\documentclass[fleqn,usenatbib]{mnras}

\usepackage{newtxtext,newtxmath}

\usepackage[T1]{fontenc}

\DeclareRobustCommand{\VAN}[3]{#2}
\let\VANthebibliography\thebibliography
\def\thebibliography{\DeclareRobustCommand{\VAN}[3]{##3}\VANthebibliography}

\usepackage{graphicx}	
\usepackage{amsmath}	
\usepackage{longtable}  

\usepackage{caption}
\usepackage{subcaption}

\usepackage{booktabs}
\usepackage{threeparttable}

\title[Multi-phase morphology and kinematics at $z=7.31$]{REBELS-25: multi-phase morphology and kinematics at $z=7.31$}

\author[Lucie E. Rowland]{
Lucie E. Rowland$^{1}$\thanks{E-mail: lrowland@strw.leidenuniv.nl},
Hiddo S. B. Algera$^{2}$,
Jacqueline Hodge$^{1}$,
Mauro Stefanon$^{3,4}$,
Rychard Bouwens$^{1}$,
\newauthor Manuel Aravena$^{5,6}$,
Lucy Astles$^{7}$,
Karin Cescon$^{1}$,
Elisabete da Cunha$^{8}$,
Ilse de Looze$^{9}$,
Andrea Ferrara$^{10}$,
\newauthor Rebecca Fisher$^{11}$,
Yoshinobu Fudamoto$^{12}$,
Thomas Herard-Demanche$^{1}$,
Hanae Inami$^{13}$,
Mahsa Kohandel$^{10}$,
\newauthor Lena Komarova$^{3,4}$,
Andrés Laza Ramos$^{3,4}$,
Themiya Nanayakkara$^{14}$,
Katherine Ormerod$^{7}$,
Andrea Pallottini$^{10,15}$,
\newauthor Siân Phillips$^{3,4}$,
Sander Schouws$^{1}$,
Piyush Sharda$^{1}$,
Renske Smit$^{7}$,
Paul van der Werf$^{1}$
\\
$^{1}$ Leiden Observatory, Leiden University, P.O. Box 9513, 2300 RA Leiden, The Netherlands \\
$^{2}$ Institute of Astronomy and Astrophysics, Academia Sinica, 11F of Astronomy-Mathematics Building, No.1, Section 4, Roosevelt Rd, Taipei 106319, Taiwan, R.O.C. \\
$^{3}$ Departament d’Astronomia i Astrofísica, Universitat de València, C. Dr Moliner 50, E-46100 Burjassot, València, Spain \\
$^{4}$ Unidad Asociada CSIC ‘Grupo de Astrofísica Extragaláctica y Cosmología’ (Instituto de Física de Cantabria – Universitat de València), Spain \\
$^{5}$ Instituto de Estudios Astrofísicos, Facultad de Ingeniería y Ciencias, Universidad Diego Portales, Av. Ejército 441, Santiago, Chile \\
$^{6}$ Millennium Nucleus for Galaxies (MINGAL) \\
$^{7}$ Astrophysics Research Institute, Liverpool John Moores University, Liverpool, UK \\
$^{8}$ International Centre for Radio Astronomy Research, University of Western Australia, ICRAR M468, 35 Stirling Hwy, Crawley 6009, Western Australia, Australia \\
$^{9}$ Sterrenkundig Observatorium, Ghent University, Ghent, Belgium \\
$^{10}$ Scuola Normale Superiore, Piazza dei Cavalieri 7, 56126 Pisa, Italy \\
$^{11}$ Jodrell Bank Centre for Astrophysics, Department of Physics and Astronomy, University of Manchester, Manchester, UK \\
$^{12}$ Center for Frontier Science, Chiba University, 1-33 Yayoi-cho, Inage-ku, Chiba 263-8522, Japan \\
$^{13}$ Hiroshima Astrophysical Science Center, Hiroshima University, Hiroshima 739-8526, Japan \\
$^{14}$ Sydney Institute for Astronomy, School of Physics, The University of Sydney, Sydney, NSW 2006, Australia \\
$^{15}$ Dipartimento di Fisica ``Enrico Fermi'', Università di Pisa, Pisa, Italy
}
\date{Accepted XXX. Received YYY; in original form \today}

\pubyear{\the\year{}}

\begin{document}
\label{firstpage}
\pagerange{\pageref{firstpage}--\pageref{lastpage}}
\maketitle

\begin{abstract}
We present a spatially resolved, multi-wavelength study of the massive, star-forming galaxy REBELS-25 at a redshift of $z=7.31$. We combine new high-resolution ALMA [O \textsc{iii}]88$\upmu$m observations with resolution-matched [C \textsc{ii}], dust continuum, and \textit{JWST}/NIRSpec IFU spectroscopy, providing a $\lesssim1$ kpc view of the morphology, interstellar medium (ISM) conditions, and multi-tracer gas kinematics of one of the most mature galaxies known in the reionisation era. We find differing morphologies from the rest-frame UV to far-infrared (FIR) emission, with the UV and optical emission appearing clumpy and irregular, whereas the FIR emission is well-described by near-exponential disc profiles, with [C \textsc{ii}] being the most extended. Comparing the resolved UV and FIR emission, we find that obscured star formation contributes $\sim55$--$98$\% of the total star formation rate across the galaxy, demonstrating that dust obscuration strongly shapes the observed UV and optical morphology. Resolved ionisation-line diagnostics show no significant variation across the source at $\sim1$ kpc resolution, consistent with broadly similar ISM conditions among the identified regions. Kinematic modelling reveals that both the warm ionised gas, traced by [O \textsc{iii}]88$\upmu$m, and the colder neutral gas, traced by [C \textsc{ii}], share the same large-scale rotating structure and are dynamically cold, with ratios of ordered-to-random motion, $V/\sigma$, of $\sim11$ and 4.5, respectively, although we find evidence for non-circular motions that are not well-described by a simple rotating disc. Overall, these results further support a picture in which REBELS-25 hosts a dusty, chemically enriched, and dynamically cold ISM already in place at $z=7.31$.
\end{abstract}




\section{Introduction}
\label{sec:intro}

In recent years, the \textit{James Webb Space Telescope} (\textit{JWST}) and the Atacama Large Millimeter/submillimeter Array (ALMA) have significantly expanded our view of galaxy evolution in the early Universe. Together, these observatories have made it possible to trace the gas, dust, and stars in galaxies well into the Epoch of Reionisation (EoR), even out to redshifts $z\gtrsim14$ (\citealt{schouws_detection_2025,carniani_eventful_2025,naidu_cosmic_2026}). As a result, there is now growing evidence that galaxies assembled and evolved more rapidly than anticipated prior to the advent of \textit{JWST} (see e.g. recent reviews by  \citealt{matthee_jwst_2025,herrera-camus_early_2026}). Among the most surprising findings are signs of maturity in galaxies within the first Gyr of cosmic time, including galaxies that have already assembled Milky Way-like stellar masses (e.g., \citealt{glazebrook_massive_2024, xiao_panoramic_2025}), chemically enriched their interstellar medium (ISM) to a significant fraction of solar metallicity (\citealt{shapley_aurora_2025,schouws_detection_2025,rowland_rebels-ifu_2026,castellano_investigating_2026}), and/or developed disc-like morphologies (\citealt{ferreira_panic_2022,robertson_morpheus_2023,kuhn_jwst_2024,xiao_panoramic_2025,wang_2025}) with rotationally-supported gas (\citealt{ parlanti_alma_2023, neeleman_alma_2023, nelson_ionized_2024, rowland_rebels-25_2024, de_graaff_ionised_2024, fujimoto_primordial_2025}). How such apparently mature systems assembled and evolved so rapidly, however, remains an open question.

Addressing this question requires spatially resolved observations across multiple wavelengths, capable of tracing the structure, kinematics, and physical properties of the different components of these early galaxies. With its near- and mid-infrared capabilities, \textit{JWST} has enabled systematic studies of the rest-frame UV and optical emission of galaxies at $z > 3$, providing insight into their stellar populations and ionised gas properties at near-to-sub-kpc resolution (e.g. \citealt{matthee_jwst_2025, adamo_2025}). At longer wavelengths, ALMA traces the far-infrared (FIR) emission of high-$z$ galaxies associated with ionised, neutral, and molecular gas, as well as dust. This includes FIR fine-structure lines and continuum emission, which can reveal obscured components otherwise inaccessible at shorter wavelengths (e.g. \citealt{Hodge_2020, Pozzi_2021, inami_alma_2022, sommovigo_alma_2022, Decarli_2025}). Accounting for this obscured emission is essential for building a complete picture of star formation and galaxy evolution, and remains important even during the EoR, where substantial obscured star formation fractions and dust reservoirs have been inferred for some systems (e.g., \citealt{watson_dusty_2015, tamura_detection_2019,schouws_significant_2022,algera_alma_2023,van_leeuwen_characterising_2024,inami_alma_2022,bowler_alma_2024,bakx_2025,fisher_rebels-ifu_2026, rajulal_2026}).

This multi-wavelength approach has also become particularly important for studies of galaxy kinematics. Much of our previous understanding of galaxy kinematics at high redshift has been based on rest-frame optical emission lines tracing warm ionised gas, such as H$\upalpha$ and [O \textsc{iii}]$\lambda5007$, which are accessible from the ground only out to $z\sim3$. These studies found that galaxies become increasingly turbulent with redshift (e.g. \citealt{wisnioski_kmos3d_2015,simons_2017,ubler_evolution_2019}), exhibiting lower rotational support at earlier times and contributing to the prevailing view that well-ordered discs emerged at later cosmic times. More recently, \textit{JWST} and ALMA have begun to challenge this picture, with \textit{JWST} observations extending ionised-gas kinematic studies to higher redshifts and revealing some rotating discs at early times (e.g. \citealt{nelson_ionized_2024,de_graaff_ionised_2024,jones_ga-nifs_2025,danhaive_dawn_2025}), while ALMA observations of FIR lines tracing colder gas phases have revealed dynamically cold discs out to $z\gtrsim4$, in some cases with substantially lower velocity dispersions and a higher degree of rotational support than previously expected (e.g., \citealt{rizzo_dynamically_2020,neeleman_cold_2020,tsukui_2021,roman-oliveira_regular_2023,pope_2023,rowland_rebels-25_2024}).

Early disc formation also remains debated among theoretical models, with some cosmological simulations predicting that rotationally supported discs become common only at $z\lesssim3$ (e.g. \citealt{HaywardHopkins2017,pillepich_first_2019}), while others have demonstrated that dynamically cold discs can form at substantially earlier epochs (e.g. \citealt{kohandel_2020,kretschmer_origin_2022,kohandel_dynamically_2024,bhagwat_spice_2024}). Recent simulations have further shown that different gas phases can exhibit systematically different kinematics (e.g., \citealt{ejdetjarn_giant_2022,kohandel_dynamically_2024, he_dynamically_2026, casavecchia_new_2026}), complicating direct comparisons between observations and theoretical predictions. Determining whether apparent differences in dynamical state reflect genuine galaxy evolution or the gas phase being traced therefore requires direct comparisons of multiple ISM phases within the same systems.

Whilst multi-wavelength observations are now available for a growing number of high-$z$ galaxies on global scales, spatially resolved studies combining rest-frame UV, optical, and FIR emission remain limited to only a handful of sources, particularly those that enable direct kinematic comparisons between multiple gas tracers within the same galaxy (e.g., \citealt{fujimoto_primordial_2025,jones_ga-nifs_2025,parlanti_ga-nifs_2025}). With the new observations presented in this work, REBELS-25 is now one such system.

REBELS-25 is a massive star-forming galaxy at $z=7.3065\pm0.0001$ (\citealt{hygate_alma_2023}) from the REBELS (Reionisation Era Bright Emission Line Survey) ALMA Large Programme, with a stellar mass of $M_\star\simeq2\times10^9\mathrm{M_{\odot}}$ (\citealt{rowland_rebels-ifu_2026}; Stefanon et al. in prep.), a dynamical mass of $M_{\mathrm{dyn}}\simeq10^{11}\mathrm{M_{\odot}}$ (\citealt{rowland_rebels-25_2024}), and SFR$_{\mathrm{UV+IR}}=67^{+54}_{-20}~\mathrm{M_{\odot}}$ yr$^{-1}$ (\citealt{fisher_rebels-ifu_2026}). Previous resolved kinematic observations have already revealed a highly rotationally supported disc (\citealt{rowland_rebels-25_2024}), making it an interesting target for a multi-phase kinematic study. It also exhibits a number of other properties indicative of a relatively evolved system, including a substantial dust reservoir (\citealt{algera_accurate_2024}), near-solar metallicity and relatively low ionisation parameter (\citealt{algera_cold_2024,algera_rebels-ifu_2025,rowland_rebels-ifu_2026}), together with a large molecular gas reservoir traced by CO(3--2) and CO(7--6) (\citealt{Cescon_2026})\footnote{There is currently no strong evidence that REBELS-25 hosts an active galactic nucleus (AGN), with no indication of broad emission lines or extreme line ratios (\citealt{rowland_rebels-ifu_2026}). We therefore assume throughout this work that the observed emission is dominated by star formation, while noting that a weak or obscured AGN cannot be completely ruled out.}. Together, these properties also make REBELS-25 a particularly valuable case study for investigating how mass, metals, dust, and ordered structure can assemble within only $\sim700$ Myr of the Big Bang.

In this work, we present new high-resolution ($\sim$1 kpc) ALMA observations of [O \textsc{iii}]88$\upmu$m and the underlying FIR continuum emission in REBELS-25, which we combine with the existing resolved [C \textsc{ii}]158$\upmu$m (hereafter [C \textsc{ii}]) and dust continuum observations from \cite{rowland_rebels-25_2024} and \textit{JWST}/NIRSpec (Near-Infrared Spectrograph) IFU (integral field unit) data (\citealt{rowland_rebels-ifu_2025,fisher_rebels-ifu_2025,algera_rebels-ifu_2025,rowland_rebels-ifu_2026,komarova_rebels-ifu_2026, fisher_rebels-ifu_2026}; Stefanon et al. in prep). The [C \textsc{ii}] and [O \textsc{iii}]88$\upmu$m observations allow us to compare the sub-kpc morphology and kinematics of different components of the ISM. Whilst [C \textsc{ii}] emission can arise from ionised, neutral atomic, and molecular gas, theoretical models and observations of local and some high-$z$ galaxies (e.g., \citealt{hollenbach_photodissociation_1999,wolfire_neutral_2003,ferrara_physical_2019,wolfire_2022,fudamoto_alma_2025}), including [O \textsc{i}] observations of REBELS-25 itself (\citealt{fudamoto_alma_2025}), suggest that a substantial fraction of the [C \textsc{ii}] emission originates from neutral atomic gas regions, whereas [O \textsc{iii}]88$\upmu$m primarily traces highly ionised gas. Together with the \textit{JWST}/NIRSpec IFU and FIR continuum data, these observations also enable a resolved study of the ISM properties across REBELS-25, including spatial variations in dust obscuration, metallicity, and the [O \textsc{iii}]88$\upmu$m/[C \textsc{ii}] luminosity ratio. This allows us to test whether its integrated properties conceal significant internal variations on $\sim1$ kpc scales, and provides further insight into the assembly and enrichment of the galaxy.

The structure of this paper is as follows: in Section \ref{sec:data}, we describe the ALMA and \textit{JWST}/NIRSpec IFU data. In Section \ref{sec:morphology}, we present the multi-wavelength morphology from these observations, and in Section \ref{sec:dust obscuration} we quantify the degree of dust obscuration across the disc. We then present a resolved analysis of its emission line ratios and related ISM properties in Section \ref{sec: resolved IFU}. In Section \ref{sec:kinematics}, we present the [O \textsc{iii}]88$\upmu$m kinematic fitting results, and compare to the [C \textsc{ii}] kinematics in Section \ref{sec:kinematics comparison}.  In Section \ref{sec:discussions}, we discuss these findings and place REBELS-25 into context with other galaxies with comparable multi-wavelength coverage. Finally, we summarise our conclusions in Section \ref{sec:conclusions}.

Throughout this work, we adopt a standard $\Lambda$CDM cosmology with Hubble constant $H_0=70$ km s$^{-1}$ Mpc$^{-1}$, matter density $\Omega_m=0.3$ and vacuum energy $\Omega_{\Lambda}=0.7$. At the redshift of REBELS-25, this corresponds to a luminosity distance of 72519 Mpc and physical angular scale of 5.09552 pkpc/arcsec. We also assume a \cite{chabrier_galactic_2003} initial mass function (IMF) and a solar oxygen abundance value of $12+\log\mathrm{(O/H)}=8.69$ (\citealt{asplund_chemical_2009}).

\section{Data} 
\label{sec:data}

\begin{figure*}
    \centering
    
    \includegraphics[width=0.9\textwidth]{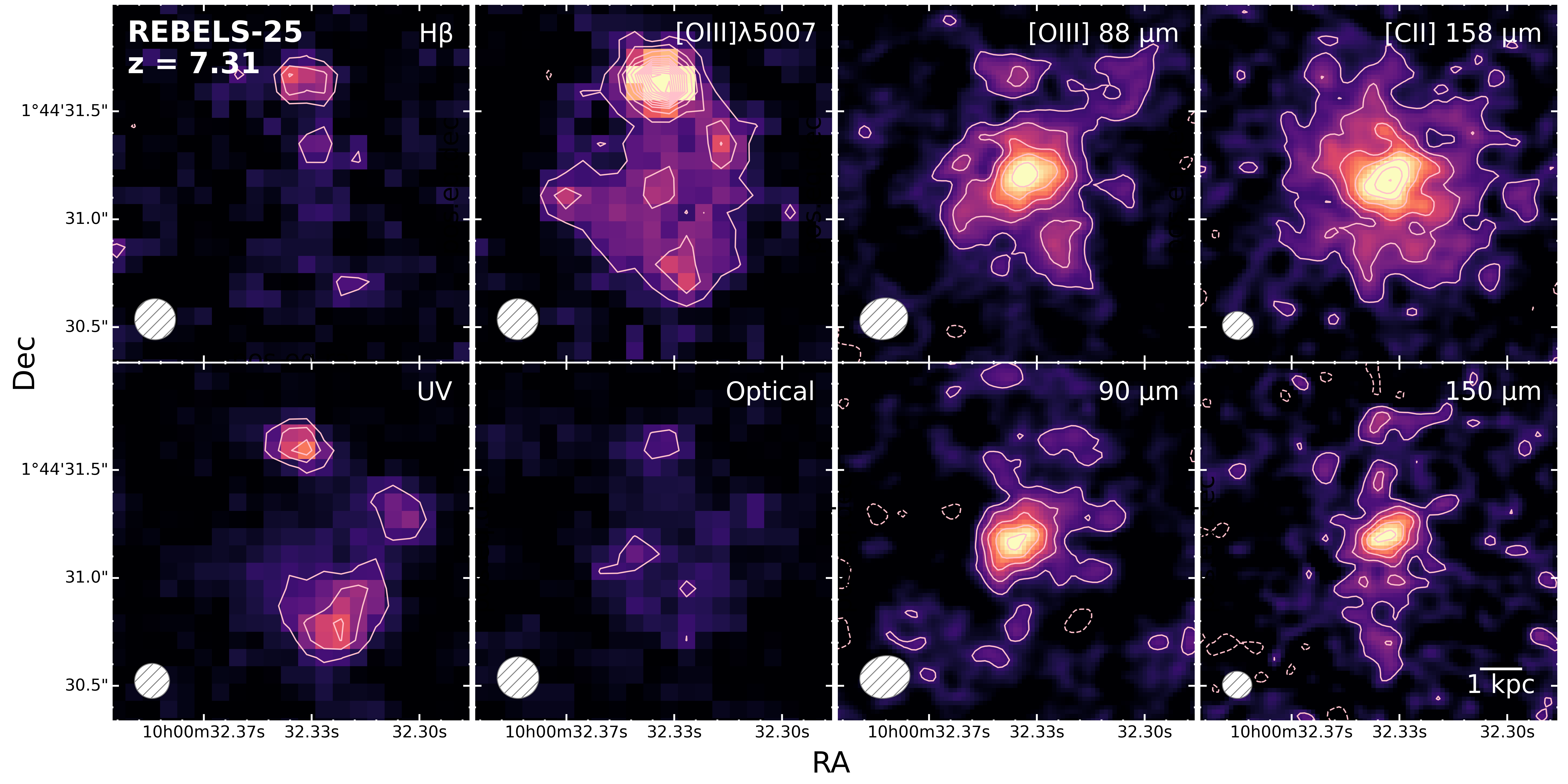}
    \caption{We present observed emission maps from the multi-wavelength ALMA and \textit{JWST} data analysed in this work. The first two columns are from the \textit{JWST}/NIRSpec IFU observations detailed in Section \ref{sec:jwst data}, and the two right-most columns are from the high resolution ($\lesssim1$ kpc) ALMA observations in Bands 6 and 8. From left-to-right, the top row depicts  H$\upbeta$, [O \textsc{iii}]$\lambda$5007, [O \textsc{iii}]88$\upmu$m and [C \textsc{ii}] emission line maps, produced by integrating $\pm1$ FWHM around the detected emission line in the corresponding cubes. In the bottom row, from left-to-right, we show the rest-frame UV, optical, 90$\upmu$m and 150$\upmu$m continuum imaging from collapsed channels, excluding emission lines. The UV map is collapsed over the range 1215.67-3700 \AA~ in the rest frame and the optical map over the range 3700-6340 \AA, masking any detected emission lines. All these maps are presented at their observed resolution, which varies from $\sim0.13$ to 0.22 arcsec. The corresponding PSF sizes are depicted by grey ellipses in the bottom left corners of each map. The contours shown start from ($-$)$2 \sigma_{\mathrm{RMS}}$, and (decrease) increase in increments of $2 \sigma_{\mathrm{RMS}}$. All maps show a 1.75 arcsec $\times$ 1.75 arcsec cutout around the centroid of the [C \textsc{ii}] emission. From these data, it is clear the morphology of REBELS-25  changes significantly across different wavelengths, and in particular the UV/optical and FIR emission are spatially offset.}
    \label{fig:line maps}
\end{figure*}

REBELS-25, also known as UVISTA-Y-003 or UVISTA-Y3, was selected for the REBELS ALMA LP based on its prior detections in deep near-infrared imaging with COSMOS/UltraVISTA (\citealt{scoville_cosmic_2007}) and \textit{Spitzer}/IRAC (analysed in \citealt{stefanon_brightest_2019}). These observations identified it as a candidate UV-luminous, massive $z>8$ merger system. As discussed in Section~\ref{sec:intro}, subsequent ALMA observations confirmed it to be a $z = 7.3065 \pm 0.0001$ rotating disc galaxy (\citealt{hygate_alma_2023, rowland_rebels-25_2024}).

REBELS-25 has since been targeted by a number of facilities, including \textit{HST} in F160W (\citealt{stefanon_brightest_2019, hygate_alma_2023}), in six bands with ALMA (\citealt{schouws_significant_2022,bouwens_reionization_2022, hygate_alma_2023, algera_accurate_2024, rowland_rebels-25_2024, Cescon_2026}), the VLA (Karl G. Jansky Very Large Array; \citealt{Cescon_2026}), \textit{JWST}/NIRSpec IFU (\citealt{rowland_rebels-ifu_2025,fisher_rebels-ifu_2025,algera_rebels-ifu_2025,rowland_rebels-ifu_2026,komarova_rebels-ifu_2026, fisher_rebels-ifu_2026}; Stefanon et al. in prep), and five bands with \textit{JWST}/NIRCam (Near Infrared Camera; Cescon et al. in prep, Schouws et al. in prep). 

In this work, we focus on a subset of these observations that enable a spatially resolved, multi-wavelength analysis at approximately matched resolution. Specifically, we present new ALMA Band 8 observations of the [O \textsc{iii}]88$\upmu$m line and underlying dust continuum at $\sim$0.2 arcsec ($\sim$1 kpc), which we compare to the high-resolution ($\sim$0.14 arcsec, $\sim$710 pc) [C \textsc{ii}] and 150$\upmu$m continuum data from \cite{rowland_rebels-25_2024}, and to \textit{JWST}/NIRSpec IFU prism observations, with a PSF (point spread function) FWHM (full width at half maximum) ranging from $\sim$0.16 to 0.21 arcsec from observed wavelengths  $\mathrm{\lambda_{obs}} \sim$ 0.5 to 5.3 $\upmu$m. Although higher-resolution imaging from \textit{JWST}/NIRCam is newly available, we do not include it here, as the NIRSpec IFU data already enables a joint analysis of key emission lines and stellar continuum over the full NIRSpec wavelength range at a comparable resolution to the ALMA data. Analysis of the NIRCam data will be the focus of forthcoming studies (Cescon et al. in prep; Astles et al. in prep).

\subsection{ALMA data}
\label{sec:alma data}

\begin{table*}
\centering
\caption{Summary of the ALMA Band 6 and Band 8 data used in this work.}
\label{tab:alma data summary}
\begin{tabular}{lcccccccc}
\hline
Data & Band & BMAJ & BMIN & BPA & RMS & Channel width & ALMA PID & Reference \\
     &      & (arcsec) & (arcsec) & (deg) & ($\upmu$Jy beam$^{-1}$) & (km s$^{-1}$) & & \\
\hline
{[C \textsc{ii}] cube} & 6 & 0.14 & 0.13 & 80.4 & 110 & 15.4 & 2021.1.01603.S & \citet{rowland_rebels-25_2024}\\
150$\upmu$m continuum & 6 & 0.14 & 0.13 & 83.4 & 5.7 & -- & 2021.1.01603.S & \citet{rowland_rebels-25_2024}\\
{[O \textsc{iii}]88$\upmu$m cube} & 8 & 0.22 & 0.19 & -74.6 & 281 & 28.7 & 2022.1.01324.S & This work \\
90$\upmu$m continuum & 8 & 0.23 & 0.20 & -75.9 & 33 & -- & 2022.1.01324.S & This work \\
\hline
\end{tabular}
\end{table*}

We make use of high spatial resolution ALMA Band 6 (\citealt{ediss_alma_2004,kerr_alma_2004}) and Band 8 (\citealt{shan_development_2005,sekimoto_development_2008,sekimoto_evaluation_2009})  observations of REBELS-25, covering the [C \textsc{ii}] and [O \textsc{iii}]88$\upmu$m emission lines, respectively, as well as the underlying FIR continua. The key observational properties of these datasets are summarised in Table \ref{tab:alma data summary}.

The reduction and imaging of the Band 6 data are described in detail in \cite{rowland_rebels-25_2024}, and we refer to that paper for full details. Where necessary, we convolve the higher-resolution Band 6 data to the coarsest resolution used in this work, which corresponds to the Band 8 rest-frame $\lambda \sim 90~\mu$m continuum map with a beam size of $0.23$ arcsec $\times$ $0.20$ arcsec (FWHM $\sim 0.21$ arcsec, or $\sim1.1$ kpc). Details of the beam-matching procedure are given in Section \ref{sec:psf matching}.

The Band 8 data, which trace the [O \textsc{iii}]88$\upmu$m line and underlying FIR dust continuum, were obtained in ALMA Cycle 9 (project ID 2022.1.01324.S, PI: H. Algera) using the C-6 configuration, achieving a maximum recoverable scale of $\sim 2.4$ arcsec with an on-source integration time of $\sim$2.8 hours. These observations were reduced using the standard ALMA pipeline in \texttt{CASA} (\citealt{mcmullin_casa_2007, casa_team_casa_2022}) v6.5.4.9, and no additional flagging or calibration was required beyond the pipeline products. 

As we aim to make comparisons with the Band 6 data products, we follow the same imaging procedures used in \cite{rowland_rebels-25_2024}. In summary, the [O \textsc{iii}]88$\upmu$m line cube was generated using \texttt{uvcontsub} for continuum subtraction, excluding channels within twice the expected FWHM of the line (where we use the [C \textsc{ii}] FWHM from \citealt{hygate_alma_2023}; 316 km s$^{-1}$)\footnote{The [O \textsc{iii}]88$\upmu$m FWHM subsequently measured from the final data products is $295\pm34$ km s$^{-1}$; \citealt{algera_rebels-ifu_2025}.}. Imaging was performed with \texttt{tclean} in `cube' mode, using automasking optimised for long-baseline data. The final cube was cleaned down to $2\times$ the channel RMS (root mean square) using multi-scale \texttt{CLEAN} with deconvolution scales of 0, 1$\times$, and 3$\times$ the beam size, enabling recovery of both compact and extended emission. Continuum imaging ($\lambda_{\mathrm{rest}}\sim90\mu$m) was performed using line-free channels in `mfs' mode, with the same automasking and cleaning strategies. Both the Band 6 and Band 8 data have a native channel width of $\sim 5$ km s$^{-1}$, but we note that we adopt a wider channel binning of 28.7 km s$^{-1}$ (five channels per bin) for the [O \textsc{iii}]88$\upmu$m cube, compared to 15.4 km s$^{-1}$ (three channels per bin) for the [C \textsc{ii}]158$\upmu$m cube, in order to enhance the signal-to-noise (S/N) per channel given the lower sensitivity of the Band 8 data.\footnote{Using a uniform channel width across both cubes does not significantly impact the main kinematic results presented in this work.}  

Both natural and Briggs-weighted imaging were produced (the latter with robust = 0.5), but we adopt the naturally weighted imaging for all subsequent analysis due to its higher sensitivity and S/N across the source, as in \cite{rowland_rebels-25_2024}. The final beam sizes and RMS values for both line and continuum images are listed in Table \ref{tab:alma data summary}.  Lower resolution data are available for this source in both Band 6 and Band 8, covering both the [C \textsc{ii}] and [O \textsc{iii}]88$\upmu$m emission lines. Combining the low and high resolution observations could marginally increase the S/N of the final data products. However, combining observations obtained with very different array configurations produces a highly non-Gaussian synthesised beam and therefore introduces an incorrect flux scaling in the default imaging products (the so-called JvM effect, named after \citealt{jorsater_high_1995}). In \cite{rowland_rebels-25_2024}, we opted not to combine these observations, and we follow the same approach in this work for the Band 8 data. Additionally, the low-resolution Band 8 observations are considerably shallower than the higher resolution data, with only $\sim10$ minutes on source, and therefore contribute negligibly to the overall S/N. The lower S/N also makes the low resolution Band 8 data more susceptible to contamination from an atmospheric absorption feature at $\sim408.28$ GHz, which affects the recovered line profile. As discussed by \cite{algera_rebels-ifu_2025} when comparing to \cite{algera_cold_2024}, fitting the low-resolution spectrum results in a narrower FWHM and a blueshifted line centroid relative to the new Cycle 9 observations. We therefore use only the high-resolution [O \textsc{iii}]88$\upmu$m data throughout this work. We further discuss the impact of the atmospheric feature in Appendix \ref{appendix:atmospheric feature}.

Even when considering the high-resolution measurement sets alone, a minor JvM effect can still be present (e.g., \citealt{czekala_molecules_2021}). We therefore test all the analysis presented in this work on the high-resolution data both with and without applying the JvM correction from \cite{czekala_molecules_2021}. This correction, in summary, involves calculating the ratio between the clean and dirty beam volumes, $\epsilon$, rescaling the residual image by this factor and then adding the residual image to the convolved, cleaned model image. For the Band 6 and Band 8 data, we compute $\epsilon=0.37$ and 0.45 at the centroids of the [C \textsc{ii}] and [O \textsc{iii}]88$\mu$m emission lines, respectively. This correction factor is beam- and therefore frequency- dependent, but we find that it varies by $<0.2\%$ across the observed frequency ranges, so the same $\epsilon$ is applied to the respective full linecubes.

Details of the tests on both JvM-corrected and uncorrected data are given in Appendix \ref{appendix:jvm correction}, but overall we find that the relatively small JvM effect in the high-resolution data does not alter the main conclusions from Sections \ref{sec:morphology} and \ref{sec:kinematics}. Since applying the JvM correction reduces the RMS of the background by a factor $\epsilon$, meaning that the corrected RMS no longer accurately reflects the sensitivity to point-like sources, in Sections \ref{sec:morphology} and \ref{sec:kinematics} we present the non-JvM-corrected data for visualisation purposes, with contour levels corresponding to the true point-source sensitivity. However, the JvM effect does have a measurable impact on the FIR fluxes presented in Section \ref{sec:resolved ism}. In the worst cases, the uncorrected data overestimates the binned flux by a factor of $\sim1.8$. As these fluxes are compared directly across multiple wavelengths, this can significantly impact the interpretation. We therefore present only JvM-corrected measurements in Section \ref{sec:resolved ism}.

\subsection{\textit{JWST} data}
\label{sec:jwst data}

REBELS-25 was observed for a total on-source time of 1750 seconds with \textit{JWST}/NIRSpec IFU in the prism mode as part of the Cycle 1 ‘REBELS-IFU’ program (\citealt{rowland_rebels-ifu_2025,algera_rebels-ifu_2025,fisher_rebels-ifu_2025,fisher_rebels-ifu_2026,rowland_rebels-ifu_2026,komarova_rebels-ifu_2026,algera_rebels-ifu_2026}), with 11 sources (including REBELS-25) in GO-1626, PI M. Stefanon, and one additional source in GO-2659, PI J. Weaver. The NIRSpec prism mode offers low spectral resolution ($\mathcal{R} \sim 100$) across an observed wavelength range of 0.6--5.3 $\upmu$m, corresponding to a rest-frame coverage of $\sim$602--6340 \AA~ at the redshift of REBELS-25. The full details of the IFU cube data reduction are presented in Stefanon et al. (in prep.). In summary, the standard \textit{JWST} pipeline was supplemented with custom outlier detection to mask hot edge pixels and cosmic ray hits, followed by subtraction of a two-dimensional interpolated background. The final data cube adopts a spatial sampling of 0.08 arcsec per pixel.

In the emission line analysis presented in \cite{rowland_rebels-ifu_2026}, UV continuum emission and the [O \textsc{ii}]$\lambda\lambda$3727,9, [Ne \textsc{iii}]$\lambda$3869, H$\upgamma$, H$\upbeta$ and [O \textsc{iii}]$\lambda\lambda4959,5007$ emission lines (spectrally unresolved) are successfully detected at S/N $>3$ in the integrated NIRSpec spectrum of this source. Where necessary, we adopt the integrated line fluxes and global properties from \cite{rowland_rebels-ifu_2026} in our analysis.

The NIRSpec IFU covers a 3 arcsec × 3 arcsec field of view, with a wavelength-dependent PSF FWHM ranging from $\sim 0.15$ arcsec at $\lambda_{\mathrm{obs}}\sim0.96 \upmu$m to $\sim 0.21$ arcsec at $\lambda_{\mathrm{obs}}\sim5.3 \upmu$m, with a median value of $\sim0.17$ arcsec ($=0.88$ kpc at $z=7.31$). For the spatially resolved analysis presented in Sections \ref{sec:dust obscuration} and \ref{sec: resolved IFU}, we PSF-match the full NIRSpec IFU cube to the coarser Band 8 beam, with details described below in Section \ref{sec:psf matching}.

\subsection{Astrometry correction}

\label{sec:astrometry}

As we aim to make spatially resolved comparisons between the \textit{JWST} and ALMA data, we applied an astrometric correction to the NIRSpec IFU cube to ensure a more accurate spatial alignment. Due to the small NIRSpec IFU field of view, there are no additional  bright sources within the cube. We therefore aligned the cube according to the UltraVISTA DR4 imaging over the COSMOS field (\citealt{scoville_cosmic_2007}) covering this source, which was astrometry-corrected to the Gaia DR3 reference frame. For this, we first constructed a mock UVISTA H-band image from the IFU cube by applying the corresponding transmission filter. The derived shift between the mock image and the UVISTA data ($\sim0.21$ arcsec) was then applied to the full NIRSpec IFU cube. Propagating the uncertainties on the derived RA and Dec shifts gives an uncertainty on the NIRSpec IFU astrometric correction of $\sim0.04$ arcsec.

Whilst the \textit{JWST}/NIRCam imaging of REBELS-25 (from GO-6480 and GO-6036; Cescon et al. in prep.; Schouws et al. in prep.) is not analysed in this work, we used these data as an independent check on the NIRSpec IFU astrometry. To do this, we calibrated the NIRCam images to the Gaia DR3 reference frame, created mock NIRCam images from the IFU cube, and repeated the alignment. This yields a negligible shift compared to the UVISTA-based correction. We therefore find that the adopted NIRSpec IFU astrometry is sufficiently robust, within the uncertainties, for spatially resolved comparison with the ALMA data.

Next, we consider the astrometric uncertainty of the ALMA data. Following \cite{hygate_alma_2023} and \cite{cortes_alma_handbook_2026}, the nominal astrometric uncertainty of ALMA is given by $\theta = \mathrm{FWHM_{beam}}/\mathrm{SNR}/0.9$, where $\mathrm{FWHM}_{\rm beam}$ is the FWHM of the synthesized beam and SNR is the peak signal-to-noise ratio. Since the true astrometric uncertainty can be up to a factor of two larger than the nominal value (\citealt{cortes_alma_handbook_2026}), we adopt ALMA astrometric uncertainties of $\sim0.02$ arcsec for [C \textsc{ii}], $\sim0.03$ arcsec for the 150 $\upmu$m dust continuum, and $\sim0.04$ arcsec for both [O \textsc{iii}]88$\upmu$m and the 90 $\upmu$m dust continuum. Combining these with the uncertainty on the NIRSpec IFU correction gives relative astrometric uncertainties of $\sim0.05$ arcsec for comparisons with the Band 6 data, and $\sim0.06$ arcsec for comparisons with the Band 8 data.

These relative astrometric uncertainties are small compared to the ALMA beam sizes and \textit{JWST}/NIRSpec IFU PSFs (Section \ref{sec:psf matching}), and therefore do not limit the spatially resolved analysis presented in this work. Even so, they should be kept in mind when interpreting apparent spatial offsets between the different datasets (as in Section \ref{sec:morphology}), and also when comparing aperture-based spectra (as in Section \ref{sec:resolved ism}).

\subsection{PSF matching}
\label{sec:psf matching}

In addition to analysis of the data products at their native resolution, we also convolve all imaging products (the ALMA line cubes, NIRSpec IFU cube and ALMA continuum maps) to the coarsest effective resolution: the ALMA Band 8 FIR continuum beam, which has a FWHM of $0.23$ arcsec $\times$ $0.20$ arcsec (equivalent to $\sim1.1$ kpc at $z = 7.31$). For the ALMA data, we assume 2D elliptical Gaussian beams and perform the convolution in the image plane. For the \textit{JWST}/NIRSpec IFU data, the PSF has been found to vary significantly with wavelength and deviates from a simple Gaussian. We therefore adopt wavelength-dependent model PSFs generated using the \texttt{stpsf} tool, which have been validated through comparisons with archival calibration star observations (\citealt{jones_blackthunder_2026}; cf. \citealt{deugenio_fast-rotator_2024}). These models are used to construct PSF-matching kernels on a per-channel basis. Specifically, for each image requiring convolution, we generate a PSF-matching kernel using the \texttt{create\_matching\_kernel} function in the \texttt{photutils} package. The convolution itself is performed using the \texttt{convolve} function from \texttt{astropy}. For the \textit{JWST} data, the mismatch between the complex IFU PSF and a Gaussian target beam introduces artefacts at high spatial frequencies, where the Fourier transform of the PSF approaches zero and becomes dominated by noise (\citealt{aniano_common-resolution_2011}). To mitigate this, we apply a cosine-bell window function to the kernel in the Fourier domain, tapering 90\% of the array values to suppress noisy high-frequency modes and avoid introducing artefacts into the convolved images (as in e.g. \citealt{polletta_jwsts_2024}). 

At the reddest wavelengths in the NIRSpec IFU cube, the native PSF approaches the resolution of the Band 8 beam, leading to unstable kernel construction. When the PSF FWHM of the NIRSpec IFU data reaches 90\% of the target beam (at $\lambda=4.05\upmu$m), we therefore do not perform PSF matching, as the native resolution is already comparable. We have verified that all final convolved data products recover consistent integrated flux properties with the original data at its native resolution. Where direct comparisons are necessary, we reproject the final images onto the same grid (NIRSpec IFU grid with pixel scale $=0.08$ arcsec pix$^{-1}$).

\section{Morphology from the rest-UV to FIR}
\label{sec:morphology}

\subsection{Overview of the multi-wavelength emission}

In Figure \ref{fig:line maps}, we provide a multi-wavelength, $\lesssim1$ kpc-resolution view of REBELS-25 from the rest-frame UV to FIR. We show the H$\upbeta$, [O \textsc{iii}]$\lambda$5007, [O \textsc{iii}]88$\upmu$m, and [C \textsc{ii}] emission line maps, as well as UV, optical, 90$\upmu$m, and 150$\upmu$m continuum maps from the corresponding ALMA and \textit{JWST}/NIRSpec IFU data. The emission line maps are moment-0 maps integrated over a $\pm1$ FWHM range around the line centre, while the UV map is collapsed over the range 1215.67--3700 \AA~ in the rest frame and the optical map over the range 3700-6340 \AA, masking all detected emission lines. These maps reach a peak S/N of 6.2, 27.8, 11.0, 14.3, 7.2, 3.0, 10.6 and 10.1, respectively. We note that these are the observed maps at their native resolution, without any PSF-matching or correction for dust attenuation. 

From Figure \ref{fig:line maps}, it is clear that the observed morphology of REBELS-25 varies significantly with wavelength, highlighting the importance of multi-wavelength resolved observations for interpreting its structure. The emission detected by \textit{JWST} is particularly clumpy and irregular, with the brightest emission in [O \textsc{iii}]$\lambda5007$ and H$\upbeta$ originating from a prominent northern `clump' offset by $\sim$0.4 arcsec ($\sim2$ kpc) from the centre, which we hereafter define as the centroid of [C \textsc{ii}] emission as fitted in \cite{rowland_rebels-25_2024}, while the peak of the UV emission is in a southern `clump', also offset from the centre by $\sim$0.4 arcsec. Rest-frame optical continuum emission is barely detected (peak S/N $=2.96$, with only a handful of pixels exceeding 2$\sigma_{\mathrm{RMS}}$), but we find that the brightest pixel lies within one PSF FWHM of the galaxy centre. In contrast, the FIR ALMA data ([C\textsc{ii}], [O \textsc{iii}]88$\upmu$m, and the 90$\upmu$m and 150$\upmu$m FIR continuum) are all centrally concentrated, with a comparatively smoother distribution.

This clear variation across wavelength, and in particular the spatial offset between the UV/optical and the FIR emission, suggests significant dust obscuration in the central regions of the galaxy. This is also discussed in \cite{rowland_rebels-25_2024} based on \textit{HST} F160W imaging, which traces the rest-frame UV emission at a comparable wavelength range to this \textit{JWST} UV map. We explore the role of dust obscuration and attempt to quantify it in more detail in Section \ref{sec:dust obscuration}. We also use the resolved ISM diagnostics in Section \ref{sec: resolved IFU} to test whether the UV- and optical-bright regions could be distinct components, such as merging components, since these might be expected to show different physical conditions on the $\sim1$ kpc scales probed here. Alternatively, the observed clumpiness could arise from smaller-scale star-forming regions embedded within a disc (e.g. \citealt{fujimoto_primordial_2025}). Such regions would likely remain unresolved in the present data, whilst spatially varying dust attenuation could still strongly influence which parts of them are visible in the rest-frame UV and optical.

The diversity in morphology across the different maps could also reflect the fact that we are tracing different components of the ISM. This is potentially illustrated by the morphological differences among the ALMA maps, where dust obscuration should have a negligible impact. The [C \textsc{ii}] emission, often associated with neutral gas within galaxies,  appears the smoothest and most extended, although we note its higher S/N. The Band 6 and Band 8 continuum maps (tracing dusty star formation), as well as the [O \textsc{iii}]88$\upmu$m emission (tracing ionised gas), show more irregular outer structure, with hints of clumpy substructure or faint extended emission. However, these features are at low S/N and may be consistent with noise (e.g., \citealt{hodge_kiloparsec-scale_2016}), although some are potentially co-spatial with clumps in the \textit{JWST} maps (Figure \ref{fig:sersic fits}), which we discuss in more detail below. To better account for differences in the S/N, and also resolution, we present a more quantitative analysis of these data by fitting PSF-convolved Sérsic models, below.

\subsection{Sérsic fitting}
\label{sec:sersic fitting}

\begin{figure*}
\centering
\includegraphics[width=0.85\textwidth]{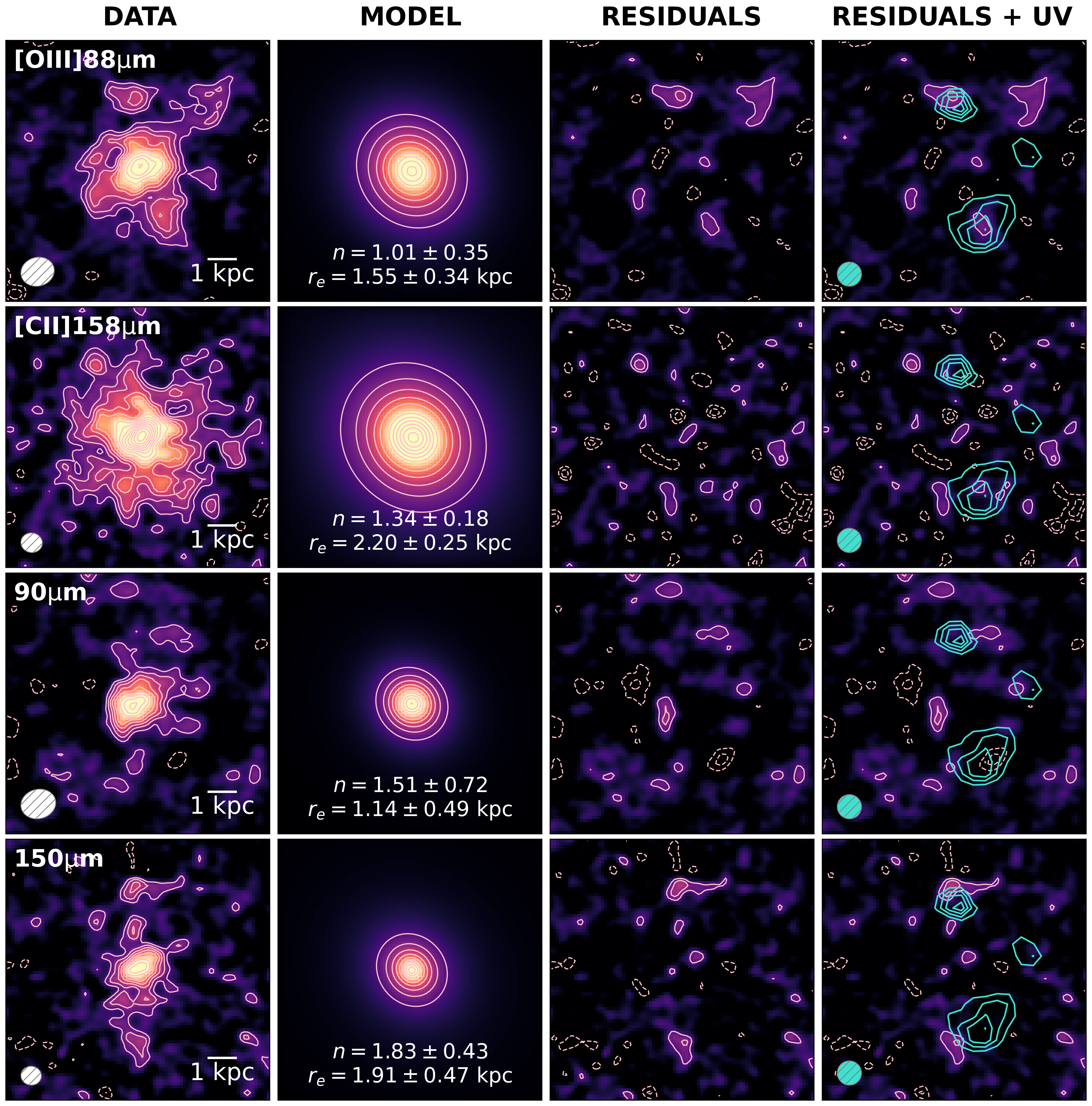}
\caption{Two-dimensional Sérsic fits to the high resolution ($\lesssim1$ kpc) ALMA emission maps. From left-to-right: observed emission map, best-fit Sérsic model, residual map, and residual map with contours from the \textit{JWST} UV map in blue. The contours show emission at $\pm 2, 3, 4, ... \sigma_{\mathrm{RMS}}$. In the `Model' column, we also include text labels with the best-fit Sérsic indices, $n$, and effective radii, $r_e$. For all fits, we fix the morphological position angle, ellipticity and centroid to the values obtained in \citet{rowland_rebels-25_2024} for the [C \textsc{ii}] emission. Overall, the FIR emission of REBELS-25 is well-described by a near-exponential disc morphology, although the FIR continuum maps may be slightly more centrally concentrated, and there may be tentative clump-like substructures that could be coincident with rest-UV emission.}
\label{fig:sersic fits}
\end{figure*}

Following \citet{rowland_rebels-25_2024}, we fit 2D Sérsic profiles to the emission maps using \texttt{PetroFit} (\citealt{geda_petrofit_2022}), which uses \texttt{Astropy Sersic2D} models convolved with the input PSFs. There exist a variety of other tools which also apply PSF-convolved Sérsic modelling, such as \texttt{GALFIT} (\citealt{peng_detailed_2010}), \texttt{statmorph} (\citealt{rodriguez-gomez_statmorph_2022}), and \texttt{PySersic} (\citealt{pasha_pysersic_2023}). In general, the  fitted Sérsic parameters across these different tools have been found to be fairly consistent, but it has been noted that \texttt{PetroFit} tends to exhibit fewer outliers (\citealt{ren_evolution_2025,song_size_2026}). Importantly, \texttt{PetroFit} also enables flexible oversampling to be applied to the input PSF and the model. This is particularly important for systems with a high Sérsic index ($n$), where the light profile rises very sharply towards the centre. Ideally, the model should therefore be estimated on an oversampled grid, particularly in the centre (e.g. \citealt{haussler_gems_2007}). In addition, the PSF can have intricate details and variations that may not be well-captured with the input image sampling. 

For the ALMA data, the PSF is already well sampled ($\sim7-10$ pixels across BMIN), so we only set a modest model oversampling factor of 4 in a box that corresponds to the central 1 kpc ($\sim10$ pixels in length) of each fitting iteration. We verified convergence by repeating fits at higher oversampling factors (up to 12) and found parameter shifts well below the uncertainties. We note that light profile fitting algorithms can systematically underestimate the uncertainties on the derived morphological parameters (e.g. \citealt{van_der_wel_structural_2012,ghosh_morphological_2023}), and so to derive conservative estimates for the uncertainties we use Markov Chain Monte Carlo (MCMC) sampling around the best-fit returned by \texttt{PetroFit}. The likelihood assumes Gaussian noise with a fixed $\sigma_{\mathrm{RMS}}$ measured in emission-free regions of the image. The reported uncertainties correspond to the 16th–84th percentile range of the posterior distributions with 32 walkers and 1200 steps, with the first 300 steps discarded.  

As discussed above, the observed emission maps from the \textit{JWST} data are not well described by a single component, and instead appear to be composed of distinct clumps. These maps are also likely to be strongly affected by dust obscuration, which we discuss in more detail in Section \ref{sec:dust obscuration}. We therefore choose to apply this one-component Sérsic fitting to the four ALMA emission maps ([O\textsc{iii}]88 $\upmu$m, [C \textsc{ii}], Band 6 continuum and Band 8 continuum) only. 

If we leave all parameters (Sérsic index, $n$, half-light radius, $r_e$, morphological position angle, PA, ellipticity, and central coordinates) free, we find that the morphological centroids are all in good agreement (all within less than one beam FWHM), but the morphological PAs and ellipticities vary significantly across these four FIR maps. For example, the ellipticity varies from 0.1 (for [C \textsc{ii}]) to 0.5 (90 $\upmu$m), and the PAs from 150$^{\circ}$ (for [O \textsc{iii}]88$\upmu$m) to 218$^{\circ}$ (for [C \textsc{ii}]), which makes it difficult to directly compare different $n$ and $r_e$ values across the different maps. Here, we consider whether these differences in PA and ellipticity could potentially be due to intrinsic substructure and/or the S/N of each observation.

In \cite{rowland_rebels-25_2024}, a central region in the inner $\sim$ 1 kpc of the [C \textsc{ii}] and 150$\upmu$m imaging was identified as a potential bar-like structure, based on the change in ellipticity and PA from elliptical isophote fits. Interestingly, the same structure can also be seen in the [O \textsc{iii}]88$\upmu$m and 90$\upmu$m data. By applying the same elliptical isophote fitting as described in \cite{rowland_rebels-25_2024}, we find that  the PA and ellipticity of the inner $\sim1$ kpc is $\sim130^{\circ}$ and $\sim0.3$, respectively, across all four ALMA maps, whereas the [C \textsc{ii}] Sersic2D fit and outer elliptical isophote fit presented in \cite{rowland_rebels-25_2024} have a PA and ellipticity of $\sim220^{\circ}$ and $0.1$ (near-circular), respectively. This could be seen as additional evidence for a bar-like component in the dust and gas emission at the centre of REBELS-25, although this component is barely resolved (1.4 - 2 beams) in the Band 6 data and is unresolved in the Band 8 data, limiting our ability to investigate this central component further.

However, it is clear from the elliptical isophotal fitting presented in \cite{rowland_rebels-25_2024}, and to some extent the Band 8 data, that the morphological PA and ellipticity in REBELS-25 changes as a function of radius. This could imply that the differences in the PAs and ellipticities derived from the Sérsic fits across the four maps likely reflect the differing S/N and radial sensitivity of each dataset. For this analysis, we therefore fix the centroid, ellipticity, and morphological PA in all fiducial Sérsic fits to the values derived from the highest-S/N data (the [C \textsc{ii}] emission, as presented in \citealt{rowland_rebels-25_2024}), which we take to provide the most reliable tracer of the global disc morphology. Under this assumption, the gas and dust in REBELS-25 are modelled as a thin disc observed close to face-on, with inclination $i=(25\pm6)^{\circ}$ (\citealt{rowland_rebels-25_2024}). We add that, when all parameters are left free, the resulting $n$ and $r_e$ values for each map are consistent within the uncertainties with the values when only $n$ and $r_e$ are left free.

We show the results from the fiducial Sérsic fitting applied to the [O\textsc{iii}]88 $\upmu$m, [C \textsc{ii}] (from \citealt{rowland_rebels-25_2024}\footnote{We note that, when fitting the [C \textsc{ii}] emission with an alternative morphology fitting tool, \texttt{CANNUBI}, in \cite{rowland_rebels-25_2024}, significant ($6\sigma_{\mathrm{RMS}}$) residual emission was found at the centre, which is not as prominent in the Sérsic fitting. However, for these \texttt{CANNUBI} fits, the model was normalised to the azimuthally averaged flux in each ring, meaning it can underestimate the peak flux at the centre, where the intensity profile is steepest.}), Band 6 continuum and Band 8 continuum emission in Figure \ref{fig:sersic fits}, where only $n$ and $r_e$ are left as free parameters. The best-fit Sérsic indices are all close to unity, implying the emission is consistent with a near-exponential disc. The [C \textsc{ii}] emission is the most extended, with an $r_e$ of $2.20\pm0.25$ kpc, and the 90$\upmu$m dust continuum is the most compact, with $r_e=1.14\pm0.49$ kpc. The dust continuum emission at 150$\upmu$m is more extended than at 90$\upmu$m, likely due to the clumpy substructures detected at S/N$\sim3$ that extend north and south from the centre. When leaving all parameters free in the Sérsic fitting, \cite{rowland_rebels-25_2024} find an even higher $r_e$ of $2.5\pm0.7$ kpc for this 150$\upmu$m imaging. This more extended emission at 150$\upmu$m than at 90$\upmu$m could suggest a colder dust temperature at the outskirts of the galaxy, but based on the 3$\sigma$ upper limits in the 90$\upmu$m imaging, the non-detection of these clumpy substructures at 90$\upmu$m is consistent with a wide range of temperatures (Section \ref{sec:dust obscuration}).

The clumpy structures in the 150$\upmu$m imaging are visible at $\sim4\sigma_{\mathrm{RMS}}$ in the residuals of the Sérsic fitting in Figure \ref{fig:sersic fits}, and we see that the north clump closely aligns with the northern clump in the UV map (which is also where the rest-frame optical emission lines peak, Figure \ref{fig:line maps}). In the [O \textsc{iii}]88$\upmu$m residual map, we also see some residual emission at $\sim3\sigma_{\mathrm{RMS}}$ that is coincident with the northern and southern clumps in the UV map (cf. Appendix \ref{appendix:atmospheric feature}). As the [O \textsc{iii}]88$\upmu$m and FIR continuum morphologies are not impacted by dust obscuration effects, we therefore consider whether these clumps could be intrinsic to the system, such as merging components or regions of intense star formation, although we caution that these clumps are detected at low significance (see e.g., \citealt{hodge_kiloparsec-scale_2016} and Appendix \ref{appendix:atmospheric feature}). If these clumps are merging components, we might expect variations in their ISM conditions, which we investigate using the multi-wavelength data in the following section, as well as kinematic signatures, which we consider in Section \ref{sec:kinematics}.

\section{Resolved ISM properties}
\label{sec:resolved ism}

\begin{table*}
\def\arraystretch{1.2}
\centering
\caption{Resolved properties of REBELS-25 derived from clump aperture spectra.}
\label{tab:rebels25_clump_properties}
\begin{tabular}{lccccc}
\hline
Property & Clump 1 & Clump 2 & Clump 3 & Clump 4 & Integrated \\
\hline
\multicolumn{6}{l}{\textbf{Dust and UV properties}} \\
$T_{\rm dust}$ (K) &
$41^{+9}_{-7}$&
32&
32&
32&
${32^{+9}_{-6}}^{\rm a}$\\
$L_{\rm IR}$ ($10^{11}L_\odot$) & $6.20^{+6.79}_{-2.78}$ & $0.97^{+0.81}_{-0.40}$ & $0.37^{+0.43}_{-0.19}$ & $0.73^{+0.65}_{-0.34}$ & ${5.0^{+5.0}_{-1.9}}^{\rm a}$ \\
SFR$_{\rm IR}$ ($M_\odot\,{\rm yr}^{-1}$) & $68.2^{+74.7}_{-30.6}$ & $10.7^{+8.9}_{-4.8}$ & $4.0^{+4.7}_{-2.1}$ & $8.0^{+7.1}_{-3.7}$ & ${55^{+54}_{-20}}^{\rm b}$ \\
$M_{\rm UV}$ (mag) & $-19.3\pm0.4$ & $-19.9\pm0.3$ & $-20.2\pm0.2$ & $-20.3\pm0.2$ & $-21.46\pm0.05^{\rm c}$ \\
SFR$_{\rm UV}$ ($M_\odot\,{\rm yr}^{-1}$) & $1.6\pm0.6$ & $2.7\pm0.6$ & $3.7\pm0.6$ & $4.0\pm0.6$ & $12\pm0.5^{\rm b}$ \\
$f_{\rm obscured}$ & $0.98^{+0.01}_{-0.02}$ & $0.81^{+0.08}_{-0.12}$ & $0.55^{+0.17}_{-0.18}$ & $0.68^{+0.12}_{-0.15}$ & $0.82\pm0.08$ \\
$\beta_{\rm UV}$ & $-1.68\pm0.52$ & $-1.17\pm0.18$ & $-1.65\pm0.18$ & $-1.39\pm0.11$ & $-1.61\pm0.09^{\rm c}$ \\
\hline
\multicolumn{6}{l}{\textbf{ALMA FIR emission-line properties}} \\
$L_{\rm [OIII]\,88}$ ($10^{9}L_\odot$) & $1.53\pm0.28$ & $0.34\pm0.17$ & $0.42\pm0.18$ & $0.41\pm0.18$ & ${3.26^{+0.49}_{-0.48}}^{\rm d}$ \\
$L_{\rm [CII]}$ ($10^{9}L_\odot$) & $0.56\pm0.08$ & $0.17\pm0.05$ & $0.18\pm0.05$ & $0.20\pm0.05$ & ${1.76^{+0.18}_{-0.19}}^{\rm d}$ \\
$L_{\rm [OIII]\,88}/L_{\rm [CII]}$ & $2.73\pm0.68$ & $2.01\pm1.15$ & $2.29\pm1.22$ & $2.02\pm1.03$ & ${1.9^{+0.4}_{-0.3}}^{\rm d}$ \\
\hline
\multicolumn{6}{l}{\textbf{\textit{JWST}/NIRSpec rest-frame optical emission-line properties}} \\
$F_{\rm [OII]}$ & $2.11\pm0.28$ & $2.25\pm0.24$ & $<0.83$ & $2.21\pm0.34$ & $6.25\pm0.74^{\rm e}$ \\
$F_{\rm [NeIII]}$ & $<0.67$ & $<1.08$ & $<0.63$ & $1.10\pm0.26$ & $2.45\pm0.63^{\rm e}$ \\
$F_{\rm H\upbeta}$ & $1.39\pm0.21$ & $1.94\pm0.22$ & $<0.58$ & $1.31\pm0.24$ & $5.19\pm0.73^{\rm e}$ \\
$F_{\rm [OIII]\,5007}$ & $4.03\pm0.19$ & $9.55\pm0.21$ & $2.45\pm0.18$ & $4.47\pm0.22$ & $21.36\pm0.63^{\rm e}$ \\
R3 & $2.89\pm0.45$ & $4.93\pm0.57$ & $>4.21$ & $3.42\pm0.65$ & $4.12\pm0.60^{\rm e}$ \\
O32 & $<2.55$ & $<5.67$ & -- & $<2.70$ & $<3.42^{\rm e}$ \\
$12+\log\mathrm{(O/H)}$ & $8.47\pm0.15$ & $8.29\pm0.15$ & -- & $8.42\pm0.15$ & $8.36\pm0.15^{\rm f}$ \\
\hline
\end{tabular}

\vspace{4pt}
\parbox{\textwidth}{\footnotesize
\textbf{Notes.} As described in the text, we use a fixed $T_{\mathrm{dust}}=32$ K with $\pm10$ K uncertainty for clumps 2, 3 and 4, where the 90$\upmu$m emission is not detected. For reasons described in the text, the sums of the clump luminosities and SFRs are not expected to be equivalent to the integrated values. For the ALMA FIR line luminosities, we note that these are derived from the JvM-corrected data products. As the correction applies a scaling of factor $\epsilon$ to the residual products from \texttt{tclean}, the $\sigma_{\mathrm{RMS}}$ of the resulting imaging no longer reflects the true sensitivity of the data. To obtain conservative estimates of the flux uncertainties, we therefore multiply the final error spectrum by a factor of $1/\epsilon$. The rest-frame optical emission-line fluxes are in units of $10^{-18}$ erg s$^{-1}$ cm$^{-2}$. R3 is defined as [O \textsc{iii}]$\lambda$5007/H$\upbeta$, and O32 is [O \textsc{iii}]$\lambda$5007/[O \textsc{ii}]$\lambda\lambda$3727,9.  The integrated properties are taken from: (a) \citet{algera_accurate_2024}; (b) \citet{fisher_rebels-ifu_2025}; (c) \citet{fisher_rebels-ifu_2026}; (d) \citet{algera_rebels-ifu_2025}; and (e) \citet{rowland_rebels-ifu_2026}. (f) The integrated oxygen abundance is re-derived using the same \citet{sanders_aurora_2025} R3 calibration as used for the clumps, in comparison to the global value reported in \citet{rowland_rebels-ifu_2026}, where the \citet{sanders_direct_2024} R3 calibration was applied.}
\end{table*}

In this section, we present a variety of resolved ISM properties from the multi-wavelength observations of REBELS-25. Specifically, we focus on spatial variations in the fraction of obscured star formation, the [O \textsc{iii}]88$\upmu$m/[C \textsc{ii}] ratio, and the oxygen abundance from rest-frame optical emission line ratios. For these investigations, we use the PSF-matched data products described in Section \ref{sec:psf matching}, which results in a resolution of $0.23\times0.20$ arcsec ($\sim1.1$ kpc). To derive resolved ISM properties, we therefore extract spectra and photometry from clump-based apertures with radii of $0.23$ arcsec, rather than on a pixel-by-pixel basis. This also helps to increase the S/N of the fainter emission features on resolved scales.  

For the purpose of further investigating the nature of the clumpy \textit{JWST} morphology, in comparison to the smoother ALMA emission, we focus on measuring emission line diagnostics for the three distinct clumps identified in the UV maps, and also a central clump (offset from the UV emission) that encompasses the majority of the FIR emission (see Figure \ref{fig:resolved dust sed}a). These diagnostics are then used to test whether the UV-bright regions exhibit strongly distinct ISM properties, as might be expected for merging components (e.g. \citealt{arribas_ga-nifs_2024,jones_2024,sugahara_rioja_2025, jones_2025, jones_2026}). However, we note that similar line ratios would not rule out intrinsic star-forming regions embedded within the disc, particularly given the $\sim1$ kpc resolution of the data. As dust obscuration is likely to contribute substantially to the differing morphologies, we focus where possible on measurements that can be considered insensitive to dust attenuation. As discussed below, obtaining dust attenuation-corrected emission on resolved scales, and even for the source as a whole, is challenging with the current data.

\subsection{Dust obscuration}
\label{sec:dust obscuration}

\begin{figure*}
\centering
\includegraphics[width=\textwidth]{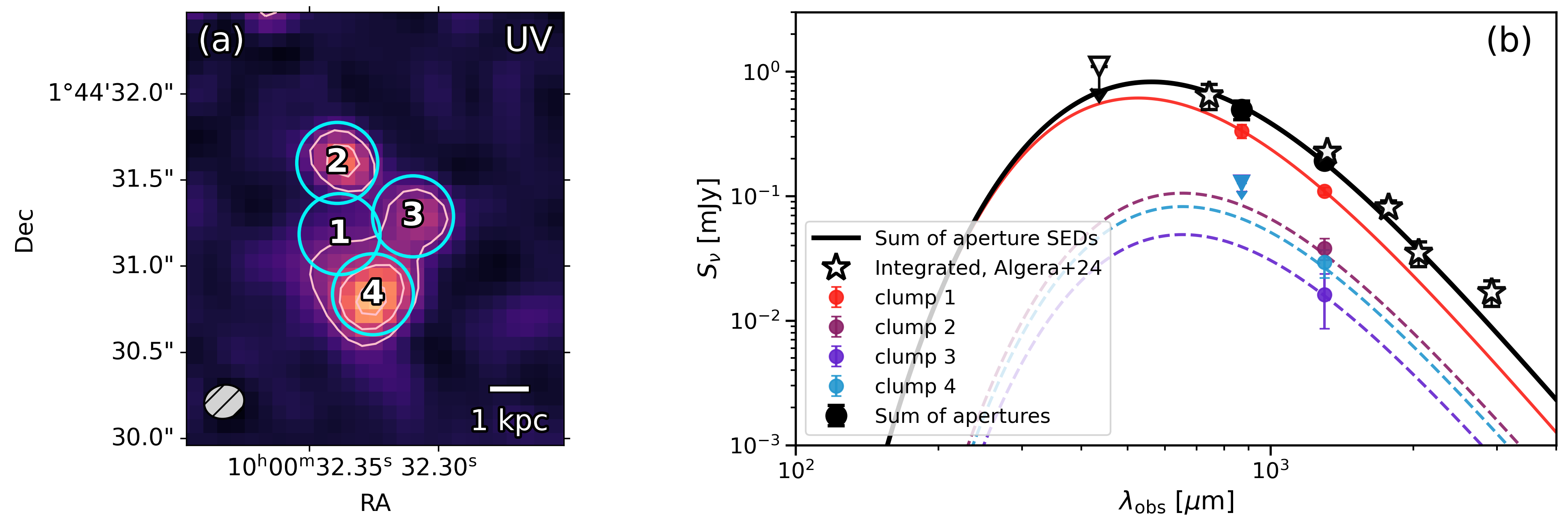}
\caption{\textit{Panel a:} We plot the PSF-matched UV map from the NIRSpec IFU cube, with the circular apertures used to extract the so-called `clump' spectra in blue and labelled 1-4. The resulting PSF is indicated by the grey ellipse in the bottom left corner. \textit{Panel b:} For each clump, we fit a modified blackbody to the Band 6 and Band 8 FIR continuum flux densities. For all fits, we assume an emissivity index, $\beta_{\mathrm{IR}}$, of 2.5, based on the integrated dust SED fit with six ALMA bands from \citet{algera_accurate_2024}. For clumps 2, 3 and 4, we use the 3$\sigma_{\mathrm{RMS}}$ upper limits for the Band 8 data (upper limits plotted with a downwards triangle) and we fix the dust temperature, $T_{\mathrm{dust}}$, to 32$\pm10$ K, also from \citet{algera_accurate_2024}. For clump 1, as we detect the continuum at S/N$>3$ in both bands, we leave $T_{\mathrm{dust}}$ as a free parameter and obtain $T_{\mathrm{dust}}=41^{+9}_{-7}$ K. We also plot the integrated continuum flux measurements from all six ALMA bands analysed in \citet{algera_accurate_2024} with white star-shaped markers. We sum each of the clump-based dust SEDs (solid black line) to ensure it is comparable to the integrated measurements from the other ALMA bands, although we note that the clump apertures have some overlapping pixels and do not cover the full extent of the dust emission, and so we caution that the summed SED should not be treated as a true representation of the global FIR emission.}
\label{fig:resolved dust sed}
\end{figure*}

To assess the role of dust obscuration in producing the observed UV-to-FIR morphology, we would ideally measure spatial variations in the dust attenuation, $A_{\mathrm{V}}$. Typical approaches include spatially resolved SED fitting and Balmer decrement maps. For REBELS-25 at $z=7.31$, however, the brightest Balmer line, H$\upalpha$, is shifted beyond the nominal wavelength range of NIRSpec. Whilst H$\upbeta$ and H$\upgamma$ are detected in the integrated spectrum (\citealt{rowland_rebels-ifu_2026}), these lines are not detected at high enough S/N to investigate spatial variations in their ratio. Even in the integrated spectrum, H$\upgamma$ is only detected at S/N$\sim3$ (and is also blended with [O \textsc{iii}]$\lambda$4363), and the resulting nebular dust attenuation ($A_\mathrm{V, neb}$) derived from H$\upbeta$/H$\upgamma$ is consistent with $A_{\mathrm{V, neb}} =0$ mag. However, from the integrated SED fitting discussed in \cite{rowland_rebels-ifu_2026} (full details in Stefanon et al. in prep), the stellar dust attenuation is $A_{\mathrm{V,*}}=0.73\pm0.04$ mag -- i.e., there should be, on average, a significant amount of dust attenuation across the galaxy.

Spatially resolved SED fitting can be performed on a pixel-by-pixel basis (e.g., \citealt{gimenez-arteaga_spatially_2023, li_alma-cristal_2024}),  or using binning methods, such as Voronoi binning to reach a S/N threshold (e.g. \citealt{abdurrouf_spatially_2023,harvey_2025}). For the NIRSpec IFU data of REBELS-25, the S/N is too low to detect extended continuum emission on a pixel-by-pixel basis, and even when binning in both spatial and spectral dimensions, the optical continuum is only detected at S/N$\lesssim3$.  A more detailed analysis using spatially resolved SED fitting with the NIRSpec IFU data is therefore limited for this source by the low S/N, but the upcoming analysis of the deeper NIRCam observations will enable a more thorough investigation (Cescon et al. in prep). 

To quantify the degree of dust obscuration across REBELS-25, we instead adopt two alternative methods. First, we compare the UV and FIR continuum emission to obtain measures of the obscured and unobscured star formation rate (SFR) in regions across the galaxy, where we use circular extraction apertures centred on the three distinct clumps in the UV map and an additional aperture at the peak of the FIR emission (Figure \ref{fig:resolved dust sed}). Secondly, we compare the degree of reddening at UV wavelengths inferred from the fitted UV slope, $\beta_{\mathrm{UV}}$, for these same regions from the NIRSpec spectra. As aforementioned, we set the radii of these apertures to the major axis FWHM of the PSF, 0.23 arcsec, which also means that they cover the bulk of the emission across the multi-wavelength observations. The apertures have some overlapping pixels, such that the extractions are not fully independent. We also test binned regions made using the \texttt{vorbin} (\citealt{cappellari_adaptive_2003,cappellari_vorbin_2012}) Python package applied to the PSF-matched UV continuum map, with a target threshold of 5$\sigma_{\mathrm{RMS}}$ as a compromise to produce enough bins to make comparisons across the source, but also to ensure that each bin is larger than the PSF. This results in five bins across the source, when using an input mask of all pixels where [C \textsc{ii}] is detected at S/N$>2$. We note that using this binning method does not significantly change the conclusions of the resolved analysis presented here, where we focus on the clump-based apertures for the fiducial comparison.

For the first method, we need to measure the UV luminosity, $L_{\mathrm{UV}}$, and the infrared luminosity, $L_{\mathrm{IR}}$, in order to derive the unobscured and obscured SFRs. For each region, we extract an integrated spectrum from the PSF-matched NIRSpec IFU cube (Figure \ref{fig:clump spectra}), and follow the methodology in \cite{fisher_rebels-ifu_2026} to obtain a measure of  $L_{\mathrm{UV}}$ for each bin. Briefly, we apply a top-hat filter of width 100 \AA~ centred at $\lambda_{\mathrm{rest}}$ = 1500 \AA~ to obtain $L_{\mathrm{UV}}$, and then we use the conversion factors given in Table 2 of \cite{fisher_rebels-ifu_2026} to derive SFR$_{\mathrm{UV}}$, assuming a \cite{chabrier_galactic_2003} initial mass function (IMF). 

To measure $L_{\mathrm{IR}}$ of each clump region, we follow the methodology of \cite{algera_accurate_2024} to fit dust SEDs using the continuum flux densities measured from the PSF-matched, JvM-corrected 90$\upmu$m and 150$\upmu$m images (Figure \ref{fig:resolved dust sed}). Specifically, we adopt the fiducial optically thin modified blackbody (MBB) model, including a correction for the cosmic microwave background (CMB) from \cite{da_cunha_effect_2013}, and use an MCMC technique to propagate the uncertainties. For all bins, we fix the dust emissivity index to $\beta_{\mathrm{IR}}=2.5$, as derived for the integrated dust SED of REBELS-25 by \cite{algera_accurate_2024}.  For bins detected at S/N$>3$ in both bands (only clump 1), we fit the dust temperature with a flat prior between $T_{\mathrm{dust},z}$ = [22.6, 150] K, and a smooth decline at temperatures above 150 K via a Gaussian of width $= 30$ K (as in \citealt{algera_accurate_2024}).  Here, 22.6 K is the temperature of the CMB at $z=7.31$. For clump 1, this results in $T_{\mathrm{dust}}=41^{+9}_{-7}$ K. For the rest of the regions that are only detected at S/N$>3$ in one band (Band 6) the dust temperature cannot be independently constrained, even with the Band 8 upper limits. For clump 2 and 3, the Band 8 continuum is detected at $\sim1.7\sigma$. If we include these as detections in the SED fit, we obtain a colder dust temperature of $\sim$ 24 K in clump 2, which is co-spatial with the northern dust clump in the Band 6 continuum imaging, and a hotter dust temperature of 58 K in clump 3, where the dust continuum is only detected at $\sim 3\sigma_{\mathrm{RMS}}$ in the Band 6 data. However, within the uncertainties these fits are also consistent with the lower bound of 22.6 K. In these cases, we therefore fix $T_{\mathrm{dust}}=32$ K, consistent with the integrated dust temperature inferred for REBELS-25 in \cite{algera_accurate_2024}, and we include an uncertainty of $\pm 10$K.

The $L_{\mathrm{IR}}$ values are then computed by integrating the best-fit MBBs over rest-frame wavelengths of 8–1000 $\upmu$m, and SFR$_{\mathrm{IR}}$ values are derived again using the conversion factors given in Table 2 of \cite{fisher_rebels-ifu_2026}. We present the derived values in Table \ref{tab:rebels25_clump_properties}, and note that the summed $L_{\mathrm{IR}}$ and SFR$_{\mathrm{IR}}$ values across the 4 clumps comes to a slightly higher total than reported in \cite{fisher_rebels-ifu_2026}. This is due to two reasons: firstly, the overlapping pixels in the circular apertures are summed more than once, and secondly, the higher temperature found for clump 1 increases its inferred $L_{\mathrm{IR}}$ as $L_{\mathrm{IR}}\propto {T_{\mathrm{dust}}}^{4+\beta_{\mathrm{IR}}}$ (e.g., \citealt{sommovigo2025}). The resolved values in Table \ref{tab:rebels25_clump_properties} are, therefore, useful mainly for relative comparisons and not as independent additive components. For comparison, if we instead assume the same dust temperature of 32 K for clump 1, we obtain $L_{\mathrm{IR}}=2.65^{+2.32}_{-1.14}\times10^{11}\mathrm{L\odot}$ and $\mathrm{SFR}_{\mathrm{IR}}=29.1^{+25.5}_{-12.6}\mathrm{M_\odot yr^{-1}}$, which does not alter the main conclusions of this section.

In \cite{algera_accurate_2024}, the integrated dust SED of REBELS-25 is well constrained by observations in six ALMA continuum bands spanning the Rayleigh–Jeans tail and the peak of the infrared emission. In contrast, our spatially resolved analysis relies on only two bands that probe wavelengths close to the SED peak, leading to strong degeneracies between $T_{\mathrm{dust}}$ and $\beta_{\mathrm{IR}}$ if both are allowed to vary. As a result of these limitations, the dust SED fitting on resolved scales is necessarily subject to additional systematic uncertainties compared to the integrated measurements. In particular, the derived values of $T_{\mathrm{dust}}$ and $L_{\mathrm{IR}}$ depend on the assumed fixed emissivity index and, for the single-band bins, on the adopted dust temperature. We therefore caution that the absolute values of $L_{\mathrm{IR}}$ and SFR$_{\mathrm{IR}}$ for individual bins should be interpreted in this context.

In Figure \ref{fig:resolved dust sed}b we show the best-fit dust SEDs for each spatial bin. In principle, the high S/N of the FIR continuum in the central region of REBELS-25 would allow a more finely binned analysis of spatial variations in dust temperature. However, in this work we adopt a coarser binning scheme in the centre in order to ensure robust measurements of both the UV and FIR emission within each bin, as our primary aim here is to compare obscured and unobscured star formation. A more detailed analysis of resolved dust temperature will be presented in future work (Astles et al. in prep).

The resulting $L_{\mathrm{IR}}$ and $L_{\mathrm{UV}}$  of each bin are given in Table \ref{tab:rebels25_clump_properties}, along with the fraction of obscured star formation, $f_{\mathrm{obs}} = \rm{SFR}_{\mathrm{IR}}/ (\rm{SFR}_{\mathrm{IR}}+\rm{SFR}_{\mathrm{UV}})$. We find that dust obscured star formation contributes between 55-98\% of the total SFR across REBELS-25. As expected, the highest $f_{\mathrm{obs}}$ is in the central clump where the dust continuum, [C \textsc{ii}], [O \textsc{iii}]88$\upmu$m, and (tentatively) the optical continuum emission all peak. This is also true if $T_{\mathrm{dust}}=32$ K is assumed for clump 1, resulting in $f_{\mathrm{obs}} = 0.95 ^{+0.03}_{-0.07}$. However, even in regions with lower S/N dust continuum and strong observed UV emission, there is still a significant amount of dust obscuration, comparable to the integrated $f_{\mathrm{obs}}$ values of other IR-luminous sources in the literature (e.g. see \citealt{fisher_rebels-ifu_2026}).

Next, we consider the amount of reddening in each region as inferred from $\beta_{\mathrm{UV}}$, which we measure by fitting a power law between the rest-frame wavelengths $\lambda_{\mathrm{rest}}=1268-2580$ \AA, where $f_\lambda \propto \lambda^{\beta_{\mathrm{UV}}}$ (following \citealt{fisher_rebels-ifu_2025}). The resulting slopes are all consistent with each other, within the uncertainties, and also with the integrated value of $\beta_{\mathrm{UV}}=-1.61$ from \cite{fisher_rebels-ifu_2025}. The limited S/N in the UV emission on resolved scales therefore likely limits an in depth comparison between resolved regions within this galaxy. However, it is interesting to note that all the inferred slopes are relatively red ($\beta_{\mathrm{UV}}\gtrsim-1.7$) in comparison to other high-$z$ galaxies. For example, from the stacked spectrum of 564 sources from JADES data release 4 (\citealt{bunker_jades_2024,deugenio_jades_2025,curtis-lake_jades_2025,scholtz_jades_2025,eisenstein_overview_2026}) at $z\sim4-7$, $\beta_{\mathrm{UV}}=-2.13$ (\citealt{isobe_jades_2026}). The range of $\beta_{\mathrm{UV}}\sim-1.2$ to $-1.7$ measured across the source therefore suggests a significant amount of reddening. Assuming an intrinsic UV slope of $\beta_0=-2.05\pm0.12$, derived from the integrated spectrum of REBELS-25 in \cite{fisher_rebels-ifu_2026}, we estimate the UV attenuation as

\begin{equation}
A_{1600} = \frac{dA_{1600}}{d\beta_{\rm UV}}\left(\beta_{\rm UV} - \beta_{\rm 0,}\right)
\end{equation}

\noindent where we assume $\frac{dA_{1600}}{d\beta_{\rm UV}}=1.44\pm0.14$ (derived for the REBELS-IFU sample in \citealt{fisher_rebels-ifu_2026}). For clump 1, the inferred attenuation is consistent with zero due to the large uncertainty on $\beta_{\rm UV}$ in this low-S/N central region, and we therefore derive an upper limit of $A_{1600}<1.30$ mag at the $1\sigma$ level. For clumps 2--4, we infer attenuation values of $A_{1600}= 1.26\pm0.32, 0.58\pm0.33$, and $0.95\pm0.26$ mag, respectively. We note that the value of $\frac{dA_{1600}}{d\beta_{\rm UV}}$ can also shift by approximately $\pm0.5$ mag, depending on the assumed attenuation curve, and that both $\beta_0$ and $\frac{dA_{1600}}{d\beta_{\rm UV}}$ are likely dependent on properties such as metallicity, stellar ages, and star formation history, and could therefore also vary spatially within a galaxy (e.g., \citealt{zanella_2021,topping_2022,topping_2024,austin_2025,saxena_2026,nakazato_clump-scale_2026}; Fisher et al. in prep.). These estimates therefore remain highly uncertain, and are only intended for relative comparisons.

Overall, this spatially resolved analysis demonstrates that dust obscuration is an important component of REBELS-25. The high obscured fractions inferred across all regions, together with the centrally concentrated FIR emission, suggest that a substantial fraction of the star formation remains hidden in the \textit{JWST} data, while the observed UV emission may preferentially trace less obscured regions towards the outskirts. Similarly, the relatively red UV slopes across the source indicate that even the UV-bright regions are still significantly attenuated. The observed clumps may therefore reflect a combination of intrinsic variations in star formation and spatial variations in dust obscuration, rather than tracing low-attenuation regions alone. At the current resolution, we cannot disentangle the relative contributions of these effects.

\subsection{Clump properties}
\label{sec: resolved IFU}

\begin{figure*}
    \centering
    \includegraphics[width=0.9\textwidth]{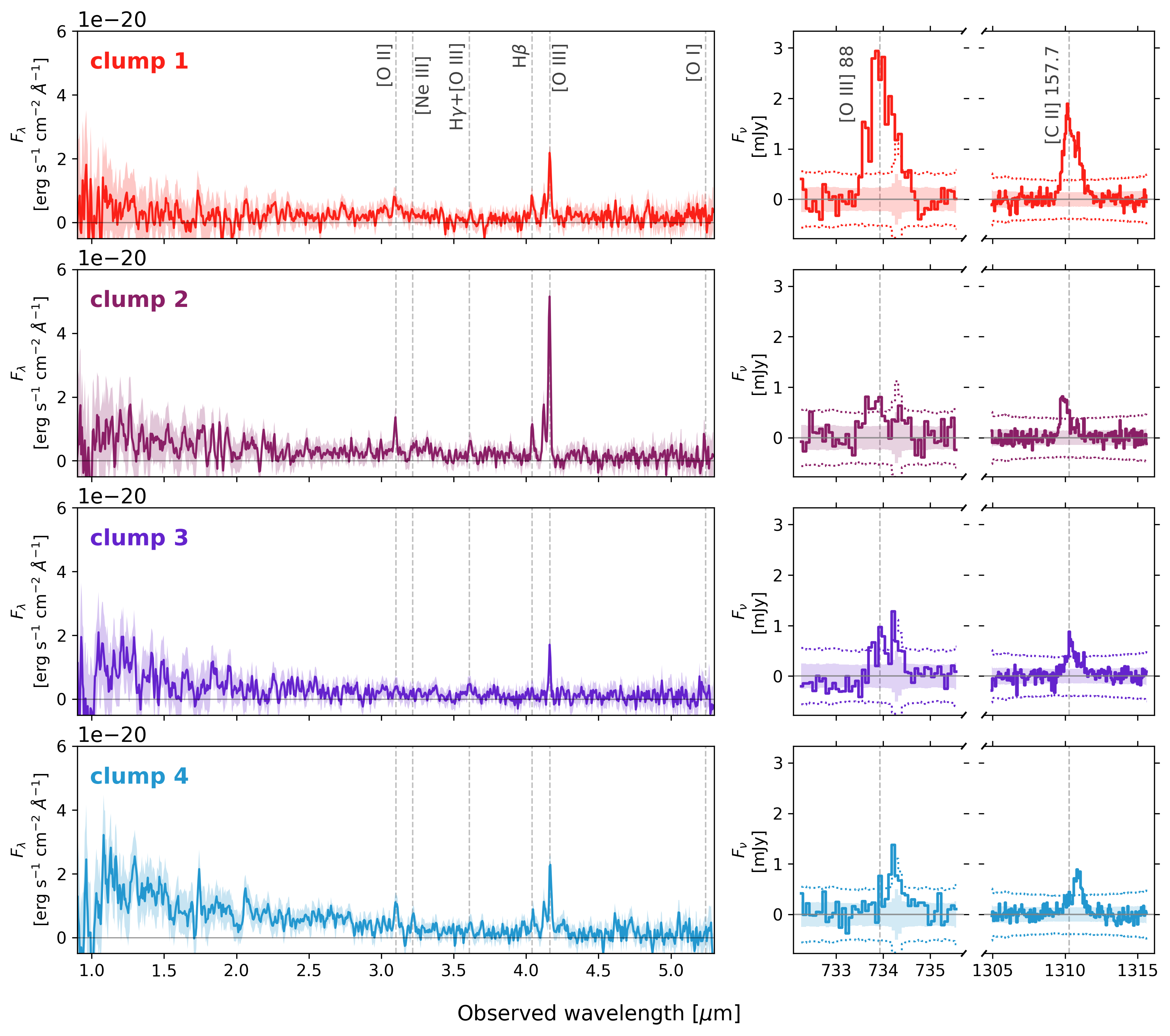}
    \caption{\textit{Left:} Prism spectra from the PSF-matched NIRSpec IFU observations, extracted from the four clump apertures shown in Figure \ref{fig:resolved dust sed}. The observed wavelengths of key rest-frame optical emission lines are indicated by vertical grey dashed lines. \textit{Right:} [O \textsc{iii}]88$\upmu$m and [C \textsc{ii}] spectra from the JvM-corrected, continuum-subtracted, beam-convolved ALMA emission line cubes for the same circular apertures. In both panels, the shaded regions depict the $1\sigma_{\mathrm{RMS}}$ uncertainties in the spectra. or the ALMA data, to which we apply the JvM correction, the dotted lines additionally show ($\sigma_{\mathrm{RMS}}/\epsilon$) as a conservative estimate of the per-channel uncertainties.}
    \label{fig:clump spectra}
\end{figure*}

To further investigate the ISM properties of these clumps, we fit key rest-frame optical emission lines to the extracted IFU spectra plotted in Figure \ref{fig:clump spectra}, applying the same methodology as in \cite{rowland_rebels-ifu_2026}. In brief, we subtract a third-order polynomial fitted to the continuum, and model emission lines as single Gaussians (including blended features in the prism spectra, such as [O \textsc{ii}]$\lambda\lambda$3727,9, [Ne \textsc{iii}]$\lambda$3869+He\textsc{i}+H$\zeta$, and H$\upgamma$+[O \textsc{iii}]$\lambda$4363). We fix the emission line widths to the empirically-derived line spread function (LSF; \citealt{rowland_rebels-ifu_2026}), and fix the ratio of [O \textsc{iii}]$\lambda$5007/[O \textsc{iii}]$\lambda$4959 to the theoretical value of 2.98 (\citealt{storey_theoretical_2000}). Due to the limited S/N of the resolved IFU spectra, the only lines detected at S/N$>3$ are [O \textsc{ii}]$\lambda\lambda$3727,9, H$\upbeta$, and [O \textsc{iii}]$\lambda$5007 (and [Ne \textsc{iii}]$\lambda$3869, for clump 4 only). We also extract spectra for each region from the [O \textsc{iii}]88$\upmu$m and [C \textsc{ii}] JvM-corrected cubes, and fit single Gaussians to obtain luminosities for each clump. 

Due to the significant uncertainties and assumptions in trying to obtain spatially resolved values of the dust attenuation, as discussed above, we do not attempt to estimate dust-corrected emission properties from the \textit{JWST} data. We therefore focus on line ratios that are insensitive to dust attenuation. 

The main ratios we use are:

\begin{equation}
\mathrm{R3}=\mathrm{[O \textsc{iii}]\lambda 5007/ H\upbeta},
\end{equation}

\begin{equation}
\mathrm{Ne3O2}=\mathrm{[Ne \textsc{iii}]\lambda3869/[O \textsc{ii}]\lambda3727,29},
\end{equation}

\begin{equation}
\mathrm{[O \textsc{iii}]/[C \textsc{ii}]}=\mathrm{[O \textsc{iii}]88\upmu m/[C \textsc{ii}]158\upmu m}.
\end{equation}

\noindent We also consider the O32 ratio, assuming that the observed ratios are upper limits when not accounting for dust attenuation:

\begin{equation}
\mathrm{O32}=\mathrm{[O \textsc{iii}]\lambda 5007/[O \textsc{ii}]\lambda3727,29}.
\end{equation}

We report the observed fluxes and ratios in Table \ref{tab:rebels25_clump_properties}. We note that for clump 3, only the [O \textsc{iii}]$\lambda$5007 line is detected in the NIRSpec spectrum at S/N$>3$, meaning a limit for the intrinsic O32 ratio cannot be constrained. We also report the values derived from integrated spectral fitting, i.e. from the full extent of observed emission, from \cite{algera_rebels-ifu_2025} (ALMA values) and \cite{rowland_rebels-ifu_2026} (\textit{JWST} values). For the \textit{JWST} line fluxes, the integrated values are close to the sum of the four clumps, since the majority of the optical nebular emission is encompassed by these regions, whereas for the ALMA data the [C \textsc{ii}] and [O \textsc{iii}] emission are more extended, meaning the summed luminosities are lower than the integrated values.

As aforementioned, the ratios R3 and [O \textsc{iii}]/[C \textsc{ii}] are relatively unaffected by dust attenuation, and we find that these ratios in general do not vary by more than a factor of $\sim2$ across each region. Clumps 2, 3, and 4 in particular are all consistent with the integrated values within the uncertainties, whereas the central clump (1) shows a slightly higher [O \textsc{iii}]/[C \textsc{ii}] and lower R3 ratio.

These line ratios likely depend on a variety of ISM properties, including but not limited to the metallicity and ionisation parameter. The R3 ratio, in particular, is an oft-used tracer of the oxygen abundance, $12+\log\mathrm{(O/H)}$. We therefore use this ratio and recent calibrations made from $z>2$ auroral line detections in \cite{sanders_aurora_2025} to derive the metallicity for each region, also reported in Table \ref{tab:rebels25_clump_properties}. Since this R3 calibration results in two metallicity values for a given R3 (a high branch and low branch solution), we use the upper limits of the O32 ratios (and, where possible, the Ne3O2 ratios) to break this degeneracy. The observed O32 ratios are low ($\lesssim5.7$), meaning that clumps 1, 2 and 4 all favour the high branch metallicity solution (clump 3 is unconstrained). As noted in \cite{rowland_rebels-ifu_2026} and discussed below in Section \ref{sec:discussions}, the R3 and O32 ratios of this source are in general lower than the $z>6$ systems in current large spectroscopic surveys, and are more comparable to sources in surveys at $1<z<3$ (e.g. MOSDEF, \citealt{kriek_mosfire_2015,shapley_mosdef_2019}; AURORA, \citealt{shapley_aurora_2025-1}), implying a relatively metal-rich ISM at very early times. Indeed, the resulting oxygen abundance values are $12+\log\mathrm{(O/H)}\sim8.29$--$8.47$ ($\sim0.4$--$0.6 ~\mathrm{Z_{\odot}}$), with the highest metallicity found in the central clump and the lowest metallicity in clump 2, where the \textit{JWST} emission lines peak (Figure \ref{fig:line maps}). With the significantly higher S/N obtained for the rest-frame optical emission lines in clump 2 from the NIRSpec IFU data, this clump is in fact likely dominating the integrated flux ratios and inferred ISM properties. However, the derived metallicities are all consistent with each other and with the integrated value within the uncertainties, suggesting no strong evidence for a metallicity gradient across the source. Similarly flat or weak metallicity gradients have been inferred for other $z\gtrsim6$ targets from both FIR (e.g. \citealt{vallini_2024}) and UV/optical measurements (e.g. \citealt{venturi_gas-phase_2024,ivey_2026}). A more detailed investigation of the resolved metallicity distribution and any possible gradient in the REBELS-IFU sample will be presented in Rowland et al. (in prep.).

These metallicity values are also subject to additional systematics, as they are based on a single optical strong-line ratio which also has a strong dependence on the ionisation parameter, and therefore the absolute values should be treated with caution (see more in depth discussion and comparison with other calibrations in \citealt{rowland_rebels-ifu_2026}). Including an emission line that also traces gas in a lower ionisation state can help overcome this degeneracy, for example the R23 ratio (= ([O \textsc{iii}]$\lambda\lambda$4959,5007 + [O \textsc{ii}]$\lambda\lambda$3727,9) / H$\upbeta$). If we assume $A_{\mathrm{V, neb}}=1.46$ mag, adopting the average $A_{\mathrm{V, neb}}/A_{\mathrm{V, *}}\sim2$ measured for the REBELS-IFU sample in \cite{fisher_rebels-ifu_2026} and the stellar dust attenuation from \cite{rowland_rebels-ifu_2025}, and use the R23 calibration, again from \cite{sanders_aurora_2025}, we obtain higher metallicities ranging from $12+\log\mathrm{(O/H)}\sim$ 8.57 to 8.65 for the clumps, and a global value of 8.53, i.e. implying near-solar metallicities already in place at $z\sim7$. Confirming whether REBELS-25 is truly enriched to near-solar metallicity, for example through the detections of temperature sensitive auroral lines, could have important implications for early galaxy evolution and chemical enrichment studies. In addition, recent studies have explored deriving more robust ISM properties from multi-phase ISM modelling by combining UV, optical and FIR emission (\citealt{harikane_jwst_2025,usui_rioja_2025,castellano_investigating_2026}). Applying similar modelling techniques to this source would strongly benefit from electron density constraints from both optical ratios (e.g. [O \textsc{ii}]$\lambda$3727/[O \textsc{ii}]$\lambda$3729) and FIR ratios (e.g. [O \textsc{iii}]52$\upmu$m/[O \textsc{iii}]88$\upmu$m). With the limited information currently available, we therefore opt not to apply these multi-phase modelling techniques, and instead only consider variations in the observed R3, O32, and [O \textsc{iii}]/[C \textsc{ii}] ratios as indications of potential changes in primarily the metallicity and/or ionisation parameter in a one-zone model assumption.

Indeed, the [O \textsc{iii}]/[C \textsc{ii}] has also been found to show some dependence on metallicity (e.g. \citealt{algera_rebels-ifu_2025}), although it likely correlates more strongly with other properties, such as the star formation `burstiness' (\citealt{algera_rebels-ifu_2025, kohandel_amaryllis_2025}), ratios of ionised-to-neutral gas (e.g. \citealt{nakazato_unveiling_2026}), and/or C/O abundances (e.g. \citealt{arata_2020,katz_nature_2022,nyhagen_theoretical_2025}). The relatively small variation in the observed [O \textsc{iii}]/[C \textsc{ii}] ratios across the clumps therefore does not point towards large spatial variations in these underlying ISM properties.

Overall, the resolved line ratios and inferred metallicities suggest that the clumps in REBELS-25 share broadly similar ISM conditions. From this we can infer that these clumps may be physically associated and embedded within the same metal-rich ISM, rather than distinct systems, such as merging  components. We note that this doesn't preclude stronger variations on smaller spatial scales, and indeed these line ratios and ISM properties can vary significantly in well-resolved sources in the local Universe (e.g., \citealt{james2016,Zhang2017,Poetrodjojo2018,monrealibero2023,fischer2025}) and in high-$z$ simulations (e.g. \citealt{vallini_high_2021,zanella_2021,nakazato_unveiling_2026}) on $\lesssim 100$ pc scales.

\section{Multi-phase gas kinematics}
\label{sec:kinematics}

While the previous sections focused on morphological differences and the role of dust obscuration in shaping the observed multi-wavelength structure of REBELS-25, the high-resolution [C \textsc{ii}] and [O \textsc{iii}]88$\upmu$m observations also enable a direct comparison of the galaxy’s gas kinematics from these different tracers. Importantly, both of these lines were observed with the same telescope at comparable spectral resolution, and at wavelengths where dust obscuration will not impact the measurements. In addition, we can further assess whether the `clumps' prominent in the rest-optical and UV emission, and tentatively also detected in the [O \textsc{iii}]88$\upmu$m emission, reflect dynamically distinct components, such as merging or outflowing components.

By modelling the warm ionised gas kinematics traced by [O \textsc{iii}]88$\upmu$m, we can also test whether the low velocity dispersion  ($\bar{\sigma}=33^{+9}_{-7}$ km s$^{-1}$) and high rotational support ($V_{\mathrm{rot,~max}}/\bar{\sigma}=11^{+6}_{-5}$) derived in the colder neutral phase (\citealt{rowland_rebels-25_2024}), which we assume [C \textsc{ii}] to be predominantly tracing, is also reflected in the warm ionised phase, or whether this source appears dynamically hotter when traced by warm ionised gas, as commonly found at lower redshifts (e.g. \citealt{levy_edge-califa_2018, ubler_evolution_2019, girard_systematic_2021}). 

We present the velocity and velocity dispersion maps of the [O \textsc{iii}]88$\upmu$m emission in Figures \ref{fig:vrot maps} and \ref{fig:vdisp maps}. These already reveal a clear large-scale velocity gradient in [O \textsc{iii}], closely matching that observed in [C \textsc{ii}]. We find no evidence for kinematically distinct components at the locations of the clumps in the \textit{JWST} data, which is also consistent with the similar [O \textsc{iii}]88$\upmu$m and [C \textsc{ii}] line profiles extracted from these regions (Figure \ref{fig:clump spectra}), suggesting that these features, if intrinsic, are embedded within the same global rotating structure.

\subsection{\texttt{3DBAROLO}}
\label{sec:barolo}

\begin{table}
\def\arraystretch{1.25}
\centering
\caption{Kinematic properties derived from the fiducial \texttt{3DBAROLO} fitting.}
\label{tab:rebels25_kinematics}
\begin{tabular}{c c c c}
\hline
Line &
$V_{\mathrm{rot, max}}$ &

$\bar{\sigma}$ &
$V_{\mathrm{rot, max}}/\sigma$ 
 \\
&
km s$^{-1}$ &

km s$^{-1}$ &

\\
\hline

 [O \textsc{iii}]88$\upmu$m & 274$^{+89}_{-87}$ &   61$^{+7}_{-8}$ & 4.5$^{+2.4}_{-1.8}$ \\

 [C \textsc{ii}] & 374$^{+86}_{-91}$ &   33$^{+9}_{-7}$ & 11$^{+6}_{-5}$ \\

\hline
\end{tabular}
\vspace{5pt}
\begin{tablenotes}
\item{\textbf{Notes:} Col. (1): emission line, col. (2): maximum rotational velocity, col. (3): average velocity dispersion, col. (4) degree of rotational support (ratio of maximum rotational velocity to average velocity dispersion). For the rotational velocity values, the uncertainty on the inclination, $i=25\pm6$, is propagated through.} 
\end{tablenotes}
\end{table}

\begin{figure*}
    \centering
    
    \includegraphics[width=\textwidth]{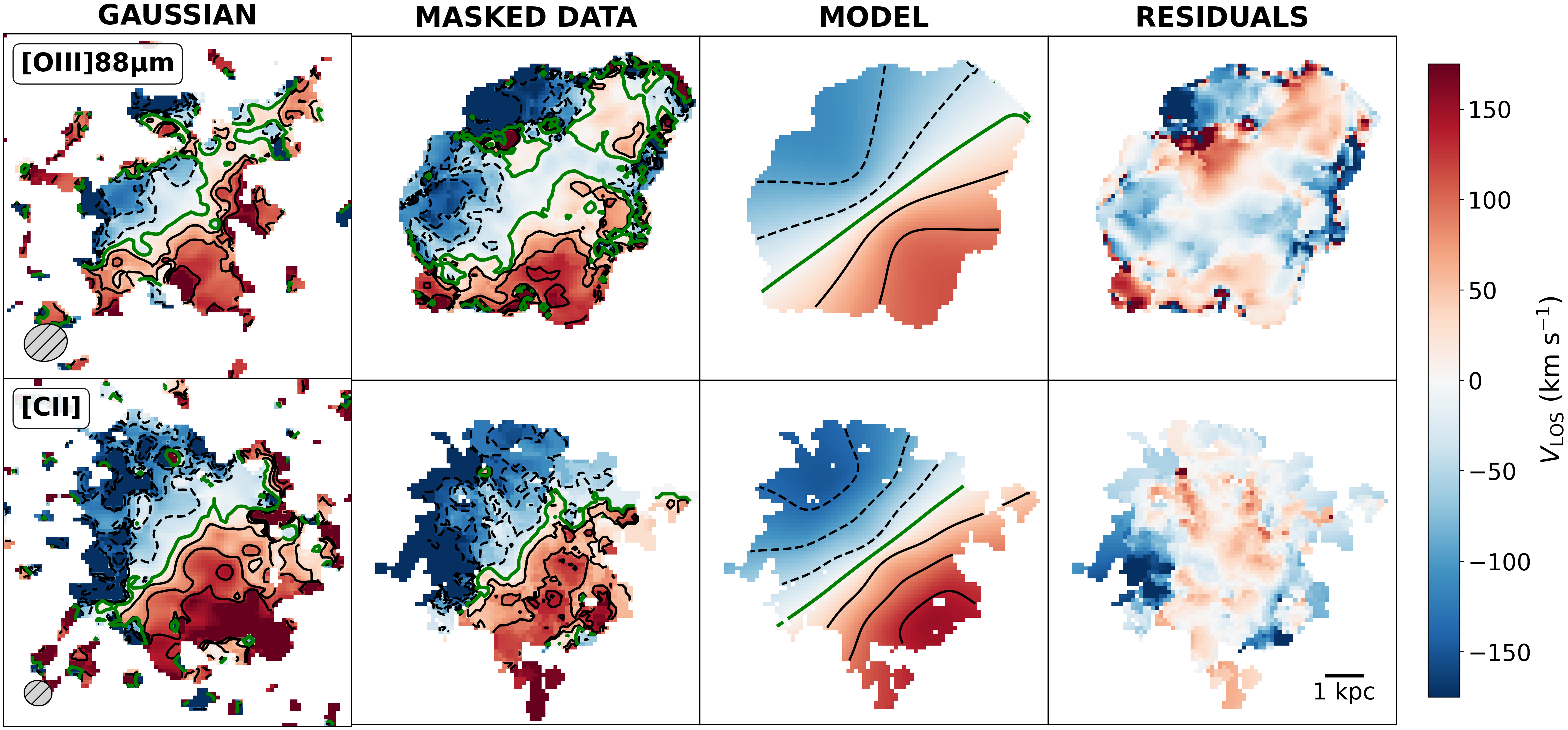}
    \caption{Velocity field maps from the $\lesssim1$ kpc-resolution ALMA observations of [O \textsc{iii}]88$\upmu$m (upper) and [C \textsc{ii}] emission (lower). From left-to-right, we show line-of-sight rotation velocity maps produced from fitting single Gaussians to each pixel, moment-1 maps from the masked cube used as an input to the \texttt{3DBAROLO} fitting, moment-1 map of the \texttt{3DBAROLO} model, and residuals from the masked moment-1 and model moment-1 maps. Isovelocity contours are plotted from $\pm180$ km s$^{-1}$ in 45 km s$^{-1}$ increments (black dashed lines for negative velocities, solid for positive) with the systemic velocity contour plotted with a thick green line. The [O \textsc{iii}]88$\upmu$m and [C \textsc{ii}] emission both appear to trace a similar rotating disc structure, albeit with significant residuals not well captured by the disc model.}
    \label{fig:vrot maps}
\end{figure*}

\begin{figure*}
    \centering
    
    \includegraphics[width=\textwidth]{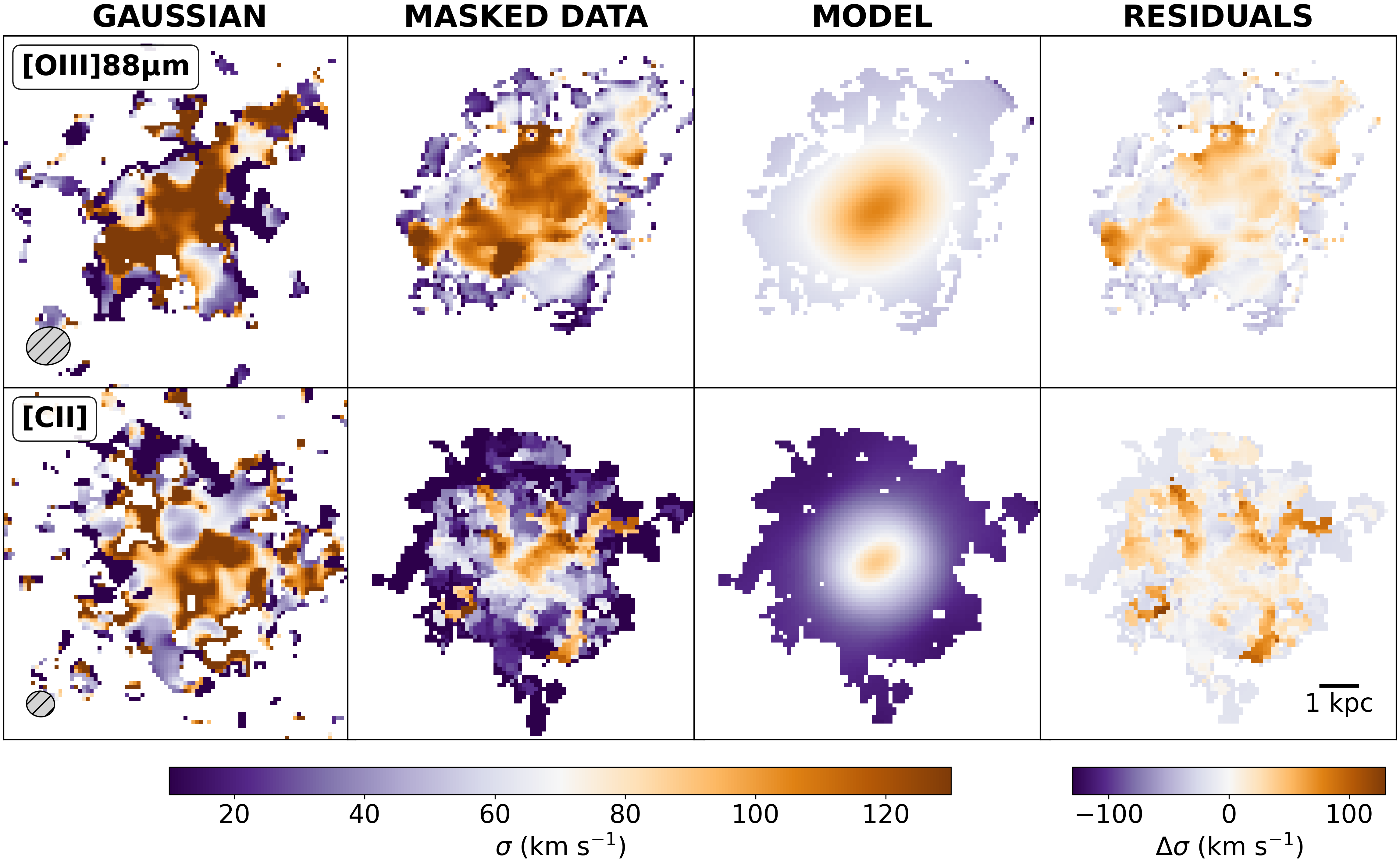}
    \caption{Velocity dispersion maps from the $\lesssim1$ kpc-resolution ALMA observations of [O \textsc{iii}]88$\upmu$m (upper) and [C \textsc{ii}] emission (lower). From left-to-right, we show dispersion maps produced from the widths of single Gaussians fitted to each pixel, moment-2 maps from the masked cube used as an input to the \texttt{3DBAROLO} fitting, moment-2 maps of the \texttt{3DBAROLO} models, and residuals from the masked data cubes and models. Due to the moderate S/N of the observations, these velocity dispersion maps are relatively noisy and should be interpreted with caution on a pixel-by-pixel basis, although both tracers show higher dispersions towards the centre and lower values at the outskirts, broadly consistent with the disc models. Note these maps are not corrected for beam smearing, which can significantly increase the observed velocity dispersion, particularly at the centre. We also add that the fitted dispersions from the Gaussian fitting to the non-masked data in the first column may be overestimated, since the data shown is not JvM-corrected (see Appendix \ref{appendix:jvm correction} for details).}
    \label{fig:vdisp maps}
\end{figure*}

\begin{figure}
    \centering
    
    \includegraphics[width=0.45\textwidth]{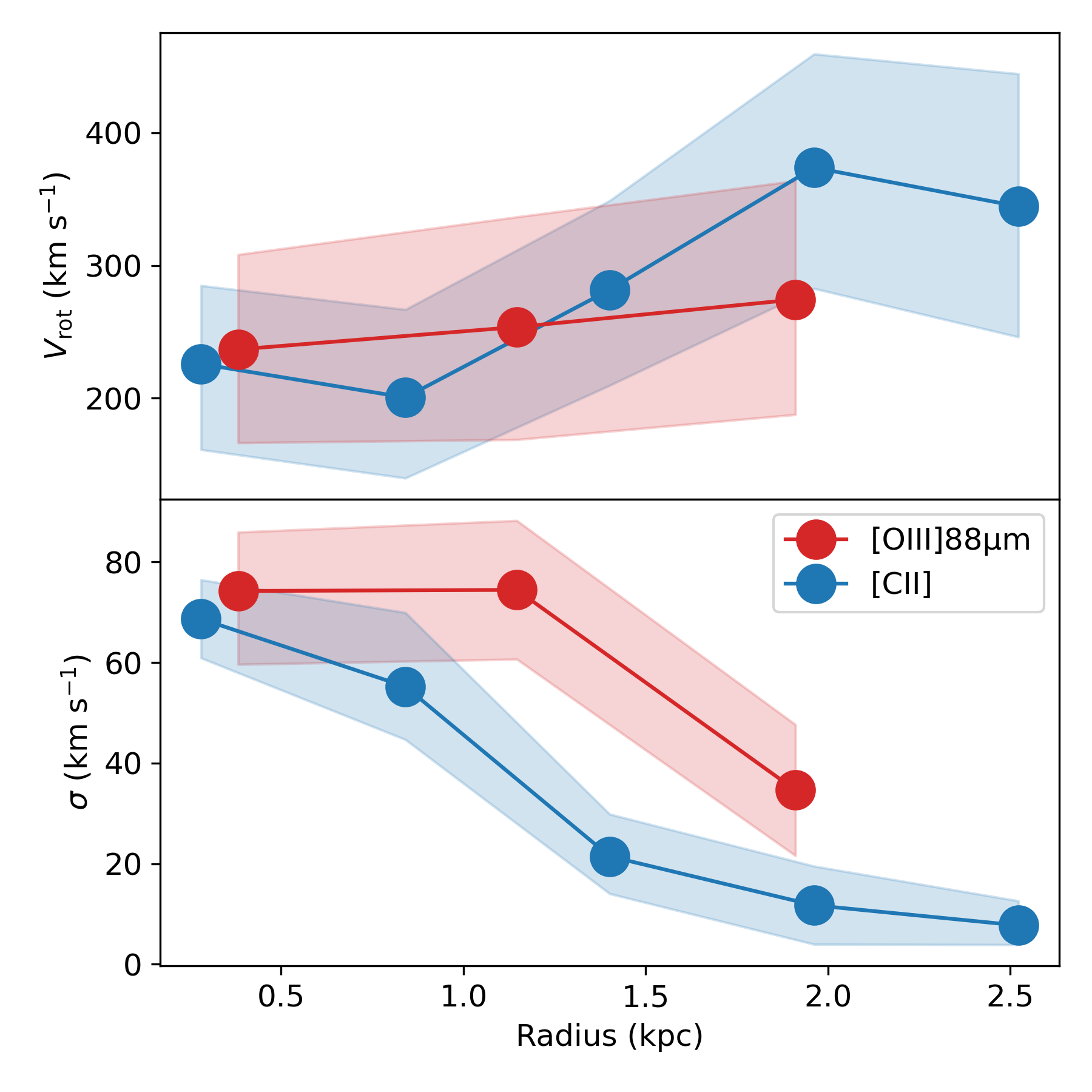}
    \caption{Rotational velocity (top; $V_{\mathrm{rot}}$) and velocity dispersion (bottom; $\sigma$) profiles of REBELS-25 from the [C \textsc{ii}] (blue) and [O \textsc{iii}]88$\upmu$m (red) kinematic modelling with \texttt{3DBAROLO}. These are the intrinsic values in each ring, corrected for inclination and beam smearing effects. The shaded region indicates the uncertainties on the fitted parameters, where for the $V_{\mathrm{rot}}$ errors we have also propagated the error on the inclination, $i=25\pm6$ degrees. As the source is found to have a low inclination, even small uncertainties in $i$ have a significant effect on the measured rotational velocity profile.}
    \label{fig:velocity profiles}
\end{figure}

\begin{figure}
    \centering
    
    \includegraphics[width=0.48\textwidth]{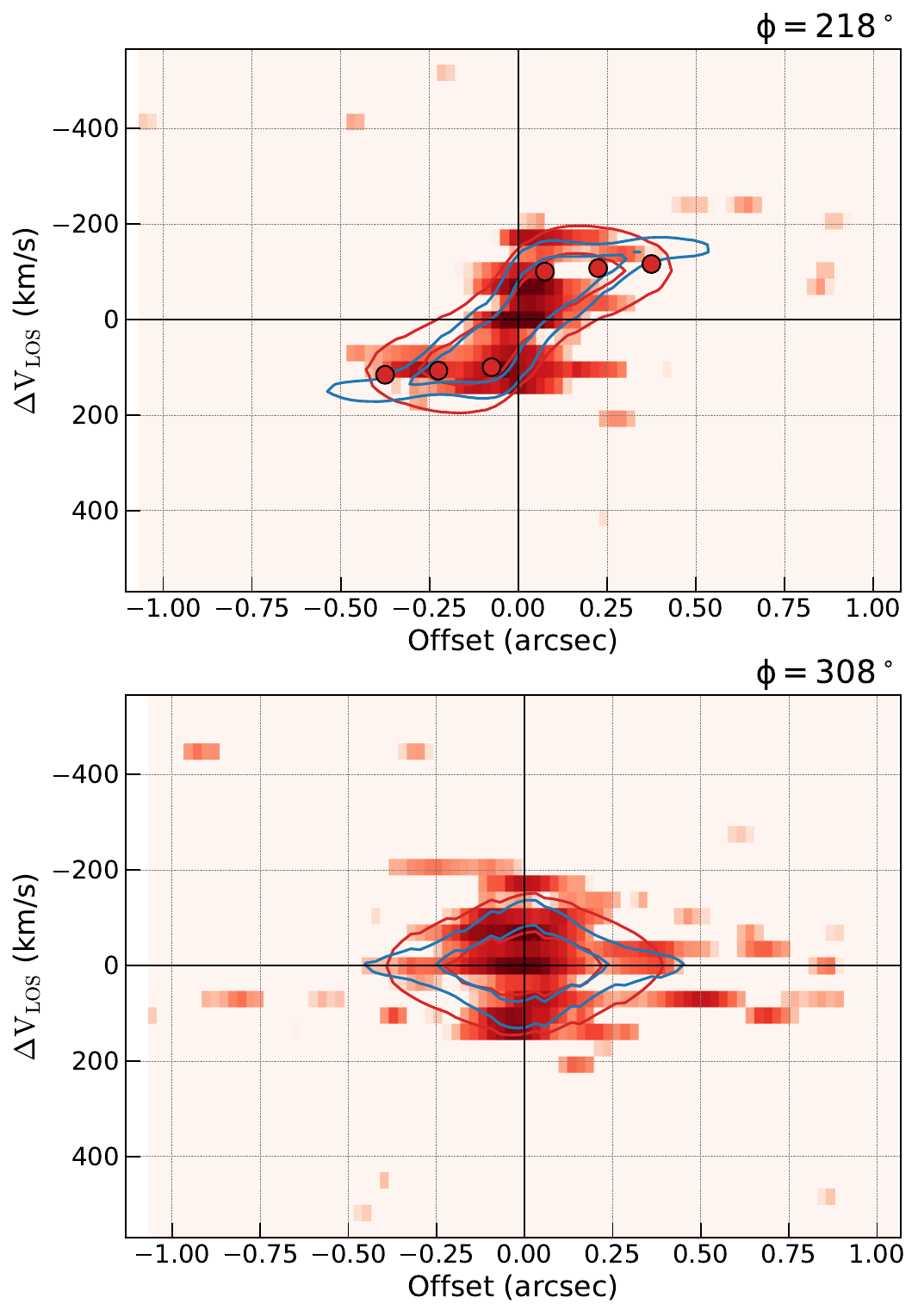}
    \caption{Position velocity diagrams (PVDs) for the [O \textsc{iii}]88$\upmu$m \texttt{3DBAROLO} fitting extracted from the kinematic major axis (218 degrees, from the [C \textsc{ii}] fitting described in \citealt{rowland_rebels-25_2024}) and minor axis. The best-fit model from the [O \textsc{iii}]88$\upmu$m fitting is shown by red contours, and we also plot the best-fit model from the [C \textsc{ii}] modelling in blue contours (2 and 4$\sigma_{\mathrm{RMS}}$). The [O \textsc{iii}]88$\upmu$m and [C \textsc{ii}] models agree well with each other, indicating that both emission lines are tracing the same overall rotating disc structure.}
    \label{fig:pvds}
\end{figure}

To enable a consistent comparison with the colder gas kinematics from [C \textsc{ii}] presented in \cite{rowland_rebels-25_2024}, we apply the same kinematic fitting methodology to the [O \textsc{iii}]88$\upmu$m cube. Specifically, we use \texttt{3DBAROLO} (v1.7; \citealt{di_teodoro_3d_2015}) to fit beam-convolved 3D tilted-ring models directly to the data. This method fits the line emission in the full 3D cube and returns best-fit morphological and kinematic parameters using a Monte Carlo sampling approach to estimate the uncertainties.

A number of alternative tools exist in the literature, including \texttt{QUBEFIT} (\citealt{neeleman_kinematics_2021}), \texttt{DysmalPy} (\citealt{lee_disk_2025}), and \texttt{GALPAK3D} (\citealt{GALPAK2015}). Recent comparison studies (\citealt{rizzo_2022,yttergren_kinematics_2025, lee_disk_2025}) show that methodological choices -- especially masking, priors, and parametric versus non-parametric assumptions -- can impact the derived kinematic properties of individual sources, particularly at modest S/N. For consistency with \cite{rowland_rebels-25_2024}, we adopt \texttt{3DBAROLO} as our fiducial method here, but also perform complementary fits using \texttt{DysmalPy} for comparison (see Appendix \ref{appendix:DysmalPy} for details).

The fiducial \texttt{3DBAROLO} setup closely follows the configuration used in \cite{rowland_rebels-25_2024}. We first extract a $3.5 \times 3.5$ arcsec sub-cube centred on the [O \textsc{iii}]88$\upmu$m emission. The radial separation of the rings is set to $\sim$80\% of the beam FWHM ($=0.168$ arcsec), which results in three rings across the detected emission. We use azimuthal normalisation for the surface brightness, fit both sides of the rotation curve simultaneously, and minimise the reduced $\chi^2$. 

For initial source identification, we set \texttt{SNRCUT} = 3, and the mask around the detected source is grown with \texttt{GROWTHCUT} = 2. However, unlike the [C \textsc{ii}] data, the [O \textsc{iii}]88$\upmu$m cube is affected by a narrow atmospheric feature near $\sim$+100 km/s, which increases the noise in those channels. Since \texttt{3DBAROLO} assumes a constant $\sigma_{\mathrm{RMS}}$ across channels when generating S/N-based masks, this can result in overestimated significance in those affected regions. To account for this, we construct a S/N cube in which the noise is computed separately for each channel using emission-free regions of the line cube. We then apply the \texttt{SEARCH} algorithm to this S/N cube using \texttt{SNRCUT} = 3 and \texttt{GROWTHCUT} = 2, and input the resulting 3D mask to \texttt{3DBAROLO}. This method is consistent with the S/N-based masking used in \cite{rowland_rebels-25_2024}, but now accounts for the non-uniform channel noise in the Band 8 data (see Appendix \ref{appendix:atmospheric feature}).

In addition, we carry out the same tests on the masking as described in Appendix C of \cite{rowland_rebels-25_2024}, where we investigate the percentage of pixels in fewer than three channels in the outer ring of the masked cube, and find that it is over 50\%. This could result in a significant number of pixels with unrealistically small values in the moment-2 map. We therefore grow the mask in the spectral direction, and ensure that every pixel is detected in at least three consecutive channels, more comparable to the [C \textsc{ii}] masking. In addition, we grow the mask in every channel by five pixels in the spatial direction so that the outermost ring is filled (see e.g.  \citealt{mancera_pina_galaxy-halo_2025,Westoby_2026}). 

As discussed in Section \ref{sec:morphology}, the morphological centre of the [O \textsc{iii}]88$\upmu$m emission is consistent with the [C \textsc{ii}] analysis, but, if left as a free parameter in the Sersic fitting, the morphological PAs are different, with the observed [O \textsc{iii}] emission in fact being more extended along the minor axis of the [C \textsc{ii}] emission. To limit the number of free parameters, we fix the morphology parameters (centroid and inclination) and the kinematic PA in \texttt{3DBAROLO} to the values derived from the [C \textsc{ii}] data in \cite{rowland_rebels-25_2024}. Inspection of the moment-1 maps (Figure \ref{fig:vrot maps}) supports the use of the same kinematic PA and centroid for [O \textsc{iii}]88$\upmu$m and [C \textsc{ii}]. We allow \texttt{3DBAROLO} to estimate the systemic velocity, finding $V_{\mathrm{sys}}=24.36$ km s$^{-1}$, consistent with the [C \textsc{ii}] value. The only parameters left free in the fitting are therefore the rotational velocity, $V_{\mathrm{rot}}$, and the velocity dispersion, $\sigma$, of each ring. We add that, since the inclination was found to be low from the [C \textsc{ii}] data in \cite{rowland_rebels-25_2024} ($i=25\pm6^{\circ}$), the inferred $V_{\mathrm{rot}}$ and $V_{\mathrm{rot, max}}/\bar{\sigma}$ values are highly sensitive to the adopted inclination. We therefore propagate the uncertainties in the inclination for these measurements.

We plot the resulting moment maps, velocity profiles and position velocity diagrams (PVDs) in Figures \ref{fig:vrot maps}, \ref{fig:vdisp maps},  \ref{fig:velocity profiles}, and \ref{fig:pvds}, respectively, with the corresponding [C \textsc{ii}] results from \cite{rowland_rebels-25_2024} also shown for comparison. This [O \textsc{iii}]88$\upmu$m fitting results in a maximum rotation velocity, $V_{\mathrm{rot, max}}$, of $274^{+89}_{-87}$ km s$^{-1}$, and an average velocity dispersion, $\bar{\sigma}$, across the three rings of $61^{+7}_{-8}$ km s$^{-1}$. This results in $V_{\mathrm{rot, max}}/\bar{\sigma} = 4.5^{+2.4}_{-1.8}$.  However, on inspection of the moment maps in Figures \ref{fig:vrot maps} and \ref{fig:vdisp maps}, we see some residuals in both the moment-1 and moment-2 maps, as also found for the [C \textsc{ii}] fits in \cite{rowland_rebels-25_2024}. This likely indicates contributions from non-circular motions to both the [O \textsc{iii}] and [C \textsc{ii}] emission, although we note that at this S/N, these features could also be attributed to noise. As there are significant residuals in the moment-2 maps, we also investigate whether \texttt{3DBAROLO} may be underestimating the velocity dispersion, below.

As discussed in \cite{neeleman_kinematics_2021}, overly severe spectral masking can lead to biased values in moment-1 and moment-2 maps, and in particular to an underestimation of the velocity dispersion. We therefore also show velocity and dispersion maps obtained by single Gaussian fits to each spaxel detected at S/N$>2$ in Figures \ref{fig:vrot maps} and \ref{fig:vdisp maps}. We indeed find higher velocity dispersion values in the single-Gaussian fitted maps, although in all maps the dispersion values decrease to $\sim30-40$ km s$^{-1}$ at the outer extent of the emission, where beam smearing is less severe. To further assess if \texttt{3DBAROLO} may be underestimating the velocity dispersion, we therefore apply fitting without spectral masking using \texttt{DysmalPy}, as described in Appendix \ref{appendix:DysmalPy}. To summarise these tests, we first model the [O \textsc{iii}]88$\upmu$m and [C \textsc{ii}] spectra with single Gaussians to input into \texttt{DysmalPy}, and we assume a constant velocity dispersion across the disc. For [O \textsc{iii}]88$\upmu$m, this yields a velocity dispersion of $\sigma=62.1\pm3.6$ km s$^{-1}$, consistent with the \texttt{3DBAROLO} value, whilst for [C \textsc{ii}] we find a higher $\sigma$ of $75.4\pm 1.5$ km s$^{-1}$. However, there is tentative evidence for a broad component in the spectra (see e.g. \citealt{hygate_alma_2023}), which, if accounted for, would result in $\sigma\sim40$km s$^{-1}$ for both [C \textsc{ii}] and [O \textsc{iii}]88$\upmu$m, as discussed in detail in Appendix \ref{appendix:DysmalPy}. We therefore caution that the absolute values of fitted kinematic parameters for individual sources should be treated with caution, and in the context of the tools they were fitted with (see e.g. the findings of \citealt{rizzo_2022,lee_disk_2025,yttergren_kinematics_2025}).

\subsection{Comparison of [O \textsc{iii}]88 and [C \textsc{ii}] kinematics}
\label{sec:kinematics comparison}

On visual inspection, the [C \textsc{ii}] and [O \textsc{iii}]88$\upmu$m velocity fields show good agreement, with comparable PVDs (Figure \ref{fig:pvds}), kinematic major axes and isovelocity contours (Figure \ref{fig:vrot maps}), and velocity profiles (Figures \ref{fig:velocity profiles}, \ref{fig:oiii DysmalPy} and \ref{fig:cii DysmalPy}). This indicates that both the warm ionised gas and colder, neutral gas components follow the same bulk motion; that is, the rotating disc of REBELS-25. Quantitatively, the velocity curves (from both the \texttt{3DBAROLO} fitting in Figure \ref{fig:velocity profiles} and the Gaussian fits in Figures \ref{fig:oiii DysmalPy} and \ref{fig:cii DysmalPy}) are also consistent within the uncertainties, although we obtain a lower $V_{\mathrm{rot, max}}$ for [O \textsc{iii}]88$\upmu$m. We also find a higher velocity dispersion from the [O \textsc{iii}]88$\upmu$m data in comparison to [C \textsc{ii}], by a factor of $\sim2$ (although see \texttt{DYSMALPY} fits in Appendix \ref{appendix:DysmalPy}).


Offsets of this kind, where the ionised gas traces systematically lower rotational velocities and higher velocity dispersions in comparison to colder gas tracers, have been reported at lower redshift and interpreted as contributions from extraplanar diffuse ionised gas (e.g. \citealt{levy_edge-califa_2018}). However, for the observations of REBELS-25 analysed here, this difference could also in part arise from the radial extent and S/N of the two tracers. The [C \textsc{ii}] emission is more extended and has higher S/N, allowing us to trace the emission out to $\sim2.5$ kpc, where there is some evidence from the PV diagrams (Figures \ref{fig:pvds} and \ref{fig:cii DysmalPy}) that the velocity profile begins to flatten and the intrinsic velocity dispersion decreases. In contrast, the [O \textsc{iii}] emission is less extended and has lower S/N, and we are only able to reliably model the kinematics out to $\lesssim2$ kpc, where the velocity may still be increasing.  We discuss whether these differences, particularly in the measured velocity dispersion, may reflect intrinsic changes in the gas phases being probed in more detail in Section \ref{sec:compilation}.

From the kinematic modelling described in this section, we overall find that REBELS-25 can be described as a dynamically cold disc in both [O \textsc{iii}] and [C \textsc{ii}] emission; i.e. in both warm ionised gas and colder neutral gas. In both cases, we recover high $V_{\mathrm{rot, max}}/\sigma$ values with \texttt{3DBAROLO}, indicating that this galaxy is already settled into a rotating disc.\footnote{The same conclusion is supported by the \texttt{DysmalPy} modelling in Appendix \ref{appendix:DysmalPy}, although it does yield a lower $V_{\mathrm{rot, max}}/\sigma$ of $\sim4$ for the [C \textsc{ii}] kinematics.} We do, however, find evidence for contributions from non-circular motions, for example in the residuals of the fitted velocity and dispersion maps, warping of the isovelocity contours, and hints of additional components in the spectra. With the current data, it is not possible to determine whether these non-circular motions are driven by outflows, inflows, minor mergers, other interactions, or noise.

\section{Discussion}
\label{sec:discussions}

\subsection{The multi-wavelength properties of REBELS-25}

In \cite{rowland_rebels-25_2024}, the high-resolution ALMA Band 6 observations of REBELS-25 revealed that its [C \textsc{ii}] emission is well-described by a near-exponential, rotating disc model, despite the apparent clumpiness of the source at rest-frame UV wavelengths. The FIR and UV emission were also found to be physically offset, indicating a significant amount of dust obscuration at the centre of the galaxy. In this work, the addition of high-resolution ALMA Band 8 data and \textit{JWST}/NIRSpec IFU observations allows us to investigate the morphology, kinematics, and ISM properties of REBELS-25 in more detail. In particular, the [O \textsc{iii}]88$\upmu$m emission traces the ionised gas without being impacted by dust attenuation, and the underlying FIR continuum at $\sim90 ~\upmu$m, in combination with the Band 6 continuum at $150 ~\upmu$m and the deeper rest-UV observations, enables us to quantify the degree of dust obscuration across the source. The optical and FIR emission lines also enable a resolved investigation of its ISM properties, allowing us to compare the FIR-bright central region with the UV- and optical-bright clumps and test whether these regions exhibit distinct underlying physical conditions.

As with the [C \textsc{ii}] emission, the [O \textsc{iii}]88$\upmu$m and FIR continuum emission are found to be centrally concentrated and well described by near-exponential Sérsic profiles, although the two FIR continuum profiles have Sérsic indices of $n\sim1.5$--$2$ (albeit with large uncertainties). The [C \textsc{ii}] emission is slightly more extended than the [O \textsc{iii}]88$\upmu$m ($r_{\mathrm{e, [C \textsc{ii}]}} = 1.4 \times r_{\mathrm{e, [O \textsc{iii}]88}}$) and FIR continuum, consistent with [C \textsc{ii}] tracing a more extended neutral gas component at these redshifts (e.g., \citealt{fudamoto_alma_2022,ikeda2025, rowland_rebels-ifu_2025}). 

The [C \textsc{ii}], [O \textsc{iii}]88$\upmu$m, and FIR continuum all peak close to the dynamical centre identified from the [C \textsc{ii}] data, whereas the brightest UV and optical emission is offset from this region. Even with the deeper \textit{JWST} data in comparison to the \textit{HST} F160W observations, the source remains undetected in individual pixels at the centre at UV wavelengths. Similar spatial offsets between UV/optical emission and IR emission have been observed in submillimetre galaxies (SMGs) and dusty star-forming galaxies (DSFGs) at lower redshifts, particularly at $z\sim2$--5 (e.g. \citealt{chen2015,hodge_kiloparsec-scale_2016,chen2017,Li_2026}) due to strong spatial variations in dust attenuation. From the combination of the FIR and UV data for REBELS-25, we find that the fraction of obscured star formation could be as high as $\sim98\%$ in the central FIR-bright region and remains dominant across the galaxy ($\gtrsim55\%$), including in regions where the observed UV emission is strongest. Our analysis of REBELS-25 therefore suggests that strong, spatially varying dust obscuration may already be present even at $z\sim7$, which is also supported by recent discoveries of `NIRCam-dark' galaxies at $z\gtrsim6$ (e.g., \citealt{Sun2026, bing2026}).

For the kinematics, both [C \textsc{ii}] and [O \textsc{iii}]88$\upmu$m show a coherent large-scale velocity gradient, with no clear evidence for dynamically distinct components at the locations of the UV/optical clumps. The kinematic major axis, systemic velocity, and rotation curves are broadly consistent between the  warm ionised ([O \textsc{iii}]88$\upmu$m) and colder neutral ([C \textsc{ii}]) gas phases, indicating that both tracers are dominated by the same rotating structure.

Within the uncertainties, the rest-frame optical line ratios and FIR line ratios are broadly consistent across the source and between the identified clumps. In particular, the comparable R3, O32 upper limits, and [O \textsc{iii}]88$\upmu$m/[C \textsc{ii}] ratios suggest that the clumps do not have dramatically different ionisation conditions, metallicities, or star-formation properties.  We stress that this does not rule out varying ISM properties on smaller spatial scales than those probed here ($\sim 1$ kpc scales), but it argues against a scenario in which the large UV/optical clumps are separate systems with substantially different ISM conditions.

Together, these observations further support the scenario where REBELS-25 is a dusty, rotationally supported disc,  with an observed UV and optical morphology that is strongly impacted by dust attenuation. Nevertheless, there are also indications that REBELS-25 is not a perfectly smooth disc. Low-significance residuals in the Sérsic modelling of the [O \textsc{iii}]88$\upmu$m and 150$\upmu$m continuum maps appear spatially coincident with some of the UV clumps. Since these FIR tracers are not strongly affected by dust obscuration, this may indicate that at least part of the clumpy substructure could be intrinsic and embedded in the large-scale rotating disc (e.g. as in \citealt{fujimoto_primordial_2025}). Similarly, the tentative central bar-like morphology, the possible broad spectral component, and the non-circular motions seen in the residuals of the kinematic modelling all suggest that the disc may be dynamically complex and still in the process of assembly.

\subsection{REBELS-25 in context}

\subsubsection{Multi-phase gas kinematics}
\label{sec:compilation}

\begin{figure*}
    \centering
    \includegraphics[width=\textwidth]{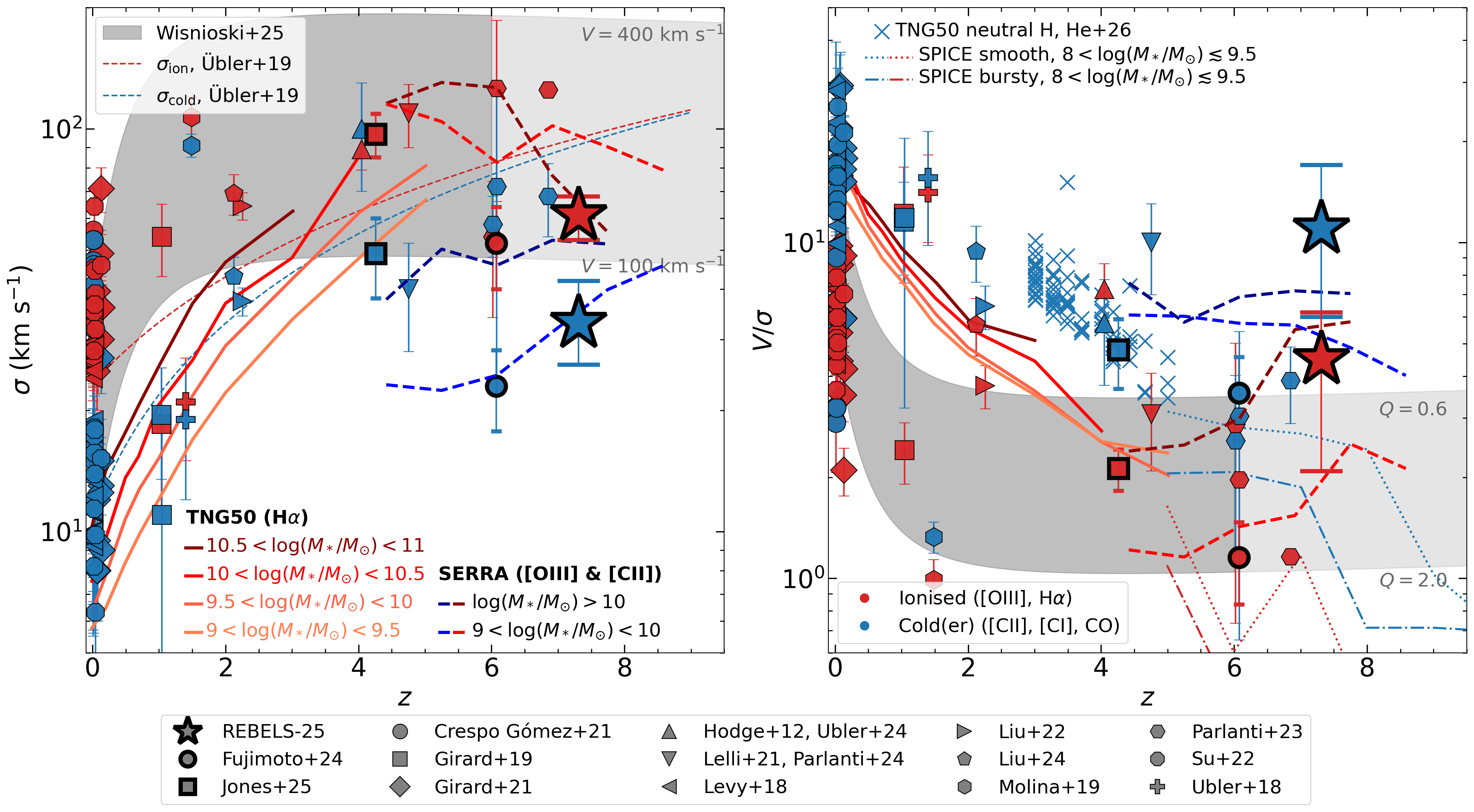} 
    \caption{\textit{Left:} Velocity dispersion as a function of redshift for warm ionised gas (red) and colder gas (blue) tracers, compiled from the literature where individual galaxies have been measured in both tracers. The grey shaded region represents the model fit from \citet{wisnioski_evolution_2025} for a stable disc with $\log(M_*/\mathrm{M_{\odot}})=9.3$ (comparable to latest estimate for REBELS-25; \citealt{rowland_rebels-ifu_2026}) and maximum rotational velocities from 100 to 400 km s$^{-1}$ , which is extrapolated beyond $z\sim6$. The thin dashed lines are the best-fit relations from \citet{ubler_evolution_2019} to ionised gas (red) and atomic $+$ molecular gas samples (blue) at $0<z<4$. We also plot the median trends from TNG50 at $0\lesssim z\lesssim5$ for H$\upalpha$ kinematics \citep{pillepich_first_2019}, with different shades of red for different mass bins (as indicated by the legend), and we plot median trends from the SERRA simulations \citep{pallottini_survey_2022,kohandel_dynamically_2024} at $4\lesssim z \lesssim 9$ for [O \textsc{iii}]88$\upmu$m (red dashed lines) and [C \textsc{ii}] (blue dashed lines). For the literature compilation, each marker corresponds to a different study, and we refer to the corresponding papers for discussion on the different samples and methods adopted. The bold markers are studies where \texttt{3DBAROLO} was also used for the kinematic modelling, as in the fiducial fits of REBELS-25. \textit{Right:} Ratio of ordered-to-random motion (as measured by the ratio of the maximum rotational velocity and the average velocity dispersion, $V/\sigma$) for this sample of galaxies with multi-phase kinematic measurements. The grey shaded region from \citet{wisnioski_evolution_2025} now shows the predicted trend for a $\log(M_*/\mathrm{M_{\odot}})=9.3$  disc with Toomre parameter $Q=0.6$ to $2.0$ (see \citealt{wisnioski_kmos3d_2015} and \citealt{wisnioski_evolution_2025} for details), which is again extrapolated beyond $z\sim6$.  We show the same trends from TNG50 and SERRA as in the left panel, but also show measurements from individual haloes in TNG50 from mock neutral hydrogen cubes with blue cross-shaped markers \citep{he_dynamically_2026}, and in addition median trends from the SPICE simulations in [C \textsc{ii}] emission \citep{bhagwat_stellar_2026} and [O \textsc{iii}]88$\upmu$m emission \citep{casavecchia_new_2026}, from their `smooth' and `bursty' feedback implementations (see \citealt{bhagwat_spice_2024} for details). We note that the ionised gas velocity dispersion values in both panels of this figure are not corrected for thermal broadening, to be more consistent with other comparisons in the literature. Overall, REBELS-25 shows relatively low $\sigma$ and high $V/\sigma$ in both [C \textsc{ii}] and [O \textsc{iii}]88$\upmu$m, characteristic of a dynamically cold disc already in place at $z=7.31$. Although the theoretical predictions span a broad range, REBELS-25 is broadly consistent with some simulations in which massive galaxies can exhibit high $V/\sigma$ ratios at early times.}
    \label{fig:sigma_redshift}
\end{figure*}

\begin{figure*}
    \centering
    \includegraphics[width=\textwidth]{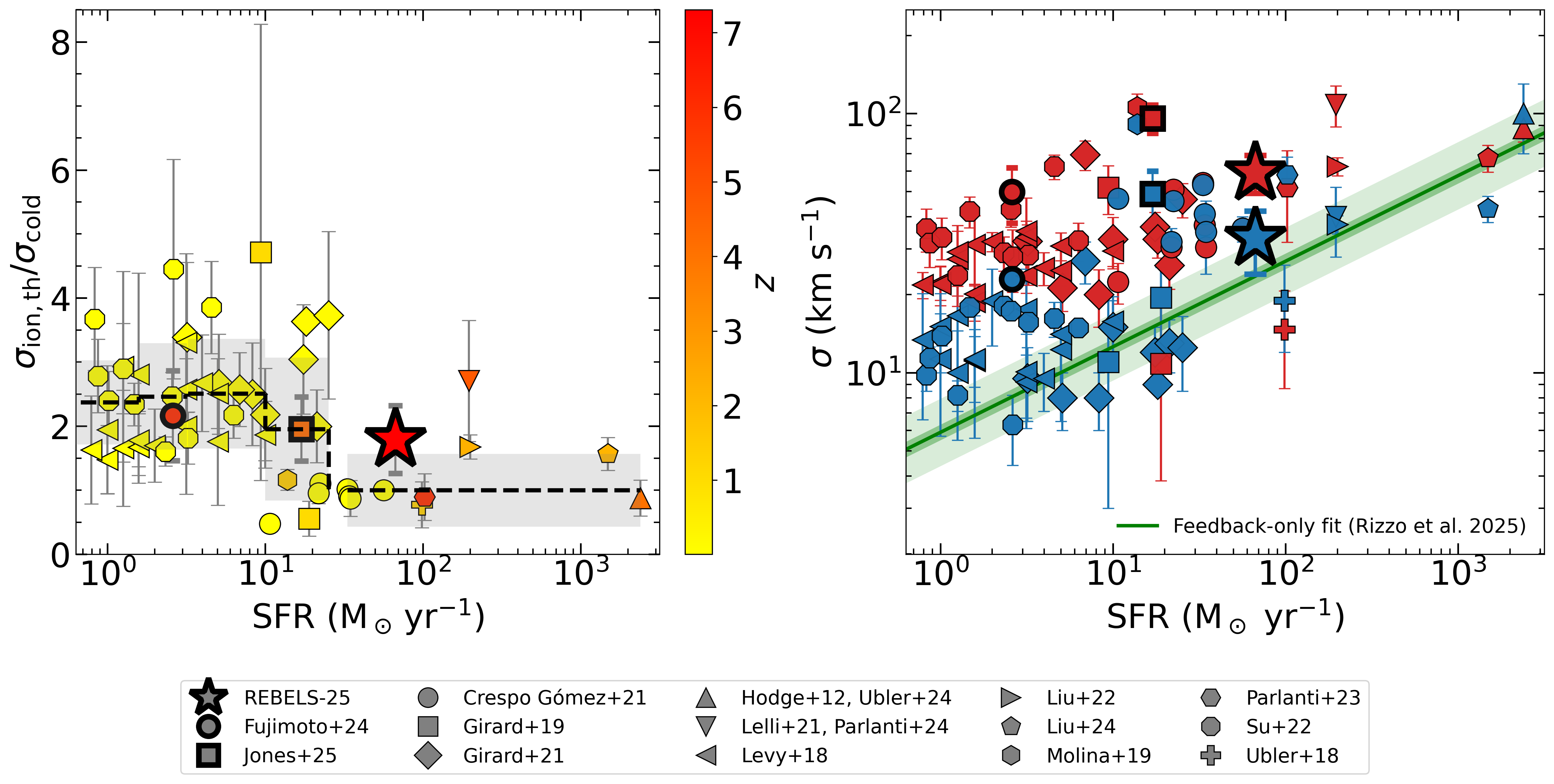} 
    \caption{\textit{Left:} Ratio of warm ionised to cold(er) gas velocity dispersion ($\sigma_{\mathrm{ion, th}} / \sigma_{\mathrm{cold}}$) as a function of total SFR for the sample compiled from the literature, where the ionised gas velocity dispersion is corrected for thermal broadening, assuming $\sigma_{\mathrm{thermal}}=15$ km s$^{-1}$. The marker shapes are the same as in Figure \ref{fig:sigma_redshift} (also with bold markers highlighting other studies that used \texttt{3DBAROLO}), but are coloured according to their redshift. We plot the running median of $\sigma_{\mathrm{ion, th}} / \sigma_{\mathrm{cold}}$ and its uncertainty in five equal size bins by the black dashed line and shaded region.
    \textit{Right:} We plot both $\sigma_{\mathrm{ion, th}}$ (red) and $\sigma_{\mathrm{cold}}$ (blue) as a function of SFR, and also plot the best-fit relation and its intrinsic scatter from \citet{rizzo_alma-alpaka_2024} (green line and shaded area). This relation is a linear fit to $\log (\sigma)$ as a function of $\log$ (SFR) for a sample of 57 discs at $z=0-5$ with velocity dispersions from colder gas tracers (CO, [C \textsc{i}], or [C \textsc{ii}]), including the \citet{girard_systematic_2021} sample, where the slope of the relation is fixed to 0.33, as expected if the turbulence in the gas is driven by supernovae feedback \citep{fraternali_fast_2021}. The compiled sample shows $\sigma_{\mathrm{ion, th}} / \sigma_{\mathrm{cold}} > 1$ in the majority of galaxies, indicating that the warm ionised gas is generally more turbulent than the colder gas phase even after correcting for thermal broadening. There is also tentative evidence that this ratio may be SFR-dependent, and/or that the SFR--$\sigma$ relation may differ between gas tracers.}
    \label{fig:sigma_ratio}
\end{figure*}

Observational and theoretical work at $z\sim0$--3 has shown that velocity dispersions tend to increase (and $V/\sigma$ correspondingly decrease) with redshift (e.g. \citealt{wisnioski_kmos3d_2015,ubler_evolution_2019, pillepich_first_2019}), with extrapolations to $z>4$ predicting that dynamically cold discs should be rare. Nonetheless, a growing number of $z>4$ galaxies have been found to be dynamically cold, most commonly from [C \textsc{ii}], [C \textsc{i}], and CO observations with ALMA (e.g. \citealt{neeleman_cold_2020,roman-oliveira_regular_2023, rizzo_alma-alpaka_2023}). One possible explanation is that these FIR emission lines probe the cooler neutral or molecular gas, which may exhibit systematically lower $\sigma$ than warm ionised gas tracers (\citealt{ levy_edge-califa_2018,ubler_evolution_2019}, but see also examples of dynamically cold high-$z$ discs from [O \textsc{iii}]$\uplambda$5007 or H$\upalpha$, e.g. \citealt{nelson_ionized_2024, de_graaff_ionised_2024, jones_ga-nifs_2025}). Alternatively, this may be partly due to sample bias, since the high-resolution spectroscopic samples currently available at $z>4$, especially with ALMA, are small and often biased towards bright, massive, or otherwise unusual galaxies. Kinematic measurements of both warm ionised and cooler neutral or molecular gas in the same galaxies are therefore needed to separate these effects. We note that, hereafter, we use $\sigma_{\mathrm{ion}}$ to denote the velocity dispersion measured from warm ionised gas tracers, including H$\upalpha$, [O \textsc{iii}]88$\upmu$m, and [O \textsc{iii}]$\lambda5007$, and $\sigma_{\mathrm{cold}}$ for measurements from cooler neutral and molecular gas tracers, including [C \textsc{ii}], [C \textsc{i}], and CO. One of the primary motivations for analysing the [O \textsc{iii}]88$\upmu$m kinematics in REBELS-25 was therefore to directly compare the warm ionised and colder gas kinematics, and determine whether the galaxy appears substantially more dispersion-dominated when observed in warm ionised gas.

Instead, we find that REBELS-25 remains strongly rotationally supported even in the ionised gas phase. Using our fiducial \texttt{3DBAROLO} modelling, the [O \textsc{iii}]88$\upmu$m emission yields a high $V/\sigma$ ratio of 4.5$^{+2.4}_{-1.8}$, consistent with a dynamically cold disc. Although the [O \textsc{iii}]88$\upmu$m velocity dispersion is moderately higher than [C \textsc{ii}], the source remains more rotationally supported with a lower velocity dispersion than expected from extrapolations of some earlier observational and theoretical predictions for this redshift (\citealt{ubler_evolution_2019} and TNG50 from \citealt{pillepich_first_2019}), as shown in Figure \ref{fig:sigma_redshift}. The same conclusion is reached when adopting the alternative \texttt{DysmalPy} modelling (Figure \ref{fig:dysmalpy results}).

The kinematics are broadly consistent with more recent theoretical work and simulations, some of which now predict the existence of dynamically cold massive discs already in place at $z>6$, and even up to $z\sim14$ (e.g. \citealt{kohandel_dynamically_2024, kohandel_amaryllis_2025}). In particular, we see in Figure \ref{fig:sigma_redshift} that the velocity dispersions agree well with the median values from the SERRA simulations (\citealt{pallottini_survey_2022,kohandel_dynamically_2024}) at $z\sim7$, both in terms of the absolute values and the relative difference between the neutral and ionised phases. These predictions were calculated from face-on synthetic [C \textsc{ii}] and [O \textsc{iii}]88$\upmu$m datacubes within a 2.5 kpc field of view, with $\sigma$ defined as the median of the moment-2 map and $V_{\rm rot}=\sqrt{G(M_\star+M_{\rm gas})/r_d}$, where $r_d$ is the gas half-mass radius.

In \cite{kohandel_amaryllis_2025}, an analog to the galaxy JADES-GS-z-14-0 ($z=14.2$, \citealt{carniani_2024,carniani_eventful_2025,schouws_deep_2025, schouws_detection_2025, scholtz_tentative_2025}) in the SERRA simulations is analysed in detail. Interestingly, by $z\sim7$, this synthetic galaxy, called Amaryllis, also has comparable properties to REBELS-25, with a dynamically cold gas disc in both [C \textsc{ii}] and [O \textsc{iii}]88$\upmu$m, a near-solar metallicity, and an [O \textsc{iii}]/[C \textsc{ii}] ratio of $\sim3$. We note its stellar mass at this redshift, $\log(M_*/\mathrm{M_{\odot}})\sim10$, is higher than the current estimate of $\log(M_*/\mathrm{M_{\odot}})\sim9.3$ for REBELS-25, but the integrated SED fit for REBELS-25 could potentially underestimate its total stellar mass due to the significant amount of dust obscuration on resolved scales (e.g., a `dust-obscuration bias', as discussed in \citealt{Li_2026}). At face value, these similarities with Amaryllis could imply that REBELS-25 is a potential descendant of the UV-luminous, `Blue Monster' galaxies observed with \textit{JWST} at $z>10$, like JADES-GS-z14-0. In addition, \cite{kohandel_amaryllis_2025} find that Amaryllis already hosts a dynamically cold gas disc by $z\sim11$, with $V/\sigma \sim 4$ and 6 in [O \textsc{iii}]88$\upmu$m and [C \textsc{ii}], respectively, and it remains disc-like up to $z\sim7$, despite undergoing mergers, bursty star formation, and strong outflows. This suggests that ordered bulk gas rotation can coexist with substantial ongoing assembly, and strong feedback effects and interactions may not necessarily disrupt the disc kinematics. Similarly, earlier numerical studies of gas-rich mergers showed that ordered discs can survive or rapidly re-form even after major merger events (\citealt{springel_2005, robertson_2008}), and some observational studies of high-redshift galaxies also find ordered rotation alongside galaxy interactions (e.g., \citealt{Westoby_2026}). This may be a comparable scenario to REBELS-25, which appears dynamically cold but may also have more complex kinematic signatures.

We also make comparisons with a recent study of massive discs in the TNG50 simulations, \cite{he_dynamically_2026} (plotted in the right panel of Figure \ref{fig:sigma_redshift}). In this study, massive ($\log (M_*/\mathrm{M_{\odot}})\gtrsim 10$) galaxies at $3<z<5$ selected from TNG50 are analysed using mock neutral hydrogen emission cubes, in order to more closely mimic ALMA [C \textsc{ii}] observations. \cite{he_dynamically_2026} find $V/\sigma$ values from these mock neutral gas observations that are broadly consistent with recent ALMA $z>4$ [C \textsc{ii}] measurements, and higher than the median values inferred from the TNG50 H$\upalpha$ kinematics (\citealt{pillepich_first_2019}), although still lower than the value derived for REBELS-25's [C \textsc{ii}] kinematics. In this study, the most extreme dynamically cold discs in TNG50 ($V/\sigma\gtrsim10$) are found to be relatively short-lived ($\lesssim200$ Myr, see also \citealt{kretschmer_origin_2022}), whilst the majority of moderately cold discs ($V/\sigma\sim5-10$) persist for longer timescales of $\sim600$ Myr. The dynamically cold discs in these simulations are also found to accrete less gas compared to other galaxies, reducing the impact of misaligned gas clumps that could disrupt the rotation.

The recent SPICE simulations (\citealt{bhagwat_spice_2024, casavecchia_new_2026, bhagwat_stellar_2026}), which also enable a direct comparison of [C \textsc{ii}] and [O \textsc{iii}]88$\upmu$m kinematics, similar to SERRA, also predict some dynamically cold [C \textsc{ii}] discs at $z\sim7$. However, their $V/\sigma$ values remain below that measured for REBELS-25 (Figure \ref{fig:sigma_redshift}). This is likely in part due to the lower stellar masses probed, with most galaxies at $z\sim7$ having $M_*<10^9\mathrm{M_{\odot}}$ (see discussions of the mass dependence of $V/\sigma$ in e.g. \citealt{rowland_rebels-25_2024}).

These simulation-based studies therefore suggest that at least some massive galaxies at high redshift may already host dynamically cold discs, potentially in both ionised and colder gas phases, even with ongoing galaxy assembly processes. However, it remains unclear whether the highest observed $V/\sigma$ values correspond to relatively transient phases of the most massive systems, and to what extent such dynamically cold discs can survive mergers, gas accretion events, and strong feedback at early times.

In addition, whilst these numerical studies suggest that dynamically cold discs may exist in both ionised and colder gas phases at high-$z$, they also predict systematic kinematic differences between the phases, with the ionised gas generally exhibiting higher velocity dispersions and lower $V/\sigma$ values. Observationally, REBELS-25 is one of only a small number of sources beyond the local Universe with resolved kinematic constraints in both a warm ionised gas phase and a cooler gas phase, allowing us to test this prediction. In Figures \ref{fig:sigma_redshift} and \ref{fig:sigma_ratio}, we compile a comparison sample from the literature with simultaneous multi-phase kinematic measurements, considering [C \textsc{ii}], [C \textsc{i}], and CO as tracers of the colder neutral or molecular gas, and H$\upalpha$, [O \textsc{iii}]$\lambda5007$, and [O \textsc{iii}]88$\upmu$m as tracers of the warm ionised gas. We find, to the best of our knowledge, an additional seven sources at $z>3$ where $\sigma$ is constrained in both phases (from \citealt{hodge_evidence_2012, girard_towards_2019, lelli_massive_2021, parlanti_alma_2023, ubler_ga-nifs_2024, parlanti_ga-nifs_2024,fujimoto_primordial_2025, jones_ga-nifs_2025}). To expand the sample, we therefore also compile sources at $z<3$ (\citealt{ubler_ionized_2018,levy_edge-califa_2018,molina_kiloparsec-scale_2019,crespo_gomez_stellar_2021,girard_systematic_2021,su_almaquest_2022,liu_600_2023,Liu_2024}), but note that this may not be a complete sample of all available measurements. 

This literature sample is inherently heterogeneous and spans a wide range of selection methods, emission line tracers, S/N, spatial resolutions, and kinematic fitting methodologies. As we demonstrated in Section \ref{sec:kinematics} (more details in Appendix \ref{appendix:DysmalPy}), the inferred kinematic parameters, in particular the velocity dispersion, can vary significantly depending on the adopted modelling technique, masking procedure, and assumptions. Individual comparisons should therefore be treated with caution, however we note that for the majority of this comparison sample the same kinematic fitting techniques are applied to both gas tracers (cf. GN20 analysed separately in \citealt{hodge_evidence_2012} and \citealt{ubler_ga-nifs_2024}, and ALESS073.1 analysed separately in \citealt{lelli_massive_2021} and \citealt{parlanti_ga-nifs_2024}). It is also worth noting that, in this sample, only REBELS-25 and the three sources from \cite{parlanti_alma_2023} have both gas phases observed with the same instrument, ALMA. In these cases, both tracers are FIR emission lines that are largely unaffected by dust attenuation, and therefore provide a cleaner comparison between the cooler and ionised gas kinematics.

Despite the ranges of sample selection and methodology adopted, almost all of the ionised gas velocity dispersions in this literature sample are higher than those measured in cooler gas phases, by an average factor of $\sim2$. This remains the case after applying a uniform correction of $\sigma_{\mathrm{thermal}}=15$ km s$^{-1}$, commonly adopted for H$\upalpha$ and intended to account for both thermal broadening and H \textsc{ii}-region expansion. Hereafter, corrected dispersions are denoted $\sigma_{\mathrm{ion,th}}$, where ${\sigma_{\mathrm{ion,th}}}^2={\sigma_{\mathrm{ion}}}^2-{\sigma_{\mathrm{thermal}}}^2$. For [O \textsc{iii}], the thermal contribution is smaller, at only $\sim2$--3 km s$^{-1}$, although broadening from bulk motions may still contribute. The uniform correction may therefore slightly over-correct the [O \textsc{iii}]-based measurements relative to H$\upalpha$, but its effect is small for the predominantly high ionised gas velocity dispersions in this sample.

From significantly larger samples of galaxies, where the ionised gas measurements and colder gas measurements are from entirely separate samples, a comparable average ratio of $\sigma_{\mathrm{ion}}/\sigma_{\mathrm{cold}}\sim2$ has also been reported in e.g. \cite{ubler_evolution_2019, levy_edge-califa_2018, wisnioski_evolution_2025, krumholz_unified_2018, rizzo_alma-alpaka_2024}, as well as in the simulations discussed above. Physically, this could suggest that ionised gas tracers, such as [O \textsc{iii}] and H$\upalpha$, are more sensitive to feedback processes, including outflows, and also extraplanar diffuse ionised gas. Indeed, \cite{ejdetjarn_giant_2022} showed in simulations that the inclusion of stellar feedback has a much stronger impact on the velocity dispersion measured in the ionised gas phase than in the neutral or molecular gas phases.

One might therefore expect the offset between ionised and neutral gas dispersions to become larger at higher SFR, where feedback effects are likely stronger. We instead observe tentative evidence for the opposite behaviour (Figure \ref{fig:sigma_ratio}). When considering the running median of $\sigma_{\mathrm{ion, th}}/\sigma_{\mathrm{cold}}$ as a function of SFR across the sample, the ratio appears to decrease from $\sim 2.5$ at lower SFRs ($\lesssim10 ~\mathrm{M_{\odot} yr ^{-1}}$) to values closer to unity at the highest SFRs. We caution that the heterogeneity of the current sample and possible systematics in the observations/methodologies may impact this apparent trend and therefore any physical interpretation. With these caveats in mind, it is nevertheless useful to consider whether the observed behaviour could be driven by the ionised and cooler gas phases following different $\sigma$--SFR trends. We therefore show the individual $\sigma_{\mathrm{ion, th}}$ and $\sigma_{\mathrm{cold}}$ measurements as a function of total SFR in the right-hand panel of Figure \ref{fig:sigma_ratio}. As a reference, we compare these measurements to the semi-analytical relation from \cite{rizzo_alma-alpaka_2024}. This relation assumes that stellar feedback is the primary driver of the gas turbulence, which fixes the slope (\citealt{fraternali_fast_2021}), whilst the normalisation is fitted to a sample of 57 discs observed in CO, [C \textsc{i}] or [C \textsc{ii}] with ALMA at high resolution from $z\sim0$ to 5.

Here, we see that our compilation of CO, [C \textsc{i}] and [C \textsc{ii}] measurements agree relatively well with this trend, including the value measured for REBELS-25 (see also discussion in \citealt{rowland_rebels-25_2024}). However, some high dispersions measured in these cold-gas tracers still lie above the feedback-only model, suggesting that other processes, such as gravitational transport and/or gas accretion (\citealt{krumholz_unified_2018,kohandel_2020,ginzburg_2022}), may also contribute in some systems (e.g. \citealt{parlanti_alma_2023}). In contrast, the majority of the ionised gas velocity dispersions are all systematically offset above the relation, especially at SFR$\lesssim10 ~\mathrm{M_{\odot}}$ yr$^{-1}$, where the median $\sigma_{\mathrm{ion, th}}/\sigma_{\mathrm{cold}}\sim2.5$. At higher SFRs, there are more $\sigma_{\mathrm{ion, th}}$ values in agreement with the relation.

A possible interpretation could be that at low-to-moderate SFRs, feedback may be more strongly coupled to the ionised gas phase, preferentially increasing the velocity dispersions of emission line tracers such as H$\upalpha$ and [O \textsc{iii}] more than lines tracing colder neutral or molecular gas (\citealt{ejdetjarn_giant_2022}). At sufficiently high SFR ($>10^2 ~\mathrm{M_{\odot}}$ yr$^{-1}$), where we tentatively see convergence between $\sigma_{\mathrm{ion, th}}$ and $\sigma_{\mathrm{cold}}$, the ISM may become sufficiently turbulent that the kinematic differences between the phases are reduced, leading to comparable dispersions across the ionised and colder gas.

Alternatively, as discussed in detail in \cite{rizzo_alma-alpaka_2024} (see also \citealt{phillips_lessons_2025}), the majority of ionised gas kinematic studies across redshifts may be subject to more systematic biases than colder gas kinematic measurements. Typically, optical instruments such as PPAK IFU, SAMI, and KMOS are used to trace warm ionised gas kinematics from rest-frame optical emission lines, such as H$\upalpha$ and [O \textsc{iii}]$\lambda$5007. Colder gas kinematics (e.g. from H \textsc{i}, CO, [C \textsc{i}], [C \textsc{ii}]) are then typically traced using interferometers such as CARMA, PdBI and ALMA, with these typically achieving much higher spectral resolution. For example, in our comparison sample, 16/57 of the sources are from \cite{levy_edge-califa_2018}, where the spectral resolution of the CO observations is $\sim3-14$ km s$^{-1}$ (smoothed to 20 km s$^{-1}$ channels) and the resolution of the H$\upalpha$ observations is 160 km s$^{-1}$. For most of the sources at $z>3$, \textit{JWST}/NIRSpec IFU is used to trace the ionised kinematics, and ALMA is used for the colder gas kinematics. The native velocity resolution for these ALMA data are of order $R\sim$ 50,000-60,000, whereas even the highest resolution grating of NIRSpec has $R\sim2700$. The typically poorer spectral resolution for ionised gas kinematics could therefore bias the inferred velocity dispersion values high. From these rest-frame optical studies, further uncertainties could also be introduced by substantial dust obscuration at these wavelengths. In this sense, the observations of REBELS-25 are unique compared to the majority of similar studies: both the warm ionised gas and colder gas were traced using FIR emission lines observed with ALMA, enabling a more robust kinematic comparison at similar spectral resolutions and without the effects of dust attenuation. The fact that $\sigma_{\mathrm{ion, th}}/\sigma_{\mathrm{cold}}=1.9^{+0.8}_{-0.6}$ for REBELS-25 from the \texttt{3DBAROLO} fitting could still imply that there may be intrinsic differences in the kinematics and therefore the drivers of turbulence in different gas phases, although we caution that with the \texttt{DYSMALPY} fits in Appendix \ref{appendix:DysmalPy}, $\sigma_{\mathrm{ion, th}}/\sigma_{\mathrm{cold}}\sim0.8$. Larger samples of galaxies observed at matched spatial and spectral resolution across multiple gas tracers, spanning a wide range of galaxy properties and redshifts, would be essential to determine if these offsets are universal, and to establish their physical origin.

\subsubsection{ISM properties at high-$z$}

\begin{figure*}
\centering
\includegraphics[width=\textwidth]{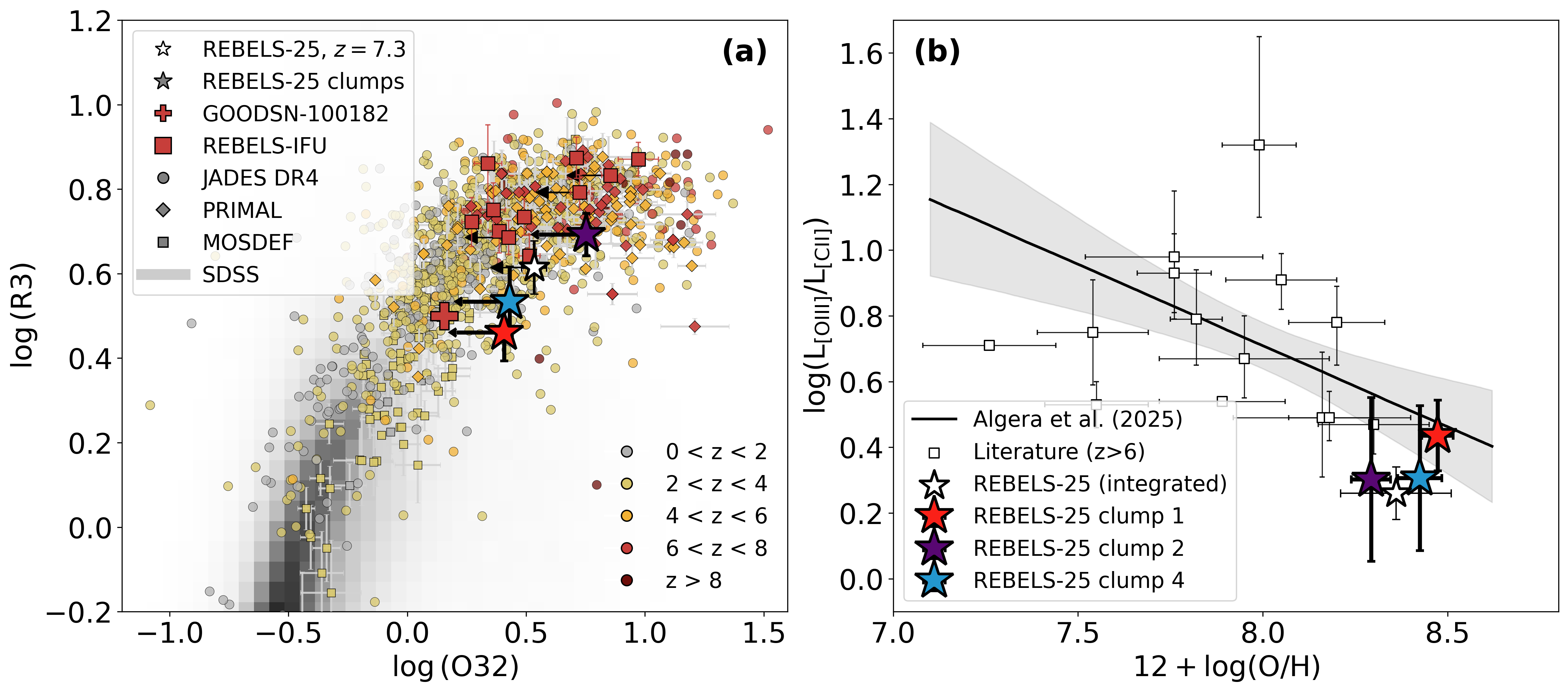}
\caption{\textit{Panel a:} We plot the emission line ratios R3 (a common tracer of metallicity) against O32 (primarily traces the ionisation parameter) of galaxies from a range of surveys at $0\lesssim z \lesssim 10$, in comparison to the integrated ratios for REBELS-25 (white star-shaped marker) and clumps 1, 2 and 4, where the necessary emission lines are detected at S/N$>3$. Here we see that, in particular for clumps 1 and 4, the R3 and O32 ratios are lower than the typical $z>6$ population (but comparable to the $z=6.73$ source, GOODSN-100182, in \citealt{shapley_aurora_2025}), implying a high metallicity and low ionisation parameter across the source. We note that the O32 ratio for REBELS-25 (and some other REBELS-IFU sources, square-shaped markers) are upper limits, since at $z>7$ the H$\upalpha$ emission line moves beyond the nominal wavelength range of NIRSpec, and so we do not apply a reddening correction to this ratio, as discussed in \citet{rowland_rebels-ifu_2026}. For the REBELS-25 values, the size of the arrows in the x-direction indicates the $\sim-0.2$ dex shift in $\log$(O32) if an $A_{\mathrm{V, neb}}=2\times A_{\mathrm{V, *}}\simeq1.46$ mag is assumed (see text).   \textit{Panel b:} [O \textsc{iii}]/[C \textsc{ii}] ratio as a function of metallicity for $z>6$ sources observed with ALMA (see compilation in \citealt{algera_rebels-ifu_2025}). Where possible, we recalculate the metallicities based on the R3 calibration from \citet{sanders_aurora_2025}. The best-fit relation from \citet{algera_rebels-ifu_2025} is plotted with a black line, with a grey shaded region to depict the $1\sigma$ uncertainties. The integrated and clump-based [O \textsc{iii}]/[C \textsc{ii}] values for REBELS-25 are amongst the lowest in the high-$z$ literature, implying more evolved ISM conditions.}
\label{fig:clump ratios}
\end{figure*}

Thanks to \textit{JWST}, an increasingly large number of spectroscopic surveys have revealed the ISM properties of $z>4$ galaxy populations for the first time (e.g. \citealt{arellano-cordova_first_2022,schaerer_first_2022,curti_chemical_2023, nakajima_jwst_2023,trump_physical_2023,brinchmann_high-z_2023,sun_first_2023}). For the majority of these high-$z$ systems, these spectra have revealed extreme emission line ratios when compared to local galaxies, indicative of metal-poor environments, recent bursty star formation, and high ionisation parameters. The integrated and resolved analysis of REBELS-25 presented here and in previous works (\citealt{ algera_rebels-ifu_2025,rowland_rebels-ifu_2026}) therefore provides a surprising contrast, instead pointing towards a system that is more chemically evolved, with emission line ratios and properties more comparable to $z<3$ systems. This is illustrated in Figure \ref{fig:clump ratios}a, where we compare the integrated and clump-based R3 and O32 ratios of REBELS-25 to galaxies from spectroscopic surveys spanning $0\lesssim z \lesssim10$, including JADES data release 4 (\citealt{bunker_jades_2024,deugenio_jades_2025,curtis-lake_jades_2025,scholtz_jades_2025,eisenstein_overview_2026}), PRIMAL (\citealt{Heintz_2025}), MOSDEF (\citealt{kriek_mosfire_2015,shapley_mosdef_2019}), and the Sloan Digital Sky Survey (SDSS) Data Release 8 (DR8; \citealt{aihara_2011}).

From the integrated analyses presented in \cite{rowland_rebels-ifu_2026}, the optical emission line ratios of REBELS-25 indicate that it already hosts an enriched ISM, with a metallicity of $\gtrsim50\% ~\mathrm{Z_{\odot}}$ based on the recent strong-line calibrations of \cite{sanders_aurora_2025}. The resolved, clump-based analysis presented here suggests that it is similarly enriched across the disc, even in clumps $\sim 0.4$ arcsec ($\sim 2$ kpc) from the centre. The R3 and O32 ratios observed are lower than typical high-$z$ systems, for example, \cite{isobe_jades_2026} find R3$=5.7\pm0.08$ and O32$=6.31\pm0.19$ from a stacked spectrum of 564 sources at redshifts $z=4-7$. The observed R3 ratios for REBELS-25, R3$\simeq$2.9-4.9, are in fact more consistent with the median value of the AURORA survey at $z=1.4-2.7$ (R3$\sim$4.4, \citealt{shapley_aurora_2025-1}) and the median value of the MOSDEF survey at $z\sim2.3$ (R3$\sim$2.9, \citealt{shapley_mosdef_2019}). This finding is also comparable to the study of GOODSN-100182, a $z=6.73$ galaxy identified from the AURORA survey as having surprisingly mature ISM properties in \cite{shapley_aurora_2025}.

Similarly to the emission line ratios in the rest-frame optical, ratios of FIR fine-structure lines can also offer insight into a variety of ISM conditions. In recent years, there have been multiple ALMA observations targeting both [O \textsc{iii}]88$\upmu$m and [C \textsc{ii}] in high-$z$ galaxies (e.g. see recent review of \citealt{smit_alma_2026}). Several studies have since investigated the global [O \textsc{iii}]/[C \textsc{ii}] ratio as a probe of ISM conditions, both from observational (e.g. \citealt{harikane_large_2020,witstok_dual_2022,algera_cold_2024,algera_rebels-ifu_2025,hagimoto_compact_2025}) and theoretical (e.g. \citealt{arata_2020,katz_nature_2022,pallottini_survey_2022, nyhagen_theoretical_2025,kohandel_amaryllis_2025, nakazato_unveiling_2026}) perspectives. At $z>6$, this ratio has been found to be as high as $\gtrsim2$--$20$ (e.g., \citealt{inoue_detection_2016, laporte_absence_2019, carniani_missing_2020, harikane_large_2020, witstok_dual_2022, ren_updated_2023, algera_rebels-ifu_2025, schouws_deep_2025}; cf. \citealt{hashimoto_detections_2019}), whereas typical star-forming galaxies in the local Universe tend to show ratios of $0.04$--1 (e.g., \citealt{de_looze_applicability_2014,diaz-santos_herschelpacs_2017}), although similarly high values can be found in some local low-metallicity dwarf galaxies, reaching $\gtrsim1$ and in some cases up to $\sim10$ (e.g., \citealt{cormier2015,kumari_study_2024}). The origin of these elevated ratios is still under debate, but could result from several factors, including higher ionisation parameters, lower metallicities, enhanced oxygen-to-carbon ratios, and lower neutral gas covering fractions, amongst others.

Interestingly, REBELS-25 exhibits amongst the lowest global [O \textsc{iii}] / [C \textsc{ii}] ratios observed so far at $z>6$ (\citealt{algera_rebels-ifu_2025}; although also see e.g. \citealt{walter2018,marrone2018}), with an integrated ratio of $1.9^{+0.4}_{-0.3}$. This is likely attributed to its high metallicity ($12+\log(\mathrm{O/H})=8.36\pm0.15$) and relatively low [O \textsc{iii}]$\lambda$5007 equivalent width (EW$_{[\mathrm{O \textsc{iii}}]\lambda5007}=511\pm56$ \AA), suggestive of it being a more evolved and less bursty source (\citealt{algera_cold_2024, algera_rebels-ifu_2025}). From the resolved analysis in Section \ref{sec: resolved IFU}, the clump-based ratios are slightly higher ($\sim2.0-2.8$), but still amongst the lowest [O \textsc{iii}]/[C \textsc{ii}] ratios from the literature (Figure \ref{fig:clump ratios}b; see compilation in \citealt{algera_rebels-ifu_2025,nakazato_unveiling_2026}). The integrated ratio is likely lower because, as shown in Section \ref{sec:morphology}, the [C \textsc{ii}] emission is $\sim1.4\times$ more extended than [O \textsc{iii}]88$\upmu$m emission, and extends beyond the region covered by these clump apertures.

Taken together, the optical and FIR emission line ratios analysed in this work suggest that REBELS-25 has less extreme ionising conditions than many galaxies in current reionisation-era spectroscopic samples. The resolved analysis further suggests that these less extreme and more evolved ISM conditions extend across the galaxy. However, we note that, in particular for the optical line ratios, we may be biased towards less obscured outer regions of the galaxy, rather than the obscured centre where the FIR maps indicate that much of the star formation is concentrated. In addition, we see tentative evidence that the centre of the galaxy is more metal-enriched, however the clump-based metallicities are consistent within the uncertainties.

\section{Summary \& Conclusions}
\label{sec:conclusions}

In this work, we have presented a multi-wavelength, $\lesssim1$ kpc view of the massive, $z=7.31$ star-forming galaxy REBELS-25, combining new high spatial-resolution [O \textsc{iii}]88$\upmu$m ALMA observations with resolution-matched [C \textsc{ii}] and \textit{JWST}/NIRSpec IFU data. Together, these data provide a detailed view of the morphology, ISM conditions, and multi-phase gas kinematics of a surprisingly evolved galaxy $\sim700$ Myr after the Big Bang.

With this wealth of data, we first focused our analysis on further investigating the nature of REBELS-25, which our previous study of the [C \textsc{ii}] observations identified as potentially the most distant dynamically cold disc galaxy observed to-date (\citealt{rowland_rebels-25_2024}). As discussed throughout this work, such a mature system at $z=7.31$ is surprising given expectations for more dispersion-dominated galaxy kinematics at this epoch. At the same time, the offset between the \textit{HST} rest-frame UV and FIR emission already suggested a more complex picture, where patchy dust obscuration and/or clumpy substructure may play an important role in shaping the observed morphology. The new \textit{JWST}/NIRSpec IFU and ALMA Band 8 observations allow us to investigate this in greater detail. We therefore carried out a detailed, spatially resolved investigation of the UV-to-FIR morphology, ISM properties, and multi-phase gas kinematics of REBELS-25, the main results of which are:

\begin{itemize}
    \item REBELS-25 exhibits dramatically different morphologies from the UV to FIR. The \textit{JWST} rest-frame UV and optical nebular emission is highly clumpy and irregular, with multiple bright clumps offset by $\sim2$ kpc from the peak of each of the FIR emission maps ([C \textsc{ii}], [O \textsc{iii}]88$\upmu$m, 90$\upmu$m continuum and 150$\upmu$m continuum). By contrast, these FIR emission maps are well-described by a single near-exponential disc profile ($n\sim1$--$1.8$).
    
    \item Of these FIR emission profiles, the [C \textsc{ii}] emission of REBELS-25 is the most extended ($r_e\sim2.2$ kpc), whilst the 90$\upmu$m continuum is the most compact ($r_e\sim1.1$ kpc), as expected if [C \textsc{ii}] traces primarily a more extended neutral gas component, whereas the dust continuum and [O \textsc{iii}] emission primarily trace star forming regions. 
    
    \item Dust obscuration appears to play a major role in shaping the observed UV and optical morphology of REBELS-25. By comparing resolved UV and FIR emission across four clump apertures, we find that obscured star formation contributes approximately 55--98\% of the total SFR throughout the galaxy, even in the most UV-luminous clumps. The central FIR-bright region is the most heavily obscured, indicating that much of the star formation near the dynamical centre is hidden at rest-frame UV and optical wavelengths. 

    \item The resolved emission line ratios and ISM properties show very little variation at this resolution. From the different binning methods applied, key emission line ratios, such as R3 and [O \textsc{iii}]/[C \textsc{ii}], vary by less than a factor of $\sim2$ across the galaxy, with most regions having consistent ratios. This implies that the ISM conditions, such as the metallicity and ionisation parameter, are broadly consistent across the source. This argues against the UV-bright regions being separate systems with substantially different ISM conditions, but does not exclude physical star-forming structures embedded within a common, chemically enriched ($\gtrsim50\% ~\mathrm{Z_{\odot}}$) disc.

    \item From the kinematic analysis, we find that the warm ionised gas (traced by [O \textsc{iii}]88$\upmu$m) and colder neutral gas (traced by [C \textsc{ii}]) share the same large-scale rotating structure, with a comparable kinematic major axis and velocity profile. We find no compelling evidence for kinematically distinct components associated with the UV/optical clumps, although we cannot completely rule this out at the current resolution and S/N.

    \item REBELS-25 remains dynamically cold, even in the ionised gas. Kinematic modelling with \texttt{3DBAROLO} yields high rotational support in both phases, with $V_{\mathrm{rot, max}}/\sigma \simeq 4.5$ for [O \textsc{iii}]88$\upmu$m and $\simeq11$ for [C \textsc{ii}]. Although the ionised gas shows moderately higher velocity dispersions than the colder gas, both tracers are consistent with a rotationally-supported disc already in place at $z = 7.31$.

    \item We also see evidence for non-circular motions in both the [C \textsc{ii}] and [O \textsc{iii}]88$\upmu$m emission, which could indicate contributions from inflows, outflows, non-axisymmetric structure, and/or minor interactions.

\end{itemize}

Overall, the analysis presented here demonstrates that dust obscuration strongly shapes the observed UV-to-FIR morphology of REBELS-25. However, it remains likely that intrinsic clumpy substructure and/or non-axisymmetric features also contribute to the observed morphology and kinematics.  In particular, the tentative FIR substructures (some of which may be spatially coincident with clumps in the UV/optical emission) and also the evidence for non-circular motions indicate that REBELS-25 is not a perfectly smooth rotating disc.

We next placed the observed properties of REBELS-25 into the broader context of both observational and theoretical studies spanning low and high redshift. From these comparisons, we find that:

\begin{itemize}
    \item The high degree of rotational support in REBELS-25 in both warm and colder gas tracers is broadly consistent with some recent observational and theoretical works at $z\gtrsim4$, which find that at least some of the most massive galaxies at these epochs may already host dynamically cold discs in multiple tracers. However, these high $V/\sigma$ phases could be short-lived ($\sim200$ Myr; \citealt{he_dynamically_2026}) and therefore rare, or they may persist for longer timescales (e.g., \citealt{kohandel_amaryllis_2025}).

    \item Comparisons between warm ionised and cooler gas kinematics of the same galaxies up to $z\sim7$ support  findings at lower redshifts, where velocity dispersions from ionised gas tracers are systematically higher by a factor of $\sim2$. Given the heterogeneity of current datasets, it remains unclear if these differences reflect intrinsic changes in the kinematics probed,  or if this could be dominated by systematics in the measurements. 

    \item The optical and FIR emission line ratios of REBELS-25 indicate that it is more chemically evolved compared to the reionisation-era population from current large spectroscopic surveys. In particular, the R3, O32, and FIR [O \textsc{iii}]/[C \textsc{ii}] ratios imply a relatively high metallicity and low ionisation parameter.
\end{itemize}

From these comparisons, it is therefore evident that REBELS-25 is both dynamically and chemically evolved, and hence is a remarkably mature galaxy for its epoch. This may challenge earlier expectations of high-$z$ galaxies as compact, metal-poor systems dominated by turbulent motion from pristine gas accretion, major mergers and violent feedback effects. This analysis of REBELS-25 also presents a key case study for the importance of obtaining spatially resolved multi-wavelength observations. The dramatically different morphologies observed from the UV to FIR demonstrate that rest-frame UV and optical emission alone may provide an incomplete picture of early galaxy structure and properties.

However, it is important to note that REBELS-25 is a single galaxy, and perhaps an exceptionally rare and luminous one. Extending such detailed, resolved studies to more representative galaxy populations at these epochs, and to the even earlier systems now being uncovered by \textit{JWST} at $z>10$, will require substantial improvements in sensitivity to FIR line and continuum emission. Future facilities and upgrades, such as those envisioned by ALMA2040\footnote{\url{https://www.euroalma2040.com/}} (\citealt{Facchini_2025}), will be crucial for tracing the build-up of dust, metals, and ordered structure across the broader high-$z$ galaxy population.

\section*{Acknowledgements}

LER would like to thank their PhD thesis reading committee for their helpful comments and suggestions on this work. HSBA gratefully acknowledges support from Academia Sinica through grant AS-PD-1141-M01-2. JH acknowledges support from the ERC Consolidator Grant 101088676 (VOYAJ).  MA is supported by FONDECYT grant number 1252054, and gratefully acknowledges support from ANID Basal Project FB210003,  ANID MILENIO NCN2024\_112 and ANID + Vinculaci\'on Internacional + FOVI250261.

This work is based in part on observations made with the NASA/ESA/CSA \textit{\textit{JWST}}. The data were obtained from the Mikulski Archive for Space Telescopes at the Space Telescope Science Institute, which is operated by the Association of Universities for Research in Astronomy, Inc., under NASA contract NAS 5-03127 for \textit{\textit{JWST}}. These observations are associated with programme \#1626.

This paper also makes use of the following ALMA data:

\noindent ADS/JAO.ALMA\#2021.1.01603.S,
\noindent ADS/JAO.ALMA\#2022.1.01324.S .

ALMA is a partnership of ESO (representing its member states), NSF (USA) and NINS (Japan), together with NRC (Canada), MOST and ASIAA (Taiwan), and KASI (Republic of Korea), in cooperation with the Republic of Chile. The Joint ALMA Observatory is operated by ESO, AUI/NRAO and NAOJ.

\section*{Data Availability}

All data utilised in this work is available upon reasonable request to the corresponding authors.
 



\bibliographystyle{mnras}
\bibliography{o3_88_bib_v3_journal_abbrev_no_adsnotes} 

@article{smit_alma_2026,
  author  = {Smit, R. and Bowler, A. A.},
  title   = {The {ALMA} View of High-Redshift Galaxy Formation},
  journal = {Annual Review of Astronomy and Astrophysics},
  year    = {2026},
  volume  = {64},
  pages   = {493--535},
  doi     = {10.1146/annurev-astro-052722-104242}
}

@ARTICLE{tsukui_2021,
       author = {{Tsukui}, Takafumi and {Iguchi}, Satoru},
        title = "{Spiral morphology in an intensely star-forming disk galaxy more than 12 billion years ago}",
      journal = {Science},
         year = 2021,
        month = jun,
       volume = {372},
       number = {6547},
        pages = {1201-1205},
          doi = {10.1126/science.abe9680},
archivePrefix = {arXiv},
       eprint = {2108.02206},
 primaryClass = {astro-ph.GA},
       
}

@ARTICLE{pope_2023,
       author = {{Pope}, Alexandra and {McKinney}, Jed and {Kamieneski}, Patrick and {Battisti}, Andrew and {Aretxaga}, Itziar and {Brammer}, Gabriel and {Diego}, Jose M. and {Hughes}, David H. and {Keller}, Erica and {Marchesini}, Danilo and {Mizener}, Andrew and {Monta{\~n}a}, Alfredo and {Murphy}, Eric and {Whitaker}, Katherine E. and {Wilson}, Grant and {Yun}, Min},
        title = "{ALMA Reveals a Stable Rotating Gas Disk in a Paradoxical Low-mass, Ultradusty Galaxy at z = 4.274}",
      journal = {\apjl},
         year = 2023,
        month = jul,
       volume = {951},
       number = {2},
          eid = {L46},
        pages = {L46},
          doi = {10.3847/2041-8213/acdf5a},
archivePrefix = {arXiv},
       eprint = {2306.10450},
 primaryClass = {astro-ph.GA},
       
}

@ARTICLE{saxena_2026,
       author = {{Saxena}, Aayush and {Cameron}, Alex J. and {Katz}, Harley and {Bunker}, Andrew J. and {Chevallard}, Jacopo and {D'Eugenio}, Francesco and {Arribas}, Santiago and {Bhatawdekar}, Rachana and {Boyett}, Kristan and {Cargile}, Phillip A. and {Carniani}, Stefano and {Charlot}, St{\'e}phane and {Curti}, Mirko and {Curtis-Lake}, Emma and {Hainline}, Kevin and {Ji}, Zhiyuan and {Johnson}, Benjamin D. and {Jones}, Gareth C. and {Kumari}, Nimisha and {Laseter}, Isaac and {Maseda}, Michael V. and {Robertson}, Brant and {Simmonds}, Charlotte and {Tacchella}, Sandro and {{\"U}bler}, Hannah and {Williams}, Christina C. and {Willott}, Chris and {Witstok}, Joris and {Zhu}, Yongda},
        title = "{Hitting the slopes: a spectroscopic view of UV continuum slopes of galaxies reveals a reddening at z > 9.5}",
      journal = {\mnras},
         year = 2026,
        month = jun,
       volume = {548},
       number = {4},
          eid = {stag808},
        pages = {stag808},
          doi = {10.1093/mnras/stag808},
archivePrefix = {arXiv},
       eprint = {2411.14532},
 primaryClass = {astro-ph.GA},
      
}

@ARTICLE{austin_2025,
       author = {{Austin}, Duncan and {Conselice}, Christopher J. and {Adams}, Nathan J. and {Harvey}, Thomas and {Duan}, Qiao and {Trussler}, James and {Li}, Qiong and {Juod{\v{z}}balis}, Ignas and {Ormerod}, Katherine and {Ferreira}, Leonardo and {Westcott}, Lewi and {Harris}, Honor and {Wilkins}, Stephen M. and {Bhatawdekar}, Rachana and {Caruana}, Joseph and {Coe}, Dan and {Cohen}, Seth H. and {Driver}, Simon P. and {D'Silva}, Jordan C.~J. and {Frye}, Brenda and {Furtak}, Lukas J. and {Grogin}, Norman A. and {Hathi}, Nimish P. and {Holwerda}, Benne W. and {Jansen}, Rolf A. and {Koekemoer}, Anton M. and {Marshall}, Madeline A. and {Nonino}, Mario and {Ortiz}, III, Rafael and {Pirzkal}, Nor and {Robotham}, Aaron and {Ryan}, Jr., Russell E. and {Summers}, Jake and {Willmer}, Christopher N.~A. and {Windhorst}, Rogier A. and {Yan}, Haojing and {Zackrisson}, Erik},
        title = "{EPOCHS. III. Unbiased UV Continuum Slopes at 6.5 < z < 13 from Combined PEARLS GTO and Public JWST/NIRCam Imaging}",
      journal = {\apj},
         year = 2025,
        month = dec,
       volume = {995},
       number = {1},
          eid = {43},
        pages = {43},
          doi = {10.3847/1538-4357/ae07db},
archivePrefix = {arXiv},
       eprint = {2404.10751},
 primaryClass = {astro-ph.GA},
      
}

@ARTICLE{topping_2022,
       author = {{Topping}, Michael W. and {Stark}, Daniel P. and {Endsley}, Ryan and {Plat}, Adele and {Whitler}, Lily and {Chen}, Zuyi and {Charlot}, St{\'e}phane},
        title = "{Searching for Extremely Blue UV Continuum Slopes at z = 7-11 in JWST/NIRCam Imaging: Implications for Stellar Metallicity and Ionizing Photon Escape in Early Galaxies}",
      journal = {\apj},
         year = 2022,
        month = dec,
       volume = {941},
       number = {2},
          eid = {153},
        pages = {153},
          doi = {10.3847/1538-4357/aca522},
archivePrefix = {arXiv},
       eprint = {2208.01610},
 primaryClass = {astro-ph.GA},
       
}

@ARTICLE{topping_2024,
       author = {{Topping}, Michael W. and {Stark}, Daniel P. and {Endsley}, Ryan and {Whitler}, Lily and {Hainline}, Kevin and {Johnson}, Benjamin D. and {Robertson}, Brant and {Tacchella}, Sandro and {Chen}, Zuyi and {Alberts}, Stacey and {Baker}, William M. and {Bunker}, Andrew J. and {Carniani}, Stefano and {Charlot}, Stephane and {Chevallard}, Jacopo and {Curtis-Lake}, Emma and {DeCoursey}, Christa and {Egami}, Eiichi and {Eisenstein}, Daniel J. and {Ji}, Zhiyuan and {Maiolino}, Roberto and {Williams}, Christina C. and {Willmer}, Christopher N.~A. and {Willott}, Chris and {Witstok}, Joris},
        title = "{The UV continuum slopes of early star-forming galaxies in JADES}",
      journal = {\mnras},
         year = 2024,
        month = apr,
       volume = {529},
       number = {4},
        pages = {4087-4103},
          doi = {10.1093/mnras/stae800},
archivePrefix = {arXiv},
       eprint = {2307.08835},
 primaryClass = {astro-ph.GA},
       
}

@ARTICLE{adamo_2025,
       author = {{Adamo}, Angela and {Atek}, Hakim and {Bagley}, Micaela B. and {Ba{\~n}ados}, Eduardo and {Barrow}, Kirk S.~S. and {Berg}, Danielle A. and {Bezanson}, Rachel and {Brada{\v{c}}}, Maru{\v{s}}a and {Brammer}, Gabriel and {Carnall}, Adam C. and {Chisholm}, John and {Coe}, Dan and {Dayal}, Pratika and {Eisenstein}, Daniel J. and {Eldridge}, Jan J. and {Ferrara}, Andrea and {Fujimoto}, Seiji and {Graaff}, Anna de and {Habouzit}, Melanie and {Hutchison}, Taylor A. and {Kartaltepe}, Jeyhan S. and {Kassin}, Susan A. and {Kriek}, Mariska and {Labb{\'e}}, Ivo and {Maiolino}, Roberto and {Marques-Chaves}, Rui and {Maseda}, Michael V. and {Mason}, Charlotte and {Matthee}, Jorryt and {McQuinn}, Kristen B.~W. and {Meynet}, Georges and {Naidu}, Rohan P. and {Oesch}, Pascal A. and {Pentericci}, Laura and {P{\'e}rez-Gonz{\'a}lez}, Pablo G. and {Rigby}, Jane R. and {Roberts-Borsani}, Guido and {Schaerer}, Daniel and {Shapley}, Alice E. and {Stark}, Daniel P. and {Stiavelli}, Massimo and {Strom}, Allison L. and {Vanzella}, Eros and {Wang}, Feige and {Wilkins}, Stephen M. and {Williams}, Christina C. and {Willott}, Chris J. and {Wylezalek}, Dominika and {Nota}, Antonella},
        title = "{The first billion years according to JWST}",
      journal = {Nature Astronomy},
         year = 2025,
        month = aug,
       volume = {9},
        pages = {1134-1147},
          doi = {10.1038/s41550-025-02624-5},
archivePrefix = {arXiv},
       eprint = {2405.21054},
 primaryClass = {astro-ph.GA},
       
}

@ARTICLE{harvey_2025,
       author = {{Harvey}, Thomas and {Conselice}, Christopher J. and {Adams}, Nathan J. and {Austin}, Duncan and {Li}, Qiong and {Rusakov}, Vadim and {Westcott}, Lewi and {Goolsby}, Caio M. and {Lovell}, Christopher C. and {Cochrane}, Rachel K. and {Vijayan}, Aswin P. and {Trussler}, James},
        title = "{Behind the spotlight: a systematic assessment of outshining using NIRCam medium bands in the JADES Origins Field}",
      journal = {\mnras},
         year = 2025,
        month = oct,
       volume = {542},
       number = {4},
        pages = {2998-3027},
          doi = {10.1093/mnras/staf1396},
archivePrefix = {arXiv},
       eprint = {2504.05244},
 primaryClass = {astro-ph.GA},
       
}

@ARTICLE{bakx_2025,
       author = {{Bakx}, T.~J.~L.~C. and {Sommovigo}, Laura and {Tamura}, Yoichi and {Smit}, Renske and {Ferrara}, Andrea and {Algera}, Hiddo and {Aalto}, Susanne and {Bossion}, Duncan and {Carniani}, Stefano and {Esmerian}, Clarke and {Hagimoto}, Masato and {Hashimoto}, Takuya and {Hatsukade}, Bunyo and {Ibar}, Edo and {Inami}, Hanae and {Inoue}, Akio K. and {Knudsen}, Kirsten and {Laporte}, Nicolas and {Mawatari}, Ken and {Molina}, Juan and {Nyman}, Gunnar and {Okamoto}, Takashi and {Pallottini}, Andrea and {Sameera}, W.~M.~C. and {Umehata}, Hideki and {Vlemmings}, Wouter and {Yoshida}, Naoki},
        title = "{A warm ultraluminous infrared galaxy just 600 million years after the big bang}",
      journal = {\mnras},
         year = 2025,
        month = dec,
       volume = {544},
       number = {2},
        pages = {1502-1513},
          doi = {10.1093/mnras/staf1714},
archivePrefix = {arXiv},
       eprint = {2511.08327},
 primaryClass = {astro-ph.GA},
       
}

@ARTICLE{carniani_2024,
       author = {{Carniani}, Stefano and {Hainline}, Kevin and {D'Eugenio}, Francesco and {Eisenstein}, Daniel J. and {Jakobsen}, Peter and {Witstok}, Joris and {Johnson}, Benjamin D. and {Chevallard}, Jacopo and {Maiolino}, Roberto and {Helton}, Jakob M. and {Willott}, Chris and {Robertson}, Brant and {Alberts}, Stacey and {Arribas}, Santiago and {Baker}, William M. and {Bhatawdekar}, Rachana and {Boyett}, Kristan and {Bunker}, Andrew J. and {Cameron}, Alex J. and {Cargile}, Phillip A. and {Charlot}, St{\'e}phane and {Curti}, Mirko and {Curtis-Lake}, Emma and {Egami}, Eiichi and {Giardino}, Giovanna and {Isaak}, Kate and {Ji}, Zhiyuan and {Jones}, Gareth C. and {Kumari}, Nimisha and {Maseda}, Michael V. and {Parlanti}, Eleonora and {P{\'e}rez-Gonz{\'a}lez}, Pablo G. and {Rawle}, Tim and {Rieke}, George and {Rieke}, Marcia and {Del Pino}, Bruno Rodr{\'\i}guez and {Saxena}, Aayush and {Scholtz}, Jan and {Smit}, Renske and {Sun}, Fengwu and {Tacchella}, Sandro and {{\"U}bler}, Hannah and {Venturi}, Giacomo and {Williams}, Christina C. and {Willmer}, Christopher N.~A.},
        title = "{Spectroscopic confirmation of two luminous galaxies at a redshift of 14}",
      journal = {\nat},
         year = 2024,
        month = sep,
       volume = {633},
       number = {8029},
        pages = {318-322},
          doi = {10.1038/s41586-024-07860-9},
archivePrefix = {arXiv},
       eprint = {2405.18485},
 primaryClass = {astro-ph.GA},
       
}

@ARTICLE{Heintz_2025,
       author = {{Heintz}, K.~E. and {Brammer}, G.~B. and {Watson}, D. and {Oesch}, P.~A. and {Keating}, L.~C. and {Hayes}, M.~J. and {Abdurro'uf} and {Arellano-C{\'o}rdova}, K.~Z. and {Carnall}, A.~C. and {Christiansen}, C.~R. and {Cullen}, F. and {Dav{\'e}}, R. and {Dayal}, P. and {Ferrara}, A. and {Finlator}, K. and {Fynbo}, J.~P.~U. and {Flury}, S.~R. and {Gelli}, V. and {Gillman}, S. and {Gottumukkala}, R. and {Gould}, K. and {Greve}, T.~R. and {Hardin}, S.~E. and {Hsiao}, T.~Y.-Y. and {Hutter}, A. and {Jakobsson}, P. and {Killi}, M. and {Khosravaninezhad}, N. and {Laursen}, P. and {Lee}, M.~M. and {Magdis}, G.~E. and {Matthee}, J. and {Naidu}, R.~P. and {Narayanan}, D. and {Pollock}, C. and {Prescott}, M.~K.~M. and {Rusakov}, V. and {Shuntov}, M. and {Sneppen}, A. and {Smit}, R. and {Tanvir}, N.~R. and {Terp}, C. and {Toft}, S. and {Valentino}, F. and {Vijayan}, A.~P. and {Weaver}, J.~R. and {Wise}, J.~H. and {Witstok}, J.},
        title = "{The JWST-PRIMAL archival survey: A JWST/NIRSpec reference sample for the physical properties and Lyman-{\ensuremath{\alpha}} absorption and emission of {\ensuremath{\sim}}600 galaxies at z = 5.0 {\ensuremath{-}} 13.4}",
      journal = {\aap},
         year = 2025,
        month = jan,
       volume = {693},
          eid = {A60},
        pages = {A60},
          doi = {10.1051/0004-6361/202450243},
archivePrefix = {arXiv},
       eprint = {2404.02211},
 primaryClass = {astro-ph.GA},
       
}

@ARTICLE{aihara_2011,
       author = {{Aihara}, Hiroaki and {Allende Prieto}, Carlos and {An}, Deokkeun and {Anderson}, Scott F. and {Aubourg}, {\'E}ric and {Balbinot}, Eduardo and {Beers}, Timothy C. and {Berlind}, Andreas A. and {Bickerton}, Steven J. and {Bizyaev}, Dmitry and {Blanton}, Michael R. and {Bochanski}, John J. and {Bolton}, Adam S. and {Bovy}, Jo and {Brandt}, W.~N. and {Brinkmann}, J. and {Brown}, Peter J. and {Brownstein}, Joel R. and {Busca}, Nicolas G. and {Campbell}, Heather and {Carr}, Michael A. and {Chen}, Yanmei and {Chiappini}, Cristina and {Comparat}, Johan and {Connolly}, Natalia and {Cortes}, Marina and {Croft}, Rupert A.~C. and {Cuesta}, Antonio J. and {da Costa}, Luiz N. and {Davenport}, James R.~A. and {Dawson}, Kyle and {Dhital}, Saurav and {Ealet}, Anne and {Ebelke}, Garrett L. and {Edmondson}, Edward M. and {Eisenstein}, Daniel J. and {Escoffier}, Stephanie and {Esposito}, Massimiliano and {Evans}, Michael L. and {Fan}, Xiaohui and {Femen{\'\i}a Castell{\'a}}, Bruno and {Font-Ribera}, Andreu and {Frinchaboy}, Peter M. and {Ge}, Jian and {Gillespie}, Bruce A. and {Gilmore}, G. and {Gonz{\'a}lez Hern{\'a}ndez}, Jonay I. and {Gott}, J. Richard and {Gould}, Andrew and {Grebel}, Eva K. and {Gunn}, James E. and {Hamilton}, Jean-Christophe and {Harding}, Paul and {Harris}, David W. and {Hawley}, Suzanne L. and {Hearty}, Frederick R. and {Ho}, Shirley and {Hogg}, David W. and {Holtzman}, Jon A. and {Honscheid}, Klaus and {Inada}, Naohisa and {Ivans}, Inese I. and {Jiang}, Linhua and {Johnson}, Jennifer A. and {Jordan}, Cathy and {Jordan}, Wendell P. and {Kazin}, Eyal A. and {Kirkby}, David and {Klaene}, Mark A. and {Knapp}, G.~R. and {Kneib}, Jean-Paul and {Kochanek}, C.~S. and {Koesterke}, Lars and {Kollmeier}, Juna A. and {Kron}, Richard G. and {Lampeitl}, Hubert and {Lang}, Dustin and {Le Goff}, Jean-Marc and {Lee}, Young Sun and {Lin}, Yen-Ting and {Long}, Daniel C. and {Loomis}, Craig P. and {Lucatello}, Sara and {Lundgren}, Britt and {Lupton}, Robert H. and {Ma}, Zhibo and {MacDonald}, Nicholas and {Mahadevan}, Suvrath and {Maia}, Marcio A.~G. and {Makler}, Martin and {Malanushenko}, Elena and {Malanushenko}, Viktor and {Mandelbaum}, Rachel and {Maraston}, Claudia and {Margala}, Daniel and {Masters}, Karen L. and {McBride}, Cameron K. and {McGehee}, Peregrine M. and {McGreer}, Ian D. and {M{\'e}nard}, Brice and {Miralda-Escud{\'e}}, Jordi and {Morrison}, Heather L. and {Mullally}, F. and {Muna}, Demitri and {Munn}, Jeffrey A. and {Murayama}, Hitoshi and {Myers}, Adam D. and {Naugle}, Tracy and {Neto}, Angelo Fausti and {Nguyen}, Duy Cuong and {Nichol}, Robert C. and {O'Connell}, Robert W. and {Ogando}, Ricardo L.~C. and {Olmstead}, Matthew D. and {Oravetz}, Daniel J. and {Padmanabhan}, Nikhil and {Palanque-Delabrouille}, Nathalie and {Pan}, Kaike and {Pandey}, Parul and {P{\^a}ris}, Isabelle and {Percival}, Will J. and {Petitjean}, Patrick and {Pfaffenberger}, Robert and {Pforr}, Janine and {Phleps}, Stefanie and {Pichon}, Christophe and {Pieri}, Matthew M. and {Prada}, Francisco and {Price-Whelan}, Adrian M. and {Raddick}, M. Jordan and {Ramos}, Beatriz H.~F. and {Reyl{\'e}}, C{\'e}line and {Rich}, James and {Richards}, Gordon T. and {Rix}, Hans-Walter and {Robin}, Annie C. and {Rocha-Pinto}, Helio J. and {Rockosi}, Constance M. and {Roe}, Natalie A. and {Rollinde}, Emmanuel and {Ross}, Ashley J. and {Ross}, Nicholas P. and {Rossetto}, Bruno M. and {S{\'a}nchez}, Ariel G. and {Sayres}, Conor and {Schlegel}, David J. and {Schlesinger}, Katharine J. and {Schmidt}, Sarah J. and {Schneider}, Donald P. and {Sheldon}, Erin and {Shu}, Yiping and {Simmerer}, Jennifer and {Simmons}, Audrey E. and {Sivarani}, Thirupathi and {Snedden}, Stephanie A. and {Sobeck}, Jennifer S. and {Steinmetz}, Matthias and {Strauss}, Michael A. and {Szalay}, Alexander S. and {Tanaka}, Masayuki and {Thakar}, Aniruddha R. and {Thomas}, Daniel and {Tinker}, Jeremy L. and {Tofflemire}, Benjamin M. and {Tojeiro}, Rita and {Tremonti}, Christy A. and {Vandenberg}, Jan and {Vargas Maga{\~n}a}, M. and {Verde}, Licia and {Vogt}, Nicole P. and {Wake}, David A. and {Wang}, Ji and {Weaver}, Benjamin A. and {Weinberg}, David H. and {White}, Martin and {White}, Simon D.~M. and {Yanny}, Brian and {Yasuda}, Naoki and {Yeche}, Christophe and {Zehavi}, Idit},
        title = "{The Eighth Data Release of the Sloan Digital Sky Survey: First Data from SDSS-III}",
      journal = {\apjs},
         year = 2011,
        month = apr,
       volume = {193},
       number = {2},
          eid = {29},
        pages = {29},
          doi = {10.1088/0067-0049/193/2/29},
archivePrefix = {arXiv},
       eprint = {1101.1559},
 primaryClass = {astro-ph.IM},
       
}

@ARTICLE{ivey_2026,
       author = {{Ivey}, L.~R. and {Scholtz}, J. and {Danhaive}, A.~L. and {Koudmani}, S. and {Jones}, G.~C. and {Maiolino}, R. and {Curti}, M. and {D'Eugenio}, F. and {Tacchella}, S. and {Baker}, W.~M. and {Arribas}, S. and {Charlot}, S. and {Eisenstein}, D. and {Ji}, Z. and {Koller}, M. and {Laporte}, N. and {Perna}, M. and {Pusk{\'a}s}, D. and {Robertson}, B. and {Sijacki}, D. and {Trussler}, J.~A.~A. and {Witten}, C.},
        title = "{Exploring spatially resolved metallicities, dynamics, and outflows in low-mass galaxies at z {\ensuremath{\sim}} 7.6}",
      journal = {\mnras},
         year = 2026,
        month = mar,
       volume = {546},
       number = {3},
          eid = {stag094},
        pages = {stag094},
          doi = {10.1093/mnras/stag094},
archivePrefix = {arXiv},
       eprint = {2507.14936},
 primaryClass = {astro-ph.GA},
       
}

@ARTICLE{jones_2026,
       author = {{Jones}, Gareth C. and {Bowler}, Rebecca A.~A. and {Bunker}, Andrew J. and {Curti}, Mirko and {Arribas}, Santiago and {Carniani}, Stefano and {Charlot}, Stephane and {Perna}, Michele and {Rodr{\'\i}guez Del Pino}, Bruno and {{\"U}bler}, Hannah and {Willott}, Chris J. and {Chevallard}, Jacopo and {Cresci}, Giovanni and {Parlanti}, Eleonora and {Scholtz}, Jan and {Venturi}, Giacomo},
        title = "{GA-NIFS: interstellar medium properties and tidal interactions in the evolved massive merging system B14-65666 at z = 7.152}",
      journal = {\mnras},
         year = 2026,
        month = apr,
       volume = {547},
       number = {2},
          eid = {stag336},
        pages = {stag336},
          doi = {10.1093/mnras/stag336},
archivePrefix = {arXiv},
       eprint = {2412.15027},
 primaryClass = {astro-ph.GA},
       
}

@ARTICLE{jones_2025,
       author = {{Jones}, Gareth C. and {Bunker}, Andrew J. and {Telikova}, Kseniia and {Arribas}, Santiago and {Carniani}, Stefano and {Charlot}, Stephane and {D'Eugenio}, Francesco and {Maiolino}, Roberto and {Perna}, Michele and {Rodr{\'\i}guez Del Pino}, Bruno and {{\"U}bler}, Hannah and {Willott}, Chris and {Aravena}, Manuel and {B{\"o}ker}, Torsten and {Cresci}, Giovanni and {Curti}, Mirko and {Gonz{\'a}lez-L{\'o}pez}, Jorge and {Herrera-Camus}, Rodrigo and {Lamperti}, Isabella and {Parlanti}, Eleonora and {P{\'e}rez-Gonz{\'a}lez}, Pablo G. and {Villanueva}, Vicente},
        title = "{GA-NIFS: witnessing the complex assembly of a star-forming system at z = 5.7}",
      journal = {\mnras},
         year = 2025,
        month = jul,
       volume = {540},
       number = {4},
        pages = {3311-3329},
          doi = {10.1093/mnras/staf899},
archivePrefix = {arXiv},
       eprint = {2405.12955},
 primaryClass = {astro-ph.GA},
       
}

@ARTICLE{jones_2024,
       author = {{Jones}, Gareth C. and {{\"U}bler}, Hannah and {Perna}, Michele and {Arribas}, Santiago and {Bunker}, Andrew J. and {Carniani}, Stefano and {Charlot}, Stephane and {Maiolino}, Roberto and {Del Pino}, Bruno Rodr{\'\i}guez and {Willott}, Chris and {Bowler}, Rebecca A.~A. and {B{\"o}ker}, Torsten and {Cameron}, Alex J. and {Chevallard}, Jacopo and {Cresci}, Giovanni and {Curti}, Mirko and {D'Eugenio}, Francesco and {Kumari}, Nimisha and {Saxena}, Aayush and {Scholtz}, Jan and {Venturi}, Giacomo and {Witstok}, Joris},
        title = "{GA-NIFS: JWST/NIRSpec integral field unit observations of HFLS3 reveal a dense galaxy group at z {\ensuremath{\sim}} 6.3}",
      journal = {\aap},
         year = 2024,
        month = feb,
       volume = {682},
          eid = {A122},
        pages = {A122},
          doi = {10.1051/0004-6361/202347838},
archivePrefix = {arXiv},
       eprint = {2308.16620},
 primaryClass = {astro-ph.GA},
       
}

@ARTICLE{vallini_2024,
       author = {{Vallini}, Livia and {Witstok}, Joris and {Sommovigo}, Laura and {Pallottini}, Andrea and {Ferrara}, Andrea and {Carniani}, Stefano and {Kohandel}, Mahsa and {Smit}, Renske and {Gallerani}, Simona and {Gruppioni}, Carlotta},
        title = "{Spatially resolved Kennicutt-Schmidt relation at z {\ensuremath{\approx}} 7 and its connection with the interstellar medium properties}",
      journal = {\mnras},
         year = 2024,
        month = jan,
       volume = {527},
       number = {1},
        pages = {10-22},
          doi = {10.1093/mnras/stad3150},
archivePrefix = {arXiv},
       eprint = {2309.07957},
 primaryClass = {astro-ph.GA},
       
}

@ARTICLE{zanella_2021,
       author = {{Zanella}, A. and {Pallottini}, A. and {Ferrara}, A. and {Gallerani}, S. and {Carniani}, S. and {Kohandel}, M. and {Behrens}, C.},
        title = "{Early galaxy growth: mergers or gravitational instability?}",
      journal = {\mnras},
         year = 2021,
        month = jan,
       volume = {500},
       number = {1},
        pages = {118-137},
          doi = {10.1093/mnras/staa2776},
archivePrefix = {arXiv},
       eprint = {2009.03927},
 primaryClass = {astro-ph.GA},
       
}

@ARTICLE{wolfire_2022,
       author = {{Wolfire}, Mark G. and {Vallini}, Livia and {Chevance}, M{\'e}lanie},
        title = "{Photodissociation and X-Ray-Dominated Regions}",
      journal = {\araa},
         year = 2022,
        month = aug,
       volume = {60},
        pages = {247-318},
          doi = {10.1146/annurev-astro-052920-010254},
archivePrefix = {arXiv},
       eprint = {2202.05867},
 primaryClass = {astro-ph.GA},
       
}

@ARTICLE{rizzo_2022,
       author = {{Rizzo}, F. and {Kohandel}, M. and {Pallottini}, A. and {Zanella}, A. and {Ferrara}, A. and {Vallini}, L. and {Toft}, S.},
        title = "{Dynamical characterization of galaxies up to z {\ensuremath{\sim}} 7}",
      journal = {\aap},
         year = 2022,
        month = nov,
       volume = {667},
          eid = {A5},
        pages = {A5},
          doi = {10.1051/0004-6361/202243582},
archivePrefix = {arXiv},
       eprint = {2204.05325},
 primaryClass = {astro-ph.GA},
       
}

@ARTICLE{wang_2025,
       author = {{Wang}, Weichen and {Cantalupo}, Sebastiano and {Pensabene}, Antonio and {Galbiati}, Marta and {Travascio}, Andrea and {Steidel}, Charles C. and {Maseda}, Michael V. and {Pezzulli}, Gabriele and {de Beer}, Stephanie and {Fossati}, Matteo and {Fumagalli}, Michele and {Gallego}, Sofia G. and {Lazeyras}, Titouan and {Mackenzie}, Ruari and {Matthee}, Jorryt and {Nanayakkara}, Themiya and {Quadri}, Giada},
        title = "{A giant disk galaxy two billion years after the Big Bang}",
      journal = {Nature Astronomy},
         year = 2025,
        month = may,
       volume = {9},
        pages = {710-719},
          doi = {10.1038/s41550-025-02500-2},
archivePrefix = {arXiv},
       eprint = {2409.17956},
 primaryClass = {astro-ph.GA},
       
}

@ARTICLE{arata_2020,
       author = {{Arata}, Shohei and {Yajima}, Hidenobu and {Nagamine}, Kentaro and {Abe}, Makito and {Khochfar}, Sadegh},
        title = "{Starbursting [O III] emitters and quiescent [C II] emitters in the reionization era}",
      journal = {\mnras},
         year = 2020,
        month = nov,
       volume = {498},
       number = {4},
        pages = {5541-5556},
          doi = {10.1093/mnras/staa2809},
archivePrefix = {arXiv},
       eprint = {2001.01853},
 primaryClass = {astro-ph.GA},
       
}

@ARTICLE{kohandel_2020,
       author = {{Kohandel}, M. and {Pallottini}, A. and {Ferrara}, A. and {Carniani}, S. and {Gallerani}, S. and {Vallini}, L. and {Zanella}, A. and {Behrens}, C.},
        title = "{Velocity dispersion in the interstellar medium of early galaxies}",
      journal = {\mnras},
         year = 2020,
        month = nov,
       volume = {499},
       number = {1},
        pages = {1250-1265},
          doi = {10.1093/mnras/staa2792},
archivePrefix = {arXiv},
       eprint = {2009.05049},
 primaryClass = {astro-ph.GA},
       
}

@article{ubler_evolution_2019,
	title = {The {Evolution} and {Origin} of {Ionized} {Gas} {Velocity} {Dispersion} from z ∼ 2.6 to z ∼ 0.6 with {KMOS3D}},
	volume = {880},
	issn = {0004-637X},
	url = {https://ui.adsabs.harvard.edu/abs/2019ApJ...880...48U},
	doi = {10.3847/1538-4357/ab27cc},
	urldate = {2025-02-24},
	journal = {\apj},
	publisher = {IOP},
	author = {Übler, H. and Genzel, R. and Wisnioski, E. and Förster Schreiber, N. M. and Shimizu, T. T. and Price, S. H. and Tacconi, L. J. and Belli, S. and Wilman, D. J. and Fossati, M. and Mendel, J. T. and Davies, R. L. and Beifiori, A. and Bender, R. and Brammer, G. B. and Burkert, A. and Chan, J. and Davies, R. I. and Fabricius, M. and Galametz, A. and Herrera-Camus, R. and Lang, P. and Lutz, D. and Momcheva, I. G. and Naab, T. and Nelson, E. J. and Saglia, R. P. and Tadaki, K. and van Dokkum, P. G. and Wuyts, S.},
	month = jul,
	year = {2019},
	pages = {48},
}

@article{girard_systematic_2021,
	title = {Systematic {Difference} between {Ionized} and {Molecular} {Gas} {Velocity} {Dispersions} in z ∼ 1–2 {Disks} and {Local} {Analogs}},
	volume = {909},
	issn = {0004-637X},
	url = {https://dx.doi.org/10.3847/1538-4357/abd5b9},
	doi = {10.3847/1538-4357/abd5b9},
	language = {en},
	number = {1},
	urldate = {2025-02-24},
	journal = {\apj},
	publisher = {The American Astronomical Society},
	author = {Girard, M. and Fisher, D. B. and Bolatto, A. D. and Abraham, R. and Bassett, R. and Glazebrook, K. and Herrera-Camus, R. and Jiménez, E. and Lenkić, L. and Obreschkow, D.},
	month = mar,
	year = {2021},
	pages = {12},
}

@ARTICLE{walter2018,
       author = {{Walter}, Fabian and {Riechers}, Dominik and {Novak}, Mladen and {Decarli}, Roberto and {Ferkinhoff}, Carl and {Venemans}, Bram and {Ba{\~n}ados}, Eduardo and {Bertoldi}, Frank and {Carilli}, Chris and {Fan}, Xiaohui and {Farina}, Emanuele and {Mazzucchelli}, Chiara and {Neeleman}, Marcel and {Rix}, Hans-Walter and {Strauss}, Michael A. and {Uzgil}, Bade and {Wang}, Ran},
        title = "{No Evidence for Enhanced [O III] 88 {\ensuremath{\mu}}m Emission in a z {\ensuremath{\sim}} 6 Quasar Compared to Its Companion Starbursting Galaxy}",
      journal = {\apjl},
         year = 2018,
        month = dec,
       volume = {869},
       number = {2},
          eid = {L22},
        pages = {L22},
          doi = {10.3847/2041-8213/aaf4fa},
archivePrefix = {arXiv},
       eprint = {1811.12836},
 primaryClass = {astro-ph.GA},
       
}

@ARTICLE{marrone2018,
       author = {{Marrone}, D.~P. and {Spilker}, J.~S. and {Hayward}, C.~C. and {Vieira}, J.~D. and {Aravena}, M. and {Ashby}, M.~L.~N. and {Bayliss}, M.~B. and {B{\'e}thermin}, M. and {Brodwin}, M. and {Bothwell}, M.~S. and {Carlstrom}, J.~E. and {Chapman}, S.~C. and {Chen}, Chian-Chou and {Crawford}, T.~M. and {Cunningham}, D.~J.~M. and {De Breuck}, C. and {Fassnacht}, C.~D. and {Gonzalez}, A.~H. and {Greve}, T.~R. and {Hezaveh}, Y.~D. and {Lacaille}, K. and {Litke}, K.~C. and {Lower}, S. and {Ma}, J. and {Malkan}, M. and {Miller}, T.~B. and {Morningstar}, W.~R. and {Murphy}, E.~J. and {Narayanan}, D. and {Phadke}, K.~A. and {Rotermund}, K.~M. and {Sreevani}, J. and {Stalder}, B. and {Stark}, A.~A. and {Strandet}, M.~L. and {Tang}, M. and {Wei{\ss}}, A.},
        title = "{Galaxy growth in a massive halo in the first billion years of cosmic history}",
      journal = {\nat},
         year = 2018,
        month = jan,
       volume = {553},
       number = {7686},
        pages = {51-54},
          doi = {10.1038/nature24629},
archivePrefix = {arXiv},
       eprint = {1712.03020},
 primaryClass = {astro-ph.GA},
       
}

@ARTICLE{cormier2015,
       author = {{Cormier}, D. and {Madden}, S.~C. and {Lebouteiller}, V. and {Abel}, N. and {Hony}, S. and {Galliano}, F. and {R{\'e}my-Ruyer}, A. and {Bigiel}, F. and {Baes}, M. and {Boselli}, A. and {Chevance}, M. and {Cooray}, A. and {De Looze}, I. and {Doublier}, V. and {Galametz}, M. and {Hughes}, T. and {Karczewski}, O. {\L}. and {Lee}, M.-Y. and {Lu}, N. and {Spinoglio}, L.},
        title = "{The Herschel Dwarf Galaxy Survey. I. Properties of the low-metallicity ISM from PACS spectroscopy}",
      journal = {\aap},
         year = 2015,
        month = jun,
       volume = {578},
          eid = {A53},
        pages = {A53},
          doi = {10.1051/0004-6361/201425207},
archivePrefix = {arXiv},
       eprint = {1502.03131},
 primaryClass = {astro-ph.GA},
       adsurl = {https://ui.adsabs.harvard.edu/abs/2015A&A...578A..53C},
}

@ARTICLE{robertson_2008,
       author = {{Robertson}, Brant E. and {Bullock}, James S.},
        title = "{High-Redshift Galaxy Kinematics: Constraints on Models of Disk Formation}",
      journal = {\apjl},
         year = 2008,
        month = sep,
       volume = {685},
       number = {1},
        pages = {L27},
          doi = {10.1086/592329},
archivePrefix = {arXiv},
       eprint = {0808.1100},
 primaryClass = {astro-ph},
       adsurl = {https://ui.adsabs.harvard.edu/abs/2008ApJ...685L..27R},
}

@ARTICLE{springel_2005,
       author = {{Springel}, Volker and {Hernquist}, Lars},
        title = "{Formation of a Spiral Galaxy in a Major Merger}",
      journal = {\apjl},
         year = 2005,
        month = mar,
       volume = {622},
       number = {1},
        pages = {L9-L12},
          doi = {10.1086/429486},
archivePrefix = {arXiv},
       eprint = {astro-ph/0411379},
 primaryClass = {astro-ph},
       adsurl = {https://ui.adsabs.harvard.edu/abs/2005ApJ...622L...9S},
}

@ARTICLE{bing2026,
       author = {{Bing}, Longji and {Oliver}, Seb and {Xiao}, Mengyuan and {Lagache}, Guilaine and {Adscheid}, Sylvia and {Liu}, Daizhong and {Magnelli}, Benjamin and {Neri}, Roberto and {Dessauges-Zavadsky}, Miroslava and {Koekemoer}, Anton M. and {Franco}, Maximilien and {Jin}, Shuowen and {Cooper}, Olivia R. and {Faisst}, Andreas L. and {Casey}, Catilin M. and {Kartaltepe}, Jeyhan S. and {Akins}, Hollis and {Beelen}, Alexandre and {Elbaz}, David and {Gillman}, Steven and {Harish}, Santosh and {Long}, Arianna S. and {McCracken}, Henry Joy and {Oesch}, Pascal and {Paquereau}, Louise and {Ponthieu}, Nicolas and {Rhodes}, Jason and {Robertson}, Brant and {Sanders}, David B. and {Shuntov}, Marko and {Wilkins}, Stephen},
        title = "{An almost NIRCam-dark dusty star-forming galaxy at z = 6.63}",
      journal = {\mnras},
         year = 2026,
        month = jun,
       volume = {549},
       number = {2},
          eid = {stag846},
        pages = {stag846},
          doi = {10.1093/mnras/stag846},
archivePrefix = {arXiv},
       eprint = {2511.08672},
 primaryClass = {astro-ph.GA},
       adsurl = {https://ui.adsabs.harvard.edu/abs/2026MNRAS.549ag846B},
}

@ARTICLE{Sun2026,
       author = {{Sun}, Fengwu and {Yang}, Jinyi and {Wang}, Feige and {Eisenstein}, Daniel J. and {Decarli}, Roberto and {Fan}, Xiaohui and {Rieke}, George H. and {Ba{\~n}ados}, Eduardo and {Bosman}, Sarah E.~I. and {Cai}, Zheng and {Champagne}, Jaclyn B. and {Colina}, Luis and {D'Eugenio}, Francesco and {Fudamoto}, Yoshinobu and {Li}, Mingyu and {Lin}, Xiaojing and {Liu}, Weizhe and {Lyu}, Jianwei and {Mazzucchelli}, Chiara and {Jin}, Xiangyu and {Jun}, Hyunsung D. and {Tee}, Wei Leong and {Wu}, Yunjing and {Zhang}, Huanian},
        title = "{The Identification of Two JWST/NIRCam-dark Starburst Galaxies at z = 6.6 with ALMA}",
      journal = {\apj},
         year = 2026,
        month = jun,
       volume = {1003},
       number = {2},
          eid = {206},
        pages = {206},
          doi = {10.3847/1538-4357/ae66e3},
archivePrefix = {arXiv},
       eprint = {2506.06418},
 primaryClass = {astro-ph.GA},
       adsurl = {https://ui.adsabs.harvard.edu/abs/2026ApJ..1003..206S},
}

@ARTICLE{HaywardHopkins2017,
       author = {{Hayward}, Christopher C. and {Hopkins}, Philip F.},
        title = "{How stellar feedback simultaneously regulates star formation and drives outflows}",
      journal = {\mnras},
         year = 2017,
        month = feb,
       volume = {465},
       number = {2},
        pages = {1682-1698},
          doi = {10.1093/mnras/stw2888},
archivePrefix = {arXiv},
       eprint = {1510.05650},
 primaryClass = {astro-ph.GA},
       adsurl = {https://ui.adsabs.harvard.edu/abs/2017MNRAS.465.1682H},
}

@ARTICLE{simons_2017,
       author = {{Simons}, Raymond C. and {Kassin}, Susan A. and {Weiner}, Benjamin J. and {Faber}, Sandra M. and {Trump}, Jonathan R. and {Heckman}, Timothy M. and {Koo}, David C. and {Pacifici}, Camilla and {Primack}, Joel R. and {Snyder}, Gregory F. and {de la Vega}, Alexander},
        title = "{z {\ensuremath{\sim}} 2: An Epoch of Disk Assembly}",
      journal = {\apj},
         year = 2017,
        month = jul,
       volume = {843},
       number = {1},
          eid = {46},
        pages = {46},
          doi = {10.3847/1538-4357/aa740c},
archivePrefix = {arXiv},
       eprint = {1705.03474},
 primaryClass = {astro-ph.GA},
       adsurl = {https://ui.adsabs.harvard.edu/abs/2017ApJ...843...46S},
}

@ARTICLE{Li_2026,
       author = {{Li}, J. and {da Cunha}, E. and {Hodge}, J.~A. and {Smail}, I. and {Kendrew}, S. and {battisti}, A. and {Cracraft}, M. and {Boogaard}, L.~A. and {Brandt}, W.~N. and {Chen}, C.-C. and {Cox}, P. and {Knudsen}, K.~K. and {Liao}, C.-L. and {Calistro Rivera}, G. and {Rybak}, M. and {Swinbank}, A.~M. and {van der Werf}, P. and {Walter}, F. and {Weiss}, A. and {Westoby}, B.~A.},
        title = "{ALESS--JWST: Dust-driven Morphologies and Hidden Stellar Mass in $z\sim3$ Sub-millimeter Galaxies}",
      journal = {arXiv e-prints},
         year = 2026,
        month = apr,
          eid = {arXiv:2604.13408},
        pages = {arXiv:2604.13408},
          doi = {10.48550/arXiv.2604.13408},
archivePrefix = {arXiv},
       eprint = {2604.13408},
 primaryClass = {astro-ph.GA},
       adsurl = {https://ui.adsabs.harvard.edu/abs/2026arXiv260413408L},
}

@ARTICLE{Pozzi_2021,
       author = {{Pozzi}, F. and {Calura}, F. and {Fudamoto}, Y. and {Dessauges-Zavadsky}, M. and {Gruppioni}, C. and {Talia}, M. and {Zamorani}, G. and {Bethermin}, M. and {Cimatti}, A. and {Enia}, A. and {Khusanova}, Y. and {Decarli}, R. and {Le F{\`e}vre}, O. and {Capak}, P. and {Cassata}, P. and {Faisst}, A.~L. and {Yan}, L. and {Schaerer}, D. and {Silverman}, J. and {Bardelli}, S. and {Boquien}, M. and {Enia}, A. and {Narayanan}, D. and {Ginolfi}, M. and {Hathi}, N.~P. and {Jones}, G.~C. and {Koekemoer}, A.~M. and {Lemaux}, B.~C. and {Loiacono}, F. and {Maiolino}, R. and {Riechers}, D.~A. and {Rodighiero}, G. and {Romano}, M. and {Vallini}, L. and {Vergani}, D. and {Zucca}, E.},
        title = "{The ALPINE-ALMA [CII] survey. Dust mass budget in the early Universe}",
      journal = {\aap},
         year = 2021,
        month = sep,
       volume = {653},
          eid = {A84},
        pages = {A84},
          doi = {10.1051/0004-6361/202040258},
archivePrefix = {arXiv},
       eprint = {2105.14789},
 primaryClass = {astro-ph.GA},
       adsurl = {https://ui.adsabs.harvard.edu/abs/2021A&A...653A..84P},
}

@ARTICLE{Hodge_2020,
       author = {{Hodge}, J.~A. and {da Cunha}, E.},
        title = "{High-redshift star formation in the Atacama large millimetre/submillimetre array era}",
      journal = {R. Soc. Open Sci.},
         year = 2020,
        month = dec,
       volume = {7},
       number = {12},
          eid = {200556},
        pages = {200556},
          doi = {10.1098/rsos.200556},
archivePrefix = {arXiv},
       eprint = {2004.00934},
 primaryClass = {astro-ph.GA},
       adsurl = {https://ui.adsabs.harvard.edu/abs/2020RSOS....700556H},
}

@ARTICLE{Decarli_2025,
       author = {{Decarli}, Roberto and {D{\'\i}az-Santos}, Tanio},
        title = "{Infrared fine-structure lines at high redshift}",
      journal = {\aapr},
         year = 2025,
        month = sep,
       volume = {33},
       number = {1},
          eid = {4},
        pages = {4},
          doi = {10.1007/s00159-025-00162-7},
archivePrefix = {arXiv},
       eprint = {2509.19444},
 primaryClass = {astro-ph.GA},
       adsurl = {https://ui.adsabs.harvard.edu/abs/2025A&ARv..33....4D},
}

@ARTICLE{Facchini_2025,
       author = {{Facchini}, Stefano and {Hodge}, Jacqueline and {J{\o}rgensen}, Jes and {Schinnerer}, Eva and {Tan}, Gie Han and {Bakx}, Tom and {Baryshev}, Andrey and {Beltran}, Maite and {Boogaard}, Leindert and {Decarli}, Roberto and {D{\'\i}az Trigo}, Mar{\'\i}a and {Forbrich}, Jan and {Huggard}, Peter and {Humphreys}, Elizabeth and {Impellizeri}, Violette and {Koljonen}, Karri and {Liu}, Kuo and {Matr{\`a}}, Luca and {Pereira Santella}, Miguel and {Piccialli}, Arianna and {Popping}, Gerg{\"o} and {Querejeta}, Miguel and {Rengel}, Miriam and {Rizzo}, Francesca and {Rowland}, Lucie and {Stacey}, Hannah and {Vlemmings}, Wouter and {Walsh}, Catherine and {Wedemeyer}, Sven and {Wiedner}, Martina},
        title = "{Towards ALMA2040: An update from the European community and invitation to contribute}",
      journal = {arXiv e-prints},
         year = 2025,
        month = dec,
          eid = {arXiv:2512.15652},
        pages = {arXiv:2512.15652},
          doi = {10.48550/arXiv.2512.15652},
archivePrefix = {arXiv},
       eprint = {2512.15652},
 primaryClass = {astro-ph.IM},
       adsurl = {https://ui.adsabs.harvard.edu/abs/2025arXiv251215652F},
}

@article{ubler_ga-nifs_2024,
	title = {{GA}-{NIFS}: {NIRSpec} reveals evidence for non-circular motions and {AGN} feedback in {GN20}},
	volume = {533},
	issn = {0035-8711},
	shorttitle = {{GA}-{NIFS}},
	url = {https://ui.adsabs.harvard.edu/abs/2024MNRAS.533.4287U},
	doi = {10.1093/mnras/stae1993},
	urldate = {2025-02-25},
	journal = {\mnras},
	publisher = {OUP},
	author = {Übler, Hannah and D'Eugenio, Francesco and Perna, Michele and Arribas, Santiago and Jones, Gareth C. and Bunker, Andrew J. and Carniani, Stefano and Charlot, Stéphane and Maiolino, Roberto and Rodríguez del Pino, Bruno and Willott, Chris J. and Böker, Torsten and Cresci, Giovanni and Kumari, Nimisha and Lamperti, Isabella and Parlanti, Eleonora and Scholtz, Jan and Venturi, Giacomo},
	month = oct,
	year = {2024},
	pages = {4287--4299},
}

@article{arribas_ga-nifs_2024,
	title = {{GA}-{NIFS}: {The} core of an extremely massive protocluster at the epoch of reionisation probed with {JWST}/{NIRSpec}},
	volume = {688},
	issn = {0004-6361},
	shorttitle = {{GA}-{NIFS}},
	url = {https://ui.adsabs.harvard.edu/abs/2024A&A...688A.146A},
	doi = {10.1051/0004-6361/202348824},
	urldate = {2025-02-25},
	journal = {\aap},
	publisher = {EDP},
	author = {Arribas, Santiago and Perna, Michele and Rodríguez Del Pino, Bruno and Lamperti, Isabella and D'Eugenio, Francesco and Pérez-González, Pablo G. and Jones, Gareth C. and Crespo Gómez, Alejandro and Curti, Mirko and Lim, Seunghwan and Álvarez-Márquez, Javier and Bunker, Andrew J. and Carniani, Stefano and Charlot, Stéphane and Jakobsen, Peter and Maiolino, Roberto and Übler, Hannah and Willott, Chris J. and Böker, Torsten and Chevallard, Jacopo and Circosta, Chiara and Cresci, Giovanni and Kumari, Nimisha and Parlanti, Eleonora and Scholtz, Jan and Venturi, Giacomo and Witstok, Joris},
	month = aug,
	year = {2024},
	pages = {A146},
}

@article{lelli_massive_2021,
	title = {A massive stellar bulge in a regularly rotating galaxy 1.2 billion years after the {Big} {Bang}},
	volume = {371},
	url = {https://www.science.org/doi/10.1126/science.abc1893},
	doi = {10.1126/science.abc1893},
	number = {6530},
	urldate = {2025-02-25},
	journal = {\sci},
	publisher = {American Association for the Advancement of Science},
	author = {Lelli, Federico and Di Teodoro, Enrico M. and Fraternali, Filippo and Man, Allison W. S. and Zhang, Zhi-Yu and De Breuck, Carlos and Davis, Timothy A. and Maiolino, Roberto},
	month = feb,
	year = {2021},
	pages = {713--716},
}

@article{rowland_rebels-25_2024,
	title = {{REBELS}-25: discovery of a dynamically cold disc galaxy at z = 7.31},
	volume = {535},
	issn = {0035-8711},
	shorttitle = {{REBELS}-25},
	url = {https://ui.adsabs.harvard.edu/abs/2024MNRAS.535.2068R},
	doi = {10.1093/mnras/stae2217},
	urldate = {2025-02-25},
	journal = {\mnras},
	publisher = {OUP},
	author = {Rowland, Lucie E. and Hodge, Jacqueline and Bouwens, Rychard and Piña, Pavel E. Mancera and Hygate, Alexander and Algera, Hiddo and Aravena, Manuel and Bowler, Rebecca and da Cunha, Elisabete and Dayal, Pratika and Ferrara, Andrea and Herard-Demanche, Thomas and Inami, Hanae and van Leeuwen, Ivana and de Looze, Ilse and Oesch, Pascal and Pallottini, Andrea and Phillips, Siân and Rybak, Matus and Schouws, Sander and Smit, Renske and Sommovigo, Laura and Stefanon, Mauro and van der Werf, Paul},
	month = dec,
	year = {2024},
	pages = {2068--2091},
}

@ARTICLE{fischer2025,
       author = {{Fischer}, C. and {Madden}, S.~C. and {Krabbe}, A. and {Polles}, F.~L. and {Fadda}, D. and {Tarantino}, E. and {Galliano}, F. and {Chen}, C.-H.~R. and {Abel}, N. and {Beck}, {\'A}. and {Belloir}, L. and {Bigiel}, F. and {Bolatto}, A. and {Chevance}, M. and {Colditz}, S. and {Fischer}, N. and {Green}, A. and {Hughes}, A. and {Indebetouw}, R. and {Iserlohe}, C. and {Ka{\'z}mierczak-Barthel}, M. and {Klein}, R. and {Lambert-Huyghe}, A. and {Lebouteiller}, V. and {Mikheeva}, E. and {Poglitsch}, A. and {Ramambason}, L. and {Reach}, W. and {Rubio}, M. and {Vacca}, W. and {Wong}, T. and {Zinnecker}, H.},
        title = "{LMC$^{+}$: Large-scale mapping of [C II] and [O III] in the LMC molecular ridge: I. Dataset and line ratio analyses}",
      journal = {\aap},
         year = 2025,
        month = oct,
       volume = {702},
          eid = {A273},
        pages = {A273},
          doi = {10.1051/0004-6361/202555833},
archivePrefix = {arXiv},
       eprint = {2509.09417},
 primaryClass = {astro-ph.GA},
       adsurl = {https://ui.adsabs.harvard.edu/abs/2025A&A...702A.273F},
}

@article{ubler_ionized_2018,
	title = {Ionized and {Molecular} {Gas} {Kinematics} in a z = 1.4 {Star}-forming {Galaxy}*},
	volume = {854},
	issn = {2041-8205},
	url = {https://dx.doi.org/10.3847/2041-8213/aaacfa},
	doi = {10.3847/2041-8213/aaacfa},
	language = {en},
	number = {2},
	urldate = {2025-02-25},
	journal = {\apjl},
	publisher = {The American Astronomical Society},
	author = {Übler, H. and Genzel, R. and Tacconi, L. J. and Schreiber, N. M. Förster and Neri, R. and Contursi, A. and Belli, S. and Nelson, E. J. and Lang, P. and Shimizu, T. T. and Davies, R. and Herrera-Camus, R. and Lutz, D. and Plewa, P. M. and Price, S. H. and Schuster, K. and Sternberg, A. and Tadaki, K. and Wisnioski, E. and Wuyts, S.},
	month = feb,
	year = {2018},
	pages = {L24},
}

@article{rizzo_alma-alpaka_2024,
	title = {The {ALMA}-{ALPAKA} survey - {II}. {Evolution} of turbulence in galaxy disks across cosmic time: {Difference} between cold and warm gas},
	volume = {689},
	copyright = {© The Authors 2024},
	issn = {0004-6361, 1432-0746},
	shorttitle = {The {ALMA}-{ALPAKA} survey - {II}. {Evolution} of turbulence in galaxy disks across cosmic time},
	url = {https://www.aanda.org/articles/aa/abs/2024/09/aa50455-24/aa50455-24.html},
	doi = {10.1051/0004-6361/202450455},
	language = {en},
	urldate = {2025-02-25},
	journal = {\aap},
	publisher = {EDP Sciences},
	author = {Rizzo, F. and Bacchini, C. and Kohandel, M. and Mascolo, L. Di and Fraternali, F. and Roman-Oliveira, F. and Zanella, A. and Popping, G. and Valentino, F. and Magdis, G. and Whitaker, K.},
	month = sep,
	year = {2024},
	pages = {A273},
}

@article{levy_edge-califa_2018,
	title = {The {EDGE}-{CALIFA} {Survey}: {Molecular} and {Ionized} {Gas} {Kinematics} in {Nearby} {Galaxies}},
	volume = {860},
	issn = {0004-637X},
	shorttitle = {The {EDGE}-{CALIFA} {Survey}},
	url = {https://dx.doi.org/10.3847/1538-4357/aac2e5},
	doi = {10.3847/1538-4357/aac2e5},
	language = {en},
	number = {2},
	urldate = {2025-02-27},
	journal = {\apj},
	publisher = {The American Astronomical Society},
	author = {Levy, Rebecca C. and Bolatto, Alberto D. and Teuben, Peter and Sánchez, Sebastián F. and Barrera-Ballesteros, Jorge K. and Blitz, Leo and Colombo, Dario and García-Benito, Rubén and Herrera-Camus, Rodrigo and Husemann, Bernd and Kalinova, Veselina and Lan, Tian and Leung, Gigi Y. C. and Mast, Damián and Utomo, Dyas and Ven, Glenn van de and Vogel, Stuart N. and Wong, Tony},
	month = jun,
	year = {2018},
	pages = {92},
}

@article{polletta_jwsts_2024,
	title = {{JWST}’s {PEARLS}: {Resolved} study of the stellar and dust components in starburst galaxies at cosmic noon},
	volume = {690},
	copyright = {https://creativecommons.org/licenses/by/4.0},
	issn = {0004-6361, 1432-0746},
	shorttitle = {{JWST}’s {PEARLS}},
	url = {https://www.aanda.org/10.1051/0004-6361/202450671},
	doi = {10.1051/0004-6361/202450671},
	language = {en},
	urldate = {2025-03-04},
	journal = {\aap},
	author = {Polletta, M. and Frye, B. L. and Garuda, N. and Willner, S. P. and Berta, S. and Kneissl, R. and Dole, H. and Jansen, R. A. and Lehnert, M. D. and Cohen, S. H. and Summers, J. and Windhorst, R. A. and D’Silva, J. C. J. and Koekemoer, A. M. and Coe, D. and Conselice, C. J. and Driver, S. P. and Grogin, N. A. and Marshall, M. A. and Nonino, M. and Ortiz Iii, R. and Pirzkal, N. and Robotham, A. and Ryan, R. E. and Willmer, C. N. A. and Yan, H. and Arumugam, V. and Cheng, C. and Gim, H. B. and Hathi, N. P. and Holwerda, B. and Kamieneski, P. and Keel, W. C. and Li, J. and Pascale, M. and Rottgering, H. and Smith, B. M. and Yun, M. S.},
	month = oct,
	year = {2024},
	pages = {A285},
}

@misc{danhaive_dawn_2025,
	title = {The dawn of disks: unveiling the turbulent ionised gas kinematics of the galaxy population at \$z{\textbackslash}sim4-6\$ with {JWST}/{NIRCam} grism spectroscopy},
	shorttitle = {The dawn of disks},
	url = {http://arxiv.org/abs/2503.21863},
	doi = {10.48550/arXiv.2503.21863},
	urldate = {2025-03-31},
	publisher = {arXiv},
	author = {Danhaive, A. Lola and Tacchella, Sandro and Übler, Hannah and Graaff, Anna de and Egami, Eiichi and Johnson, Benjamin D. and Sun, Fengwu and Arribas, Santiago and Bunker, Andrew J. and Carniani, Stefano and Jones, Gareth C. and Maiolino, Roberto and McClymont, William and Parlanti, Eleonora and Simmonds, Charlotte and Villanueva, Natalia C. and Baker, William M. and Jaffe, Daniel T. and Eisenstein, Daniel and Hainline, Kevin and Helton, Jakob M. and Ji, Zhiyuan and Lin, Xiaojing and Puskás, Dávid and Rieke, Marcia and Rinaldi, Pierluigi and Robertson, Brant and Scholz, Jan and Williams, Christina C. and Willmer, Christopher N. A.},
	month = mar,
	year = {2025},
	note = {arXiv:2503.21863 [astro-ph]},
}

@ARTICLE{Zhang2017,
       author = {{Zhang}, Kai and {Yan}, Renbin and {Bundy}, Kevin and {Bershady}, Matthew and {Haffner}, L. Matthew and {Walterbos}, Ren{\'e} and {Maiolino}, Roberto and {Tremonti}, Christy and {Thomas}, Daniel and {Drory}, Niv and {Jones}, Amy and {Belfiore}, Francesco and {S{\'a}nchez}, Sebastian F. and {Diamond-Stanic}, Aleksandar M. and {Bizyaev}, Dmitry and {Nitschelm}, Christian and {Andrews}, Brett and {Brinkmann}, Jon and {Brownstein}, Joel R. and {Cheung}, Edmond and {Li}, Cheng and {Law}, David R. and {Roman Lopes}, Alexandre and {Oravetz}, Daniel and {Pan}, Kaike and {Storchi Bergmann}, Thaisa and {Simmons}, Audrey},
        title = "{SDSS-IV MaNGA: the impact of diffuse ionized gas on emission-line ratios, interpretation of diagnostic diagrams and gas metallicity measurements}",
      journal = {\mnras},
         year = 2017,
        month = apr,
       volume = {466},
       number = {3},
        pages = {3217-3243},
          doi = {10.1093/mnras/stw3308},
archivePrefix = {arXiv},
       eprint = {1612.02000},
 primaryClass = {astro-ph.GA},
       adsurl = {https://ui.adsabs.harvard.edu/abs/2017MNRAS.466.3217Z},
}

@ARTICLE{chen2017,
       author = {{Chen}, Chian-Chou and {Hodge}, J.~A. and {Smail}, Ian and {Swinbank}, A.~M. and {Walter}, Fabian and {Simpson}, J.~M. and {Calistro Rivera}, Gabriela and {Bertoldi}, F. and {Brandt}, W.~N. and {Chapman}, S.~C. and {da Cunha}, Elisabete and {Dannerbauer}, H. and {De Breuck}, C. and {Harrison}, C.~M. and {Ivison}, R.~J. and {Karim}, A. and {Knudsen}, K.~K. and {Wardlow}, J.~L. and {Wei{\ss}}, A. and {van der Werf}, P.~P.},
        title = "{A Spatially Resolved Study of Cold Dust, Molecular Gas, H II Regions, and Stars in the z = 2.12 Submillimeter Galaxy ALESS67.1}",
      journal = {\apj},
         year = 2017,
        month = sep,
       volume = {846},
       number = {2},
          eid = {108},
        pages = {108},
          doi = {10.3847/1538-4357/aa863a},
archivePrefix = {arXiv},
       eprint = {1708.08937},
 primaryClass = {astro-ph.GA},
       adsurl = {https://ui.adsabs.harvard.edu/abs/2017ApJ...846..108C},
}

@ARTICLE{chen2015,
       author = {{Chen}, Chian-Chou and {Smail}, Ian and {Swinbank}, A.~M. and {Simpson}, J.~M. and {Ma}, Cheng-Jiun and {Alexander}, D.~M. and {Biggs}, A.~D. and {Brandt}, W.~N. and {Chapman}, S.~C. and {Coppin}, K.~E.~K. and {Danielson}, A.~L.~R. and {Dannerbauer}, H. and {Edge}, A.~C. and {Greve}, T.~R. and {Ivison}, R.~J. and {Karim}, A. and {Menten}, Karl M. and {Schinnerer}, E. and {Walter}, F. and {Wardlow}, J.~L. and {Wei{\ss}}, A. and {van der Werf}, P.~P.},
        title = "{An ALMA Survey of Submillimeter Galaxies in the Extended Chandra Deep Field South: Near-infrared Morphologies and Stellar Sizes}",
      journal = {\apj},
         year = 2015,
        month = feb,
       volume = {799},
       number = {2},
          eid = {194},
        pages = {194},
          doi = {10.1088/0004-637X/799/2/194},
archivePrefix = {arXiv},
       eprint = {1412.0668},
 primaryClass = {astro-ph.GA},
       adsurl = {https://ui.adsabs.harvard.edu/abs/2015ApJ...799..194C},
}

@ARTICLE{james2016,
       author = {{James}, Bethan L. and {Auger}, Matthew and {Aloisi}, Alessandra and {Calzetti}, Daniela and {Kewley}, Lisa},
        title = "{Resolving Ionization and Metallicity on Parsec Scales across Mrk 71 with HST-WFC3}",
      journal = {\apj},
         year = 2016,
        month = jan,
       volume = {816},
       number = {1},
          eid = {40},
        pages = {40},
          doi = {10.3847/0004-637X/816/1/40},
archivePrefix = {arXiv},
       eprint = {1510.02447},
 primaryClass = {astro-ph.GA},
       adsurl = {https://ui.adsabs.harvard.edu/abs/2016ApJ...816...40J},
}

@ARTICLE{monrealibero2023,
       author = {{Monreal-Ibero}, Ana and {Weilbacher}, Peter M. and {Micheva}, Genoveva and {Kollatschny}, Wolfram and {Maseda}, Michael},
        title = "{UM 462, a local green pea galaxy analogue under the MUSE magnifying glass}",
      journal = {\aap},
         year = 2023,
        month = jun,
       volume = {674},
          eid = {A210},
        pages = {A210},
          doi = {10.1051/0004-6361/202345891},
archivePrefix = {arXiv},
       eprint = {2304.06096},
 primaryClass = {astro-ph.GA},
       adsurl = {https://ui.adsabs.harvard.edu/abs/2023A&A...674A.210M},
}

@ARTICLE{Poetrodjojo2018,
       author = {{Poetrodjojo}, Henry and {Groves}, Brent and {Kewley}, Lisa J. and {Medling}, Anne M. and {Sweet}, Sarah M. and {van de Sande}, Jesse and {Sanchez}, Sebastian F. and {Bland-Hawthorn}, Joss and {Brough}, Sarah and {Bryant}, Julia J. and {Cortese}, Luca and {Croom}, Scott M. and {L{\'o}pez-S{\'a}nchez}, {\'A}ngel R. and {Richards}, Samuel N. and {Zafar}, Tayyaba and {Lawrence}, Jon S. and {Lorente}, Nuria P.~F. and {Owers}, Matt S. and {Scott}, Nicholas},
        title = "{The SAMI Galaxy Survey: Spatially resolved metallicity and ionization mapping}",
      journal = {\mnras},
         year = 2018,
        month = oct,
       volume = {479},
       number = {4},
        pages = {5235-5265},
          doi = {10.1093/mnras/sty1782},
archivePrefix = {arXiv},
       eprint = {1807.01522},
 primaryClass = {astro-ph.GA},
       adsurl = {https://ui.adsabs.harvard.edu/abs/2018MNRAS.479.5235P},
}

@article{sugahara_rioja_2025,
	title = {{RIOJA}. {Complex} {Dusty} {Starbursts} in a {Major} {Merger} {B14}-65666 at z = 7.15},
	volume = {981},
	issn = {0004-637X},
	url = {https://ui.adsabs.harvard.edu/abs/2025ApJ...981..135S},
	doi = {10.3847/1538-4357/adb02a},
	urldate = {2025-05-19},
	journal = {\apj},
	publisher = {IOP},
	author = {Sugahara, Yuma and Álvarez-Márquez, Javier and Hashimoto, Takuya and Colina, Luis and Inoue, Akio K. and Costantin, Luca and Fudamoto, Yoshinobu and Mawatari, Ken and Ren, Yi W. and Arribas, Santiago and Bakx, Tom J. L. C. and Blanco-Prieto, Carmen and Ceverino, Daniel and Crespo Gómez, Alejandro and Hagimoto, Masato and Hashigaya, Takeshi and Marques-Chaves, Rui and Matsuo, Hiroshi and Nakazato, Yurina and Pereira-Santaella, Miguel and Tamura, Yoichi and Usui, Mitsutaka and Yoshida, Naoki},
	month = mar,
	year = {2025},
	pages = {135},
}

@article{ejdetjarn_giant_2022,
	title = {From giant clumps to clouds – {III}. {The} connection between star formation and turbulence in the {ISM}},
	volume = {514},
	issn = {0035-8711},
	url = {https://doi.org/10.1093/mnras/stac1414},
	doi = {10.1093/mnras/stac1414},
	number = {1},
	urldate = {2025-08-13},
	journal = {\mnras},
	author = {Ejdetjärn, Timmy and Agertz, Oscar and Östlin, Göran and Renaud, Florent and Romeo, Alessandro B},
	month = jul,
	year = {2022},
	pages = {480--496},
}

@article{harikane_large_2020,
	title = {Large {Population} of {ALMA} {Galaxies} at z{\textgreater}6 with {Very} {High} [{OIII}]88um to [{CII}]158um {Flux} {Ratios}: {Evidence} of {Extremely} {High} {Ionization} {Parameter} or {PDR} {Deficit}?},
	volume = {896},
	issn = {0004-637X, 1538-4357},
	shorttitle = {Large {Population} of {ALMA} {Galaxies} at z{\textgreater}6 with {Very} {High} [{OIII}]88um to [{CII}]158um {Flux} {Ratios}},
	url = {http://arxiv.org/abs/1910.10927},
	doi = {10.3847/1538-4357/ab94bd},
	number = {2},
	urldate = {2025-08-19},
	journal = {\apj},
	author = {Harikane, Yuichi and Ouchi, Masami and Inoue, Akio K. and Matsuoka, Yoshiki and Tamura, Yoichi and Bakx, Tom and Fujimoto, Seiji and Moriwaki, Kana and Ono, Yoshiaki and Nagao, Tohru and Tadaki, Ken-ichi and Kojima, Takashi and Shibuya, Takatoshi and Egami, Eiichi and Ferrara, Andrea and Gallerani, Simona and Hashimoto, Takuya and Kohno, Kotaro and Matsuda, Yuichi and Matsuo, Hiroshi and Pallottini, Andrea and Sugahara, Yuma and Vallini, Livia},
	month = jun,
	year = {2020},
	note = {arXiv:1910.10927 [astro-ph]},
	pages = {93},
}

@article{su_almaquest_2022,
	title = {The {ALMaQUEST} {Survey}. {VIII}. {What} {Causes} the {Discrepancy} in the {Velocity} between the {CO} and {H$\alpha$} {Rotation} {Curves} in {Galaxies}?},
	volume = {934},
	issn = {0004-637X},
	url = {https://ui.adsabs.harvard.edu/abs/2022ApJ...934..173S},
	doi = {10.3847/1538-4357/ac77fd},
	urldate = {2025-09-04},
	journal = {\apj},
	author = {Su, Yung-Chau and Lin, Lihwai and Pan, Hsi-An and López Cobá, Carlos and Hsieh, Bau-Ching and Sánchez, Sebastián F. and Thorp, Mallory D. and Bureau, Martin and Ellison, Sara L.},
	month = aug,
	year = {2022},
	pages = {173},
}

@article{rizzo_alma-alpaka_2023,
	title = {The {ALMA}-{ALPAKA} survey. {I}. {High}-resolution {CO} and [{CI}] kinematics of star-forming galaxies at z = 0.5-3.5},
	volume = {679},
	issn = {0004-6361},
	url = {https://ui.adsabs.harvard.edu/abs/2023A&A...679A.129R},
	doi = {10.1051/0004-6361/202346444},
	urldate = {2026-01-26},
	journal = {\aap},
	publisher = {EDP},
	author = {Rizzo, F. and Roman-Oliveira, F. and Fraternali, F. and Frickmann, D. and Valentino, F. M. and Brammer, G. and Zanella, A. and Kokorev, V. and Popping, G. and Whitaker, K. E. and Kohandel, M. and Magdis, G. E. and Di Mascolo, L. and Ikeda, R. and Jin, S. and Toft, S.},
	month = nov,
	year = {2023},
	pages = {A129},
}

@article{hollenbach_photodissociation_1999,
	title = {Photodissociation regions in the interstellar medium of galaxies},
	volume = {71},
	issn = {0034-6861},
	url = {https://ui.adsabs.harvard.edu/abs/1999RvMP...71..173H},
	doi = {10.1103/RevModPhys.71.173},
	urldate = {2026-02-23},
	journal = {Rev. Mod. Phys.},
	publisher = {APS},
	author = {Hollenbach, D. J. and Tielens, A. G. G. M.},
	month = jan,
	year = {1999},
	pages = {173--230},
}

@article{wolfire_neutral_2003,
	title = {Neutral {Atomic} {Phases} of the {Interstellar} {Medium} in the {Galaxy}},
	volume = {587},
	issn = {0004-637X},
	url = {https://ui.adsabs.harvard.edu/abs/2003ApJ...587..278W},
	doi = {10.1086/368016},
	urldate = {2026-02-23},
	journal = {\apj},
	publisher = {IOP},
	author = {Wolfire, Mark G. and McKee, Christopher F. and Hollenbach, David and Tielens, A. G. G. M.},
	month = apr,
	year = {2003},
	pages = {278--311},
}

@article{ferrara_physical_2019,
	title = {A physical model for [{C} {II}] line emission from galaxies},
	volume = {489},
	issn = {0035-8711},
	url = {https://ui.adsabs.harvard.edu/abs/2019MNRAS.489....1F},
	doi = {10.1093/mnras/stz2031},
	urldate = {2026-02-23},
	journal = {\mnras},
	publisher = {OUP},
	author = {Ferrara, A. and Vallini, L. and Pallottini, A. and Gallerani, S. and Carniani, S. and Kohandel, M. and Decataldo, D. and Behrens, C.},
	month = oct,
	year = {2019},
	pages = {1--12},
}

@article{algera_accurate_2024,
	title = {Accurate simultaneous constraints on the dust mass, temperature, and emissivity index of a galaxy at redshift 7.31},
	volume = {533},
	issn = {0035-8711},
	url = {https://ui.adsabs.harvard.edu/abs/2024MNRAS.533.3098A},
	doi = {10.1093/mnras/stae1994},
	urldate = {2026-02-23},
	journal = {\mnras},
	publisher = {OUP},
	author = {Algera, Hiddo S. B. and Inami, Hanae and De Looze, Ilse and Ferrara, Andrea and Hirashita, Hiroyuki and Aravena, Manuel and Bakx, Tom and Bouwens, Rychard and Bowler, Rebecca A. A. and Da Cunha, Elisabete and Dayal, Pratika and Fudamoto, Yoshinobu and Hodge, Jacqueline and Hygate, Alexander and van Leeuwen, Ivana and Nanayakkara, Themiya and Palla, Marco and Pallottini, Andrea and Rowland, Lucie and Smit, Renske and Sommovigo, Laura and Stefanon, Mauro and Vijayan, Aswin P. and van der Werf, Paul},
	month = sep,
	year = {2024},
	pages = {3098--3113},
}

@article{algera_cold_2024,
	title = {Cold dust and low [{O} {III}]/[{C} {II}] ratios: an evolved star-forming population at redshift 7},
	volume = {527},
	issn = {0035-8711},
	shorttitle = {Cold dust and low [{O} {III}]/[{C} {II}] ratios},
	url = {https://ui.adsabs.harvard.edu/abs/2024MNRAS.527.6867A},
	doi = {10.1093/mnras/stad3111},
	urldate = {2026-02-23},
	journal = {\mnras},
	publisher = {OUP},
	author = {Algera, Hiddo S. B. and Inami, Hanae and Sommovigo, Laura and Fudamoto, Yoshinobu and Schneider, Raffaella and Graziani, Luca and Dayal, Pratika and Bouwens, Rychard and Aravena, Manuel and da Cunha, Elisabete and Ferrara, Andrea and Hygate, Alexander P. S. and van Leeuwen, Ivana and De Looze, Ilse and Palla, Marco and Pallottini, Andrea and Smit, Renske and Stefanon, Mauro and Topping, Michael and van der Werf, Paul P.},
	month = jan,
	year = {2024},
	pages = {6867--6887},
}

@article{algera_rebels-ifu_2026,
	title = {{REBELS}-{IFU}: dust build-up in massive galaxies at redshift 7},
	volume = {545},
	issn = {0035-8711},
	shorttitle = {{REBELS}-{IFU}},
	url = {https://ui.adsabs.harvard.edu/abs/2026MNRAS.545f1897A},
	doi = {10.1093/mnras/staf1897},
	urldate = {2026-02-23},
	journal = {\mnras},
	publisher = {OUP},
	author = {Algera, Hiddo S. B. and Rowland, Lucie and Stefanon, Mauro and Palla, Marco and Sommovigo, Laura and Inami, Hanae and Bouwens, Rychard and Aravena, Manuel and Bowler, Rebecca A. A. and Dayal, Pratika and De Looze, Ilse and Ferrara, Andrea and Fisher, Rebecca and Graziani, Luca and Gulis, Cindy and Heintz, Kasper and Hodge, Jacqueline and Laza-Ramos, Andrés and van Leeuwen, Ivana and Pallottini, Andrea and Phillips, Siân and Schouws, Sander and Smit, Renske and Stark, Daniel P. and van der Werf, Paul},
	month = jan,
	year = {2026},
	pages = {staf1897},
}

@article{hygate_alma_2023,
	title = {The {ALMA} {REBELS} {Survey}: discovery of a massive, highly star-forming, and morphologically complex {ULIRG} at z = 7.31},
	volume = {524},
	issn = {0035-8711},
	shorttitle = {The {ALMA} {REBELS} {Survey}},
	url = {https://ui.adsabs.harvard.edu/abs/2023MNRAS.524.1775H},
	doi = {10.1093/mnras/stad1212},
	urldate = {2026-02-23},
	journal = {\mnras},
	publisher = {OUP},
	author = {Hygate, A. P. S. and Hodge, J. A. and da Cunha, E. and Rybak, M. and Schouws, S. and Inami, H. and Stefanon, M. and Graziani, L. and Schneider, R. and Dayal, P. and Bouwens, R. J. and Smit, R. and Bowler, R. A. A. and Endsley, R. and Gonzalez, V. and Oesch, P. A. and Stark, D. P. and Algera, H. S. B. and Aravena, M. and Barrufet, L. and Ferrara, A. and Fudamoto, Y. and Hilhorst, J. H. A. and De Looze, I. and Nanayakkara, T. and Pallottini, A. and Riechers, D. A. and Sommovigo, L. and Topping, M. W. and van der Werf, P.},
	month = sep,
	year = {2023},
	pages = {1775--1795},
}

@article{bouwens_reionization_2022,
	title = {Reionization {Era} {Bright} {Emission} {Line} {Survey}: {Selection} and {Characterization} of {Luminous} {Interstellar} {Medium} {Reservoirs} in the z {\textgreater} 6.5 {Universe}},
	volume = {931},
	issn = {0004-637X},
	shorttitle = {Reionization {Era} {Bright} {Emission} {Line} {Survey}},
	url = {https://ui.adsabs.harvard.edu/abs/2022ApJ...931..160B},
	doi = {10.3847/1538-4357/ac5a4a},
	urldate = {2026-02-23},
	journal = {\apj},
	publisher = {IOP},
	author = {Bouwens, R. J. and Smit, R. and Schouws, S. and Stefanon, M. and Bowler, R. and Endsley, R. and Gonzalez, V. and Inami, H. and Stark, D. and Oesch, P. and Hodge, J. and Aravena, M. and da Cunha, E. and Dayal, P. and de Looze, I. and Ferrara, A. and Fudamoto, Y. and Graziani, L. and Li, C. and Nanayakkara, T. and Pallottini, A. and Schneider, R. and Sommovigo, L. and Topping, M. and van der Werf, P. and Algera, H. and Barrufet, L. and Hygate, A. and Labbé, I. and Riechers, D. and Witstok, J.},
	month = jun,
	year = {2022},
	pages = {160},
}

@article{stefanon_brightest_2019,
	title = {The {Brightest} z ≳ 8 {Galaxies} over the {COSMOS} {UltraVISTA} {Field}},
	volume = {883},
	issn = {0004-637X},
	url = {https://ui.adsabs.harvard.edu/abs/2019ApJ...883...99S},
	doi = {10.3847/1538-4357/ab3792},
	urldate = {2026-02-23},
	journal = {\apj},
	publisher = {IOP},
	author = {Stefanon, Mauro and Labbé, Ivo and Bouwens, Rychard J. and Oesch, Pascal and Ashby, Matthew L. N. and Caputi, Karina I. and Franx, Marijn and Fynbo, Johan P. U. and Illingworth, Garth D. and Le Fèvre, Olivier and Marchesini, Danilo and McCracken, Henry J. and Milvang-Jensen, Bo and Muzzin, Adam and van Dokkum, Pieter},
	month = sep,
	year = {2019},
	pages = {99},
}

@article{czekala_molecules_2021,
	title = {Molecules with {ALMA} at {Planet}-forming {Scales} ({MAPS}). {II}. {CLEAN} {Strategies} for {Synthesizing} {Images} of {Molecular} {Line} {Emission} in {Protoplanetary} {Disks}},
	volume = {257},
	issn = {0067-0049},
	url = {https://ui.adsabs.harvard.edu/abs/2021ApJS..257....2C},
	doi = {10.3847/1538-4365/ac1430},
	urldate = {2026-02-23},
	journal = {\apjs},
	publisher = {IOP},
	author = {Czekala, Ian and Loomis, Ryan A. and Teague, Richard and Booth, Alice S. and Huang, Jane and Cataldi, Gianni and Ilee, John D. and Law, Charles J. and Walsh, Catherine and Bosman, Arthur D. and Guzmán, Viviana V. and Le Gal, Romane and Öberg, Karin I. and Yamato, Yoshihide and Aikawa, Yuri and Andrews, Sean M. and Bae, Jaehan and Bergin, Edwin A. and Bergner, Jennifer B. and Cleeves, L. Ilsedore and Kurtovic, Nicolas T. and Ménard, François and Nomura, Hideko and Pérez, Laura M. and Qi, Chunhua and Schwarz, Kamber R. and Tsukagoshi, Takashi and Waggoner, Abygail R. and Wilner, David J. and Zhang, Ke},
	month = nov,
	year = {2021},
	pages = {2},
}

@article{jorsater_high_1995,
	title = {High {Resolution} {Neutral} {Hydrogen} {Observations} of the {Barred} {Spiral} {Galaxy} {NGC} 1365},
	volume = {110},
	issn = {0004-6256},
	url = {https://ui.adsabs.harvard.edu/abs/1995AJ....110.2037J},
	doi = {10.1086/117668},
	urldate = {2026-02-23},
	journal = {\aj},
	publisher = {IOP},
	author = {Jorsater, Steven and van Moorsel, Gustaaf A.},
	month = nov,
	year = {1995},
	pages = {2037},
}

@article{nakazato_unveiling_2026,
	title = {Unveiling the {Ionized} and {Neutral} {ISM} at z {\textgreater} 10: {The} {Origin} of [{O} {III}] /[{C} {II}] {Ratios} from a {Subparsec} {Resolution} {Radiative} {Transfer} {Simulation}},
	volume = {998},
	issn = {0004-637X},
	shorttitle = {Unveiling the {Ionized} and {Neutral} {ISM} at z {\textgreater} 10},
	url = {https://ui.adsabs.harvard.edu/abs/2026ApJ...998...26N},
	doi = {10.3847/1538-4357/ae2fee},
	urldate = {2026-02-23},
	journal = {\apj},
	publisher = {IOP},
	author = {Nakazato, Yurina and Sugimura, Kazuyuki and Inoue, Akio K. and Ricotti, Massimo},
	month = feb,
	year = {2026},
	pages = {26},
}

@article{inoue_detection_2016,
	title = {Detection of an oxygen emission line from a high-redshift galaxy in the reionization epoch},
	volume = {352},
	issn = {0036-8075},
	url = {https://ui.adsabs.harvard.edu/abs/2016Sci...352.1559I},
	doi = {10.1126/science.aaf0714},
	urldate = {2026-02-23},
	journal = {\sci},
	author = {Inoue, Akio K. and Tamura, Yoichi and Matsuo, Hiroshi and Mawatari, Ken and Shimizu, Ikkoh and Shibuya, Takatoshi and Ota, Kazuaki and Yoshida, Naoki and Zackrisson, Erik and Kashikawa, Nobunari and Kohno, Kotaro and Umehata, Hideki and Hatsukade, Bunyo and Iye, Masanori and Matsuda, Yuichi and Okamoto, Takashi and Yamaguchi, Yuki},
	month = jun,
	year = {2016},
	pages = {1559--1562},
}

@article{hashimoto_detections_2019,
	title = {Detections of [{O} {III}] 88 $\mu$m in two quasars in the reionization epoch},
	volume = {71},
	issn = {0004-6264},
	url = {https://ui.adsabs.harvard.edu/abs/2019PASJ...71..109H},
	doi = {10.1093/pasj/psz094},
	urldate = {2026-02-23},
	journal = {\pasj},
	publisher = {OUP},
	author = {Hashimoto, Takuya and Inoue, Akio K. and Tamura, Yoichi and Matsuo, Hiroshi and Mawatari, Ken and Yamaguchi, Yuki},
	month = dec,
	year = {2019},
	pages = {109},
}

@article{laporte_absence_2019,
	title = {The absence of [{C} {II}] 158 $\mu$m emission in spectroscopically confirmed galaxies at z {\textgreater} 8},
	volume = {487},
	issn = {0035-8711},
	url = {https://ui.adsabs.harvard.edu/abs/2019MNRAS.487L..81L},
	doi = {10.1093/mnrasl/slz094},
	urldate = {2026-02-23},
	journal = {\mnras},
	publisher = {OUP},
	author = {Laporte, N. and Katz, H. and Ellis, R. S. and Lagache, G. and Bauer, F. E. and Boone, F. and Inoue, A. K. and Hashimoto, T. and Matsuo, H. and Mawatari, K. and Tamura, Y.},
	month = jul,
	year = {2019},
	pages = {L81--L85},
}

@article{witstok_dual_2022,
	title = {Dual constraints with {ALMA}: new [{O} {III}] 88-$\mu$m and dust-continuum observations reveal the {ISM} conditions of luminous {LBGs} at z   7},
	volume = {515},
	issn = {0035-8711},
	shorttitle = {Dual constraints with {ALMA}},
	url = {https://ui.adsabs.harvard.edu/abs/2022MNRAS.515.1751W},
	doi = {10.1093/mnras/stac1905},
	urldate = {2026-02-23},
	journal = {\mnras},
	publisher = {OUP},
	author = {Witstok, Joris and Smit, Renske and Maiolino, Roberto and Kumari, Nimisha and Aravena, Manuel and Boogaard, Leindert and Bouwens, Rychard and Carniani, Stefano and Hodge, Jacqueline A. and Jones, Gareth C. and Stefanon, Mauro and van der Werf, Paul and Schouws, Sander},
	month = sep,
	year = {2022},
	pages = {1751--1773},
}

@article{diaz-santos_herschelpacs_2017,
	title = {A {Herschel}/{PACS} {Far}-infrared {Line} {Emission} {Survey} of {Local} {Luminous} {Infrared} {Galaxies}},
	volume = {846},
	issn = {0004-637X},
	url = {https://ui.adsabs.harvard.edu/abs/2017ApJ...846...32D},
	doi = {10.3847/1538-4357/aa81d7},
	urldate = {2026-02-23},
	journal = {\apj},
	publisher = {IOP},
	author = {Díaz-Santos, T. and Armus, L. and Charmandaris, V. and Lu, N. and Stierwalt, S. and Stacey, G. and Malhotra, S. and van der Werf, P. P. and Howell, J. H. and Privon, G. C. and Mazzarella, J. M. and Goldsmith, P. F. and Murphy, E. J. and Barcos-Muñoz, L. and Linden, S. T. and Inami, H. and Larson, K. L. and Evans, A. S. and Appleton, P. and Iwasawa, K. and Lord, S. and Sanders, D. B. and Surace, J. A.},
	month = sep,
	year = {2017},
	pages = {32},
}

@article{schouws_detection_2025,
	title = {Detection of [{O} {III}]88 $\mu$m in {JADES}-{GS}-z14-0 at z = 14.1793},
	volume = {988},
	issn = {0004-637X},
	url = {https://ui.adsabs.harvard.edu/abs/2025ApJ...988...19S},
	doi = {10.3847/1538-4357/adbf1b},
	urldate = {2026-05-07},
	journal = {\apj},
	publisher = {IOP},
	author = {Schouws, Sander and Bouwens, Rychard J. and Ormerod, Katherine and Smit, Renske and Algera, Hiddo and Sommovigo, Laura and Hodge, Jacqueline and Ferrara, Andrea and Oesch, Pascal A. and Rowland, Lucie E. and van Leeuwen, Ivana and Stefanon, Mauro and Herard-Demanche, Thomas and Fudamoto, Yoshinobu and Röttgering, Huub and van der Werf, Paul},
	month = jul,
	year = {2025},
	pages = {19},
}

@article{carniani_eventful_2025,
	title = {The eventful life of a luminous galaxy at z = 14: metal enrichment, feedback, and low gas fraction?},
	volume = {696},
	issn = {0004-6361},
	shorttitle = {The eventful life of a luminous galaxy at z = 14},
	url = {https://ui.adsabs.harvard.edu/abs/2025A&A...696A..87C},
	doi = {10.1051/0004-6361/202452451},
	urldate = {2026-05-07},
	journal = {\aap},
	publisher = {EDP},
	author = {Carniani, Stefano and D'Eugenio, Francesco and Ji, Xihan and Parlanti, Eleonora and Scholtz, Jan and Sun, Fengwu and Venturi, Giacomo and Bakx, Tom J. L. C. and Curti, Mirko and Maiolino, Roberto and Tacchella, Sandro and Zavala, Jorge A. and Hainline, Kevin and Witstok, Joris and Johnson, Benjamin D. and Alberts, Stacey and Bunker, Andrew J. and Charlot, Stéphane and Eisenstein, Daniel J. and Helton, Jakob M. and Jakobsen, Peter and Kumari, Nimisha and Robertson, Brant and Saxena, Aayush and Übler, Hannah and Williams, Christina C. and Willmer, Christopher N. A. and Willott, Chris},
	month = apr,
	year = {2025},
	pages = {A87},
}

@article{naidu_cosmic_2026,
	title = {A {Cosmic} {Miracle}: {A} {Remarkably} {Luminous} {Galaxy} at zspec = 14.44 {Confirmed} with {JWST}},
	volume = {9},
	issn = {2565-6120},
	shorttitle = {A {Cosmic} {Miracle}},
	url = {https://ui.adsabs.harvard.edu/abs/2026OJAp....956033N},
	doi = {10.33232/001c.156033},
	urldate = {2026-05-07},
	journal = {Open J. Astrophys.},
	author = {Naidu, Rohan P. and Oesch, Pascal A. and Brammer, Gabriel and Weibel, Andrea and Li, Yijia and Matthee, Jorryt and Chisolm, John and Pollock, Clara L. and Heintz, Kasper E. and Johnson, Benjamin D. and Shen, Xuejian and Hviding, Raphael E. and Leja, Joel and Tacchella, Sandro and Ganguly, Arpita and Witten, Callum and Atek, Hakim and Belli, Siro and Bose, Sownak and Bouwens, Rychard and Dayal, Pratika and Decarli, Roberto and de Graaff, Anna and Fudamoto, Yoshinobu and Giovinazzo, Emma and Greene, Jenny E. and Illingworth, Garth and Inoue, Akio K. and Kane, Sarah G. and Labbe, Ivo and Leonova, Ecaterina and Marques-Chaves, Rui and Meyer, Roman A. and Nelson, Erica J. and Roberts-Borsani, Guido and Schaerer, Daniel and Simcoe, Robert A. and Stefanon, Mauro and Sugahara, Yuma and Toft, Sune and van der Wel, Arjen and van Dokkum, Pieter and Walter, Fabian and Watson, Darrach and Weaver, John R. and Whitaker, Katherine E.},
	month = jan,
	year = {2026},
	pages = {56033},
}

@article{glazebrook_massive_2024,
	title = {A massive galaxy that formed its stars at z ≈ 11},
	volume = {628},
	issn = {0028-0836},
	url = {https://ui.adsabs.harvard.edu/abs/2024Natur.628..277G},
	doi = {10.1038/s41586-024-07191-9},
	urldate = {2026-05-07},
	journal = {\nat},
	author = {Glazebrook, Karl and Nanayakkara, Themiya and Schreiber, Corentin and Lagos, Claudia and Kawinwanichakij, Lalitwadee and Jacobs, Colin and Chittenden, Harry and Brammer, Gabriel and Kacprzak, Glenn G. and Labbe, Ivo and Marchesini, Danilo and Marsan, Z. Cemile and Oesch, Pascal A. and Papovich, Casey and Remus, Rhea-Silvia and Tran, Kim-Vy H. and Esdaile, James and Chandro-Gomez, Angel},
	month = apr,
	year = {2024},
	pages = {277--281},
}

@article{xiao_panoramic_2025,
	title = {{PANORAMIC}: {Discovery} of an ultra-massive grand-design spiral galaxy at z ∼ 5.2},
	volume = {696},
	issn = {0004-6361},
	shorttitle = {{PANORAMIC}},
	url = {https://ui.adsabs.harvard.edu/abs/2025A&A...696A.156X},
	doi = {10.1051/0004-6361/202453487},
	urldate = {2026-05-07},
	journal = {\aap},
	publisher = {EDP},
	author = {Xiao, Mengyuan and Williams, Christina C. and Oesch, Pascal A. and Elbaz, David and Dessauges-Zavadsky, Miroslava and Marques-Chaves, Rui and Bing, Longji and Ji, Zhiyuan and Weibel, Andrea and Bezanson, Rachel and Brammer, Gabriel and Casey, Caitlin and Cloonan, Aidan P. and Daddi, Emanuele and Dayal, Pratika and Faisst, Andreas L. and Franx, Marijn and Glazebrook, Karl and Hutter, Anne and Kartaltepe, Jeyhan S. and Labbe, Ivo and Lagache, Guilaine and Lim, Seunghwan and Magnelli, Benjamin and Martinez, Felix and Maseda, Michael V. and Nanayakkara, Themiya and Schaerer, Daniel and Whitaker, Katherine E.},
	month = apr,
	year = {2025},
	pages = {A156},
}

@article{shapley_aurora_2025,
	title = {The {AURORA} {Survey}: {An} {Extraordinarily} {Mature}, {Star}-forming {Galaxy} at z ∼ 7},
	volume = {981},
	issn = {0004-637X},
	shorttitle = {The {AURORA} {Survey}},
	url = {https://ui.adsabs.harvard.edu/abs/2025ApJ...981..167S},
	doi = {10.3847/1538-4357/adaf98},
	urldate = {2026-05-07},
	journal = {\apj},
	publisher = {IOP},
	author = {Shapley, Alice E. and Sanders, Ryan L. and Topping, Michael W. and Reddy, Naveen A. and Pahl, Anthony J. and Oesch, Pascal A. and Berg, Danielle A. and Bouwens, Rychard J. and Brammer, Gabriel and Carnall, Adam C. and Cullen, Fergus and Davé, Romeel and Dunlop, James S. and Ellis, Richard S. and Förster Schreiber, N. M. and Furlanetto, Steven R. and Glazebrook, Karl and Illingworth, Garth D. and Jones, Tucker and Kriek, Mariska and McLeod, Derek J. and McLure, Ross J. and Narayanan, Desika and Pettini, Max and Schaerer, Daniel and Stark, Daniel P. and Steidel, Charles C. and Tang, Mengtao and Clarke, Leonardo and Donnan, Callum T. and Kehoe, Emily},
	month = mar,
	year = {2025},
	pages = {167},
}

@misc{algera_rebels-ifu_2025,
	title = {{REBELS}-{IFU}: on the origin of the elevated [{OIII}]/[{CII}] ratios in the early {Universe}},
	shorttitle = {{REBELS}-{IFU}},
	url = {https://ui.adsabs.harvard.edu/abs/2025arXiv250916071A},
	doi = {10.48550/arXiv.2509.16071},
	urldate = {2026-05-07},
	publisher = {arXiv},
	author = {Algera, Hiddo and Rowland, Lucie and Smit, Renske and Fisher, Rebecca and Ramambason, Lise and Kumari, Nimisha and Vallini, Livia and Inami, Hanae and Nanayakkara, Themiya and Stefanon, Mauro and Aravena, Manuel and Bakx, Tom and Bouwens, Rychard and Bowler, Rebecca and Cescon, Karin and Chen, Chian-Chou and Dayal, Pratika and De Looze, Ilse and Ferrara, Andrea and Fudamoto, Yoshinobu and Komarova, Lena and van Leeuwen, Ivana and Ormerod, Katherine and Schouws, Sander and Sommovigo, Laura and Vijayan, Aswin and Wang, Wei-Hao and van der Werf, Paul and Witstok, Joris},
	month = sep,
	year = {2025},
}

@article{inami_alma_2022,
	title = {The {ALMA} {REBELS} {Survey}: dust continuum detections at z {\textgreater} 6.5},
	volume = {515},
	issn = {0035-8711},
	shorttitle = {The {ALMA} {REBELS} {Survey}},
	url = {https://ui.adsabs.harvard.edu/abs/2022MNRAS.515.3126I},
	doi = {10.1093/mnras/stac1779},
	urldate = {2026-05-07},
	journal = {\mnras},
	publisher = {OUP},
	author = {Inami, Hanae and Algera, Hiddo S. B. and Schouws, Sander and Sommovigo, Laura and Bouwens, Rychard and Smit, Renske and Stefanon, Mauro and Bowler, Rebecca A. A. and Endsley, Ryan and Ferrara, Andrea and Oesch, Pascal and Stark, Daniel and Aravena, Manuel and Barrufet, Laia and da Cunha, Elisabete and Dayal, Pratika and De Looze, Ilse and Fudamoto, Yoshinobu and Gonzalez, Valentino and Graziani, Luca and Hodge, Jacqueline A. and Hygate, Alexander P. S. and Nanayakkara, Themiya and Pallottini, Andrea and Riechers, Dominik A. and Schneider, Raffaella and Topping, Michael and van der Werf, Paul},
	month = sep,
	year = {2022},
	pages = {3126--3143},
}

@article{watson_dusty_2015,
	title = {A dusty, normal galaxy in the epoch of reionization},
	volume = {519},
	issn = {0028-0836},
	url = {https://ui.adsabs.harvard.edu/abs/2015Natur.519..327W},
	doi = {10.1038/nature14164},
	urldate = {2026-05-07},
	journal = {\nat},
	author = {Watson, Darach and Christensen, Lise and Knudsen, Kirsten Kraiberg and Richard, Johan and Gallazzi, Anna and Michałowski, Michał Jerzy},
	month = mar,
	year = {2015},
	pages = {327--330},
}

@article{tamura_detection_2019,
	title = {Detection of the {Far}-infrared [{O} {III}] and {Dust} {Emission} in a {Galaxy} at {Redshift} 8.312: {Early} {Metal} {Enrichment} in the {Heart} of the {Reionization} {Era}},
	volume = {874},
	issn = {0004-637X},
	shorttitle = {Detection of the {Far}-infrared [{O} {III}] and {Dust} {Emission} in a {Galaxy} at {Redshift} 8.312},
	url = {https://ui.adsabs.harvard.edu/abs/2019ApJ...874...27T},
	doi = {10.3847/1538-4357/ab0374},
	urldate = {2026-05-07},
	journal = {\apj},
	publisher = {IOP},
	author = {Tamura, Yoichi and Mawatari, Ken and Hashimoto, Takuya and Inoue, Akio K. and Zackrisson, Erik and Christensen, Lise and Binggeli, Christian and Matsuda, Yuichi and Matsuo, Hiroshi and Takeuchi, Tsutomu T. and Asano, Ryosuke S. and Sunaga, Kaho and Shimizu, Ikkoh and Okamoto, Takashi and Yoshida, Naoki and Lee, Minju M. and Shibuya, Takatoshi and Taniguchi, Yoshiaki and Umehata, Hideki and Hatsukade, Bunyo and Kohno, Kotaro and Ota, Kazuaki},
	month = mar,
	year = {2019},
	pages = {27},
}

@ARTICLE{rowland_rebels-ifu_2025,
       author = {{Rowland}, Lucie E. and {Heintz}, Kasper E. and {Algera}, Hiddo and {Stefanon}, Mauro and {Hodge}, Jacqueline and {Bouwens}, Rychard and {Aravena}, Manuel and {Astles}, Lucy and {da Cunha}, Elisabete and {Dayal}, Pratika and {Ferrara}, Andrea and {Fisher}, Rebecca and {Gonz{\'a}lez}, Valentino and {Inami}, Hanae and {Komarova}, Lena and {de Looze}, Ilse and {Nanayakkara}, Themiya and {Ormerod}, Katherine and {Pallottini}, Andrea and {Pollock}, Clara L. and {Smit}, Renske and {van der Werf}, Paul and {Witstok}, Joris},
        title = "{REBELS-IFU: Linking damped Lyman-{\ensuremath{\alpha}} absorption to [C II] emission and dust content in the Epoch of Reonisation}",
      journal = {\aap},
         year = 2026,
        month = jun,
       volume = {710},
          eid = {A374},
        pages = {A374},
          doi = {10.1051/0004-6361/202557654},
archivePrefix = {arXiv},
       eprint = {2510.11351},
 primaryClass = {astro-ph.GA},
       
}

@ARTICLE{rajulal_2026,
       author = {{Rajulal}, Gitanjali and {Algera}, Hiddo S.~B. and {Sugahara}, Yuma and {Bakx}, Tom J.~L.~C. and {Hashimoto}, Takuya and {Arai}, Suzuka and {Inoue}, Akio K. and {Mitsuhashi}, Ikki and {Aravena}, Manuel and {Cescon}, Karin and {Chen}, Chian-Chou and {da Cunha}, Elisabete and {Dayal}, Pratika and {De Looze}, Ilse and {Faisst}, Andreas and {Fudamoto}, Yoshinobu and {Herrera-Camus}, Rodrigo and {Inami}, Hanae and {Koekemoer}, Anton M. and {Mizukoshi}, Shoichiro and {Romano}, Michael and {Rowland}, Lucie and {Schouws}, Sander and {Smit}, Renske and {Vallini}, Livia and {Wang}, Wei-Hao and {Zamorani}, Giovanni and {Zanella}, Anita},
        title = "{Breathing Fire: Hot Dust in the Big Three Dragons at z = 7.15}",
      journal = {arXiv e-prints},
         year = 2026,
        month = aug,
          eid = {arXiv:2608.08015},
        pages = {arXiv:2608.08015},
          doi = {10.48550/arXiv.2608.08015},
archivePrefix = {arXiv},
       eprint = {2608.08015},
 primaryClass = {astro-ph.GA},
       
}

@ARTICLE{sommovigo2025,
       author = {{Sommovigo}, Laura and {Algera}, Hiddo},
        title = "{Realistic multitemperature dust: how well can we constrain the dust properties of high-redshift galaxies?}",
      journal = {\mnras},
         year = 2025,
        month = jul,
       volume = {540},
       number = {4},
        pages = {3693-3708},
          doi = {10.1093/mnras/staf897},
archivePrefix = {arXiv},
       eprint = {2505.20105},
 primaryClass = {astro-ph.GA},
       adsurl = {https://ui.adsabs.harvard.edu/abs/2025MNRAS.540.3693S},
}

@ARTICLE{ginzburg_2022,
       author = {{Ginzburg}, Omri and {Dekel}, Avishal and {Mandelker}, Nir and {Krumholz}, Mark R.},
        title = "{The evolution of turbulent galactic discs: gravitational instability, feedback, and accretion}",
      journal = {\mnras},
         year = 2022,
        month = jul,
       volume = {513},
       number = {4},
        pages = {6177-6195},
          doi = {10.1093/mnras/stac1324},
archivePrefix = {arXiv},
       eprint = {2202.12331},
 primaryClass = {astro-ph.GA},
       adsurl = {https://ui.adsabs.harvard.edu/abs/2022MNRAS.513.6177G},
}

@manual{cortes_alma_handbook_2026,
author       = {Cortes, P. C. and Asaki, Y. and Cerrigone, L. and Kneissl, R. and Messias, H. and Carpenter, J. and Kameno, S. and Loomis, R. and Vila-Vilaro, B. and Immer, K. and Plunkett, A. and Law, J. and Stoehr, F. and Saini, K. and Hales, A.},
title        = {{ALMA Technical Handbook}},
year         = {2026},
number       = {ALMA Doc. 13.3},
version      = {1.0},
doi          = {10.5281/zenodo.18793803},
isbn         = {978-3-923524-66-2}
}

@ARTICLE{Cescon_2026,
       author = {{Cescon}, Karin and {Hodge}, Jacqueline A. and {Boogaard}, Leindert A. and {Algera}, Hiddo S.~B. and {Rowland}, Lucie E. and {Riechers}, Dominik A. and {Smit}, Renske and {De Looze}, Ilse and {Bouwens}, Rychard and {van der Werf}, Paul and {Aravena}, Manuel and {da Cunha}, Elisabete and {Dayal}, Pratika and {Ferrara}, Andrea and {Fisher}, Rebecca and {Inami}, Hanae and {Mancera Pi{\~n}a}, Pavel E. and {Oesch}, Pascal A. and {Pallottini}, Andrea and {Rybak}, Matus and {Schouws}, Sander and {Sommovigo}, Laura and {Stefanon}, Mauro and {Vallini}, Livia},
        title = "{Direct detection of cool molecular gas in a star-forming galaxy at $z=7.31$}",
      journal = {arXiv e-prints},
         year = 2026,
        month = jun,
          eid = {arXiv:2606.13393},
        pages = {arXiv:2606.13393},
archivePrefix = {arXiv},
       eprint = {2606.13393},
 primaryClass = {astro-ph.GA},
       adsurl = {https://ui.adsabs.harvard.edu/abs/2026arXiv260613393C},
}

@ARTICLE{Westoby_2026,
       author = {{Westoby}, B.~A. and {Hodge}, J.~A. and {Sharda}, P. and {Mancera Pi{\~n}a}, P.~E. and {Rybak}, M. and {da Cunha}, E. and {Li}, J. and {Smail}, I. and {Swinbank}, A.~M. and {Battisti}, A. and {Boogaard}, L.~A. and {Brandt}, W.~N. and {Calistro Rivera}, G. and {Chen}, C.-C. and {Cox}, P. and {Cracraft}, M. and {Dannerbauer}, H. and {Decarli}, R. and {Greve}, T.~R. and {Kendrew}, S. and {Knudsen}, K. and {Liao}, C.-L. and {van Marrewijk}, J. and {Nayak}, O. and {Neeleman}, M. and {Rowland}, L.~E. and {Schinnerer}, E. and {Walter}, F. and {Wardlow}, J.~L. and {Weiss}, A. and {van der Werf}, P.},
        title = "{Investigating the role of turbulence in the interstellar medium in $z\sim3$ dusty star-forming galaxies using kpc-resolution ALMA dust and gas maps}",
      journal = {arXiv e-prints},
         year = 2026,
        month = jun,
          eid = {arXiv:2606.11444},
        pages = {arXiv:2606.11444},
          doi = {10.48550/arXiv.2606.11444},
archivePrefix = {arXiv},
       eprint = {2606.11444},
 primaryClass = {astro-ph.GA},
       adsurl = {https://ui.adsabs.harvard.edu/abs/2026arXiv260611444W},
}

@ARTICLE{Liu_2024,
       author = {{Liu}, Daizhong and {F{\"o}rster Schreiber}, Natascha M. and {Harrington}, Kevin C. and {Lee}, Lilian L. and {Kamieneski}, Patrick S. and {Davies}, Richard I. and {Lutz}, Dieter and {Renzini}, Alvio and {Wuyts}, Stijn and {Tacconi}, Linda J. and {Genzel}, Reinhard and {Burkert}, Andreas and {Herrera-Camus}, Rodrigo and {Alcalde Pampliega}, Bel{\'e}n and {Vishwas}, Amit and {Kaasinen}, Melanie and {Wang}, Q. Daniel and {Jim{\'e}nez-Andrade}, Eric F. and {Lowenthal}, James and {Foo}, Nicholas and {Frye}, Brenda L. and {Shangguan}, Jinyi and {Cao}, Yixian and {Agapito}, Guido and {Agudo Berbel}, Alex and {Barfety}, Capucine and {Baruffolo}, Andrea and {Berman}, Derek and {Black}, Martin and {Bonaglia}, Marco and {Briguglio}, Runa and {Carbonaro}, Luca and {Chapman}, Lee and {Chen}, Jianhang and {Cikota}, Aleksandar and {Concas}, Alice and {Cooper}, Olivia and {Cresci}, Giovanni and {Dallilar}, Yigit and {Deysenroth}, Matthias and {Di Antonio}, Ivan and {Di Cianno}, Amico and {Di Rico}, Gianluca and {Doelman}, David and {Dolci}, Mauro and {Eisenhauer}, Frank and {Espejo}, Juan and {Esposito}, Simone and {Fantinel}, Daniela and {Ferruzzi}, Debora and {Feuchtgruber}, Helmut and {Gao}, Xiaofeng and {Garcia Diaz}, Carlos and {Gillessen}, Stefan and {Grani}, Paolo and {Hartl}, Michael and {Henry}, David and {Huber}, Heinrich and {Jolly}, Jean-Baptiste and {Keller}, Christoph U. and {Kenworthy}, Matthew and {Kravchenko}, Kateryna and {Lee}, Minju M. and {Lightfoot}, John and {Lunney}, David and {Macintosh}, Mike and {Mannucci}, Filippo and {Ott}, Thomas and {Pascale}, Massimo and {Pastras}, Stavros and {Pearson}, David and {Puglisi}, Alfio and {Pulsoni}, Claudia and {Rabien}, Sebastian and {Rau}, Christian and {Riccardi}, Armando and {Salasnich}, Bernardo and {Shimizu}, Taro and {Snik}, Frans and {Sturm}, Eckhard and {Taylor}, William and {Valentini}, Angelo and {Waring}, Christopher and {Wiezorrek}, Erich and {Xompero}, Marco and {Yun}, Min S.},
        title = "{Detailed study of a rare hyperluminous rotating disk in an Einstein ring 10 billion years ago}",
      journal = {Nat. Astron.},
         year = 2024,
        month = sep,
       volume = {8},
        pages = {1181-1194},
          doi = {10.1038/s41550-024-02296-7},
       adsurl = {https://ui.adsabs.harvard.edu/abs/2024NatAs...8.1181L},
}

@ARTICLE{ikeda2025,
       author = {{Ikeda}, Ryota and {Tadaki}, Ken-ichi and {Mitsuhashi}, Ikki and {Aravena}, Manuel and {De Looze}, Ilse and {F{\"o}rster Schreiber}, Natascha M. and {Gonz{\'a}lez-L{\'o}pez}, Jorge and {Herrera-Camus}, Rodrigo and {Spilker}, Justin and {Barcos-Mu{\~n}oz}, Loreto and {Bowler}, Rebecca A.~A. and {Calistro Rivera}, Gabriela and {da Cunha}, Elisabete and {Davies}, Rebecca and {D{\'\i}az-Santos}, Tanio and {Ferrara}, Andrea and {Killi}, Meghana and {Lee}, Lilian L. and {Li}, Juno and {Lutz}, Dieter and {Posses}, Ana and {Smit}, Renske and {Solimano}, Manuel and {Telikova}, Kseniia and {{\"U}bler}, Hannah and {Veilleux}, Sylvain and {Villanueva}, Vicente},
        title = "{The ALMA-CRISTAL Survey: Spatial extent of [CII] line emission in star-forming galaxies at z = 4{\ensuremath{-}}6}",
      journal = {\aap},
         year = 2025,
        month = jan,
       volume = {693},
          eid = {A237},
        pages = {A237},
          doi = {10.1051/0004-6361/202451811},
archivePrefix = {arXiv},
       eprint = {2408.03374},
 primaryClass = {astro-ph.GA},
       adsurl = {https://ui.adsabs.harvard.edu/abs/2025A&A...693A.237I},
}

@ARTICLE{GALPAK2015,
       author = {{Bouch{\'e}}, N. and {Carfantan}, H. and {Schroetter}, I. and {Michel-Dansac}, L. and {Contini}, T.},
        title = "{GalPak$^{3D}$: A Bayesian Parametric Tool for Extracting Morphokinematics of Galaxies from 3D Data}",
      journal = {\aj},
         year = 2015,
        month = sep,
       volume = {150},
       number = {3},
          eid = {92},
        pages = {92},
          doi = {10.1088/0004-6256/150/3/92},
archivePrefix = {arXiv},
       eprint = {1501.06586},
 primaryClass = {astro-ph.IM},
       adsurl = {https://ui.adsabs.harvard.edu/abs/2015AJ....150...92B},
}

@article{de_graaff_ionised_2024,
	title = {Ionised gas kinematics and dynamical masses of z ≳ 6 galaxies from {JADES}/{NIRSpec} high-resolution spectroscopy},
	volume = {684},
	issn = {0004-6361},
	url = {https://ui.adsabs.harvard.edu/abs/2024A&A...684A..87D},
	doi = {10.1051/0004-6361/202347755},
	urldate = {2026-05-07},
	journal = {\aap},
	publisher = {EDP},
	author = {de Graaff, Anna and Rix, Hans-Walter and Carniani, Stefano and Suess, Katherine A. and Charlot, Stéphane and Curtis-Lake, Emma and Arribas, Santiago and Baker, William M. and Boyett, Kristan and Bunker, Andrew J. and Cameron, Alex J. and Chevallard, Jacopo and Curti, Mirko and Eisenstein, Daniel J. and Franx, Marijn and Hainline, Kevin and Hausen, Ryan and Ji, Zhiyuan and Johnson, Benjamin D. and Jones, Gareth C. and Maiolino, Roberto and Maseda, Michael V. and Nelson, Erica and Parlanti, Eleonora and Rawle, Tim and Robertson, Brant and Tacchella, Sandro and Übler, Hannah and Williams, Christina C. and Willmer, Christopher N. A. and Willott, Chris},
	month = apr,
	year = {2024},
	pages = {A87},
}

@article{scholtz_tentative_2025,
	title = {Tentative rotation in a galaxy at z {\textasciitilde} 14 with {ALMA}},
	volume = {544},
	issn = {0035-8711},
	url = {https://ui.adsabs.harvard.edu/abs/2025MNRAS.544L.113S},
	doi = {10.1093/mnrasl/slaf109},
	urldate = {2026-05-07},
	journal = {\mnras},
	publisher = {OUP},
	author = {Scholtz, J. and Parlanti, E. and Carniani, S. and Kohandel, M. and Sun, F. and Danhaive, A. L. and Maiolino, R. and Arribas, S. and Bhatawdekar, R. and Bunker, A. J. and Charlot, S. and D'Eugenio, F. and Ferrara, A. and Ji, Z. and Jones, Gareth C. and Rinaldi, P. and Robertson, B. and Pallottini, A. and Shivaei, I. and Sun, Y. and Tacchella, S. and Übler, H. and Venturi, G.},
	month = nov,
	year = {2025},
	pages = {L113--L120},
}

@article{scoville_cosmic_2007,
	title = {The {Cosmic} {Evolution} {Survey} ({COSMOS}): {Overview}},
	volume = {172},
	issn = {0067-0049},
	shorttitle = {The {Cosmic} {Evolution} {Survey} ({COSMOS})},
	url = {https://ui.adsabs.harvard.edu/abs/2007ApJS..172....1S},
	doi = {10.1086/516585},
	urldate = {2026-05-07},
	journal = {\apjs},
	publisher = {IOP},
	author = {Scoville, N. and Aussel, H. and Brusa, M. and Capak, P. and Carollo, C. M. and Elvis, M. and Giavalisco, M. and Guzzo, L. and Hasinger, G. and Impey, C. and Kneib, J.-P. and LeFevre, O. and Lilly, S. J. and Mobasher, B. and Renzini, A. and Rich, R. M. and Sanders, D. B. and Schinnerer, E. and Schminovich, D. and Shopbell, P. and Taniguchi, Y. and Tyson, N. D.},
	month = sep,
	year = {2007},
	pages = {1--8},
}

@article{jones_blackthunder_2026,
	title = {{BlackTHUNDER}: {Shedding} light on a dormant and extreme little red dot at z = 8.50},
	volume = {546},
	issn = {0035-8711},
	shorttitle = {{BlackTHUNDER}},
	url = {https://ui.adsabs.harvard.edu/abs/2026MNRAS.546ag115J},
	doi = {10.1093/mnras/stag115},
	urldate = {2026-05-07},
	journal = {\mnras},
	publisher = {OUP},
	author = {Jones, Gareth C. and Übler, Hannah and Maiolino, Roberto and Ji, Xihan and Marconi, Alessandro and D'Eugenio, Francesco and Arribas, Santiago and Bunker, Andrew J. and Carniani, Stefano and Charlot, Stéphane and Cresci, Giovanni and Inayoshi, Kohei and Isobe, Yuki and Juodžbalis, Ignas and Mazzolari, Giovanni and Pérez-González, Pablo G. and Perna, Michele and Schneider, Raffaella and Scholtz, Jan and Tacchella, Sandro},
	month = mar,
	year = {2026},
	pages = {stag115},
}

@article{deugenio_jades_2025,
	title = {{JADES} {Data} {Release} 3: {NIRSpec}/{Microshutter} {Assembly} {Spectroscopy} for 4000 {Galaxies} in the {GOODS} {Fields}},
	volume = {277},
	issn = {0067-0049},
	shorttitle = {{JADES} {Data} {Release} 3},
	url = {https://ui.adsabs.harvard.edu/abs/2025ApJS..277....4D},
	doi = {10.3847/1538-4365/ada148},
	urldate = {2026-05-07},
	journal = {\apjs},
	publisher = {IOP},
	author = {D'Eugenio, Francesco and Cameron, Alex J. and Scholtz, Jan and Carniani, Stefano and Willott, Chris J. and Curtis-Lake, Emma and Bunker, Andrew J. and Parlanti, Eleonora and Maiolino, Roberto and Willmer, Christopher N. A. and Jakobsen, Peter and Robertson, Brant E. and Johnson, Benjamin D. and Tacchella, Sandro and Cargile, Phillip A. and Rawle, Tim and Arribas, Santiago and Chevallard, Jacopo and Curti, Mirko and Egami, Eiichi and Eisenstein, Daniel J. and Kumari, Nimisha and Looser, Tobias J. and Rieke, Marcia J. and Rodríguez Del Pino, Bruno and Saxena, Aayush and Übler, Hannah and Venturi, Giacomo and Witstok, Joris and Baker, William M. and Bhatawdekar, Rachana and Bonaventura, Nina and Boyett, Kristan and Charlot, Stephane and Danhaive, A. Lola and Hainline, Kevin N. and Hausen, Ryan and Helton, Jakob M. and Ji, Xihan and Ji, Zhiyuan and Jones, Gareth C. and Juodžbalis, Ignas and Maseda, Michael V. and Pérez-González, Pablo G. and Perna, Michele and Puskás, Dávid and Shivaei, Irene and Silcock, Maddie S. and Simmonds, Charlotte and Smit, Renske and Sun, Fengwu and Villanueva, Natalia C. and Williams, Christina C. and Zhu, Yongda},
	month = mar,
	year = {2025},
	pages = {4},
}

@article{lee_alma-cristal_2025,
	title = {The {ALMA}-{CRISTAL} survey: {Resolved} kinematic studies of main sequence star-forming galaxies at 4 {\textless} z {\textless} 6},
	volume = {701},
	issn = {0004-6361},
	shorttitle = {The {ALMA}-{CRISTAL} survey},
	url = {https://ui.adsabs.harvard.edu/abs/2025A&A...701A.260L},
	doi = {10.1051/0004-6361/202555362},
	urldate = {2026-05-07},
	journal = {\aap},
	publisher = {EDP},
	author = {Lee, Lilian L. and Förster Schreiber, Natascha M. and Herrera-Camus, Rodrigo and Liu, Daizhong and Price, Sedona H. and Genzel, Reinhard and Tacconi, Linda J. and Lutz, Dieter and Davies, Ric and Naab, Thorsten and Übler, Hannah and Aravena, Manuel and Assef, Roberto J. and Barcos-Muñoz, Loreto and Bowler, Rebecca A. A. and Burkert, Andreas and Chen, Jianhang and Davies, Rebecca L. and De Looze, Ilse and Diaz-Santos, Tanio and González-López, Jorge and Ikeda, Ryota and Mitsuhashi, Ikki and Posses, Ana and Relaño Pastor, Mónica and Renzini, Alvio and Solimano, Manuel and Spilker, Justin S. and Sternberg, Amiel and Tadaki, Kenichi and Telikova, Kseniia and Veilleux, Sylvain and Villanueva, Vicente},
	month = sep,
	year = {2025},
	pages = {A260},
}

@article{di_teodoro_3d_2015,
	title = {{3D} {BAROLO}: a new {3D} algorithm to derive rotation curves of galaxies},
	volume = {451},
	issn = {0035-8711},
	shorttitle = {{3D} {BAROLO}},
	url = {https://ui.adsabs.harvard.edu/abs/2015MNRAS.451.3021D},
	doi = {10.1093/mnras/stv1213},
	urldate = {2026-05-07},
	journal = {\mnras},
	publisher = {OUP},
	author = {Di Teodoro, E. M. and Fraternali, F.},
	month = aug,
	year = {2015},
	pages = {3021--3033},
}

@article{wisnioski_evolution_2025,
	title = {Evolution of gas velocity dispersion in discs from z ∼ 8 to z ∼ 0.5},
	volume = {544},
	issn = {0035-8711},
	url = {https://ui.adsabs.harvard.edu/abs/2025MNRAS.544.2777W},
	doi = {10.1093/mnras/staf1606},
	urldate = {2026-05-07},
	journal = {\mnras},
	publisher = {OUP},
	author = {Wisnioski, E. and Mendel, J. T. and Leaman, R. and Tsukui, T. and Übler, H. and Förster Schreiber, N. M.},
	month = dec,
	year = {2025},
	pages = {2777--2794},
}

@article{wisnioski_kmos3d_2015,
	title = {The {KMOS3D} {Survey}: {Design}, {First} {Results}, and the {Evolution} of {Galaxy} {Kinematics} from 0.7 {\textless}= z {\textless}= 2.7},
	volume = {799},
	issn = {0004-637X},
	shorttitle = {The {KMOS3D} {Survey}},
	url = {https://ui.adsabs.harvard.edu/abs/2015ApJ...799..209W},
	doi = {10.1088/0004-637X/799/2/209},
	urldate = {2026-05-07},
	journal = {\apj},
	publisher = {IOP},
	author = {Wisnioski, E. and Förster Schreiber, N. M. and Wuyts, S. and Wuyts, E. and Bandara, K. and Wilman, D. and Genzel, R. and Bender, R. and Davies, R. and Fossati, M. and Lang, P. and Mendel, J. T. and Beifiori, A. and Brammer, G. and Chan, J. and Fabricius, M. and Fudamoto, Y. and Kulkarni, S. and Kurk, J. and Lutz, D. and Nelson, E. J. and Momcheva, I. and Rosario, D. and Saglia, R. and Seitz, S. and Tacconi, L. J. and van Dokkum, P. G.},
	month = feb,
	year = {2015},
	pages = {209},
}

@article{crespo_gomez_stellar_2021,
	title = {Stellar kinematics in the nuclear regions of nearby {LIRGs} with {VLT}-{SINFONI}. {Comparison} with gas phases and implications for dynamical mass estimations},
	volume = {650},
	issn = {0004-6361},
	url = {https://ui.adsabs.harvard.edu/abs/2021A&A...650A.149C},
	doi = {10.1051/0004-6361/202039472},
	urldate = {2026-05-07},
	journal = {\aap},
	publisher = {EDP},
	author = {Crespo Gómez, A. and Piqueras López, J. and Arribas, S. and Pereira-Santaella, M. and Colina, L. and Rodríguez del Pino, B.},
	month = jun,
	year = {2021},
	pages = {A149},
}

@article{hodge_evidence_2012,
	title = {Evidence for a {Clumpy}, {Rotating} {Gas} {Disk} in a {Submillimeter} {Galaxy} at z = 4},
	volume = {760},
	issn = {0004-637X},
	url = {https://ui.adsabs.harvard.edu/abs/2012ApJ...760...11H},
	doi = {10.1088/0004-637X/760/1/11},
	urldate = {2026-05-07},
	journal = {\apj},
	publisher = {IOP},
	author = {Hodge, J. A. and Carilli, C. L. and Walter, F. and de Blok, W. J. G. and Riechers, D. and Daddi, E. and Lentati, L.},
	month = nov,
	year = {2012},
	pages = {11},
}

@article{nelson_ionized_2024,
	title = {Ionized {Gas} {Kinematics} with {FRESCO}: {An} {Extended}, {Massive}, {Rapidly} {Rotating} {Galaxy} at z = 5.4},
	volume = {976},
	issn = {0004-637X},
	shorttitle = {Ionized {Gas} {Kinematics} with {FRESCO}},
	url = {https://ui.adsabs.harvard.edu/abs/2024ApJ...976L..27N},
	doi = {10.3847/2041-8213/ad7b17},
	urldate = {2026-05-07},
	journal = {\apj},
	publisher = {IOP},
	author = {Nelson, Erica and Brammer, Gabriel and Giménez-Arteaga, Clara and Oesch, Pascal A. and Naidu, Rohan P. and Übler, Hannah and Matharu, Jasleen and Shapley, Alice E. and Whitaker, Katherine E. and Wisnioski, Emily and Förster Schreiber, Natascha M. and Smit, Renske and van Dokkum, Pieter and Chisholm, John and Endsley, Ryan and Hartley, Abigail I. and Gibson, Justus and Giovinazzo, Emma and Illingworth, Garth and Labbe, Ivo and Maseda, Michael V. and Matthee, Jorryt and Covelo Paz, Alba and Price, Sedona H. and Reddy, Naveen A. and Shivaei, Irene and Weibel, Andrea and Wuyts, Stijn and Xiao, Mengyuan and Alberts, Stacey and Baker, William M. and Bunker, Andrew J. and Cameron, Alex J. and Charlot, Stephane and Eisenstein, Daniel J. and de Graaff, Anna and Ji, Zhiyuan and Johnson, Benjamin D. and Jones, Gareth C. and Maiolino, Roberto and Robertson, Brant and Sandles, Lester and Suess, Katherine A. and Tacchella, Sandro and Williams, Christina C. and Witstok, Joris},
	month = dec,
	year = {2024},
	pages = {L27},
}

@article{fujimoto_primordial_2025,
	title = {Primordial rotating disk composed of at least 15 dense star-forming clumps at cosmic dawn},
	volume = {9},
	issn = {2397-3366},
	url = {https://ui.adsabs.harvard.edu/abs/2025NatAs...9.1553F},
	doi = {10.1038/s41550-025-02592-w},
	urldate = {2026-05-07},
	journal = {Nat. Astron.},
	author = {Fujimoto, S. and Ouchi, M. and Kohno, K. and Valentino, F. and Giménez-Arteaga, C. and Brammer, G. B. and Furtak, L. J. and Kohandel, M. and Oguri, M. and Pallottini, A. and Richard, J. and Zitrin, A. and Bauer, F. E. and Boylan-Kolchin, M. and Dessauges-Zavadsky, M. and Egami, E. and Finkelstein, S. L. and Ma, Z. and Smail, I. and Watson, D. and Hutchison, T. A. and Rigby, J. R. and Welch, B. D. and Ao, Y. and Bradley, L. D. and Caminha, G. B. and Caputi, K. I. and Espada, D. and Endsley, R. and Fudamoto, Y. and González-López, J. and Hatsukade, B. and Koekemoer, A. M. and Kokorev, V. and Laporte, N. and Lee, M. and Magdis, G. E. and Ono, Y. and Rizzo, F. and Shibuya, T. and Shimasaku, K. and Sun, F. and Toft, S. and Umehata, H. and Wang, T. and Yajima, H.},
	month = oct,
	year = {2025},
	pages = {1553--1567},
}

@article{girard_towards_2019,
	title = {Towards sub-kpc scale kinematics of molecular and ionized gas of star-forming galaxies at z ∼ 1},
	volume = {631},
	issn = {0004-6361},
	url = {https://ui.adsabs.harvard.edu/abs/2019A&A...631A..91G},
	doi = {10.1051/0004-6361/201935896},
	urldate = {2026-05-07},
	journal = {\aap},
	publisher = {EDP},
	author = {Girard, M. and Dessauges-Zavadsky, M. and Combes, F. and Chisholm, J. and Patrício, V. and Richard, J. and Schaerer, D.},
	month = nov,
	year = {2019},
	pages = {A91},
}

@article{parlanti_ga-nifs_2025,
	title = {{GA}-{NIFS}: {Multiphase} analysis of a star-forming galaxy at z ∼ 5.5},
	volume = {695},
	issn = {0004-6361},
	shorttitle = {{GA}-{NIFS}},
	url = {https://ui.adsabs.harvard.edu/abs/2025A&A...695A...6P},
	doi = {10.1051/0004-6361/202451692},
	urldate = {2026-05-07},
	journal = {\aap},
	publisher = {EDP},
	author = {Parlanti, Eleonora and Carniani, Stefano and Venturi, Giacomo and Herrera-Camus, Rodrigo and Arribas, Santiago and Bunker, Andrew J. and Charlot, Stéphane and D'Eugenio, Francesco and Maiolino, Roberto and Perna, Michele and Übler, Hannah and Böker, Torsten and Cresci, Giovanni and Curti, Mirko and Jones, Gareth C. and Lamperti, Isabella and Pérez-González, Pablo G. and Del Pino, Bruno Rodríguez and Zamora, Sandra},
	month = mar,
	year = {2025},
	pages = {A6},
}

@article{parlanti_alma_2023,
	title = {{ALMA} hints at the presence of turbulent disk galaxies at z {\textgreater} 5},
	volume = {673},
	issn = {0004-6361},
	url = {https://ui.adsabs.harvard.edu/abs/2023A&A...673A.153P},
	doi = {10.1051/0004-6361/202245603},
	urldate = {2026-05-07},
	journal = {\aap},
	publisher = {EDP},
	author = {Parlanti, E. and Carniani, S. and Pallottini, A. and Cignoni, M. and Cresci, G. and Kohandel, M. and Mannucci, F. and Marconi, A.},
	month = may,
	year = {2023},
	pages = {A153},
}

@article{parlanti_ga-nifs_2024,
	title = {{GA}-{NIFS}: {Early}-stage feedback in a heavily obscured active galactic nucleus at z = 4.76},
	volume = {684},
	issn = {0004-6361},
	shorttitle = {{GA}-{NIFS}},
	url = {https://ui.adsabs.harvard.edu/abs/2024A&A...684A..24P},
	doi = {10.1051/0004-6361/202347914},
	urldate = {2026-05-07},
	journal = {\aap},
	publisher = {EDP},
	author = {Parlanti, Eleonora and Carniani, Stefano and Übler, Hannah and Venturi, Giacomo and Circosta, Chiara and D'Eugenio, Francesco and Arribas, Santiago and Bunker, Andrew J. and Charlot, Stéphane and Lützgendorf, Nora and Maiolino, Roberto and Perna, Michele and Rodríguez Del Pino, Bruno and Willott, Chris J. and Böker, Torsten and Cameron, Alex J. and Chevallard, Jacopo and Cresci, Giovanni and Jones, Gareth C. and Kumari, Nimisha and Lamperti, Isabella and Scholtz, Jan},
	month = apr,
	year = {2024},
	pages = {A24},
}

@article{liu_600_2023,
	title = {An  600 pc {View} of the {Strongly} {Lensed}, {Massive} {Main}-sequence {Galaxy} {J0901}: {A} {Baryon}-dominated, {Thick} {Turbulent} {Rotating} {Disk} with a {Clumpy} {Cold} {Gas} {Ring} at z = 2.259},
	volume = {942},
	issn = {0004-637X},
	shorttitle = {An  600 pc {View} of the {Strongly} {Lensed}, {Massive} {Main}-sequence {Galaxy} {J0901}},
	url = {https://ui.adsabs.harvard.edu/abs/2023ApJ...942...98L},
	doi = {10.3847/1538-4357/aca46b},
	urldate = {2026-05-07},
	journal = {\apj},
	publisher = {IOP},
	author = {Liu, Daizhong and Förster Schreiber, N. M. and Genzel, R. and Lutz, D. and Price, S. H. and Lee, L. L. and Baker, Andrew J. and Burkert, A. and Coogan, R. T. and Davies, R. I. and Davies, R. L. and Herrera-Camus, R. and Kodama, Tadayuki and Lee, Minju M. and Nestor, A. and Pulsoni, C. and Renzini, A. and Sharon, Chelsea E. and Shimizu, T. T. and Tacconi, L. J. and Tadaki, Ken-ichi and Übler, H.},
	month = jan,
	year = {2023},
	pages = {98},
}

@article{molina_kiloparsec-scale_2019,
	title = {The kiloparsec-scale gas kinematics in two star-forming galaxies at z ∼ 1.47 seen with {ALMA} and {VLT}-{SINFONI}},
	volume = {487},
	issn = {0035-8711},
	url = {https://ui.adsabs.harvard.edu/abs/2019MNRAS.487.4856M},
	doi = {10.1093/mnras/stz1643},
	urldate = {2026-05-07},
	journal = {\mnras},
	publisher = {OUP},
	author = {Molina, J. and Ibar, Edo and Smail, I. and Swinbank, A. M. and Villard, E. and Escala, A. and Sobral, D. and Hughes, T. M.},
	month = aug,
	year = {2019},
	pages = {4856--4869},
}

@article{kuhn_jwst_2024,
	title = {{JWST} {Reveals} a {Surprisingly} {High} {Fraction} of {Galaxies} {Being} {Spiral}-like at 0.5 ≤ z ≤ 4},
	volume = {968},
	issn = {0004-637X},
	url = {https://ui.adsabs.harvard.edu/abs/2024ApJ...968L..15K},
	doi = {10.3847/2041-8213/ad43eb},
	urldate = {2026-05-11},
	journal = {\apj},
	publisher = {IOP},
	author = {Kuhn, Vicki and Guo, Yicheng and Martin, Alec and Bayless, Julianna and Gates, Ellie and Puleo, AJ},
	month = jun,
	year = {2024},
	pages = {L15},
}

@article{ferreira_panic_2022,
	title = {Panic! at the {Disks}: {First} {Rest}-frame {Optical} {Observations} of {Galaxy} {Structure} at z {\textgreater} 3 with {JWST} in the {SMACS} 0723 {Field}},
	volume = {938},
	issn = {0004-637X},
	shorttitle = {Panic! at the {Disks}},
	url = {https://ui.adsabs.harvard.edu/abs/2022ApJ...938L...2F},
	doi = {10.3847/2041-8213/ac947c},
	urldate = {2026-05-11},
	journal = {\apj},
	publisher = {IOP},
	author = {Ferreira, Leonardo and Adams, Nathan and Conselice, Christopher J. and Sazonova, Elizaveta and Austin, Duncan and Caruana, Joseph and Ferrari, Fabricio and Verma, Aprajita and Trussler, James and Broadhurst, Tom and Diego, Jose and Frye, Brenda L. and Pascale, Massimo and Wilkins, Stephen M. and Windhorst, Rogier A. and Zitrin, Adi},
	month = oct,
	year = {2022},
	pages = {L2},
}

@article{robertson_morpheus_2023,
	title = {Morpheus {Reveals} {Distant} {Disk} {Galaxy} {Morphologies} with {JWST}: {The} {First} {AI}/{ML} {Analysis} of {JWST} {Images}},
	volume = {942},
	issn = {0004-637X},
	shorttitle = {Morpheus {Reveals} {Distant} {Disk} {Galaxy} {Morphologies} with {JWST}},
	url = {https://ui.adsabs.harvard.edu/abs/2023ApJ...942L..42R},
	doi = {10.3847/2041-8213/aca086},
	urldate = {2026-05-11},
	journal = {\apj},
	publisher = {IOP},
	author = {Robertson, Brant E. and Tacchella, Sandro and Johnson, Benjamin D. and Hausen, Ryan and Alabi, Adebusola B. and Boyett, Kristan and Bunker, Andrew J. and Carniani, Stefano and Egami, Eiichi and Eisenstein, Daniel J. and Hainline, Kevin N. and Helton, Jakob M. and Ji, Zhiyuan and Kumari, Nimisha and Lyu, Jianwei and Maiolino, Roberto and Nelson, Erica J. and Rieke, Marcia J. and Shivaei, Irene and Sun, Fengwu and Übler, Hannah and Williams, Christina C. and Willmer, Christopher N. A. and Witstok, Joris},
	month = jan,
	year = {2023},
	pages = {L42},
}

@article{rizzo_dynamically_2020,
	title = {A dynamically cold disk galaxy in the early {Universe}},
	volume = {584},
	issn = {0028-0836},
	url = {https://ui.adsabs.harvard.edu/abs/2020Natur.584..201R},
	doi = {10.1038/s41586-020-2572-6},
	urldate = {2026-05-11},
	journal = {\nat},
	author = {Rizzo, F. and Vegetti, S. and Powell, D. and Fraternali, F. and McKean, J. P. and Stacey, H. R. and White, S. D. M.},
	month = aug,
	year = {2020},
	pages = {201--204},
}

@article{neeleman_cold_2020,
	title = {A cold, massive, rotating disk galaxy 1.5 billion years after the {Big} {Bang}},
	volume = {581},
	issn = {0028-0836},
	url = {https://ui.adsabs.harvard.edu/abs/2020Natur.581..269N},
	doi = {10.1038/s41586-020-2276-y},
	urldate = {2026-05-11},
	journal = {\nat},
	author = {Neeleman, Marcel and Prochaska, J. Xavier and Kanekar, Nissim and Rafelski, Marc},
	month = may,
	year = {2020},
	pages = {269--272},
}

@article{neeleman_alma_2023,
	title = {{ALMA} 400 pc {Imaging} of a z = 6.5 {Massive} {Warped} {Disk} {Galaxy}},
	volume = {958},
	issn = {0004-637X},
	url = {https://ui.adsabs.harvard.edu/abs/2023ApJ...958..132N},
	doi = {10.3847/1538-4357/ad05d2},
	urldate = {2026-05-11},
	journal = {\apj},
	publisher = {IOP},
	author = {Neeleman, Marcel and Walter, Fabian and Decarli, Roberto and Drake, Alyssa B. and Eilers, Anna-Christina and Meyer, Romain A. and Venemans, Bram P.},
	month = dec,
	year = {2023},
	pages = {132},
}

@article{neeleman_kinematics_2021,
	title = {The {Kinematics} of z ≳ 6 {Quasar} {Host} {Galaxies}},
	volume = {911},
	issn = {0004-637X},
	url = {https://ui.adsabs.harvard.edu/abs/2021ApJ...911..141N},
	doi = {10.3847/1538-4357/abe70f},
	urldate = {2026-05-11},
	journal = {\apj},
	publisher = {IOP},
	author = {Neeleman, Marcel and Novak, Mladen and Venemans, Bram P. and Walter, Fabian and Decarli, Roberto and Kaasinen, Melanie and Schindler, Jan-Torge and Bañados, Eduardo and Carilli, Chris L. and Drake, Alyssa B. and Fan, Xiaohui and Rix, Hans-Walter},
	month = apr,
	year = {2021},
	pages = {141},
}

@article{van_leeuwen_characterising_2024,
	title = {Characterising the contribution of dust-obscured star formation at z ≳ 5 using 18 serendipitously identified [{C} {II}] emitters},
	volume = {534},
	issn = {0035-8711},
	url = {https://ui.adsabs.harvard.edu/abs/2024MNRAS.534.2062V},
	doi = {10.1093/mnras/stae2171},
	urldate = {2026-05-11},
	journal = {\mnras},
	publisher = {OUP},
	author = {van Leeuwen, I. F. and Bouwens, R. J. and van der Werf, P. P. and Hodge, J. A. and Schouws, S. and Stefanon, M. and Algera, H. S. B. and Aravena, M. and Boogaard, L. A. and Bowler, R. A. A. and da Cunha, E. and Dayal, P. and Decarli, R. and Gonzalez, V. and Inami, H. and de Looze, I. and Sommovigo, L. and Venemans, B. P. and Walter, F. and Barrufet, L. and Ferrara, A. and Graziani, L. and Hygate, A. P. S. and Oesch, P. and Palla, M. and Rowland, L. and Schneider, R.},
	month = nov,
	year = {2024},
	pages = {2062--2085},
}

@article{algera_alma_2023,
	title = {The {ALMA} {REBELS} survey: the dust-obscured cosmic star formation rate density at redshift 7},
	volume = {518},
	issn = {0035-8711},
	shorttitle = {The {ALMA} {REBELS} survey},
	url = {https://ui.adsabs.harvard.edu/abs/2023MNRAS.518.6142A},
	doi = {10.1093/mnras/stac3195},
	urldate = {2026-05-11},
	journal = {\mnras},
	publisher = {OUP},
	author = {Algera, Hiddo S. B. and Inami, Hanae and Oesch, Pascal A. and Sommovigo, Laura and Bouwens, Rychard J. and Topping, Michael W. and Schouws, Sander and Stefanon, Mauro and Stark, Daniel P. and Aravena, Manuel and Barrufet, Laia and da Cunha, Elisabete and Dayal, Pratika and Endsley, Ryan and Ferrara, Andrea and Fudamoto, Yoshinobu and Gonzalez, Valentino and Graziani, Luca and Hodge, Jacqueline A. and Hygate, Alexander P. S. and de Looze, Ilse and Nanayakkara, Themiya and Schneider, Raffaella and van der Werf, Paul P.},
	month = feb,
	year = {2023},
	pages = {6142--6157},
}

@article{deugenio_fast-rotator_2024,
	title = {A fast-rotator post-starburst galaxy quenched by supermassive black-hole feedback at z = 3},
	volume = {8},
	issn = {2397-3366},
	url = {https://ui.adsabs.harvard.edu/abs/2024NatAs...8.1443D},
	doi = {10.1038/s41550-024-02345-1},
	urldate = {2026-05-11},
	journal = {Nat. Astron.},
	author = {D'Eugenio, Francesco and Pérez-González, Pablo G. and Maiolino, Roberto and Scholtz, Jan and Perna, Michele and Circosta, Chiara and Übler, Hannah and Arribas, Santiago and Böker, Torsten and Bunker, Andrew J. and Carniani, Stefano and Charlot, Stephane and Chevallard, Jacopo and Cresci, Giovanni and Curtis-Lake, Emma and Jones, Gareth C. and Kumari, Nimisha and Lamperti, Isabella and Looser, Tobias J. and Parlanti, Eleonora and Rix, Hans-Walter and Robertson, Brant and Rodríguez Del Pino, Bruno and Tacchella, Sandro and Venturi, Giacomo and Willott, Chris J.},
	month = nov,
	year = {2024},
	pages = {1443--1456},
}

@article{komarova_rebels-ifu_2026,
	title = {{REBELS}-{IFU}: {Spatially} {Resolved} {Ionizing} {Photon} {Production} {Efficiencies} of 12 {Bright} {Galaxies} in the {Epoch} of {Reionization}},
	volume = {1000},
	issn = {0004-637X},
	shorttitle = {{REBELS}-{IFU}},
	url = {https://ui.adsabs.harvard.edu/abs/2026ApJ..1000..228K},
	doi = {10.3847/1538-4357/ae474f},
	urldate = {2026-05-11},
	journal = {\apj},
	publisher = {IOP},
	author = {Komarova, Lena and Stefanon, Mauro and Laza-Ramos, Andrés and Algera, Hiddo S. B. and Aravena, Manuel and Bouwens, Rychard and Bowler, Rebecca and da Cunha, Elisabete and Dayal, Pratika and Ferrara, Andrea and Fisher, Rebecca and Nanayakkara, Themiya and Rowland, Lucie E. and Schouws, Sander and Smit, Renske and Sommovigo, Laura and Stark, Daniel P. and van der Werf, Paul},
	month = apr,
	year = {2026},
	pages = {228},
}

@article{fisher_rebels-ifu_2025,
	title = {{REBELS}-{IFU}: dust attenuation curves of 12 massive galaxies at z ≃ 7},
	volume = {539},
	issn = {0035-8711},
	shorttitle = {{REBELS}-{IFU}},
	url = {https://ui.adsabs.harvard.edu/abs/2025MNRAS.539..109F},
	doi = {10.1093/mnras/staf485},
	urldate = {2026-05-11},
	journal = {\mnras},
	publisher = {OUP},
	author = {Fisher, R. and Bowler, R. A. A. and Stefanon, M. and Rowland, L. E. and Algera, H. S. B. and Aravena, M. and Bouwens, R. and Dayal, P. and Ferrara, A. and Fudamoto, Y. and Gulis, C. and Hodge, J. A. and Inami, H. and Ormerod, K. and Pallottini, A. and Phillips, S. G. and Sartorio, N. S. and Schouws, S. and Smit, R. and Sommovigo, L. and Stark, D. P. and van der Werf, P. P.},
	month = may,
	year = {2025},
	pages = {109--126},
}

@article{bowler_alma_2024,
	title = {The {ALMA} {REBELS} survey: obscured star formation in massive {Lyman}-break galaxies at z= 4-8 revealed by the {IRX}-$\beta$ and {M}⋆ relations},
	volume = {527},
	issn = {0035-8711},
	shorttitle = {The {ALMA} {REBELS} survey},
	url = {https://ui.adsabs.harvard.edu/abs/2024MNRAS.527.5808B},
	doi = {10.1093/mnras/stad3578},
	urldate = {2026-05-11},
	journal = {\mnras},
	publisher = {OUP},
	author = {Bowler, R. A. A. and Inami, H. and Sommovigo, L. and Smit, R. and Algera, H. S. B. and Aravena, M. and Barrufet, L. and Bouwens, R. and da Cunha, E. and Cullen, F. and Dayal, P. and De Looze, I. and Dunlop, J. S. and Fudamoto, Y. and Mauerhofer, V. and McLure, R. J. and Stefanon, M. and Schneider, R. and Ferrara, A. and Graziani, L. and Hodge, J. A. and Nanayakkara, T. and Palla, M. and Schouws, S. and Stark, D. P. and van der Werf, P. P.},
	month = jan,
	year = {2024},
	pages = {5808--5828},
}

@article{fudamoto_alma_2022,
	title = {The {ALMA} {REBELS} {Survey}: {Average} [{C} {II}] 158 $\mu$m {Sizes} of {Star}-forming {Galaxies} from z   7 to z   4},
	volume = {934},
	issn = {0004-637X},
	shorttitle = {The {ALMA} {REBELS} {Survey}},
	url = {https://ui.adsabs.harvard.edu/abs/2022ApJ...934..144F},
	doi = {10.3847/1538-4357/ac7a47},
	urldate = {2026-05-11},
	journal = {\apj},
	publisher = {IOP},
	author = {Fudamoto, Y. and Smit, R. and Bowler, R. A. A. and Oesch, P. A. and Bouwens, R. and Stefanon, M. and Inami, H. and Endsley, R. and Gonzalez, V. and Schouws, S. and Stark, D. and Algera, H. S. B. and Aravena, M. and Barrufet, L. and da Cunha, E. and Dayal, P. and Ferrara, A. and Graziani, L. and Hodge, J. A. and Hygate, A. P. S. and Inoue, A. K. and Nanayakkara, T. and Pallottini, A. and Pizzati, E. and Schneider, R. and Sommovigo, L. and Sugahara, Y. and Topping, M. and van der Werf, P. and Bethermin, M. and Cassata, P. and Dessauges-Zavadsky, M. and Ibar, E. and Faisst, A. L. and Fujimoto, S. and Ginolfi, M. and Hathi, N. and Jones, G. C. and Pozzi, F. and Schaerer, D.},
	month = aug,
	year = {2022},
	pages = {144},
}

@article{sommovigo_alma_2022,
	title = {The {ALMA} {REBELS} {Survey}: cosmic dust temperature evolution out to z   7},
	volume = {513},
	issn = {0035-8711},
	shorttitle = {The {ALMA} {REBELS} {Survey}},
	url = {https://ui.adsabs.harvard.edu/abs/2022MNRAS.513.3122S},
	doi = {10.1093/mnras/stac302},
	urldate = {2026-05-11},
	journal = {\mnras},
	publisher = {OUP},
	author = {Sommovigo, L. and Ferrara, A. and Pallottini, A. and Dayal, P. and Bouwens, R. J. and Smit, R. and da Cunha, E. and De Looze, I. and Bowler, R. A. A. and Hodge, J. and Inami, H. and Oesch, P. and Endsley, R. and Gonzalez, V. and Schouws, S. and Stark, D. and Stefanon, M. and Aravena, M. and Graziani, L. and Riechers, D. and Schneider, R. and van der Werf, P. and Algera, H. and Barrufet, L. and Fudamoto, Y. and Hygate, A. P. S. and Labbé, I. and Li, Y. and Nanayakkara, T. and Topping, M.},
	month = jul,
	year = {2022},
	pages = {3122--3135},
}

@article{aniano_common-resolution_2011,
	title = {Common-{Resolution} {Convolution} {Kernels} for {Space}- and {Ground}-{Based} {Telescopes}},
	volume = {123},
	issn = {0004-6280},
	url = {https://ui.adsabs.harvard.edu/abs/2011PASP..123.1218A},
	doi = {10.1086/662219},
	urldate = {2026-05-11},
	journal = {\pasp},
	publisher = {IOP},
	author = {Aniano, G. and Draine, B. T. and Gordon, K. D. and Sandstrom, K.},
	month = oct,
	year = {2011},
	pages = {1218},
}

@article{geda_petrofit_2022,
	title = {{PetroFit}: {A} {Python} {Package} for {Computing} {Petrosian} {Radii} and {Fitting} {Galaxy} {Light} {Profiles}},
	volume = {163},
	issn = {0004-6256},
	shorttitle = {{PetroFit}},
	url = {https://ui.adsabs.harvard.edu/abs/2022AJ....163..202G},
	doi = {10.3847/1538-3881/ac5908},
	urldate = {2026-05-11},
	journal = {\aj},
	publisher = {IOP},
	author = {Geda, Robel and Crawford, Steven M. and Hunt, Lucas and Bershady, Matthew and Tollerud, Erik and Randriamampandry, Solohery},
	month = may,
	year = {2022},
	pages = {202},
}

@article{ren_evolution_2025,
	title = {The {Evolution} of the {Size} and {Merger} {Fraction} of {Submillimeter} {Galaxies} across 1 {\textless} z ≲ 6 as {Observed} by {JWST}},
	volume = {982},
	issn = {0004-637X},
	url = {https://ui.adsabs.harvard.edu/abs/2025ApJ...982..200R},
	doi = {10.3847/1538-4357/adb961},
	urldate = {2026-05-11},
	journal = {\apj},
	publisher = {IOP},
	author = {Ren, Jian and Liu, F. S. and Li, Nan and Zhao, Pinsong and Cui, Qifan and Song, Qi and Li, Yubin and Mo, Hao and Yesuf, Hassen M. and Wang, Weichen and An, Fangxia and Zheng, Xian Zhong},
	month = apr,
	year = {2025},
	pages = {200},
}

@article{song_size_2026,
	title = {The {Size} {Evolution} and the {Size}─{Mass} {Relation} of {Ly$\alpha$} {Emitters} across 3 ≲ z{\textless} 7 as {Observed} by {JWST}},
	volume = {997},
	issn = {0004-637X},
	url = {https://ui.adsabs.harvard.edu/abs/2026ApJ...997..126S},
	doi = {10.3847/1538-4357/ae24e4},
	urldate = {2026-05-11},
	journal = {\apj},
	publisher = {IOP},
	author = {Song, Qi and Liu, F. S. and Ren, Jian and Zhao, Pinsong and Cui, Qifan and Li, Yubin and Mo, Hao and Luo, Yuchong and Wang, Guanghuan and Li, Nan and Yesuf, Hassen M. and Wang, Weichen and Zhang, Xin and Meng, Xianmin and Fu, Mingxiang and Zhang, Bingqing and Ling, Chenxiaoji},
	month = jan,
	year = {2026},
	pages = {126},
}

@article{haussler_gems_2007,
	title = {{GEMS}: {Galaxy} {Fitting} {Catalogs} and {Testing} {Parametric} {Galaxy} {Fitting} {Codes}: {GALFIT} and {GIM2D}},
	volume = {172},
	issn = {0067-0049},
	shorttitle = {{GEMS}},
	url = {https://ui.adsabs.harvard.edu/abs/2007ApJS..172..615H},
	doi = {10.1086/518836},
	urldate = {2026-05-11},
	journal = {\apjs},
	publisher = {IOP},
	author = {Häussler, Boris and McIntosh, Daniel H. and Barden, Marco and Bell, Eric F. and Rix, Hans-Walter and Borch, Andrea and Beckwith, Steven V. W. and Caldwell, John A. R. and Heymans, Catherine and Jahnke, Knud and Jogee, Shardha and Koposov, Sergey E. and Meisenheimer, Klaus and Sánchez, Sebastian F. and Somerville, Rachel S. and Wisotzki, Lutz and Wolf, Christian},
	month = oct,
	year = {2007},
	pages = {615--633},
}

@article{van_der_wel_structural_2012,
	title = {Structural {Parameters} of {Galaxies} in {CANDELS}},
	volume = {203},
	issn = {0067-0049},
	url = {https://ui.adsabs.harvard.edu/abs/2012ApJS..203...24V},
	doi = {10.1088/0067-0049/203/2/24},
	urldate = {2026-05-11},
	journal = {\apjs},
	publisher = {IOP},
	author = {van der Wel, A. and Bell, E. F. and Häussler, B. and McGrath, E. J. and Chang, Yu-Yen and Guo, Yicheng and McIntosh, D. H. and Rix, H.-W. and Barden, M. and Cheung, E. and Faber, S. M. and Ferguson, H. C. and Galametz, A. and Grogin, N. A. and Hartley, W. and Kartaltepe, J. S. and Kocevski, D. D. and Koekemoer, A. M. and Lotz, J. and Mozena, M. and Peth, M. A. and Peng, Chien Y.},
	month = dec,
	year = {2012},
	pages = {24},
}

@article{ghosh_morphological_2023,
	title = {Morphological {Parameters} and {Associated} {Uncertainties} for 8 {Million} {Galaxies} in the {Hyper} {Suprime}-{Cam} {Wide} {Survey}},
	volume = {953},
	issn = {0004-637X},
	url = {https://ui.adsabs.harvard.edu/abs/2023ApJ...953..134G},
	doi = {10.3847/1538-4357/acd546},
	urldate = {2026-05-11},
	journal = {\apj},
	publisher = {IOP},
	author = {Ghosh, Aritra and Urry, C. Megan and Mishra, Aayush and Perreault-Levasseur, Laurence and Natarajan, Priyamvada and Sanders, David B. and Nagai, Daisuke and Tian, Chuan and Cappelluti, Nico and Kartaltepe, Jeyhan S. and Powell, Meredith C. and Rau, Amrit and Treister, Ezequiel},
	month = aug,
	year = {2023},
	pages = {134},
}

@article{rodriguez-gomez_statmorph_2022,
	title = {statmorph: {Non}-parametric morphological diagnostics of galaxy images},
	shorttitle = {statmorph},
	url = {https://ui.adsabs.harvard.edu/abs/2022ascl.soft01010R},
	urldate = {2026-05-11},
	journal = {ASCL},
	author = {Rodriguez-Gomez, Vicente and Lotz, Jennifer and Snyder, Greg},
	month = jan,
	year = {2022},
	pages = {ascl:2201.010},
}

@article{pasha_pysersic_2023,
	title = {pysersic: {A} {Python} package for determining galaxy structural properties via {Bayesian} inference, accelerated with jax},
	volume = {8},
	shorttitle = {pysersic},
	url = {https://ui.adsabs.harvard.edu/abs/2023JOSS....8.5703P},
	doi = {10.21105/joss.05703},
	urldate = {2026-05-11},
	journal = {J. Open Source Softw.},
	author = {Pasha, Imad and Miller, Tim B.},
	month = sep,
	year = {2023},
	pages = {5703},
}

@article{peng_detailed_2010,
	title = {Detailed {Decomposition} of {Galaxy} {Images}. {II}. {Beyond} {Axisymmetric} {Models}},
	volume = {139},
	issn = {0004-6256},
	url = {https://ui.adsabs.harvard.edu/abs/2010AJ....139.2097P},
	doi = {10.1088/0004-6256/139/6/2097},
	urldate = {2026-05-11},
	journal = {\aj},
	publisher = {IOP},
	author = {Peng, Chien Y. and Ho, Luis C. and Impey, Chris D. and Rix, Hans-Walter},
	month = jun,
	year = {2010},
	pages = {2097--2129},
}

@article{hodge_kiloparsec-scale_2016,
	title = {Kiloparsec-scale {Dust} {Disks} in {High}-redshift {Luminous} {Submillimeter} {Galaxies}},
	volume = {833},
	issn = {0004-637X},
	url = {https://ui.adsabs.harvard.edu/abs/2016ApJ...833..103H},
	doi = {10.3847/1538-4357/833/1/103},
	urldate = {2026-05-11},
	journal = {\apj},
	publisher = {IOP},
	author = {Hodge, J. A. and Swinbank, A. M. and Simpson, J. M. and Smail, I. and Walter, F. and Alexander, D. M. and Bertoldi, F. and Biggs, A. D. and Brandt, W. N. and Chapman, S. C. and Chen, C. C. and Coppin, K. E. K. and Cox, P. and Dannerbauer, H. and Edge, A. C. and Greve, T. R. and Ivison, R. J. and Karim, A. and Knudsen, K. K. and Menten, K. M. and Rix, H.-W. and Schinnerer, E. and Wardlow, J. L. and Weiss, A. and van der Werf, P.},
	month = dec,
	year = {2016},
	pages = {103},
}

@article{da_cunha_effect_2013,
	title = {On the {Effect} of the {Cosmic} {Microwave} {Background} in {High}-redshift ({Sub}-)millimeter {Observations}},
	volume = {766},
	issn = {0004-637X},
	url = {https://ui.adsabs.harvard.edu/abs/2013ApJ...766...13D},
	doi = {10.1088/0004-637X/766/1/13},
	urldate = {2026-05-11},
	journal = {\apj},
	publisher = {IOP},
	author = {da Cunha, Elisabete and Groves, Brent and Walter, Fabian and Decarli, Roberto and Weiss, Axel and Bertoldi, Frank and Carilli, Chris and Daddi, Emanuele and Elbaz, David and Ivison, Rob and Maiolino, Roberto and Riechers, Dominik and Rix, Hans-Walter and Sargent, Mark and Smail, Ian},
	month = mar,
	year = {2013},
	pages = {13},
}

@article{isobe_jades_2026,
	title = {{JADES}: the chemical enrichment pattern of distant galaxies ─ $\alpha$ enhancement, silicon depletion, and iron enhancement},
	volume = {547},
	issn = {0035-8711},
	shorttitle = {{JADES}},
	url = {https://ui.adsabs.harvard.edu/abs/2026MNRAS.547ag123I},
	doi = {10.1093/mnras/stag123},
	urldate = {2026-05-11},
	journal = {\mnras},
	publisher = {OUP},
	author = {Isobe, Yuki and Maiolino, Roberto and Ji, Xihan and D'Eugenio, Francesco and Simmonds, Charlotte and Scholtz, Jan and Juodžbalis, Ignas and Saxena, Aayush and Witstok, Joris and Kobayashi, Chiaki and Vanni, Irene and Salvadori, Stefania and Watanabe, Kuria and Monty, Stephanie and Belokurov, Vasily and Feltre, Anna and McClymont, William and Tacchella, Sandro and Curti, Mirko and Übler, Hannah and Charlot, Stéphane and Bunker, Andrew J. and Chevallard, Jacopo and Curtis-Lake, Emma and Kumari, Nimisha and Rinaldi, Pierluigi and Robertson, Brant and Williams, Christina C. and Willott, Chris},
	month = apr,
	year = {2026},
	pages = {stag123},
}

@misc{curtis-lake_jades_2025,
	title = {{JADES} {Data} {Release} 4 {Paper} {I}: {Sample} {Selection}, {Observing} {Strategy} and {Redshifts} of the complete spectroscopic sample},
	shorttitle = {{JADES} {Data} {Release} 4 {Paper} {I}},
	url = {https://ui.adsabs.harvard.edu/abs/2025arXiv251001033C},
	doi = {10.48550/arXiv.2510.01033},
	urldate = {2026-05-11},
	publisher = {arXiv},
	author = {Curtis-Lake, Emma and Cameron, Alex J. and Bunker, Andrew J. and Scholtz, Jan and Carniani, Stefano and Parlanti, Eleonora and D'Eugenio, Francesco and Jakobsen, Peter and Willmer, Christopher N. A. and Arribas, Santiago and Baker, William M. and Charlot, Stéphane and Chevallard, Jacopo and Circosta, Chiara and Curti, Mirko and Eisenstein, Daniel J. and Hainline, Kevin and Ji, Zhiyuan and Johnson, Benjamin D. and Jones, Gareth C. and Maiolino, Roberto and Maseda, Michael V. and Pérez-González, Pablo G. and Rawle, Tim and Rieke, Marcia and Rinaldi, Pierluigi and Robertson, Brant and Rodrígez Del Pino, Bruno and Saxena, Aayush and Shivaei, Irene and Smit, Renske and Tacchella, Sandro and Übler, Hannah and Venturi, Giacomo and Williams, Christina C. and Willott, Chris and Duan, Qiao},
	month = oct,
	year = {2025},
}

@misc{scholtz_jades_2025,
	title = {{JADES} {Data} {Release} 4 -- {Paper} {II}: {Data} reduction, analysis and emission-line fluxes of the complete spectroscopic sample},
	shorttitle = {{JADES} {Data} {Release} 4 -- {Paper} {II}},
	url = {https://ui.adsabs.harvard.edu/abs/2025arXiv251001034S},
	doi = {10.48550/arXiv.2510.01034},
	urldate = {2026-05-11},
	publisher = {arXiv},
	author = {Scholtz, J. and Carniani, S. and Parlanti, E. and D'Eugenio, F. and Curtis-Lake, E. and Jakobsen, P. and Bunker, A. J. and Cameron, A. J. and Arribas, S. and Baker, W. M. and Charlot, S. and Chevellard, J. and Circosta, C. and Curti, M. and Duan, Q. and Eisenstein, D. J. and Hainline, K. and Ji, Z. and Johnson, B. D. and Jones, G. C. and Kumari, N. and Maiolino, R. and Maseda, M. V. and Perna, M. and Pérez-González, P. G. and Rawle, T. and Rieke, M. and Rinaldi, P. and Robertson, B. and Saxena, A. and Shivaei, I. and Silcock, M. S. and Sun, Y. and Rodríguez Del Pino, B. and Tacchella, S. and Übler, H. and Venturi, G. and Williams, C. C. and Willmer, C. N. A. and Willott, C. and Witstok, J.},
	month = oct,
	year = {2025},
}

@article{eisenstein_overview_2026,
	title = {Overview of the {JWST} {Advanced} {Deep} {Extragalactic} {Survey} ({JADES})},
	volume = {283},
	issn = {0067-0049},
	url = {https://ui.adsabs.harvard.edu/abs/2026ApJS..283....6E},
	doi = {10.3847/1538-4365/ae3163},
	urldate = {2026-05-11},
	journal = {\apjs},
	publisher = {IOP},
	author = {Eisenstein, Daniel J. and Willott, Chris and Alberts, Stacey and Arribas, Santiago and Bonaventura, Nina and Bunker, Andrew J. and Cameron, Alex J. and Carniani, Stefano and Charlot, Stephane and Curtis-Lake, Emma and D'Eugenio, Francesco and Ferruit, Pierre and Giardino, Giovanna and Hainline, Kevin and Hausen, Ryan and Jakobsen, Peter and Johnson, Benjamin D. and Maiolino, Roberto and Rauscher, Bernard J. and Rieke, Marcia and Rieke, George and Rix, Hans-Walter and Robertson, Brant and Stark, Daniel P. and Tacchella, Sandro and Williams, Christina C. and Willmer, Christopher N. A. and Baker, William M. and Baum, Stefi and Bhatawdekar, Rachana and Boyett, Kristan and Chen, Zuyi and Chevallard, Jacopo and Circosta, Chiara and Curti, Mirko and Danhaive, A. Lola and DeCoursey, Christa and Endsley, Ryan and de Graaff, Anna and Dressler, Alan and Egami, Eiichi and Helton, Jakob M. and Hviding, Raphael E. and Ji, Zhiyuan and Jones, Gareth C. and Kumari, Nimisha and Lützgendorf, Nora and Laseter, Isaac and Looser, Tobias J. and Lyu, Jianwei and Maseda, Michael V. and Nelson, Erica and Parlanti, Eleonora and Perna, Michele and Puskás, Dávid and Rawle, Tim and Rodríguez Del Pino, Bruno and Rujopakarn, Wiphu and Sandles, Lester and Saxena, Aayush and Scholtz, Jan and Sharpe, Katherine and Shivaei, Irene and Silcock, Maddie S. and Simmonds, Charlotte and Skarbinski, Maya and Smit, Renske and Stone, Meredith and Suess, Katherine A. and Sun, Fengwu and Tang, Mengtao and Topping, Michael W. and Übler, Hannah and Villanueva, Natalia C. and Wallace, Imaan E. B. and Whitler, Lily and Witstok, Joris and Woodrum, Charity},
	month = mar,
	year = {2026},
	pages = {6},
}

@article{bunker_jades_2024,
	title = {{JADES} {NIRSpec} initial data release for the {Hubble} {Ultra} {Deep} {Field}: {Redshifts} and line fluxes of distant galaxies from the deepest {JWST} {Cycle} 1 {NIRSpec} multi-object spectroscopy},
	volume = {690},
	issn = {0004-6361},
	shorttitle = {{JADES} {NIRSpec} initial data release for the {Hubble} {Ultra} {Deep} {Field}},
	url = {https://ui.adsabs.harvard.edu/abs/2024A&A...690A.288B},
	doi = {10.1051/0004-6361/202347094},
	urldate = {2026-05-11},
	journal = {\aap},
	publisher = {EDP},
	author = {Bunker, Andrew J. and Cameron, Alex J. and Curtis-Lake, Emma and Jakobsen, Peter and Carniani, Stefano and Curti, Mirko and Witstok, Joris and Maiolino, Roberto and D'Eugenio, Francesco and Looser, Tobias J. and Willott, Chris and Bonaventura, Nina and Hainline, Kevin and Übler, Hannah and Willmer, Christopher N. A. and Saxena, Aayush and Smit, Renske and Alberts, Stacey and Arribas, Santiago and Baker, William M. and Baum, Stefi and Bhatawdekar, Rachana and Bowler, Rebecca A. A. and Boyett, Kristan and Charlot, Stephane and Chen, Zuyi and Chevallard, Jacopo and Circosta, Chiara and DeCoursey, Christa and de Graaff, Anna and Egami, Eiichi and Eisenstein, Daniel J. and Endsley, Ryan and Ferruit, Pierre and Giardino, Giovanna and Hausen, Ryan and Helton, Jakob M. and Hviding, Raphael E. and Ji, Zhiyuan and Johnson, Benjamin D. and Jones, Gareth C. and Kumari, Nimisha and Laseter, Isaac and Lützgendorf, Nora and Maseda, Michael V. and Nelson, Erica and Parlanti, Eleonora and Perna, Michele and Rauscher, Bernard J. and Rawle, Tim and Rix, Hans-Walter and Rieke, Marcia and Robertson, Brant and Rodríguez Del Pino, Bruno and Sandles, Lester and Scholtz, Jan and Sharpe, Katherine and Skarbinski, Maya and Stark, Daniel P. and Sun, Fengwu and Tacchella, Sandro and Topping, Michael W. and Villanueva, Natalia C. and Wallace, Imaan E. B. and Williams, Christina C. and Woodrum, Charity},
	month = oct,
	year = {2024},
	pages = {A288},
}

@article{shapley_aurora_2025-1,
	title = {The {AURORA} {Survey}: {A} {New} {Era} of {Emission}-line {Diagrams} with {JWST}/{NIRSpec}},
	volume = {980},
	issn = {0004-637X},
	shorttitle = {The {AURORA} {Survey}},
	url = {https://ui.adsabs.harvard.edu/abs/2025ApJ...980..242S},
	doi = {10.3847/1538-4357/adad68},
	urldate = {2026-05-11},
	journal = {\apj},
	publisher = {IOP},
	author = {Shapley, Alice E. and Sanders, Ryan L. and Topping, Michael W. and Reddy, Naveen A. and Berg, Danielle A. and Bouwens, Rychard J. and Brammer, Gabriel and Carnall, Adam C. and Cullen, Fergus and Davé, Romeel and Dunlop, James S. and Ellis, Richard S. and Förster Schreiber, N. M. and Furlanetto, Steven R. and Glazebrook, Karl and Illingworth, Garth D. and Jones, Tucker and Kriek, Mariska and McLeod, Derek J. and McLure, Ross J. and Narayanan, Desika and Oesch, Pascal and Pahl, Anthony J. and Pettini, Max and Schaerer, Daniel and Stark, Daniel P. and Steidel, Charles C. and Tang, Mengtao and Clarke, Leonardo and Donnan, Callum T. and Kehoe, Emily},
	month = feb,
	year = {2025},
	pages = {242},
}

@article{shapley_mosdef_2019,
	title = {The {MOSDEF} {Survey}: {Sulfur} {Emission}-line {Ratios} {Provide} {New} {Insights} into {Evolving} {Interstellar} {Medium} {Conditions} at {High} {Redshift}},
	volume = {881},
	issn = {0004-637X},
	shorttitle = {The {MOSDEF} {Survey}},
	url = {https://ui.adsabs.harvard.edu/abs/2019ApJ...881L..35S},
	doi = {10.3847/2041-8213/ab385a},
	urldate = {2026-05-11},
	journal = {\apj},
	publisher = {IOP},
	author = {Shapley, Alice E. and Sanders, Ryan L. and Shao, Peng and Reddy, Naveen A. and Kriek, Mariska and Coil, Alison L. and Mobasher, Bahram and Siana, Brian and Shivaei, Irene and Freeman, William R. and Azadi, Mojegan and Price, Sedona H. and Leung, Gene C. K. and Fetherolf, Tara and de Groot, Laura and Zick, Tom and Fornasini, Francesca M. and Barro, Guillermo},
	month = aug,
	year = {2019},
	pages = {L35},
}

@article{kriek_mosfire_2015,
	title = {The {MOSFIRE} {Deep} {Evolution} {Field} ({MOSDEF}) {Survey}: {Rest}-frame {Optical} {Spectroscopy} for {\textasciitilde}1500 {H}-selected {Galaxies} at 1.37 {\textless} z {\textless} 3.8},
	volume = {218},
	issn = {0067-0049},
	shorttitle = {The {MOSFIRE} {Deep} {Evolution} {Field} ({MOSDEF}) {Survey}},
	url = {https://ui.adsabs.harvard.edu/abs/2015ApJS..218...15K},
	doi = {10.1088/0067-0049/218/2/15},
	urldate = {2026-05-11},
	journal = {\apjs},
	publisher = {IOP},
	author = {Kriek, Mariska and Shapley, Alice E. and Reddy, Naveen A. and Siana, Brian and Coil, Alison L. and Mobasher, Bahram and Freeman, William R. and de Groot, Laura and Price, Sedona H. and Sanders, Ryan and Shivaei, Irene and Brammer, Gabriel B. and Momcheva, Ivelina G. and Skelton, Rosalind E. and van Dokkum, Pieter G. and Whitaker, Katherine E. and Aird, James and Azadi, Mojegan and Kassis, Marc and Bullock, James S. and Conroy, Charlie and Davé, Romeel and Kereš, Dušan and Krumholz, Mark},
	month = jun,
	year = {2015},
	pages = {15},
}

@article{nyhagen_theoretical_2025,
	title = {A theoretical investigation of far-infrared fine structure lines at z {\textgreater} 6 and of the origin of the [{O} {III}]88 $\mu$m/[{C} {II}]158 $\mu$m enhancement},
	volume = {702},
	issn = {0004-6361},
	url = {https://ui.adsabs.harvard.edu/abs/2025A&A...702A.260N},
	doi = {10.1051/0004-6361/202452718},
	urldate = {2026-05-11},
	journal = {\aap},
	publisher = {EDP},
	author = {Nyhagen, C. T. and Schimek, A. and Cicone, C. and Decataldo, D. and Shen, S.},
	month = oct,
	year = {2025},
	pages = {A260},
}

@article{hagimoto_compact_2025,
	title = {Compact {Ionized} {Gas} {Region} {Surrounded} by {Porous} {Neutral} {Gas} in a {Dusty} {Lyman} {Break} {Galaxy} at {Redshift} z = 8.312},
	volume = {990},
	issn = {0004-637X},
	url = {https://ui.adsabs.harvard.edu/abs/2025ApJ...990...29H},
	doi = {10.3847/1538-4357/ade87e},
	urldate = {2026-05-11},
	journal = {\apj},
	publisher = {IOP},
	author = {Hagimoto, Masato and Tamura, Yoichi and Inoue, Akio K. and Umehata, Hideki and Bakx, Tom J. L. C. and Hashimoto, Takuya and Mawatari, Ken and Sugahara, Yuma and Fudamoto, Yoshinobu and Harikane, Yuichi and Matsuo, Hiroshi and Taniguchi, Akio},
	month = sep,
	year = {2025},
	pages = {29},
}

@article{katz_nature_2022,
	title = {The nature of high [{O} {III}]88 $\mu$ m/[{C} {II}]158 $\mu$m galaxies in the epoch of reionization: {Low} carbon abundance and a top-heavy {IMF}?},
	volume = {510},
	issn = {0035-8711},
	shorttitle = {The nature of high [{O} {III}]88 $\mu$ m/[{C} {II}]158 $\mu$m galaxies in the epoch of reionization},
	url = {https://ui.adsabs.harvard.edu/abs/2022MNRAS.510.5603K},
	doi = {10.1093/mnras/stac028},
	urldate = {2026-05-11},
	journal = {\mnras},
	publisher = {OUP},
	author = {Katz, Harley and Rosdahl, Joakim and Kimm, Taysun and Garel, Thibault and Blaizot, Jérémy and Haehnelt, Martin G. and Michel-Dansac, Léo and Martin-Alvarez, Sergio and Devriendt, Julien and Slyz, Adrianne and Teyssier, Romain and Ocvirk, Pierre and Laporte, Nicolas and Ellis, Richard},
	month = mar,
	year = {2022},
	pages = {5603--5622},
}

@article{vallini_high_2021,
	title = {High [{O} {III}]/[{C} {II}] surface brightness ratios trace early starburst galaxies},
	volume = {505},
	issn = {0035-8711},
	url = {https://ui.adsabs.harvard.edu/abs/2021MNRAS.505.5543V},
	doi = {10.1093/mnras/stab1674},
	urldate = {2026-05-11},
	journal = {\mnras},
	publisher = {OUP},
	author = {Vallini, L. and Ferrara, A. and Pallottini, A. and Carniani, S. and Gallerani, S.},
	month = aug,
	year = {2021},
	pages = {5543--5553},
}

@article{chabrier_galactic_2003,
	title = {Galactic {Stellar} and {Substellar} {Initial} {Mass} {Function}},
	volume = {115},
	issn = {0004-6280},
	url = {https://ui.adsabs.harvard.edu/abs/2003PASP..115..763C},
	doi = {10.1086/376392},
	urldate = {2026-05-11},
	journal = {\pasp},
	publisher = {IOP},
	author = {Chabrier, Gilles},
	month = jul,
	year = {2003},
	pages = {763--795},
}

@article{asplund_chemical_2009,
	title = {The {Chemical} {Composition} of the {Sun}},
	volume = {47},
	issn = {0066-4146},
	url = {https://ui.adsabs.harvard.edu/abs/2009ARA&A..47..481A},
	doi = {10.1146/annurev.astro.46.060407.145222},
	urldate = {2026-05-11},
	journal = {\araa},
	author = {Asplund, Martin and Grevesse, Nicolas and Sauval, A. Jacques and Scott, Pat},
	month = sep,
	year = {2009},
	pages = {481--522},
}

@inproceedings{shan_development_2005,
	title = {Development of a 385-{500GHz} {SIS} {Mixer} for {ALMA} {Band} 8},
	url = {https://ui.adsabs.harvard.edu/abs/2005stt..conf..175S},
	urldate = {2026-05-11},
	author = {Shan, Wenlei and Asayama, Shinichiro and Kamikura, Mamoru and Noguchi, Takashi and Shi, Shengcai and Sekimoto, Yutaro},
	month = may,
	year = {2005},
	pages = {175--180},
}

@inproceedings{sekimoto_development_2008,
	title = {Development of {ALMA} {Band} 8 (385-500 {GHz}) {Cartridge}},
	url = {https://ui.adsabs.harvard.edu/abs/2008stt..conf..253S},
	urldate = {2026-05-11},
	author = {Sekimoto, Y. and Iizuko, Y. and Satou, N. and Ito, T. and Kumagai, K. and Kamikura, M. and Naruse, M. and Shan, W. L.},
	month = apr,
	year = {2008},
	pages = {253--257},
}

@inproceedings{sekimoto_evaluation_2009,
	title = {Evaluation of {ALMA} {Band} 8 {S}/{N01} {Cartridge}},
	url = {https://ui.adsabs.harvard.edu/abs/2009stt..conf....6S},
	urldate = {2026-05-11},
	author = {Sekimoto, Y. and Iizuka, Y. and Ito, T. and Kumagai, K. and Satou, N. and Kamikura, M. and Serizawa, Y. and Naruse, N. and Niizeki, Y. and Fujimoto, Y. and Shan, W. L.},
	month = apr,
	year = {2009},
	pages = {6--6},
}

@inproceedings{ediss_alma_2004,
	title = {{ALMA} {Band} 6 {Cartridge}: {Design} and {Performance}},
	shorttitle = {{ALMA} {Band} 6 {Cartridge}},
	url = {https://ui.adsabs.harvard.edu/abs/2004stt..conf..181E},
	urldate = {2026-05-11},
	author = {Ediss, G. A. and Carter, M. and Cheng, J. and Effland, J. E. and Grammer, W. and Horner, Jr., N. and Kerr, A. R. and Koller, D. and Lauria, E. F. and Morris, G. and Pan, S.-K. and Reiland, G. and Sullivan, M.},
	month = jan,
	year = {2004},
	pages = {181--188},
}

@inproceedings{kerr_alma_2004,
	title = {The {ALMA} {Band} 6 (211-275 {GHz}) {Sideband}-{Separating} {SIS} {Mixer}-{Preamp}},
	url = {https://ui.adsabs.harvard.edu/abs/2004stt..conf...55K},
	urldate = {2026-05-11},
	author = {Kerr, A. R. and Pan, S.-K. and Lauria, E. F. and Lichtenberger, A. W. and Zhang, J. and Pospieszalski, M. W. and Horner, N. and Ediss, G. A. and Effland, J. E. and Groves, R. L.},
	month = jan,
	year = {2004},
	pages = {55--61},
}

@inproceedings{mcmullin_casa_2007,
	title = {{CASA} {Architecture} and {Applications}},
	volume = {376},
	url = {https://ui.adsabs.harvard.edu/abs/2007ASPC..376..127M},
	urldate = {2026-05-11},
	author = {McMullin, J. P. and Waters, B. and Schiebel, D. and Young, W. and Golap, K.},
	month = oct,
	year = {2007},
	pages = {127},
}

@article{casa_team_casa_2022,
	title = {{CASA}, the {Common} {Astronomy} {Software} {Applications} for {Radio} {Astronomy}},
	volume = {134},
	issn = {0004-6280},
	url = {https://ui.adsabs.harvard.edu/abs/2022PASP..134k4501C},
	doi = {10.1088/1538-3873/ac9642},
	urldate = {2026-05-11},
	journal = {\pasp},
	publisher = {IOP},
	author = {{CASA Team} and Bean, Ben and Bhatnagar, Sanjay and Castro, Sandra and Donovan Meyer, Jennifer and Emonts, Bjorn and Garcia, Enrique and Garwood, Robert and Golap, Kumar and Gonzalez Villalba, Justo and Harris, Pamela and Hayashi, Yohei and Hoskins, Josh and Hsieh, Mingyu and Jagannathan, Preshanth and Kawasaki, Wataru and Keimpema, Aard and Kettenis, Mark and Lopez, Jorge and Marvil, Joshua and Masters, Joseph and McNichols, Andrew and Mehringer, David and Miel, Renaud and Moellenbrock, George and Montesino, Federico and Nakazato, Takeshi and Ott, Juergen and Petry, Dirk and Pokorny, Martin and Raba, Ryan and Rau, Urvashi and Schiebel, Darrell and Schweighart, Neal and Sekhar, Srikrishna and Shimada, Kazuhiko and Small, Des and Steeb, Jan-Willem and Sugimoto, Kanako and Suoranta, Ville and Tsutsumi, Takahiro and van Bemmel, Ilse M. and Verkouter, Marjolein and Wells, Akeem and Xiong, Wei and Szomoru, Arpad and Griffith, Morgan and Glendenning, Brian and Kern, Jeff},
	month = nov,
	year = {2022},
	pages = {114501},
}

@article{fraternali_fast_2021,
	title = {Fast rotating and low-turbulence discs at z ≃ 4.5: {Dynamical} evidence of their evolution into local early-type galaxies},
	volume = {647},
	issn = {0004-6361},
	shorttitle = {Fast rotating and low-turbulence discs at z ≃ 4.5},
	url = {https://ui.adsabs.harvard.edu/abs/2021A&A...647A.194F},
	doi = {10.1051/0004-6361/202039807},
	urldate = {2026-05-12},
	journal = {\aap},
	publisher = {EDP},
	author = {Fraternali, F. and Karim, A. and Magnelli, B. and Gómez-Guijarro, C. and Jiménez-Andrade, E. F. and Posses, A. C.},
	month = mar,
	year = {2021},
	pages = {A194},
}

@article{fisher_rebels-ifu_2026,
	title = {{REBELS}-{IFU}: {Steeply} rising star formation histories and the importance of dust obscuration in massive z ≃ 7 galaxies revealed by multiwavelength observations},
	volume = {546},
	issn = {0035-8711},
	shorttitle = {{REBELS}-{IFU}},
	url = {https://ui.adsabs.harvard.edu/abs/2026MNRAS.546ag049F},
	doi = {10.1093/mnras/stag049},
	urldate = {2026-05-12},
	journal = {\mnras},
	publisher = {OUP},
	author = {Fisher, R. and Bowler, R. A. A. and Cochrane, R. K. and Rowland, L. E. and Stefanon, M. and Algera, H. S. B. and Aravena, M. and Bouwens, R. and da Cunha, E. and Dayal, P. and Ferrara, A. and Hodge, J. A. and Inami, H. and Komarova, L. and Smit, R. and Sommovigo, L. and Stark, D. P. and van der Werf, P. P.},
	month = mar,
	year = {2026},
	pages = {stag049},
}

@article{rowland_rebels-ifu_2026,
	title = {{REBELS}-{IFU}: evidence for metal-rich massive galaxies at z ∼ 6−8},
	volume = {546},
	issn = {0035-8711},
	shorttitle = {{REBELS}-{IFU}},
	url = {https://ui.adsabs.harvard.edu/abs/2026MNRAS.546f2023R},
	doi = {10.1093/mnras/staf2023},
	urldate = {2026-05-12},
	journal = {\mnras},
	publisher = {OUP},
	author = {Rowland, Lucie E. and Stefanon, Mauro and Bouwens, Rychard and Hodge, Jacqueline and Algera, Hiddo and Fisher, Rebecca and Dayal, Pratika and Pallottini, Andrea and Stark, Daniel P. and Heintz, Kasper E. and Aravena, Manuel and Bowler, Rebecca A. A. and Cescon, Karin and Endsley, Ryan and Ferrara, Andrea and Fudamoto, Yoshinobu and Gonzalez, Valentino and Graziani, Luca and Gulis, Cindy and Herard-Demanche, Thomas and Inami, Hanae and Laza-Ramos, Andrès and van Leeuwen, Ivana and de Looze, Ilse and Nanayakkara, Themiya and Oesch, Pascal and Ormerod, Katherine and Palla, Marco and Sartorio, Nina S. and Schouws, Sander and Smit, Renske and Sommovigo, Laura and Toft, Sune and Weaver, John R. and van der Werf, Paul},
	month = feb,
	year = {2026},
	pages = {staf2023},
}

@misc{sanders_aurora_2025,
	title = {The {AURORA} {Survey}: {High}-{Redshift} {Empirical} {Metallicity} {Calibrations} from {Electron} {Temperature} {Measurements} at z=2-10},
	shorttitle = {The {AURORA} {Survey}},
	url = {https://ui.adsabs.harvard.edu/abs/2025arXiv250810099S},
	doi = {10.48550/arXiv.2508.10099},
	urldate = {2026-05-12},
	publisher = {arXiv},
	author = {Sanders, Ryan L. and Shapley, Alice E. and Topping, Michael W. and Reddy, Naveen A. and Berg, Danielle A. and Khostovan, Ali Ahmad and Bouwens, Rychard J. and Brammer, Gabriel and Carnall, Adam C. and Cullen, Fergus and Davé, Romeel and Dunlop, James S. and Ellis, Richard S. and Förster Schreiber, N. M. and Furlanetto, Steven R. and Glazebrook, Karl and Illingworth, Garth D. and Jones, Tucker and Kriek, Mariska and McLeod, Derek J. and McLure, Ross J. and Narayanan, Desika and Oesch, Pascal A. and Pahl, Anthony J. and Pettini, Max and Schaerer, Daniel and Stark, Daniel P. and Steidel, Charles C. and Tang, Mengtao and Clarke, Leonardo and Donnan, Callum T. and Kehoe, Emily},
	month = aug,
	year = {2025},
}

@article{sanders_direct_2024,
	title = {Direct {T} e-based {Metallicities} of z = 2─9 {Galaxies} with {JWST}/{NIRSpec}: {Empirical} {Metallicity} {Calibrations} {Applicable} from {Reionization} to {Cosmic} {Noon}},
	volume = {962},
	issn = {0004-637X},
	shorttitle = {Direct {T} e-based {Metallicities} of z = 2─9 {Galaxies} with {JWST}/{NIRSpec}},
	url = {https://ui.adsabs.harvard.edu/abs/2024ApJ...962...24S},
	doi = {10.3847/1538-4357/ad15fc},
	urldate = {2026-05-12},
	journal = {\apj},
	publisher = {IOP},
	author = {Sanders, Ryan L. and Shapley, Alice E. and Topping, Michael W. and Reddy, Naveen A. and Brammer, Gabriel B.},
	month = feb,
	year = {2024},
	pages = {24},
}

@article{harikane_jwst_2025,
	title = {{JWST} and {ALMA} {Joint} {Analysis} with [{O} {II}] λλ3726, 3729, [{O} {III}] λ4363, [{O} {III}] 88 $\mu$m, and [{O} {III}] 52 $\mu$m: {Multizone} {Evolution} of {Electron} {Densities} at z ∼ 0─14 and its {Impact} on {Metallicity} {Measurements}},
	volume = {993},
	issn = {0004-637X},
	shorttitle = {{JWST} and {ALMA} {Joint} {Analysis} with [{O} {II}] λλ3726, 3729, [{O} {III}] λ4363, [{O} {III}] 88 $\mu$m, and [{O} {III}] 52 $\mu$m},
	url = {https://ui.adsabs.harvard.edu/abs/2025ApJ...993..204H},
	doi = {10.3847/1538-4357/ae0e53},
	urldate = {2026-05-12},
	journal = {\apj},
	publisher = {IOP},
	author = {Harikane, Yuichi and Sanders, Ryan L. and Ellis, Richard and Jones, Tucker and Ouchi, Masami and Laporte, Nicolas and Roberts-Borsani, Guido and Katz, Harley and Nakajima, Kimihiko and Ono, Yoshiaki and Gupta, Mansi},
	month = nov,
	year = {2025},
	pages = {204},
}

@article{usui_rioja_2025,
	title = {{RIOJA}. {JWST} and {ALMA} {Unveil} the {Inhomogeneous} and {Complex} {Interstellar} {Medium} {Structure} in a {Star}-forming {Galaxy} at z = 6.81},
	volume = {991},
	issn = {0004-637X},
	url = {https://ui.adsabs.harvard.edu/abs/2025ApJ...991L..38U},
	doi = {10.3847/2041-8213/ae0574},
	urldate = {2026-05-12},
	journal = {\apj},
	publisher = {IOP},
	author = {Usui, Mitsutaka and Mawatari, Ken and Álvarez-Márquez, Javier and Hashimoto, Takuya and Sugahara, Yuma and Marques-Chaves, Rui and Inoue, Akio K. and Colina, Luis and Arribas, Santiago and Blanco-Prieto, Carmen and Nakazato, Yurina and Yoshida, Naoki and Bakx, Tom J. L. C. and Ceverino, Daniel and Costantin, Luca and Crespo Gómez, Alejandro and Hagimoto, Masato and Matsuo, Hiroshi and Osone, Wataru and Ren, Yi W. and Fudamoto, Yoshinobu and Hashigaya, Takeshi and Pereira-Santaella, Miguel and Tamura, Yoichi},
	month = oct,
	year = {2025},
	pages = {L38},
}

@article{castellano_investigating_2026,
	title = {Investigating ionising sources and the complex interstellar medium of {GHZ2} at z = 12.3},
	volume = {9},
	issn = {2565-6120},
	url = {https://ui.adsabs.harvard.edu/abs/2026OJAp....960281C},
	doi = {10.33232/001c.160281},
	urldate = {2026-05-12},
	journal = {Open J. Astrophys.},
	author = {Castellano, M. and Napolitano, L. and Moreschini, B. and Calabrò, A. and Christensen, L. and Llerena, M. and Bakx, T. J. L. C. and Belfiore, F. and Bevacqua, D. and Dickinson, M. and Fontana, A. and Gandolfi, G. and Gasparetto, T. and Marconi, A. and Mascia, S. and Merlin, E. and Morishita, T. and Nanayakkara, T. and Paris, D. and Pentericci, L. and Pérez-Díaz, B. and Roberts-Borsani, G. and Ruiz, S. Rojas and Santini, P. and Treu, T. and Vanzella, E. and Vulcani, B. and Wang, X. and Yoon, I. and Zavala, J.},
	month = apr,
	year = {2026},
	pages = {60281},
}

@article{storey_theoretical_2000,
	title = {Theoretical values for the [{OIII}] 5007/4959 line-intensity ratio and homologous cases},
	volume = {312},
	issn = {0035-8711},
	url = {https://ui.adsabs.harvard.edu/abs/2000MNRAS.312..813S},
	doi = {10.1046/j.1365-8711.2000.03184.x},
	urldate = {2026-05-12},
	journal = {\mnras},
	publisher = {OUP},
	author = {Storey, P. J. and Zeippen, C. J.},
	month = mar,
	year = {2000},
	pages = {813--816},
}

@article{yttergren_kinematics_2025,
	title = {Kinematics of synthetically observed high-z rotating discs: reliability and biases of {3D} fitting tools},
	volume = {543},
	issn = {0035-8711},
	shorttitle = {Kinematics of synthetically observed high-z rotating discs},
	url = {https://ui.adsabs.harvard.edu/abs/2025MNRAS.543.3103Y},
	doi = {10.1093/mnras/staf1489},
	urldate = {2026-05-12},
	journal = {\mnras},
	publisher = {OUP},
	author = {Yttergren, M. and Knudsen, K. K. and Molina, J. and Jones, G. C. and Kade, K. and Scholtz, J. and Bewketu Belete, A.},
	month = nov,
	year = {2025},
	pages = {3103--3122},
}

@article{mancera_pina_galaxy-halo_2025,
	title = {The galaxy-halo connection of disc galaxies over six orders of magnitude in stellar mass},
	volume = {699},
	issn = {0004-6361},
	url = {https://ui.adsabs.harvard.edu/abs/2025A&A...699A.311M},
	doi = {10.1051/0004-6361/202554381},
	urldate = {2026-05-12},
	journal = {\aap},
	publisher = {EDP},
	author = {Mancera Piña, Pavel E. and Read, Justin I. and Kim, Stacy and Marasco, Antonino and Benavides, José A. and Glowacki, Marcin and Pezzulli, Gabriele and Lagos, Claudia del P.},
	month = jul,
	year = {2025},
	pages = {A311},
}

@article{krumholz_unified_2018,
	title = {A unified model for galactic discs: star formation, turbulence driving, and mass transport},
	volume = {477},
	issn = {0035-8711},
	shorttitle = {A unified model for galactic discs},
	url = {https://ui.adsabs.harvard.edu/abs/2018MNRAS.477.2716K},
	doi = {10.1093/mnras/sty852},
	urldate = {2026-05-12},
	journal = {\mnras},
	publisher = {OUP},
	author = {Krumholz, Mark R. and Burkhart, Blakesley and Forbes, John C. and Crocker, Roland M.},
	month = jun,
	year = {2018},
	pages = {2716--2740},
}

@misc{fudamoto_alma_2025,
	title = {{ALMA} {Observations} of [{OI}]145um and [{NII}]205um {Emission} lines from {Star}-{Forming} {Galaxies} at \$z{\textbackslash}sim7\$},
	url = {https://ui.adsabs.harvard.edu/abs/2025arXiv250403831F},
	doi = {10.48550/arXiv.2504.03831},
	urldate = {2026-05-12},
	publisher = {arXiv},
	author = {Fudamoto, Yoshinobu and Inoue, Akio K. and Bouwens, Rychard and Inami, Hanae and Smit, Renske and Stark, Dan and Aravena, Manuel and Pallottini, Andrea and Hashimoto, Takuya and Oguri, Masamune and Bowler, Rebecca A. A. and da Cunha, Elisabete and Dayal, Pratika and Ferrara, Andrea and Fujimoto, Seiji and Heintz, Kasper E. and Hygate, Alexander P. S. and van Leeuwen, Ivana F. and De Looze, Ilse and Rowland, Lucie E. and Stefanon, Mauro and Sugahara, Yuma and Witstok, Joris and van der Werf, Paul P.},
	month = apr,
	year = {2025},
}

@article{roman-oliveira_regular_2023,
	title = {Regular rotation and low turbulence in a diverse sample of z ∼ 4.5 galaxies observed with {ALMA}},
	volume = {521},
	issn = {0035-8711},
	url = {https://ui.adsabs.harvard.edu/abs/2023MNRAS.521.1045R},
	doi = {10.1093/mnras/stad530},
	urldate = {2026-05-12},
	journal = {\mnras},
	publisher = {OUP},
	author = {Roman-Oliveira, Fernanda and Fraternali, Filippo and Rizzo, Francesca},
	month = may,
	year = {2023},
	pages = {1045--1065},
}

@article{matthee_jwst_2025,
	title = {{JWST} provides a new view of cosmic dawn: latest developments in studies of early galaxies},
	volume = {66},
	issn = {0010-7514},
	shorttitle = {{JWST} provides a new view of cosmic dawn},
	url = {https://ui.adsabs.harvard.edu/abs/2025ConPh..66..116M/abstract},
	doi = {10.1080/00107514.2025.2586370},
	language = {en},
	number = {1-4},
	urldate = {2026-05-13},
	journal = {Contemp. Phys.},
	author = {Matthee, Jorryt},
	month = oct,
	year = {2025},
	pages = {116},
}

@article{herrera-camus_early_2026,
	title = {The early {Universe} with {JWST} and {ALMA}},
	volume = {10},
	issn = {2397-3366},
	url = {https://ui.adsabs.harvard.edu/abs/2026NatAs..10...34H/abstract},
	doi = {10.1038/s41550-025-02726-0},
	language = {en},
	urldate = {2026-05-13},
	journal = {Nat. Astron.},
	author = {Herrera-Camus, Rodrigo and Förster Schreiber, Natascha M. and Vallini, Livia and Bouwens, Rychard and Silverman, John D.},
	month = jan,
	year = {2026},
	pages = {34--41},
}

@article{schouws_significant_2022,
	title = {Significant {Dust}-obscured {Star} {Formation} in {Luminous} {Lyman}-break {Galaxies} at z 7-8},
	volume = {928},
	issn = {0004-637X},
	url = {https://ui.adsabs.harvard.edu/abs/2022ApJ...928...31S/abstract},
	doi = {10.3847/1538-4357/ac4605},
	language = {en},
	number = {1},
	urldate = {2026-05-13},
	journal = {\apj},
	author = {Schouws, Sander and Stefanon, Mauro and Bouwens, Rychard and Smit, Renske and Hodge, Jacqueline and Labbé, Ivo and Algera, Hiddo and Boogaard, Leindert and Carniani, Stefano and Fudamoto, Yoshinobu and Holwerda, Benne W. and Illingworth, Garth D. and Maiolino, Roberto and Maseda, Michael and Oesch, Pascal and van der Werf, Paul},
	month = mar,
	year = {2022},
	pages = {31},
}

@article{li_alma-cristal_2024,
	title = {The {ALMA}-{CRISTAL} {Survey}: {Spatially} {Resolved} {Star} {Formation} {Activity} and {Dust} {Content} in 4 {\textless} z {\textless} 6 {Star}-forming {Galaxies}},
	volume = {976},
	issn = {0004-637X},
	shorttitle = {The {ALMA}-{CRISTAL} {Survey}},
	url = {https://doi.org/10.3847/1538-4357/ad7fee},
	doi = {10.3847/1538-4357/ad7fee},
	language = {en},
	number = {1},
	urldate = {2026-05-13},
	journal = {\apj},
	publisher = {The American Astronomical Society},
	author = {Li, Juno and Da Cunha, Elisabete and González-López, Jorge and Aravena, Manuel and De Looze, Ilse and Förster Schreiber, N. M. and Herrera-Camus, Rodrigo and Spilker, Justin and Tadaki, Ken-ichi and Barcos-Munoz, Loreto and Battisti, Andrew J. and Birkin, Jack E. and Bowler, Rebecca A. A. and Davies, Rebecca and Díaz-Santos, Tanio and Ferrara, Andrea and Fisher, Deanne B. and Hodge, Jacqueline and Ikeda, Ryota and Killi, Meghana and Lee, Lilian and Liu, Daizhong and Lutz, Dieter and Mitsuhashi, Ikki and Naab, Thorsten and Posses, Ana and Relaño, Monica and Solimano, Manuel and Übler, Hannah and van der Giessen, Stefan Anthony and Villanueva, Vicente},
	month = nov,
	year = {2024},
	pages = {70},
}

@article{gimenez-arteaga_spatially_2023,
	title = {Spatially {Resolved} {Properties} of {Galaxies} at 5 {\textless} z {\textless} 9 in the {SMACS} 0723 {JWST} {ERO} {Field}},
	volume = {948},
	issn = {0004-637X},
	url = {https://doi.org/10.3847/1538-4357/acc5ea},
	doi = {10.3847/1538-4357/acc5ea},
	language = {en},
	number = {2},
	urldate = {2026-05-13},
	journal = {\apj},
	publisher = {The American Astronomical Society},
	author = {Giménez-Arteaga, Clara and Oesch, Pascal A. and Brammer, Gabriel B. and Valentino, Francesco and Mason, Charlotte A. and Weibel, Andrea and Barrufet, Laia and Fujimoto, Seiji and Heintz, Kasper E. and Nelson, Erica J. and Strait, Victoria B. and Suess, Katherine A. and Gibson, Justus},
	month = may,
	year = {2023},
	pages = {126},
}

@article{abdurrouf_spatially_2023,
	title = {Spatially {Resolved} {Stellar} {Populations} of 0.3 {\textless} z {\textless} 6.0 {Galaxies} in {WHL} 0137–08 and {MACS} 0647+70 {Clusters} as {Revealed} by {JWST}: {How} {Do} {Galaxies} {Grow} and {Quench} over {Cosmic} {Time}?},
	volume = {945},
	issn = {0004-637X},
	shorttitle = {Spatially {Resolved} {Stellar} {Populations} of 0.3 {\textless} z {\textless} 6.0 {Galaxies} in {WHL} 0137–08 and {MACS} 0647+70 {Clusters} as {Revealed} by {JWST}},
	url = {https://doi.org/10.3847/1538-4357/acba06},
	doi = {10.3847/1538-4357/acba06},
	language = {en},
	number = {2},
	urldate = {2026-05-13},
	journal = {\apj},
	publisher = {The American Astronomical Society},
	author = {{Abdurro’uf} and Coe, Dan and Jung, Intae and Ferguson, Henry C. and Brammer, Gabriel and Iyer, Kartheik G. and Bradley, Larry D. and Dayal, Pratika and Windhorst, Rogier A. and Zitrin, Adi and Meena, Ashish Kumar and Oguri, Masamune and Diego, Jose M. and Kokorev, Vasily and Dimauro, Paola and Adamo, Angela and Conselice, Christopher J. and Welch, Brian and Vanzella, Eros and Hsiao, Tiger Yu-Yang and Xu, Xinfeng and Roy, Namrata and Mulcahey, Celia R.},
	month = mar,
	year = {2023},
	pages = {117},
}

@article{cappellari_vorbin_2012,
	title = {{VorBin}: {Voronoi} binning method},
	shorttitle = {{VorBin}},
	url = {https://ui.adsabs.harvard.edu/abs/2012ascl.soft11006C/abstract},
	language = {en},
	urldate = {2026-05-13},
	journal = {ASCL},
	author = {Cappellari, Michele and Copin, Yannick},
	month = nov,
	year = {2012},
	pages = {ascl:1211.006},
}

@article{cappellari_adaptive_2003,
	title = {Adaptive spatial binning of integral-field spectroscopic data using {Voronoi} tessellations},
	volume = {342},
	issn = {0035-8711},
	url = {https://ui.adsabs.harvard.edu/abs/2003MNRAS.342..345C/abstract},
	doi = {10.1046/j.1365-8711.2003.06541.x},
	language = {en},
	number = {2},
	urldate = {2026-05-13},
	journal = {\mnras},
	author = {Cappellari, Michele and Copin, Yannick},
	month = jun,
	year = {2003},
	pages = {345--354},
}

@article{lee_disk_2025,
	title = {Disk {Kinematics} at {High} {Redshift}: {DysmalPy}'s {Extension} to {3D} {Modeling} and {Comparison} with {Different} {Approaches}},
	volume = {978},
	issn = {0004-637X},
	shorttitle = {Disk {Kinematics} at {High} {Redshift}},
	url = {https://ui.adsabs.harvard.edu/abs/2025ApJ...978...14L},
	doi = {10.3847/1538-4357/ad90b5},
	urldate = {2026-05-13},
	journal = {\apj},
	publisher = {IOP},
	author = {Lee, Lilian L. and Förster Schreiber, Natascha M. and Price, Sedona H. and Liu, Daizhong and Genzel, Reinhard and Davies, Ric and Tacconi, Linda J. and Shimizu, Taro T. and Nestor Shachar, Amit and Espejo Salcedo, Juan M. and Pastras, Stavros and Wuyts, Stijn and Lutz, Dieter and Renzini, Alvio and Übler, Hannah and Herrera-Camus, Rodrigo and Sternberg, Amiel},
	month = jan,
	year = {2025},
	pages = {14},
}

@article{navarro_structure_1996,
	title = {The {Structure} of {Cold} {Dark} {Matter} {Halos}},
	volume = {462},
	issn = {0004-637X},
	url = {https://ui.adsabs.harvard.edu/abs/1996ApJ...462..563N},
	doi = {10.1086/177173},
	urldate = {2026-05-13},
	journal = {\apj},
	publisher = {IOP},
	author = {Navarro, Julio F. and Frenk, Carlos S. and White, Simon D. M.},
	month = may,
	year = {1996},
	pages = {563},
}

@article{burkert_high-redshift_2010,
	title = {High-redshift {Star}-forming {Galaxies}: {Angular} {Momentum} and {Baryon} {Fraction}, {Turbulent} {Pressure} {Effects}, and the {Origin} of {Turbulence}},
	volume = {725},
	issn = {0004-637X},
	shorttitle = {High-redshift {Star}-forming {Galaxies}},
	url = {https://ui.adsabs.harvard.edu/abs/2010ApJ...725.2324B},
	doi = {10.1088/0004-637X/725/2/2324},
	urldate = {2026-05-13},
	journal = {\apj},
	publisher = {IOP},
	author = {Burkert, A. and Genzel, R. and Bouché, N. and Cresci, G. and Khochfar, S. and Sommer-Larsen, J. and Sternberg, A. and Naab, T. and Förster Schreiber, N. and Tacconi, L. and Shapiro, K. and Hicks, E. and Lutz, D. and Davies, R. and Buschkamp, P. and Genel, S.},
	month = dec,
	year = {2010},
	pages = {2324--2332},
}

@misc{jones_ga-nifs_2025,
	title = {{GA}-{NIFS}: {A} smouldering disk galaxy undergoing ordered rotation at z=4.26},
	shorttitle = {{GA}-{NIFS}},
	url = {https://ui.adsabs.harvard.edu/abs/2025arXiv251205213J},
	doi = {10.48550/arXiv.2512.05213},
	urldate = {2026-05-13},
	publisher = {arXiv},
	author = {Jones, Gareth C. and Maiolino, Roberto and D'Eugenio, Francesco and Arribas, Santiago and Bunker, Andrew J. and Charlot, Stephane and Perna, Michele and Rodriguez del Pino, Bruno and Übler, Hannah and Böker, Torsten and Cresci, Giovanni and Lamperti, Isabella and Parlanti, Eleonora and Pascalau, Robert and Scholtz, Jan and Zamora, Sandra},
	month = dec,
	year = {2025},
}

@article{kohandel_amaryllis_2025,
	title = {Amaryllis: {A} digital twin of the earliest galaxies in the {Universe}},
	volume = {704},
	issn = {0004-6361},
	shorttitle = {Amaryllis},
	url = {https://ui.adsabs.harvard.edu/abs/2025A&A...704A..39K},
	doi = {10.1051/0004-6361/202555499},
	urldate = {2026-05-13},
	journal = {\aap},
	publisher = {EDP},
	author = {Kohandel, Mahsa and Pallottini, Andrea and Ferrara, Andrea},
	month = dec,
	year = {2025},
	pages = {A39},
}

@article{kohandel_dynamically_2024,
	title = {Dynamically cold disks in the early {Universe}: {Myth} or reality?},
	volume = {685},
	issn = {0004-6361},
	shorttitle = {Dynamically cold disks in the early {Universe}},
	url = {https://ui.adsabs.harvard.edu/abs/2024A&A...685A..72K},
	doi = {10.1051/0004-6361/202348209},
	urldate = {2026-05-13},
	journal = {\aap},
	publisher = {EDP},
	author = {Kohandel, M. and Pallottini, A. and Ferrara, A. and Zanella, A. and Rizzo, F. and Carniani, S.},
	month = may,
	year = {2024},
	pages = {A72},
}

@article{pillepich_first_2019,
	title = {First results from the {TNG50} simulation: the evolution of stellar and gaseous discs across cosmic time},
	volume = {490},
	issn = {0035-8711},
	shorttitle = {First results from the {TNG50} simulation},
	url = {https://ui.adsabs.harvard.edu/abs/2019MNRAS.490.3196P},
	doi = {10.1093/mnras/stz2338},
	urldate = {2026-05-13},
	journal = {\mnras},
	publisher = {OUP},
	author = {Pillepich, Annalisa and Nelson, Dylan and Springel, Volker and Pakmor, Rüdiger and Torrey, Paul and Weinberger, Rainer and Vogelsberger, Mark and Marinacci, Federico and Genel, Shy and van der Wel, Arjen and Hernquist, Lars},
	month = dec,
	year = {2019},
	pages = {3196--3233},
}

@article{curti_chemical_2023,
	title = {The chemical enrichment in the early {Universe} as probed by {JWST} via direct metallicity measurements at z ∼ 8},
	volume = {518},
	issn = {0035-8711},
	url = {https://ui.adsabs.harvard.edu/abs/2023MNRAS.518..425C},
	doi = {10.1093/mnras/stac2737},
	urldate = {2026-05-13},
	journal = {\mnras},
	publisher = {OUP},
	author = {Curti, Mirko and D'Eugenio, Francesco and Carniani, Stefano and Maiolino, Roberto and Sandles, Lester and Witstok, Joris and Baker, William M. and Bennett, Jake S. and Piotrowska, Joanna M. and Tacchella, Sandro and Charlot, Stephane and Nakajima, Kimihiko and Maheson, Gabriel and Mannucci, Filippo and Amiri, Amirnezam and Arribas, Santiago and Belfiore, Francesco and Bonaventura, Nina R. and Bunker, Andrew J. and Chevallard, Jacopo and Cresci, Giovanni and Curtis-Lake, Emma and Hayden-Pawson, Connor and Jones, Gareth C. and Kumari, Nimisha and Laseter, Isaac and Looser, Tobias J. and Marconi, Alessandro and Maseda, Michael V. and Scholtz, Jan and Smit, Renske and Übler, Hannah and Wallace, Imaan E. B.},
	month = jan,
	year = {2023},
	pages = {425--438},
}

@article{nakajima_jwst_2023,
	title = {{JWST} {Census} for the {Mass}-{Metallicity} {Star} {Formation} {Relations} at z = 4-10 with {Self}-consistent {Flux} {Calibration} and {Proper} {Metallicity} {Calibrators}},
	volume = {269},
	issn = {0067-0049},
	url = {https://ui.adsabs.harvard.edu/abs/2023ApJS..269...33N},
	doi = {10.3847/1538-4365/acd556},
	urldate = {2026-05-13},
	journal = {\apjs},
	publisher = {IOP},
	author = {Nakajima, Kimihiko and Ouchi, Masami and Isobe, Yuki and Harikane, Yuichi and Zhang, Yechi and Ono, Yoshiaki and Umeda, Hiroya and Oguri, Masamune},
	month = dec,
	year = {2023},
	pages = {33},
}

@article{schaerer_first_2022,
	title = {First look with {JWST} spectroscopy: {Resemblance} among z ∼ 8 galaxies and local analogs},
	volume = {665},
	issn = {0004-6361},
	shorttitle = {First look with {JWST} spectroscopy},
	url = {https://ui.adsabs.harvard.edu/abs/2022A&A...665L...4S},
	doi = {10.1051/0004-6361/202244556},
	urldate = {2026-05-13},
	journal = {\aap},
	publisher = {EDP},
	author = {Schaerer, D. and Marques-Chaves, R. and Barrufet, L. and Oesch, P. and Izotov, Y. I. and Naidu, R. and Guseva, N. G. and Brammer, G.},
	month = sep,
	year = {2022},
	pages = {L4},
}

@article{trump_physical_2023,
	title = {The {Physical} {Conditions} of {Emission}-line {Galaxies} at {Cosmic} {Dawn} from {JWST}/{NIRSpec} {Spectroscopy} in the {SMACS} 0723 {Early} {Release} {Observations}},
	volume = {945},
	issn = {0004-637X},
	url = {https://ui.adsabs.harvard.edu/abs/2023ApJ...945...35T},
	doi = {10.3847/1538-4357/acba8a},
	urldate = {2026-05-13},
	journal = {\apj},
	publisher = {IOP},
	author = {Trump, Jonathan R. and Arrabal Haro, Pablo and Simons, Raymond C. and Backhaus, Bren E. and Amorín, Ricardo O. and Dickinson, Mark and Fernández, Vital and Papovich, Casey and Nicholls, David C. and Kewley, Lisa J. and Brunker, Samantha W. and Salzer, John J. and Wilkins, Stephen M. and Almaini, Omar and Bagley, Micaela B. and Berg, Danielle A. and Bhatawdekar, Rachana and Bisigello, Laura and Buat, Véronique and Burgarella, Denis and Calabrò, Antonello and Casey, Caitlin M. and Ciesla, Laure and Cleri, Nikko J. and Cole, Justin W. and Cooper, M. C. and Cooray, Asantha R. and Costantin, Luca and Croton, Darren and Ferguson, Henry C. and Finkelstein, Steven L. and Fujimoto, Seiji and Gardner, Jonathan P. and Gawiser, Eric and Giavalisco, Mauro and Grazian, Andrea and Grogin, Norman A. and Hathi, Nimish P. and Hirschmann, Michaela and Holwerda, Benne W. and Huertas-Company, Marc and Hutchison, Taylor A. and Jogee, Shardha and Juneau, Stéphanie and Jung, Intae and Kartaltepe, Jeyhan S. and Kirkpatrick, Allison and Kocevski, Dale D. and Koekemoer, Anton M. and Lotz, Jennifer M. and Lucas, Ray A. and Magnelli, Benjamin and Matharu, Jasleen and Pérez-González, Pablo G. and Pirzkal, Nor and Rafelski, Marc and Rose, Caitlin and Seillé, Lise-Marie and Somerville, Rachel S. and Straughn, Amber N. and Tacchella, Sandro and Vanderhoof, Brittany N. and Weiner, Benjamin J. and Wuyts, Stijn and Yung, L. Y. Aaron and Zavala, Jorge A.},
	month = mar,
	year = {2023},
	pages = {35},
}

@article{arellano-cordova_first_2022,
	title = {A {First} {Look} at the {Abundance} {Pattern}-{O}/{H}, {C}/{O}, and {Ne}/{O}-in z {\textgreater} 7 {Galaxies} with {JWST}/{NIRSpec}},
	volume = {940},
	issn = {0004-637X},
	url = {https://ui.adsabs.harvard.edu/abs/2022ApJ...940L..23A},
	doi = {10.3847/2041-8213/ac9ab2},
	urldate = {2026-05-13},
	journal = {\apj},
	publisher = {IOP},
	author = {Arellano-Córdova, Karla Z. and Berg, Danielle A. and Chisholm, John and Arrabal Haro, Pablo and Dickinson, Mark and Finkelstein, Steven L. and Leclercq, Floriane and Rogers, Noah S. J. and Simons, Raymond C. and Skillman, Evan D. and Trump, Jonathan R. and Kartaltepe, Jeyhan S.},
	month = nov,
	year = {2022},
	pages = {L23},
}

@article{sun_first_2023,
	title = {First {Sample} of {H$\alpha$}+[{O} {III}]λ5007 {Line} {Emitters} at z {\textgreater} 6 {Through} {JWST}/{NIRCam} {Slitless} {Spectroscopy}: {Physical} {Properties} and {Line}-luminosity {Functions}},
	volume = {953},
	issn = {0004-637X},
	shorttitle = {First {Sample} of {H$\alpha$}+[{O} {III}]λ5007 {Line} {Emitters} at z {\textgreater} 6 {Through} {JWST}/{NIRCam} {Slitless} {Spectroscopy}},
	url = {https://ui.adsabs.harvard.edu/abs/2023ApJ...953...53S},
	doi = {10.3847/1538-4357/acd53c},
	urldate = {2026-05-13},
	journal = {\apj},
	publisher = {IOP},
	author = {Sun, Fengwu and Egami, Eiichi and Pirzkal, Nor and Rieke, Marcia and Baum, Stefi and Boyer, Martha and Boyett, Kristan and Bunker, Andrew J. and Cameron, Alex J. and Curti, Mirko and Eisenstein, Daniel J. and Gennaro, Mario and Greene, Thomas P. and Jaffe, Daniel and Kelly, Doug and Koekemoer, Anton M. and Kumari, Nimisha and Maiolino, Roberto and Maseda, Michael and Perna, Michele and Rest, Armin and Robertson, Brant E. and Schlawin, Everett and Smit, Renske and Stansberry, John and Sunnquist, Ben and Tacchella, Sandro and Williams, Christina C. and Willmer, Christopher N. A.},
	month = aug,
	year = {2023},
	pages = {53},
}

@article{brinchmann_high-z_2023,
	title = {High-z galaxies with {JWST} and local analogues - it is not only star formation},
	volume = {525},
	issn = {0035-8711},
	url = {https://ui.adsabs.harvard.edu/abs/2023MNRAS.525.2087B},
	doi = {10.1093/mnras/stad1704},
	urldate = {2026-05-13},
	journal = {\mnras},
	publisher = {OUP},
	author = {Brinchmann, Jarle},
	month = oct,
	year = {2023},
	pages = {2087--2106},
}

@article{ren_updated_2023,
	title = {Updated {Measurements} of [{O} {III}] 88 $\mu$m, [{C} {II}] 158 $\mu$m, and {Dust} {Continuum} {Emission} from a z = 7.2 {Galaxy}},
	volume = {945},
	issn = {0004-637X},
	url = {https://ui.adsabs.harvard.edu/abs/2023ApJ...945...69R},
	doi = {10.3847/1538-4357/acb8ab},
	urldate = {2026-05-13},
	journal = {\apj},
	publisher = {IOP},
	author = {Ren, Yi W. and Fudamoto, Yoshinobu and Inoue, Akio K. and Sugahara, Yuma and Tokuoka, Tsuyoshi and Tamura, Yoichi and Matsuo, Hiroshi and Kohno, Kotaro and Umehata, Hideki and Hashimoto, Takuya and Bouwens, Rychard J. and Smit, Renske and Kashikawa, Nobunari and Okamoto, Takashi and Shibuya, Takatoshi and Shimizu, Ikkoh},
	month = mar,
	year = {2023},
	pages = {69},
}

@article{kumari_study_2024,
	title = {A study of extreme {C} {III}]1908 \& [{O} {III}]88/[{C} {II}]157 emission in {Pox} 186: implications for {JWST}+{ALMA} ({FUV} + {FIR}) studies of distant galaxies},
	volume = {529},
	issn = {0035-8711},
	shorttitle = {A study of extreme {C} {III}]1908 \& [{O} {III}]88/[{C} {II}]157 emission in {Pox} 186},
	url = {https://ui.adsabs.harvard.edu/abs/2024MNRAS.529..781K},
	doi = {10.1093/mnras/stae252},
	urldate = {2026-05-13},
	journal = {\mnras},
	publisher = {OUP},
	author = {Kumari, Nimisha and Smit, Renske and Leitherer, Claus and Witstok, Joris and Irwin, Mike J. and Sirianni, Marco and Aloisi, Alessandra},
	month = apr,
	year = {2024},
	pages = {781--801},
}

@article{carniani_missing_2020,
	title = {Missing [{C} {II}] emission from early galaxies},
	volume = {499},
	issn = {0035-8711},
	url = {https://ui.adsabs.harvard.edu/abs/2020MNRAS.499.5136C},
	doi = {10.1093/mnras/staa3178},
	urldate = {2026-05-13},
	journal = {\mnras},
	publisher = {OUP},
	author = {Carniani, S. and Ferrara, A. and Maiolino, R. and Castellano, M. and Gallerani, S. and Fontana, A. and Kohandel, M. and Lupi, A. and Pallottini, A. and Pentericci, L. and Vallini, L. and Vanzella, E.},
	month = dec,
	year = {2020},
	pages = {5136--5150},
}

@article{kretschmer_origin_2022,
	title = {On the origin of surprisingly cold gas discs in galaxies at high redshift},
	volume = {510},
	issn = {0035-8711},
	url = {https://ui.adsabs.harvard.edu/abs/2022MNRAS.510.3266K},
	doi = {10.1093/mnras/stab3648},
	urldate = {2026-05-14},
	journal = {\mnras},
	publisher = {OUP},
	author = {Kretschmer, Michael and Dekel, Avishai and Teyssier, Romain},
	month = mar,
	year = {2022},
	pages = {3266--3275},
}




\appendix

\section{Atmospheric feature in Band 8 data}
\label{appendix:atmospheric feature}

\begin{figure}
    \centering
    \includegraphics[width=0.48\textwidth]{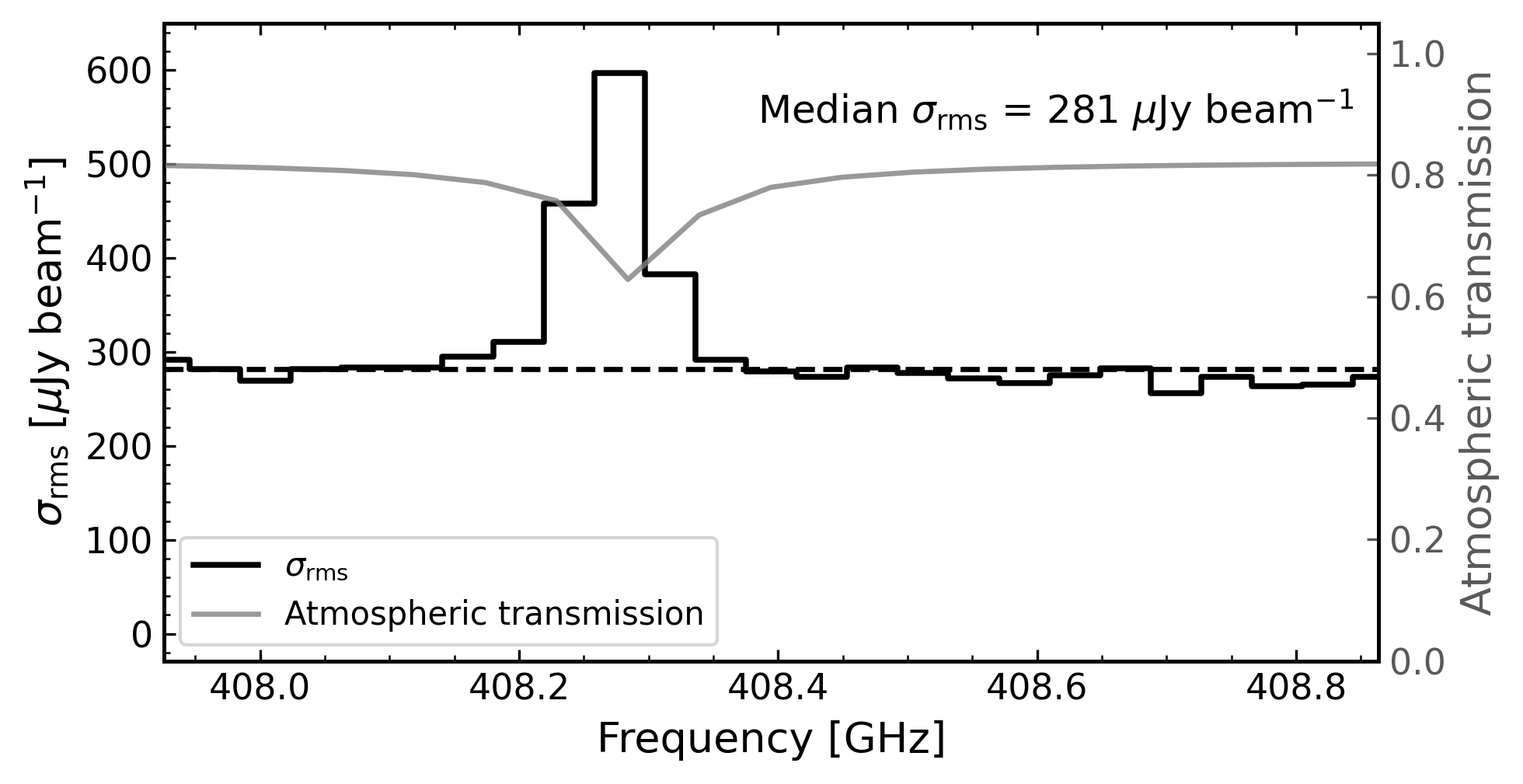}
    \caption{Per-channel $\sigma_{\mathrm{RMS}}$ spectrum (non JvM-corrected) of the ALMA Band 8 observations of REBELS-25 (black line). The grey curve shows the approximate atmospheric transmission as a function of frequency in ALMA Band 8 and is included for illustrative purposes only. An atmospheric absorption feature at $\sim408.3$ GHz coincides with an increase in the  noise by approximately a factor of two.}
    \label{fig:rms spectrum}
\end{figure}

The [O \textsc{iii}]88$\upmu$m emission of REBELS-25 coincides with an atmospheric feature at $\sim408.3$ GHz (approximately $+100$ km s$^{-1}$ relative to the centroid of the observed [O \textsc{iii}]88$\upmu$m line profile). In the channels surrounding this feature, we see a peak in the per-channel $\sigma_{\mathrm{RMS}}$, as plotted in Figure \ref{fig:rms spectrum}, where the $\sigma_{\mathrm{RMS}}$ increases by a factor of $\sim2$.  As shown in \cite{algera_rebels-ifu_2025}, the S/N of the Cycle 9 data are sufficient to recover the [O \textsc{iii}]88$\upmu$m emission even in these affected channels. However, it is still important to take this varying noise level into account. Throughout this work, we therefore use a full per-channel $\sigma_{\mathrm{RMS}}$ spectrum whenever any spectral fitting is applied. Similarly, when fitting the kinematics with \texttt{3DBAROLO}, as discussed in Section \ref{sec:barolo}, we ensure that the per-channel masking takes this $\sigma_{\mathrm{RMS}}$ spectrum into account.

\begin{figure}
    \centering
    \includegraphics[width=0.48\textwidth]{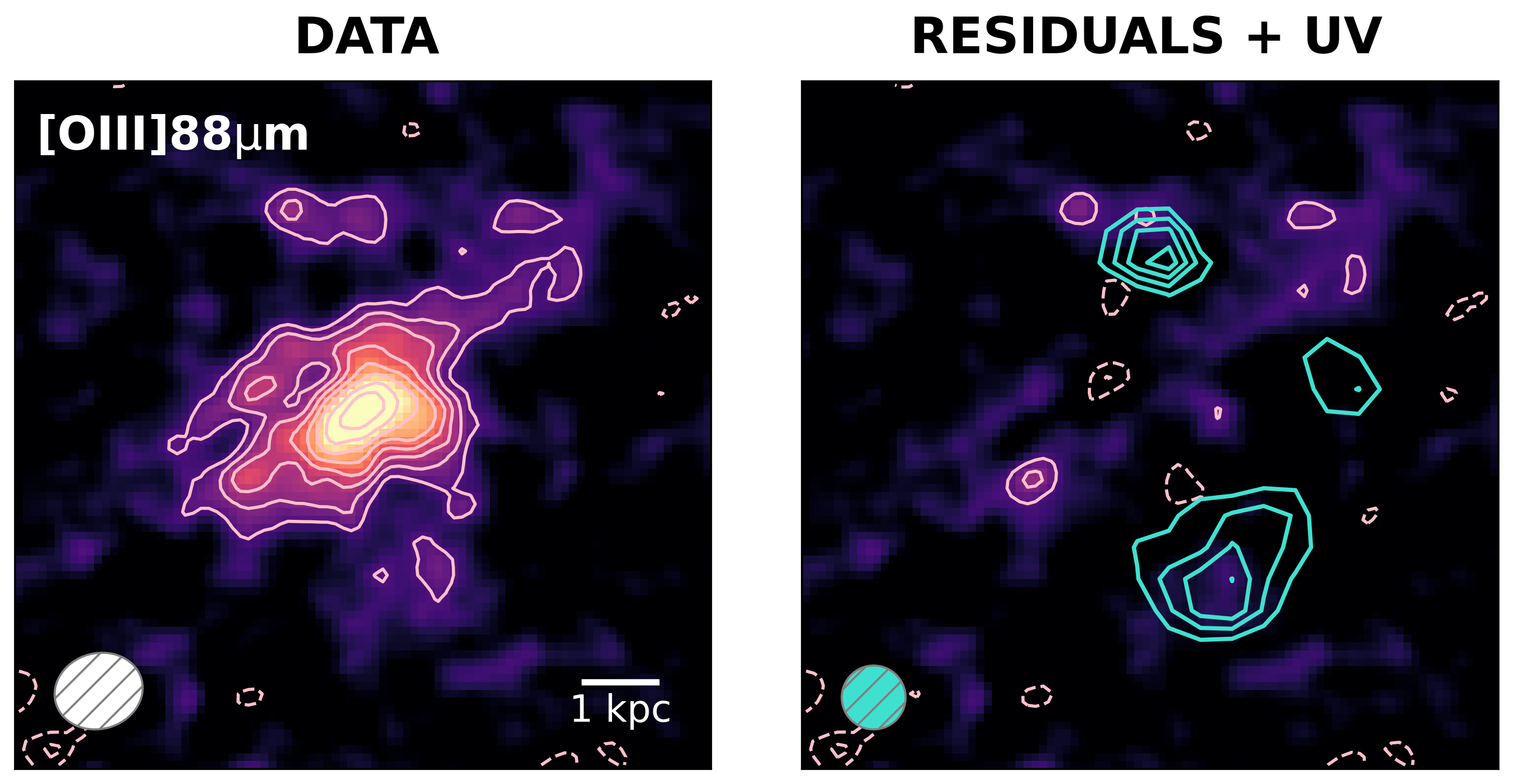}
    \caption{[O \textsc{iii}]88$\upmu$m moment-0 map (non JvM-corrected) integrated over $\pm1$ FWHM around the line centroid, where the channels affected by the atmospheric absorption feature at $\sim408.3$ GHz have been masked out. Contours show $\pm2, 3, 4... \times \sigma_{\mathrm{RMS}}$ emission (dashed for negative). Although the masking removes a small amount of emission from the receding side of the galaxy and lowers the significance of the northern clump, the inferred global morphology remains largely unchanged.}
    \label{fig:mom0 masked}
\end{figure}

We also investigate any impact of this increased noise on the moment-0 map presented in Figure \ref{fig:line maps}, and therefore the morphological fitting of the [O \textsc{iii}]88$\upmu$m data. In Figure \ref{fig:mom0 masked}, we show a new moment-0 map where the impacted channels are masked out ($\pm2$ channels from the centroid of the atmospheric absorption feature). As expected, this removes a small amount of [O \textsc{iii}]88$\upmu$m emission from the receding (southern) side of the galaxy, since the atmospheric feature lies at approximately $+100$ km s$^{-1}$ relative to the line centroid. However, it also removes some emission from regions on the northern side of the galaxy, including the northern clump, which now has a peak S/N of $\sim3$ rather than $\sim4.5$. If we re-perform the 2D Sérsic fitting described in Section \ref{sec:sersic fitting}, the derived morphological parameters ($n=1.00$, $r_e=1.31$ kpc) are consistent with results in Section \ref{sec:sersic fitting}, however we no longer see significant residuals coincident with the northern clump in the \textit{JWST} data. This suggests that the morphological parameters derived in Section \ref{sec:sersic fitting} are robust to this noise feature, but the evidence for residual clumpy substructures is less secure. Comparison between these low-significance residuals and the UV/optical clumps should therefore be treated with caution.

\section{JvM correction}
\label{appendix:jvm correction}

As discussed in Section \ref{sec:alma data}, interferometric images produced by \texttt{tclean} can systematically overestimate low-S/N emission if the volumes of the clean and dirty beams differ significantly, since the restored image is constructed by adding residual emission (in units of Jy per dirty beam), to the convolved model image (in units of Jy per clean beam). Significant sidelobes in the dirty beam can occur if observations with very different array configurations are combined, or otherwise if the \emph{uv}-plane is not well sampled by only observing with long baselines. 

For the high resolution ALMA data analysed in this work, we find that the dirty beams do exhibit moderate sidelobes that could produce this so-called JvM (\citealt{jorsater_high_1995}) effect. We calculate the volumes of the clean and dirty beams (where for each dirty beam we integrate only out to the first null, since this is the part approximated to a Gaussian during deconvolution, see e.g., \citealt{czekala_molecules_2021}), and find that these differ by factors of $\epsilon=0.37$ and 0.45 for the Band 6 and Band 8 data, respectively.

To investigate whether this residual flux scaling may have an impact on our results, we apply the JvM correction as described in \cite{czekala_molecules_2021}, whereby we multiply each of the residual images produced by \texttt{tclean} by the corresponding $\epsilon$ factors before adding them to the clean models. We call the resulting imaging the JvM-corrected data products, and we verify that the total fluxes are consistent with the fluxes obtained from imaging tapered to $>1$ arcsec (see \citealt{algera_rebels-ifu_2026}), which is an alternative method to correct for this JvM effect when obtaining integrated measurements. We then test the main analysis described in this paper both with and without applying the JvM correction.

As shown previously in \citet{rowland_rebels-25_2024}, we find that whether or not the JvM correction is applied has little impact on the kinematic parameters derived using \texttt{3DBAROLO}. For example, for [C \textsc{ii}] we obtain $V_{\mathrm{rot, max}}=342$ and $\sigma=37$ km s$^{-1}$ and for [O \textsc{iii}]88$\upmu$m we get  $V_{\mathrm{rot, max}}=256$ and $\sigma=61$ km s$^{-1}$ when using the JvM-corrected data. This negligible impact is likely because \texttt{3DBAROLO} uses spectral masking that minimise the influence of low-level residual emission. However, the uncertainties from the JvM-corrected data are likely underestimated since the residual emission is scaled down by $\epsilon$, and so the RMS no longer reflects the true sensitivity of the data. We also add that combining the low and high resolution [C \textsc{ii}] data in Appendix A of \cite{rowland_rebels-25_2024}, which increased the S/N but introduced a more significant JvM effect with $\epsilon=0.18$, resulted in a slightly higher $\sigma$ of 42 km s$^{-1}$ and lower $V/\sigma$ of 8. Based on our findings where we only change whether or not the JvM correction is applied, the difference here could instead be due to the addition of the low-resolution data rather than the JvM correction.

The effect is more significant for the \texttt{DysmalPy} modelling described in Appendix \ref{appendix:DysmalPy}. Unlike \texttt{3DBAROLO}, the \texttt{DysmalPy} analysis relies on spectral fitting without spectral masking, making it more sensitive to low-S/N residual emission. For the [C \textsc{ii}] data, omitting the JvM correction significantly increases the inferred velocity dispersion from 75 to 98 km s$^{-1}$, and decreases the baryonic mass to $\log (M_{\mathrm{bar}}[\mathrm{M_{\odot}}])=10.7$, although the dark matter fraction remains broadly unchanged. For the [O \textsc{iii}]88$\upmu$m data, the inferred velocity dispersion remains unchanged ($\sim62$ km s$^{-1}$), but the dark matter fraction significantly decreases to $f_{\mathrm{DM}}=0.04$ and the baryonic mass increases to $\log (M_{\mathrm{bar}}[\mathrm{M_{\odot}}])=10.8$. 

The JvM effect also impacts the integrated flux measurements. Repeating the clump aperture spectral fitting from Section \ref{sec:resolved ism} on the uncorrected cubes yields systematically higher line fluxes, on average by factors of approximately 1.62 and 1.56 for [C \textsc{ii}] and [O \textsc{iii}]88$\upmu$m, respectively. As a result, luminosities and derived line ratios can be biased if the correction is not applied.

For these reasons, we adopt the JvM-corrected results for the \texttt{DysmalPy} modelling in Appendix \ref{appendix:DysmalPy} and resolved spectral fitting in Section \ref{sec:resolved ism}. For the \texttt{3DBAROLO} analysis presented in Section \ref{sec:barolo}, however, we use the uncorrected cubes, following \cite{rowland_rebels-25_2024}. The \texttt{3DBAROLO} fitting relies on S/N-based spectral masking, and the JvM correction artificially rescales the residual emission, modifying both the measured S/N and the apparent noise properties of the cube. Using the uncorrected data therefore provides a more reliable estimate of the true channel-to-channel uncertainties for the masking procedure, and for the uncertainty estimation on the resulting kinematic properties. As discussed above, we find that the \texttt{3DBAROLO} parameters are consistent with and without the JvM correction, indicating that the choice of cube has a negligible impact on the final kinematic results.

The situation is slightly more complex for the morphological fitting presented in Section \ref{sec:morphology}. The uncorrected maps preserve the true noise properties but can overestimate residual flux, potentially biasing sizes high and Sérsic indices low. On the other side, the JvM-corrected maps recover the correct total flux scale but rescale the residual noise, making the emission appear artificially high-S/N and biasing fits toward more compact, centrally concentrated profiles. Indeed, when fitting the JvM-corrected data we find slightly higher $n$ and lower $r_e$ than the values presented in Section \ref{sec:morphology} (for example, $n=1.22$ and $r_e=1.51$ kpc for [O \textsc{iii}]88$\upmu$m and $n=1.58$ and $r_e=1.86$ kpc for [C \textsc{ii}]), although all values are consistent within the uncertainties. However, this doesn't significantly impact the main conclusions; the Sérsic indices are close to unity ($1<n<1.7$), suggesting a near-exponential disc, and [C \textsc{ii}] remains the most extended. An upcoming morphological analysis of this source directly in the \textit{uv}-plane will provide a more robust test of these image-plane results, avoiding potential biases introduced by the \texttt{tclean} deconvolution and residual scaling (Astles et al. in prep.).

\section{Tests with \texttt{DysmalPy}}
\label{appendix:DysmalPy}

\begin{figure*}
    \centering
    \includegraphics[width=\textwidth]{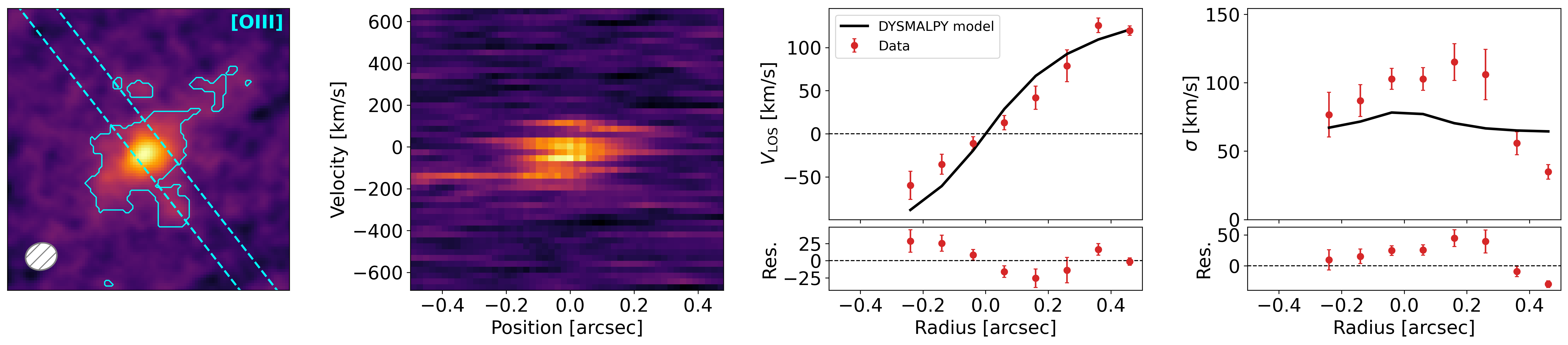}
    \caption{Plots relevant to the \texttt{DysmalPy} fitting of the JvM-corrected [O \textsc{iii}]88$\upmu$m data described in Appendix \ref{appendix:DysmalPy}. The left map shows a moment-0 image from the [O \textsc{iii}]88$\upmu$m cube. We fit a single Gaussian to the spectrum extracted from each pixel, and create a mask for pixels with an integrated S/N$>3$ from these Gaussian fits (blue contour). We then produce a position velocity diagram (PVD)  from a slit along the kinematic major axis (218 degrees) with width equal to the beam FWHM (blue dashed lines). In the third panel, we plot this extracted PVD. Spectra are then extracted column by column, with each pixel-wide column fit again with a 1D Gaussian. The resulting velocity centroids and line widths are then used to produce the line-of-sight rotational velocity (third panel) and dispersion (fourth, right-most panel). For these profiles, we take the average values in 0.5$\times$ beam FWHM-sized bins (circular markers). These profiles are then the inputs to the \texttt{DysmalPy} modelling. The resulting best-fit models are plotted as black lines.}
    \label{fig:oiii DysmalPy}
\end{figure*}

\begin{figure*}
    \centering
    \includegraphics[width=\textwidth]{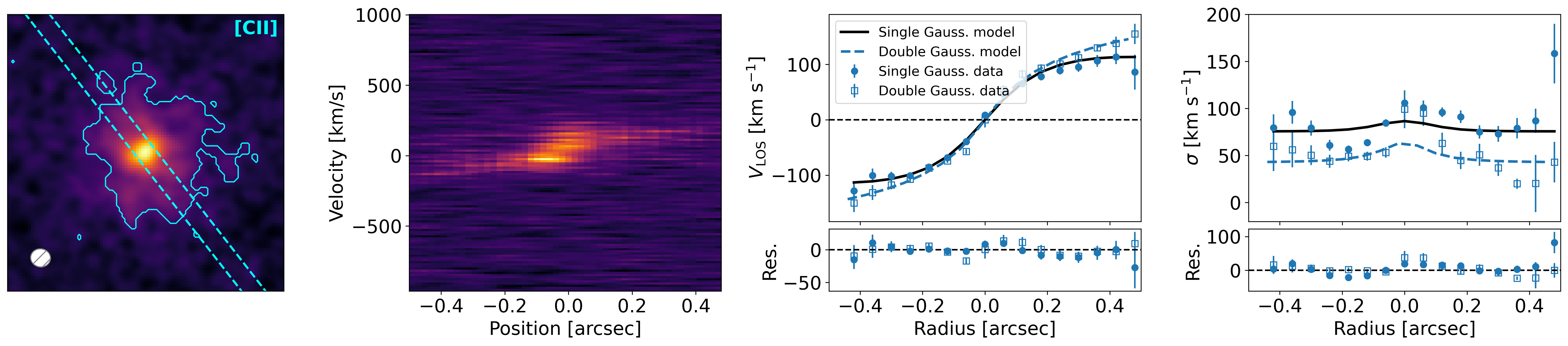}
    \caption{As with Figure \ref{fig:oiii DysmalPy}, but for the JvM-corrected [C \textsc{ii}] data. In the third and fourth panels, we also show the profiles when fitting the spectra with a double Gaussian model (empty square-shaped markers), as well as a single Gaussian (filled circle markers), and the corresponding \texttt{DysmalPy} models are plotted with dashed and solid lines, respectively.}
    \label{fig:cii DysmalPy}
\end{figure*}

\begin{figure*}
    \centering
    \includegraphics[width=\textwidth]{
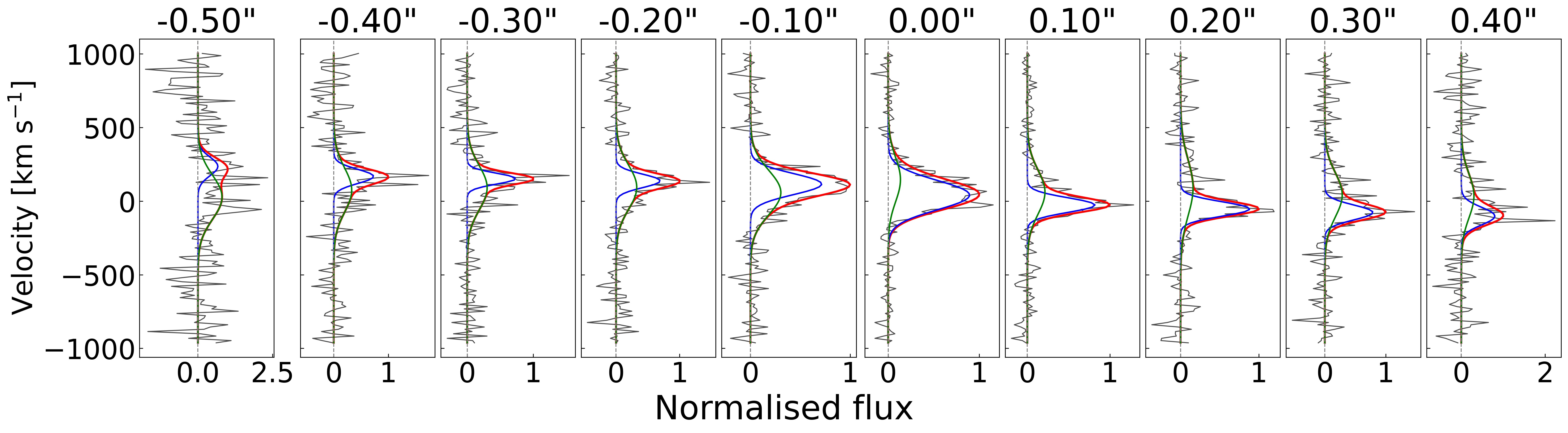}
    \caption{Example Gaussian fits to column spectra extracted from the [C \textsc{ii}] PVD, as described in the text and shown in Figure \ref{fig:cii DysmalPy}. Here we show the fits from a double Gaussian model (total model in red), with a narrow (blue) + broad (green) component. This improves the quality of the fit in comparison to a single Gaussian, according to the BIC analysis, for $\sim30\%$ of the spectra.}
    \label{fig:cii double gauss}
\end{figure*}

\begin{figure*}
    \centering
    \includegraphics[width=\textwidth]{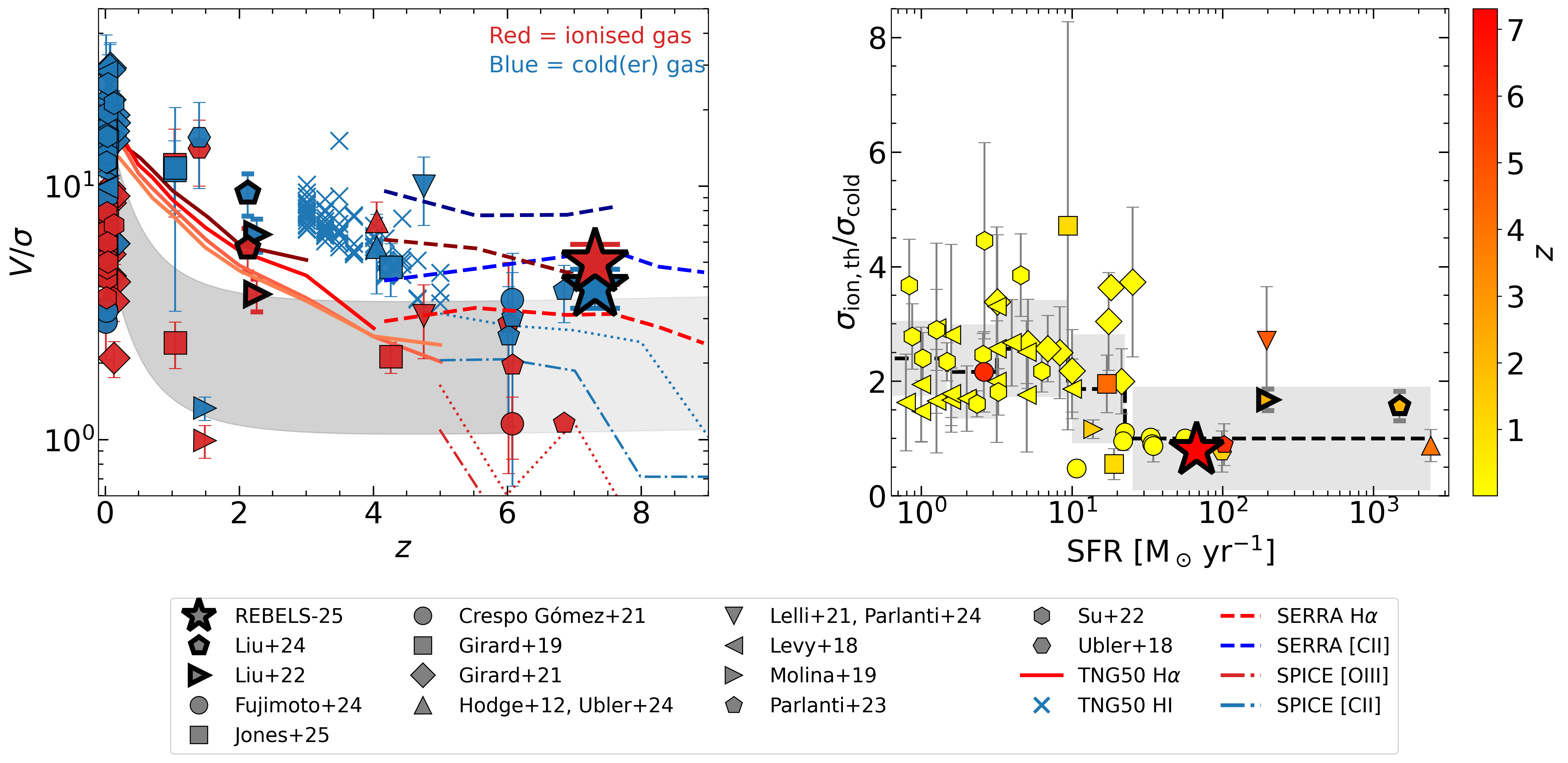}
    \caption{A recreation of some of the key figures in Section \ref{sec:compilation} (left panel of Figure \ref{fig:sigma_redshift} and left panel of Figure \ref{fig:sigma_ratio}), where we instead use the kinematic properties for REBELS-25 using \texttt{DysmalPy}. The bold markers are studies where \texttt{DysmalPy} is also used for both warm ionised gas and colder gas kinematic tracers (\citealt{liu_600_2023,Liu_2024}). We note that \texttt{DysmalPy} is also used in \citet{ubler_ga-nifs_2024}  for the H$\upalpha$ kinematics of GN20.}
    \label{fig:dysmalpy results}
\end{figure*}

We additionally model the kinematics using \texttt{DysmalPy}, an alternative forward-modelling framework that derives galaxy kinematics from a parametrised mass distribution. For this modelling with \texttt{DysmalPy}, we follow the approach described in \cite{lee_alma-cristal_2025}. This framework differs from  \texttt{3DBAROLO} in a number of ways, as described in detail in \cite{lee_disk_2025}. Most notably, it assumes parametric profiles, requires as input 1D data extracted only along the pre-identified kinematic major axis, and is used without any spectral masking.

For the tests described below, we use the JvM-corrected data cubes. As discussed in Section \ref{sec:alma data} and Appendix \ref{appendix:jvm correction}, the JvM effect can enhance low S/N emission, therefore affecting spectral fitting where no S/N masking is applied. From tests with \texttt{3DBAROLO}, where spectral masking is used, we find consistent parameters whether or not the JvM correction is applied, but for \texttt{DysmalPy} we find differing results, as discussed in more detail in Appendix \ref{appendix:jvm correction}.

Following \cite{lee_alma-cristal_2025}, we first spatially mask (note: only spatially masking is applied; no spectral masking) low-S/N spaxels ($\mathrm{S/N}<3$) and then extract a PV diagram along a slit with width equal to the beam FWHM across the kinematic major axis. We then extract velocity profiles from the PV diagram by fitting a single Gaussian profile column-by-column (i.e., collapsed emission of the velocity channels at the same position). The centroids and widths of the fitted Gaussian models are then the velocity and velocity dispersion values at the locations of each column. Finally, the extracted profiles were down-sampled by averaging to a resolution of one-half of the beam FWHM. The 1D velocity and velocity dispersion profiles then serve as an input for the dynamical modelling.

The model consists of a baryonic disc, an optional bulge component, and an NFW dark matter halo (\citealt{navarro_structure_1996}). The disc is treated as a thick oblate system, with the rotation curve derived from the mass distribution and corrected for pressure support following \cite{burkert_high-redshift_2010}, assuming a locally isotropic and radially constant velocity dispersion, $\sigma(R)=\sigma_0$. The dark matter halo virial mass is tied to the dark matter fraction within the effective radius, while the concentration is fixed using standard mass-concentration relations. For these fits, we fix the inclination to 25 degrees, and fix the disc Sérsic indices and effective radii to the values derived in Section \ref{sec:morphology} for [O \textsc{iii}]88$\upmu$m and [C \textsc{ii}]. For these fits, we do not include a bulge component, but note that including a bulge component in the fit changes $\sigma_0$ by only $\pm 5$ km s$^{^-1}$. The only free parameters left in the fitting are therefore the total baryonic mass, $M_{\mathrm{bar}}$, the dark matter fraction, $f_{\mathrm{DM}}$, and the intrinsic velocity dispersion, $\sigma_0$. 

For the [O \textsc{iii}] data, we obtain $\log (M_{\mathrm{bar}}[\mathrm{M_{\odot}}])=9.9\pm1.3$, $f_{\mathrm{DM}}=0.86\pm0.40$ and $\sigma_0=62.1\pm3.6$ km s$^{-1}$. Within a comparable radius (2 kpc) as used in the \texttt{3DBAROLO} fitting, this gives a $V_{\mathrm{max}}/\sigma \simeq 5$; consistent within the uncertainties with the value derived from \texttt{3DBAROLO}. For the [C \textsc{ii}] data, however, we obtain $\log (M_{\mathrm{bar}}[\mathrm{M_{\odot}}])=10.9\pm0.3$, $f_{\mathrm{DM}}=0.14\pm0.72$ and $\sigma_0=75.4\pm1.5$ km s$^{-1}$. Within 2.5 kpc (outer radius fit by \texttt{3DBAROLO}), this gives a $V_{\mathrm{max}}/\sigma \simeq 4$; lower than the value derived by \texttt{3DBAROLO} of $11^{+6}_{-5}$.

As with the \texttt{3DBAROLO} fits, we still see significant residuals when comparing the input data to the model, in this case from the 1D velocity and dispersion profiles. The rotation curves are relatively well-fit, particularly for [C \textsc{ii}], but the dispersion profiles are not well described by the model. This could indicate that, for REBELS-25, the velocity dispersion varies across the disc, as suggested by the \texttt{3DBAROLO} fits. However, as mentioned in Section \ref{sec:barolo}, this could also be attributed to limitations of the current data, the presence of non-circular motions, and/or that the emission is not well fitted by single Gaussian-components.

In \cite{hygate_alma_2023}, evidence for a broad component in the integrated [C \textsc{ii}] spectrum from the lower resolution LP data was identified in the range $+250$ to $+650$ km s$^{-1}$, and interpreted as either a minor merger component or outflowing gas. A search for a similar secondary component was carried out in \cite{rowland_rebels-25_2024}, using both the integrated spectrum from the higher resolution [C \textsc{ii}] data, and also beam-sized apertures across the source. No significant evidence for this secondary component was found from these extractions, however we note this was only the case when comparing a double Gaussian (comparable to a double-peaked profile typical in the integrated spectra of rotating systems) model with a model that has both a double Gaussian and an additional broad component. For this, we considered very strong evidence to be an improvement to the Bayesian Information Criterion (BIC) $>10$, and moderate evidence to be $\Delta$BIC$>2$. For the column spectra analysed in this work, we test a fit with only a single narrow component (with width constrained from 20-150 km s$^{-1}$, where $\sim150$ km s$^{-1}$ is the maximum from the single Gaussian fitting), plus an additional broad component (width constrained from 150-1000 km s$^{-1}$). We find an improvement in the BIC value in the two-component model, relative to a single Gaussian model, in 30\% of extracted spectra, mostly in pixels near the centre of the source ($\Delta \mathrm{BIC} \sim2-20$ in the spectra fit with $r_{\mathrm{offset}}\lesssim 0.2$ arcsec of the centre) with a FWHM $\sim400$ km s$^{-1}$. If we assume this secondary component in the [C \textsc{ii}] emission is real, following \cite{lee_alma-cristal_2025} (for a source in their sample also found to have a broad component in the spectra) we refit the data with \texttt{DysmalPy} using only the profiles from the narrow component. The resulting $\sigma_0$ value for [C \textsc{ii}] is then $40\pm2$ km s$^{-1}$.

We repeat this analysis for the [O \textsc{iii}] data and find $\sigma=38\pm10$ km s$^{-1}$. However, the BIC improvement for the double Gaussian fit is not significant in any of the [O \textsc{iii}] extracted spectra, which could be due to the lower S/N, with the fitted secondary components only detected at a S/N$<2$, compared to S/N$\simeq2.2-8.0$ for the [C \textsc{ii}] data.

From these tests with \texttt{DysmalPy}, we therefore find that the spectral masking applied in the \texttt{3DBAROLO} fits may be causing an underestimation of the [C \textsc{ii}] velocity dispersion. However, if the secondary component is real (see also \citealt{hygate_alma_2023}), the derived [C \textsc{ii}] velocity dispersions between the two tools would be more in agreement, indicating that the mask applied by \texttt{3DBAROLO} is effectively masking out this fainter, non-circular component. The [O \textsc{iii}] velocity dispersion estimates are in good agreement without including a secondary component, although we note that the S/N is too low to robustly confirm if there is a secondary component in the individual [O \textsc{iii}] spectra. The single Gaussian fits to the [O \textsc{iii}] data could therefore be primarily tracing the brighter, narrow component. Additionally, we find that the dispersion  profile for this source is likely not well-described by a constant velocity dispersion across the disc. For these reasons, we adopt the parameters derived from \texttt{3DBAROLO} as our fiducial estimates, but we make comparisons with results if the \texttt{DysmalPy} fits are used, below.

In addition to the tests with \texttt{3DBAROLO} and \texttt{DysmalPy} described here, we also note that when using \texttt{QUBEFIT} on the [C \textsc{ii}] data in \cite{rowland_rebels-25_2024} with a spectral masking S/N threshold of 2, we obtained $\sigma_{\mathrm{[C \textsc{ii}]}}\simeq40$ km s$^{-1}$. In \cite{parlanti_alma_2023} the low resolution data of REBELS-25 (named UVISTA-Y-003 in that work) was also fitted with a 2D kinematic fitting method without spectral masking, obtaining $\sigma_{\mathrm{[C \textsc{ii}]}}\simeq60$ km s$^{-1}$. These differences highlight that fitted velocity dispersions can depend on the modelling approach, data resolution, and masking choices, and should therefore be interpreted in the context of the specific tools and assumptions used to derive them.

In Figure \ref{fig:dysmalpy results}, we plot $V/\sigma$ against redshift and $\sigma_{\mathrm{ion, th}}/\sigma_{\mathrm{cold}}$ against SFR for the same literature comparison sample as in Figures \ref{fig:sigma_redshift} and \ref{fig:sigma_ratio}, but now for REBELS-25 we plot the kinematic parameters derived using \texttt{DysmalPy} with a single Gaussian component. Similarly to the fiducial analysis with the values derived by \texttt{3DBAROLO}, REBELS-25 remains surprisingly dynamically cold for its high redshift in both [C \textsc{ii}] and [O \textsc{iii}]88$\upmu$m, in comparison to predicted trends of lower $V/\sigma$ ratios with increasing redshift. This remains true for all the kinematic fitting techniques tested in this work and in \cite{rowland_rebels-25_2024}. However, with \texttt{DysmalPy} we find that $\sigma_{\mathrm{ion}}\lesssim\sigma_{\mathrm{cold}}$ for REBELS-25, in comparison to the \texttt{3DBAROLO} fits finding $\sigma_{\mathrm{ion}}/\sigma_{\mathrm{cold}}\sim2$. We therefore find that not only does $\sigma$ vary depending on the adopted kinematic fitting technique, but  $\sigma_{\mathrm{ion}}/\sigma_{\mathrm{cold}}$ may also be model-dependent, which is a likely caveat of the discussions presented in Section \ref{sec:compilation}.


\bsp	
\label{lastpage}
\end{document}